\documentclass[10pt,a4paper,twoside]{report}
\pdfoutput=1

\usepackage{gutthesis}
\usepackage{subcaption}

\author{Paweł Rościszewski}
\title{Optimization of hybrid parallel application execution in heterogeneous high performance computing systems considering execution time and power consumption}

\usepackage[procnames]{listings}

\usepackage[nomain]{glossaries}
\usepackage{glossary-mcols}
\newglossary[slg]{symbolslist}{syi}{syg}{Glossary}

\makeglossaries

\glstoctrue

\usepackage{amsthm}

\usepackage{tocbibind}
\renewcommand\bibname{References}

\usepackage[shortcuts]{extdash}

\colorlet{punct}{red!60!black}
\definecolor{background}{HTML}{EEEEEE}
\definecolor{delim}{RGB}{20,105,176}
\colorlet{numb}{magenta!60!black}

\lstdefinelanguage{json}{
    basicstyle=\normalfont\ttfamily,
    numbers=left,
    numberstyle=\scriptsize,
    stepnumber=1,
    numbersep=8pt,
    showstringspaces=false,
    breaklines=true,
    frame=lines,
    backgroundcolor=\color{background},
    literate=
     *{0}{{{\color{numb}0}}}{1}
      {1}{{{\color{numb}1}}}{1}
      {2}{{{\color{numb}2}}}{1}
      {3}{{{\color{numb}3}}}{1}
      {4}{{{\color{numb}4}}}{1}
      {5}{{{\color{numb}5}}}{1}
      {6}{{{\color{numb}6}}}{1}
      {7}{{{\color{numb}7}}}{1}
      {8}{{{\color{numb}8}}}{1}
      {9}{{{\color{numb}9}}}{1}
      {:}{{{\color{punct}{:}}}}{1}
      {,}{{{\color{punct}{,}}}}{1}
      {\{}{{{\color{delim}{\{}}}}{1}
      {\}}{{{\color{delim}{\}}}}}{1}
      {[}{{{\color{delim}{[}}}}{1}
      {]}{{{\color{delim}{]}}}}{1},
}

\definecolor{keywords}{RGB}{255,0,90}
\definecolor{comments}{RGB}{0,0,113}
\definecolor{red}{RGB}{160,0,0}
\definecolor{green}{RGB}{0,150,0}
\newtheorem{definition}{Definition}

\begin{document}

\pagenumbering{roman}

\includepdf{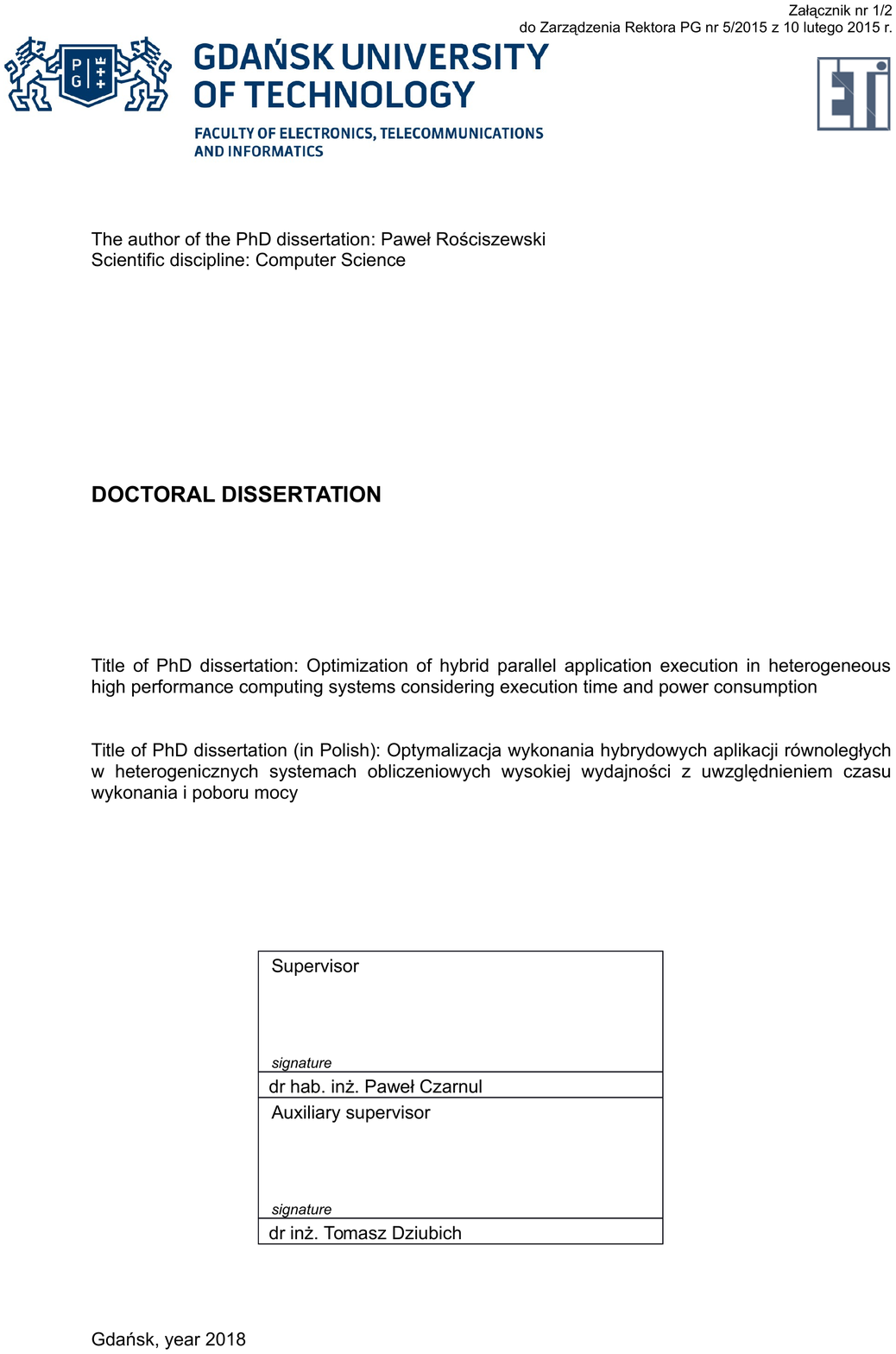}

\addcontentsline{toc}{chapter}{Abstract}
\chapter*{Abstract}

\textbf{English:}

	Many important computational problems require utilization of high performance computing (HPC) systems that consist of multi-level structures combining higher and higher numbers of devices with various characteristics. Utilizing full power of such systems requires programming parallel applications that are hybrid in two meanings: they can utilize parallelism on multiple levels at the same time and combine together programming interfaces specific for various types of computing devices.

	The main goal of parallel processing is increasing the processing performance, and therefore decreasing the application execution time. The international HPC community is targeting development of "Exascale" supercomputers (able to sustain $10^{18}$ floating point operations per second) by the year 2020. One of the main obstacles to achieving this goal is power consumption of the computing systems that exceeds the energy supply limits. New programming models and algorithms that consider this criterion are one of the key areas where significant progress is necessary in order to achieve the goal.

	The goal of the dissertation is to extract a general model of hybrid parallel application execution in heterogeneous HPC systems that is a synthesis of existing specific approaches and developing an optimization methodology for such execution aiming for minimization of the contradicting objectives of application execution time and power consumption of the utilized computing hardware. Both meanings of the application hybridity result in multiplicity of execution parameters of nontrivial interdependences and influence on the considered optimization criteria. Mapping of the application processes on computing devices has also a significant impact on these criteria.

	The dissertation consists of an Introduction, two theoretical Chapters, three empirical Chapters and a Summary. The Introduction includes motivations for the study, research problem formulation, scope, main contributions, claims and overview of the thesis. Chapter 2 contains a review of existing approaches in the area of executing, modeling and simulation of hybrid parallel applications along with examples of such applications and the meaning of their hybridity. Chapter~3 contains a critical analysis of existing approaches to parallel application optimization considering execution time and power consumption with a particular emphasis on multi-objective optimization methods, computing resource management and auto-tuning of application execution parameters.

	Chapter 4 describes five real-life parallel applications from various practical fields and five diverse computing systems that were the subject of experiments included in the dissertation. In Chapter 5, an optimization methodology of hybrid parallel application execution in heterogeneous HPC systems is proposed that consists of specific execution steps and a simulation method for fast evaluation of points in the solution search space. Chapter 6 presents results of experiments with applying the consecutive execution steps to chosen real-life applications and using the proposed optimization methodology as a whole to one application of deep neural network training for automatic speech recognition.

	 As shown in the dissertation, the execution steps specific in the context of the proposed model, including preliminary process optimization, process mapping, parameter tuning and actual execution allow to optimize execution time of hybrid parallel applications in heterogeneous high performance computing systems, while the proposed modeling and simulation method allows for fast and accurate identification of the set of optimal solutions to the problem of multi-objective execution time and power consumption optimization.
	 
\newpage

\textbf{Polish:}

	Wiele istotnych problemów obliczeniowych wymaga wykorzystania systemów obliczeniowych wysokiej wydajności, w których skład wchodzi kilkupoziomowa struktura łącząca coraz większe liczby urządzeń o różnych charakterystykach. Wykorzystanie pełnej mocy takich systemów wymaga programowania aplikacji równoległych, które są hybrydowe w dwóch znaczeniach: potrafią jednocześnie wykorzystać równoległość na wielu poziomach oraz łączą ze sobą interfejsy programistyczne charakterystyczne dla różnych typów urządzeń obliczeniowych.

	Głównym celem przetwarzania równoległego jest zwiększenie wydajności przetwarzania, a więc zmniejszenie czasu wykonania aplikacji. Międzynarodowa społeczność skupiona wokół systemów obliczeniowych wysokiej wydajności postawiła sobie za cel zbudowanie do roku 2020 superkomputerów "skali Exa", to znaczy mających możliwość wykonania $10^{18}$ operacji zmiennoprzecinkowych na sekundę. Jedną z głównych przeszkód stojących na drodze do tego celu jest pobór mocy systemów obliczeniowych przekraczający możliwości dostawy energii. Nowe modele programistyczne i algorytmy uwzględniające to kryterium są jednym z kluczowych pól, na których istotne postępy są konieczne, aby osiągnąć postawiony cel.

	Celem rozprawy jest wyodrębnienie ogólnego modelu wykonania hybrydowych aplikacji równoległych w heterogenicznych systemach obliczeniowych wysokiej wydajności będącego syntezą istniejących podejść szczegółowych oraz opracowanie metodologii optymalizacji takiego wykonania aplikacji z uwzględnieniem minimalizacji przeciwstawnych kryteriów czasu wykonania aplikacji i poboru mocy wykorzystywanego sprzętu obliczeniowego. Oba znaczenia hybrydowości aplikacji równoległych wiążą się z mnogością parametrów wykonania o nietrywialnych współzależnościach i wpływie na rozpatrywane kryteria optymalizacyjne. Istotny wpływ na owe kryteria ma także sposób mapowania procesów aplikacji na urządzenia obliczeniowe.

	Rozprawa składa się ze wstępu, dwóch rozdziałów literaturowych, trzech rozdzialów empirycznych i podsumowania. We wstępie umotywowano podjęcie badań, sformułowano problem badawczy, przedstawiono zakres, główne oryginalne osiągnięcia,  tezy oraz przegląd rozdziałów rozprawy. W rozdziale drugim zawarto przegląd istniejących rozwiązań w zakresie wykonania, modelowania i symulacji hybrydowych aplikacji równoległych oraz opisano przykłady takich aplikacji z różnych dziedzin wraz ze znaczeniem ich hybrydowości. W rozdziale trzecim dokonano krytycznej analizy istniejących podejść do optymalizacji aplikacji równoległych z uwzględnieniem czasu wykonania i poboru mocy, ze szczególnym uwzględnieniem metod optymalizacji wielokryterialnej, zarządzania zasobami obliczeniowymi oraz automatycznego strojenia parametrów wykonania aplikacji.
	
	W rozdziale czwartym opisano pięć praktycznych aplikacji równoległych z różnych dziedzin zastosowań oraz pięć różnorodnych systemów obliczeniowych wykorzystywanych w eksperymentach zawartych w rozprawie. W rozdziale piątym zaproponowano metodologię optymalizacji wykonania hybrydowych aplikacji równoległych w heterogenicznych systemach obliczeniowych wysokiej wydajności, składającą się z określonych kroków wykonania oraz metody symulacji do szybkiej ewaluacji punktów w przeszukiwanej przestrzeni rozwiązań. W rozdziale szóstym przedstawiono wyniki eksperymentów polegających na zastosowaniu poszczególnych proponowanych kroków do wybranych rzeczywistych aplikacji oraz zastosowaniu metodologii optymalizacji w całości do aplikacji treningu głębokiej sieci neuronowej do automatycznego rozpoznawania mowy.

	Jak wykazano, wykonanie specyficznych w kontekście proponowanego modelu kroków wstępnej optymalizacji procesów, mapowania procesów, strojenia parametrów i właściwego uruchomienia, pozwala na optymalizację czasu wykonania hybrydowych aplikacji równoległych w heterogenicznych systemach obliczeniowych wysokiej wydajności, a proponowana metoda modelowania i symulacji umożliwia szybkie i dokładne wyznaczenie zbioru optymalnych rozwiązań w problemie wielokryterialnej optymalizacji ich wykonania z uwzględnieniem czasu wykonania i poboru mocy.

\newpage \vspace*{8cm}
\pdfbookmark{Dedication}{dedication}
\thispagestyle{empty}
\begin{center}
  \Large \emph{To my parents Beata and Roman, who gave me \\ the rare privilege of choosing my own path}
\end{center}

\newpage
\singlespacing
\tableofcontents
\doublespacing


\newglossaryentry{flops}{
name=Flop/s,
description={- floating operations per second, measure of computer performance,},
type=symbolslist
}

\newglossaryentry{dvfs}{
name=DVFS,
description={- Dynamic Voltage and Frequency Scaling technique,},
type=symbolslist
}

\newglossaryentry{wer}{
name=WER,
description={- Word Error Rate metric of automatic speech recognition performance.},
type=symbolslist
}

\newglossaryentry{pgas}{
name=PGAS,
description={- Partitioned Global Address Space parallel programming model,},
type=symbolslist
}

\newglossaryentry{spmd}{
name=SPMD,
description={- Single Program, Multiple Data paradigm,},
type=symbolslist
}

\newglossaryentry{dac}{
name=DaC,
description={- Divide and Conquer paradigm,},
type=symbolslist
}

\newglossaryentry{openmp}{
name=OpenMP,
description={- Open Multi-Processing application programming interface,},
type=symbolslist
}

\newglossaryentry{hcs}{
name=HCS,
description={- Heterogeneous Computing System,},
type=symbolslist
}

\newglossaryentry{etc}{
name=ETC,
description={- Estimated Times to Compute matrix,},
type=symbolslist
}

\newglossaryentry{apc}{
name=APC,
description={- Average Power Consumption matrix,},
type=symbolslist
}

\newglossaryentry{ilp}{
name=ILP,
description={- Integer Linear Programming,},
type=symbolslist
}

\newglossaryentry{dcfs}{
name=DCFS,
description={- Dynamic Core and Frequency Scaling technique,},
type=symbolslist
}

\newglossaryentry{ast}{
name=AST,
description={- Abstract Syntax Tree source code representation,},
type=symbolslist
}

\newglossaryentry{ptf}{
name=PTF,
description={- Periscope Tuning Framework,},
type=symbolslist
}

\newglossaryentry{diana}{
name=DIANA,
description={- Data Intensive and Network Aware scheduling technique,},
type=symbolslist
}

\newglossaryentry{pass}{
name=PASS,
description={- Power Aware Strong Scaling model,},
type=symbolslist
}

\newglossaryentry{cct}{
name=CCT,
description={- Coflow Completion Time,},
type=symbolslist
}

\newglossaryentry{gde3}{
name=GDE3,
description={- Generalized Differential Evolution,},
type=symbolslist
}

\newglossaryentry{lan}{
name=LAN,
description={- Local Area Network,},
type=symbolslist
}

\newglossaryentry{wan}{
name=WAN,
description={- Wide Area Network,},
type=symbolslist
}

\newglossaryentry{idw}{
name=IDW,
description={- Inverse Distance Weighting interpolation method,},
type=symbolslist
}

\newglossaryentry{lstm}{
name=LSTM,
description={- Long Short-Term Memory neural network architecture,},
type=symbolslist
}

\newglossaryentry{asr}{
name=ASR,
description={- Automatic Speech Recognition,},
type=symbolslist
}

\newglossaryentry{dag}{
name=DAG,
description={- Directed Acyclic Graph,},
type=symbolslist
}

\newglossaryentry{gpu}{
name=GPU,
description={- Graphics Processing Unit,},
type=symbolslist
}

\newglossaryentry{cpu}{
name=CPU,
description={- Central Processing Unit,},
type=symbolslist
}

\newglossaryentry{sge}{
name=SGE,
description={- Sun Grid Engine computing cluster software,},
type=symbolslist
}

\newglossaryentry{hpc}{
name=HPC,
description={- High Performance Computing,},
type=symbolslist
}

\newglossaryentry{mpi}{
name=MPI,
description={- Message Passing Interface,},
type=symbolslist
}

\newglossaryentry{processes}{
name=\ensuremath{P},
description={- set of all possible \emph{parallel processes},},
type=symbolslist
}

\newglossaryentry{process}{
name=\ensuremath{p},
description={\ensuremath{\in \gls{processes}} - parallel process, a sequence of operations \ensuremath{(\gls{process}_{1}, \gls{process}_{2}, ..., \gls{process}_{|\gls{process}|})},},
type=symbolslist
}

\newglossaryentry{minrequirements}{
name=\ensuremath{\vec{rmin}_{\gls{app}}},
description={- vector of minimal process requirements of \emph{hybrid parallel application} \gls{app},},
type=symbolslist
}

\newglossaryentry{minrequirement}{
name=\ensuremath{rmin_{\gls{app}}},
description={$^j$ - minimal requirement of \emph{hybrid parallel application} \gls{app} for the number of instances of a particular \emph{parallel process} $\gls{process}^j$,},
type=symbolslist
}

\newglossaryentry{maxrequirements}{
name=\ensuremath{\vec{rmax}_{\gls{app}}},
description={- vector of maximal process requirements of \emph{hybrid parallel application} \gls{app},},
type=symbolslist
}

\newglossaryentry{maxrequirement}{
name=\ensuremath{rmax_{\gls{app}}},
description={$^j$ - maximal requirement of \emph{hybrid parallel application} \gls{app} for the number of instances of a particular \emph{parallel process} $\gls{process}^j$,},
type=symbolslist
}

\newglossaryentry{executionparameters}{
name=\ensuremath{\vec{v}},
description={- vector of application execution parameters,},
type=symbolslist
}

\newglossaryentry{feasibleexecutionparameters}{
name=\ensuremath{V_{\gls{app}, \gls{system}}},
description={- set of application execution parameter vectors feasible for \emph{hybrid parallel application} \gls{app} and \emph{heterogeneous HPC system} \gls{system},},
type=symbolslist
}

\newglossaryentry{app}{
name=\ensuremath{A},
description={- hybrid parallel application,},
type=symbolslist
}

\newglossaryentry{impl}{
name=\ensuremath{\Phi_{\gls{app}}},
description={- set of \emph{process implementations} \ensuremath{\{\gls{process}^1, \gls{process}^2, \ldots, \gls{process}^{|\gls{impl}|}\}} provided by \emph{hybrid parallel application} \gls{app},},
type=symbolslist
}


\newglossaryentry{compset}{
name=\ensuremath{\Theta},
description={- set of all possible \emph{computation operations},},
type=symbolslist
}

\newglossaryentry{comp}{
name=\ensuremath{\theta},
description={\ensuremath{\in \gls{compset}} - \emph{computation operation},},
type=symbolslist
}

\newglossaryentry{commset}{
name=\ensuremath{K},
description={- set of all possible \emph{communication operations},},
type=symbolslist
}

\newglossaryentry{comm}{
name=\ensuremath{\kappa},
description={\ensuremath{\in \gls{commset}} - \emph{communication operation},},
type=symbolslist
}

\newglossaryentry{devices}{
name=\ensuremath{D_{\gls{system}}},
description={- set of \emph{computing devices} of \emph{heterogeneous HPC system} \gls{system},},
type=symbolslist
}

\newglossaryentry{device}{
name=\ensuremath{d},
description={\ensuremath{\in \gls{devices}} - one of the computing devices available in \emph{heterogeneous HPC system} \gls{system},},
type=symbolslist
}

\newglossaryentry{networkset}{
name=\ensuremath{L_{\gls{system}}},
description={- set of network links of \emph{heterogeneous HPC system} \gls{system},},
type=symbolslist
}

\newglossaryentry{networklink}{
name=\ensuremath{l},
description={\ensuremath{\in \gls{networkset}} - one of the network links available in \emph{heterogeneous HPC system} \gls{system},},
type=symbolslist
}

\newglossaryentry{hardwarecapabilities}{
name=\ensuremath{c},
description={(\gls{device}, \gls{process}) - number of instances of process \gls{process} that device \gls{device} in \emph{heterogeneous HPC system} \gls{system} is capable of executing,},
type=symbolslist
}

\newglossaryentry{mappingset}{
name=\ensuremath{M_{\gls{app}, \gls{system}}},
description={- feasible set of process mapping functions for \emph{hybrid parallel application}
\gls{app} and \emph{heterogeneous HPC system} \gls{system},},
type=symbolslist
}

\newglossaryentry{mapping}{
name=\ensuremath{m},
description={(\gls{device}, \gls{process}) - process mapping function that determines the number of instances of \emph{parallel process} \gls{process} mapped for execution on device \gls{device},},
type=symbolslist
}

\newglossaryentry{system}{
name=\ensuremath{S},
description={(\gls{devices}, \gls{networkset}) - graph representing \emph{heterogeneous \gls{hpc} system} \gls{system},},
type=symbolslist
}

\newglossaryentry{executiontime}{
name=\ensuremath{ET},
description={(\gls{app}, \gls{system}, \gls{mapping}, \gls{executionparameters}) - execution time of \emph{hybrid parallel application} \gls{app} in \emph{heterogeneous \gls{hpc} system} \gls{system} with process mapping function \gls{mapping} and \emph{application execution parameters} \gls{executionparameters},},
type=symbolslist
}

\newglossaryentry{processexecutiontime}{
name=\ensuremath{et},
description={(\gls{process}, \gls{device}) - execution time of \emph{parallel process}
\gls{process} on \emph{device} \gls{device},},
type=symbolslist
}

\newglossaryentry{powerconsumption}{
name=\ensuremath{PC},
description={(\gls{app}, \gls{system}, \gls{mapping}, \gls{executionparameters}) - average power consumption of executing \emph{hybrid parallel application} \gls{app} in \emph{heterogeneous \gls{hpc} system} \gls{system} with process mapping function \gls{mapping} and \emph{application execution parameters} \gls{executionparameters},},
type=symbolslist
}

\newglossaryentry{powercons}{
name=\ensuremath{pc},
description={(\gls{device}, t) - power consumption of computing device \gls{device} at time t,},
type=symbolslist
}

\newglossaryentry{maxpowerconsumption}{
name=\ensuremath{MPC},
description={(\gls{app}, \gls{system}, \gls{mapping}, \gls{executionparameters}) - maximum power consumption of executing \emph{hybrid parallel application} \gls{app} in \emph{heterogeneous \gls{hpc} system} \gls{system} with process mapping function \gls{mapping} and \emph{application execution parameters} \gls{executionparameters},},
type=symbolslist
}

\newglossaryentry{deviceenergyconsumption}{
name=\ensuremath{ec},
description={(\gls{device}, \gls{app}) - total energy consumption (including while idle) of \emph{computing device} \gls{device} during the execution of \emph{hybrid parallel application} \gls{app},},
type=symbolslist
}

\newglossaryentry{commtime}{
name=\ensuremath{commtime},
description={(\gls{comm}, \gls{networklink}) - execution time of \emph{communication operation} \gls{comm} using \emph{network link} \gls{networklink},},
type=symbolslist
}

\newglossaryentry{comptime}{
name=\ensuremath{comptime},
description={(\gls{comp}, \gls{device}) - execution time of \emph{computation operation} \gls{comp} using \emph{computing device} \gls{device},},
type=symbolslist
}

\newglossaryentry{pcidle}{
name=\ensuremath{pcidle},
description={(\gls{device}) - idle power consumption of \emph{computing device} \gls{device},},
type=symbolslist
}

\newglossaryentry{pcpeak}{
name=\ensuremath{pcpeak},
description={(\gls{device}) - peak power consumption of \emph{computing device} \gls{device},},
type=symbolslist
}

\newglossaryentry{poperation}{
name=\ensuremath{poperation},
description={(\gls{device}, t) - additional power consumption of running an operation on \emph{computing device} \gls{device} at time t,},
type=symbolslist
}

\newglossaryentry{activeoperations}{
name=\ensuremath{activeoperations},
description={(\gls{device}, t) - number of operations executed on \emph{computing device} \gls{device} at time t,},
type=symbolslist
}

\newglossaryentry{ncores}{
name=\ensuremath{ncores},
description={(\gls{device}) - number of cores of \emph{computing device} \gls{device},},
type=symbolslist
}

\newglossaryentry{powerlimit}{
name=\ensuremath{PCL},
description={- power consumption limit,},
type=symbolslist
}

\newglossaryentry{paretoobjectives}{
name=\ensuremath{n},
description={- number of objectives optimized in a multi-objective optimization problem,},
type=symbolslist
}

\newglossaryentry{optimparameters}{
name=\ensuremath{x},
description={- point in the decision space of an optimization problem,},
type=symbolslist
}

\newglossaryentry{optimum}{
name=\ensuremath{\hat{x}},
description={- optimum of a single-objective optimization problem,},
type=symbolslist
}

\newglossaryentry{paretoset}{
name=\ensuremath{\mathcal{X}^*},
description={- Pareto set,},
type=symbolslist
}

\newglossaryentry{paretofront}{
name=\ensuremath{\mathcal{F}^*},
description={- Pareto front,},
type=symbolslist
}

\newglossaryentry{feasibleset}{
name=\ensuremath{\mathcal{X}},
description={- feasible set of an optimization problem,},
type=symbolslist
}

\newglossaryentry{paretofunction}{
name=\ensuremath{f},
description={- objective function of an optimization problem,},
type=symbolslist
}

\newglossaryentry{paretofunctionvector}{
name=\ensuremath{F},
description={- vector of objective functions of a multi-objective optimization problem,},
type=symbolslist
}

\printglossary[type=symbolslist, nonumberlist]

\newpage
\leavevmode\thispagestyle{empty}\newpage
\chapter{Introduction}\label{chp:introduction}
\pagenumbering{arabic}

High Performance Computing (\gls{hpc}) matters. 

It is crucial for modern scientific findings in areas such as genetics, drug discovery, 
particle physics \cite{villa_scaling_2014}, predictive models for cancer and infectious diseases
\cite{stevens_deep_2017} as well as technological development including video classification,
automatic speech recognition, language representation and many others \cite{abadi_tensorflow:_2016}.
All these fields require parallel processing. Due to its unique requirements, the
\emph{deep learning} \cite{lecun_deep_2015} method used in many scientific applications is a good
example of such an application, because it requires a combination of three different types
of parallelism \cite{stevens_deep_2017}. First, model parallelism, that is splitting the machine learning
model among multiple machines, is an example where parallelism is needed due to the problem
size not fitting into the memory of a single machine. Second, data parallelism is needed because
large quantities of data are required for the processing. In practice, finishing a contemporary
deep neural network training takes days, or even weeks \cite{you_imagenet_2017}. Discovering optimal 
results often involves a large-scale search of hyperparameters that requires utilizing the third type
of parallelism - executing multiple versions of the same training application to search a space of
thousands of model configurations \cite{stevens_deep_2017}.
Execution time of such workloads is critical and for analogous reasons, \gls{hpc} architectures are
essential for many practical applications.

Responding to these needs, because the traditional doubling of computing device clock speeds every 18-24 months
is no longer valid, hardware manufacturers replace it by doubling of cores, threads or other parallelism mechanisms \cite{shalf_exascale_2010}. This has allowed to develop supercomputers whose performance is measured in
PFlop/s, which means that they are capable of executing a multiple of $10^{15}$ \gls{flops} (floating operations per second).
For example, Sunway TaihuLight, the most powerful supercomputer according to the top500\footnote{http://top500.org/} list for November 2017, has the performance of 93.014 PFlop/s and consists of 10 649 600 cores of Sunway MPP Central Processing Units (\gls{cpu}s). The \gls{hpc} community is targeting development of supercomputers able to sustain one ExaFlop/s ($10^{18}$  \gls{flops}) by the year 2020 \cite{villa_scaling_2014}. To scale up their floating point throughput, the current generation supercomputers rely heavily on massively-parallel computing accelerators, which makes them heterogeneous in terms of utilized computing devices. For example, Tianhe-2, the second most powerful supercomputer on the aforementioned
list has a performance of 33.862 PFlop/s and consists of both Intel Xeon \gls{cpu}s and Intel Xeon Phi computing accelerators,
and the third most powerful, Piz Daint, consists of both Intel Xeon \gls{cpu}s and NVIDIA Tesla Graphics Processing Units (\gls{gpu}s). 

The primary obstacle to achieving the goal of Exascale computing is power consumption~\cite{villa_scaling_2014}.
The upper limit for a reasonable system design adopted by the Exascale Initiative Steering Committee
at the US Department of Energy is 20MW, while peak power consumption of the TaihuLight, Tianhe-2 and
Piz Daint supercomputers is 15.371, 17.808 and 2.272 MW respectively. This means that
over 20x improvement in energy efficiency is required. To achieve the goal of Exascale,
architectural improvements, circuit design and manufacturing technologies are required \cite{villa_scaling_2014},
but also software solutions such as system software, programming models, algorithms, frameworks and
compilers \cite{dongarra_international_2011}.

Given the multi-level parallelism and heterogeneity of the contemporary \gls{hpc} systems,
their efficient utilization requires developing hybrid parallel programs that have the
advantage of exploiting inter-node distributed-memory scalability and intra-node shared-memory
performance \cite{ramapantulu_approach_2015} as well as the ability to run on a heterogeneous infrastructure
\cite{ma_greengpu:_2012}, where certain computing devices can be more efficient for specific parts of the computations.
Complexity of such applications brings the requirement for automatic tuning and resource control
support to optimize the utilization of resources \cite{dongarra_international_2011}. One of the key challenges
for time- and energy-efficient execution of the applications is to determine optimal values from within a potentially high-dimensional
space of configurations~\cite{ramapantulu_approach_2015}.

\section{Motivations}

\begin{figure}[ht!]
    \centering
    \begin{subfigure}[b]{0.5\textwidth}
        \centering
                \includegraphics[scale=0.6]{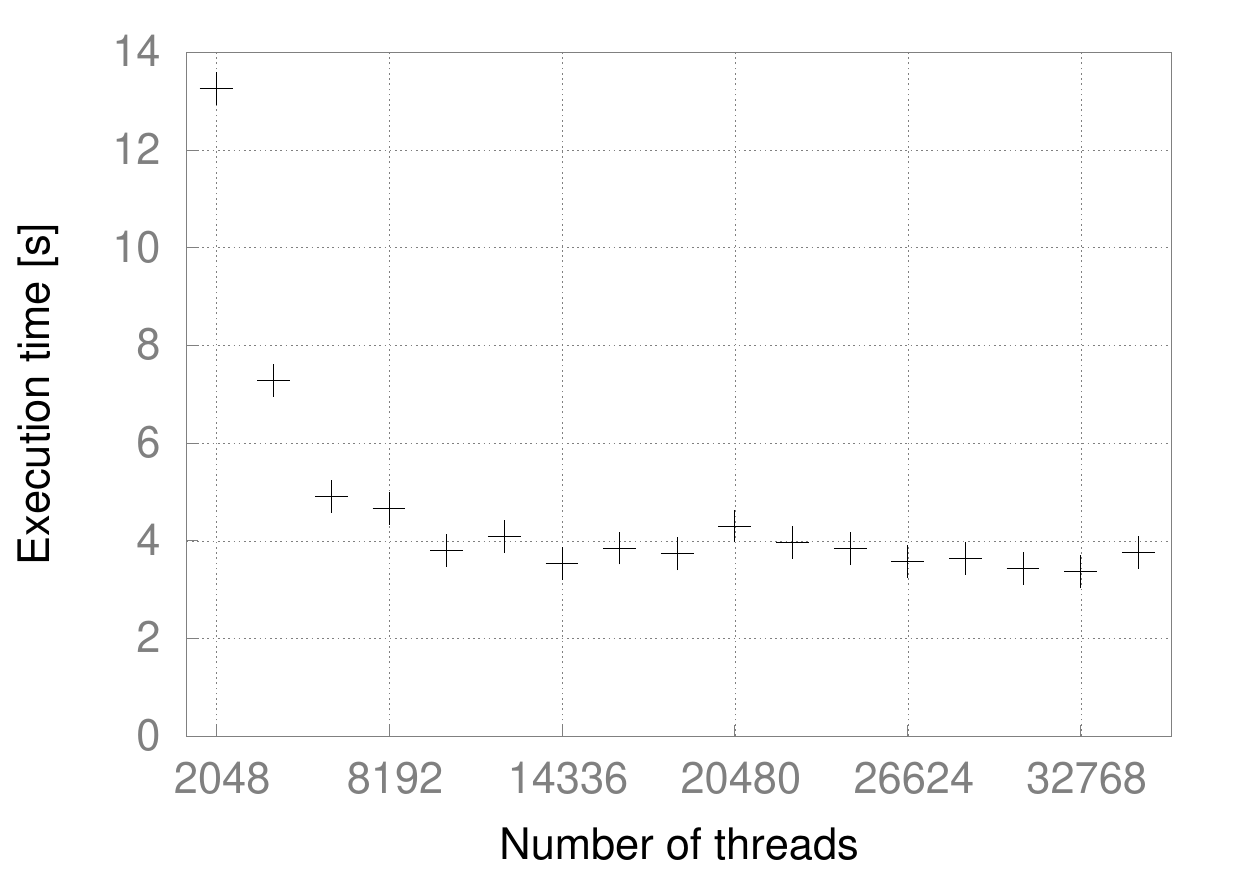}
        \caption{Regular expression matching on a \gls{gpu}\label{fig:gridregex}}
    \end{subfigure}%
    ~
    \begin{subfigure}[b]{0.5\textwidth}
        \centering
                \includegraphics[scale=0.6]{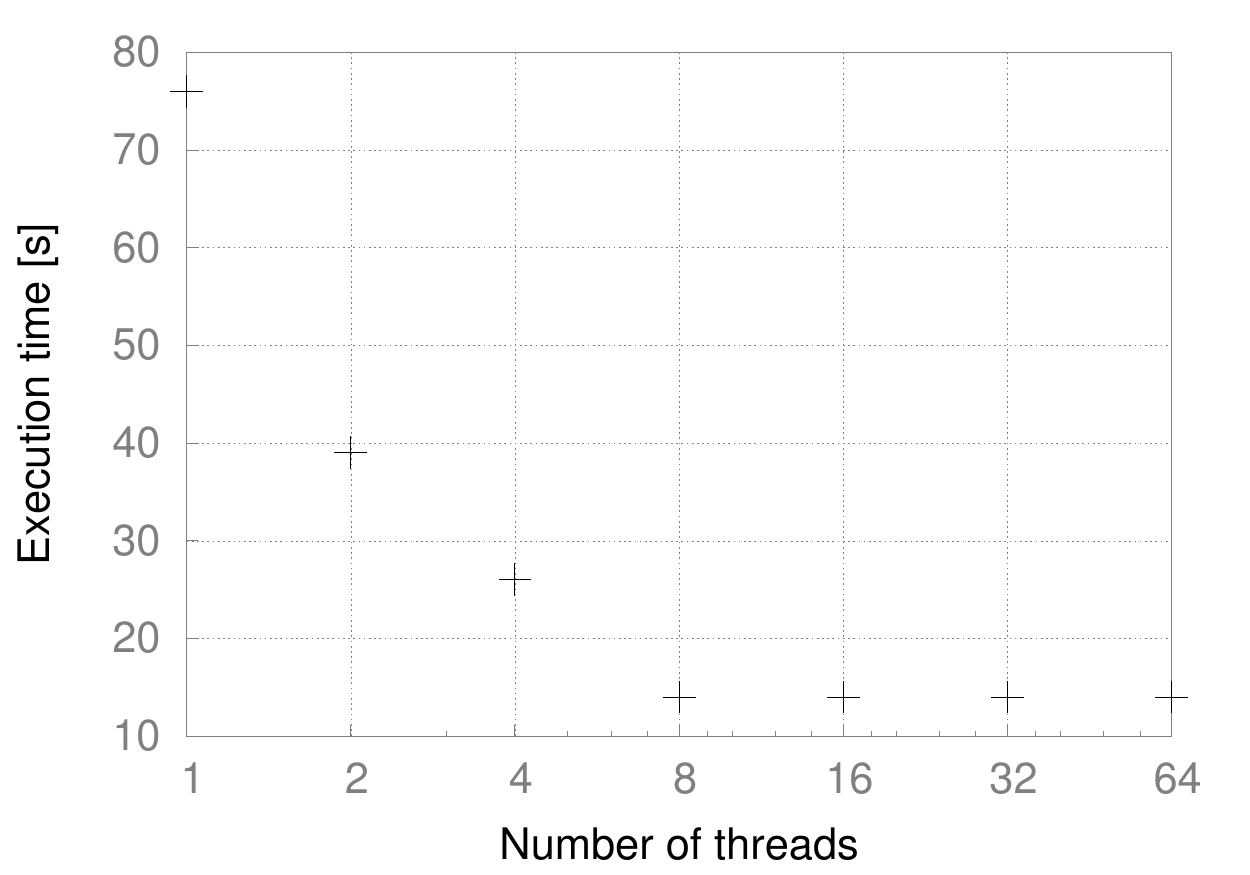}
                \caption{Geospatial interpolation on Intel Xeon Phi\label{fig:idwphigrid}}
    \end{subfigure}
    \caption{Influence of the number of used threads on execution time depending on application and device\label{fig:localthreadscomp}}
\end{figure}

The multiplicity of the possible configurations of hybrid parallel application execution
stems from various parameters of the execution and feasible allocations of hardware to
individual parts of the application. An example of an execution parameter connected with
the utilized hardware is the number of threads executed in parallel. Figure \ref{fig:localthreadscomp}
presents the influence of this parameter on execution times of a parallel regular expression matching
application (see Section \ref{sec:regex}) and a geospatial interpolation application (see Section \ref{sec:idw})
depending on the number of used threads on two different computing devices.
The first device is a GTX 480 GPU belonging to the HPC system described in Section \ref{sec:lab527}
and the second device is an Intel Xeon Phi 7120P computing accelerator belonging to the HPC system
described in Section \ref{sec:miclab}.

The charts show that the same execution parameter can take completely different values
depending on the specific application and utilized computing device - for the GPU the optimal
value is in the order of thousands of threads and for the Xeon Phi accelerator, dozens of threads.
Efficient execution of a multi-level application running in a heterogeneous HPC system may
require individual tuning of the execution parameters for each utilized device. Finding a 
single optimal value may be also non-trivial due to the presence of local optima.
 
\begin{figure}[ht!]
\begin{center}
\includegraphics[scale=0.9]{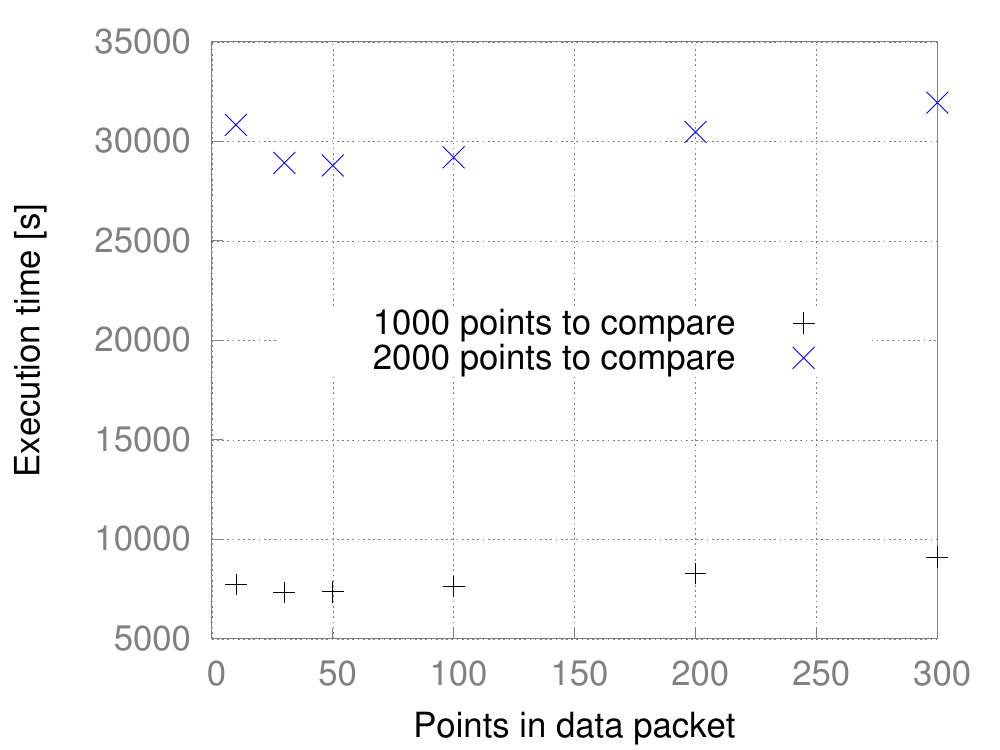}
\end{center}
\caption{Influence of data partitioning on the execution time depending on problem size \cite{czarnul_simulation_2015} \label{fig:bdsparams}}
\end{figure}

The multiple parameters of the execution may also originate from the application domain. 
An example of such parameter is size of packages into which the input data should be partitioned
for parallel execution. The application described in Section \ref{sec:big_data_sim}, given a
number of points in a high-dimensional space, computes similarity measures between the points
in parallel by distributing data packets among the available computing devices. The influence
of the number of points in such a data packet on the execution time on the HPC system described
in Section \ref{sec:lab527} for a space of 200 000 dimensions is shown in Figure \ref{fig:bdsparams}.

The chart shows that there is a certain optimal value of this parameter which differ depending
on the problem size. In the case of 1000 data points, the best partitioning
is to 30 points in data packet, but in the case of 2000 data points it is 50. There can be dozens
of such parameters in one application execution and their optimal values may depend on each other.
Indication of such parameters for specific applications in various fields and examination
of their interdependence and influence on execution time and power consumption of the application
is an important research direction. Multiple possibilities of mapping parts of the application 
to computing devices cause that the space of configurations is even more complex.

Many existing approaches focus on tuning specific execution parameters or solving a specific
scheduling problem, with regard to both execution time and power consumption optimization goals,
which are often formulated using analytical formulas. Defining a specific formula for execution
time and power consumption is often required for each individual application, which is complicated
and requires full a priori knowledge about the components of the application and their behavior.
In the light of contemporary heterogeneous and multi-level systems there is a need for analysing these
approaches, experiments with practical applications from various fields and extraction of a general
execution model and optimization methodology that includes all the important aspects and takes
into account the opposing objectives of execution time and power consumption.

In many existing approaches, evaluating an execution configuration requires running an actual
execution with this configuration. Considering high dimensionality of the configuration
space and prohibitively long execution times, there is a need for a method for fast and accurate
evaluation of configurations. A modeling and simulation method could be used for this purpose.
Research in the field of modeling and simulation of parallel application execution and 
experiments involving simulation of real applications is needed to verify if a simulation approach
can provide fast and accurate estimations of execution time and power consumption and,
thus, support optimization of practical applications.

\section{Problem Formulation}\label{sec:problemformulation}

The process optimized within this thesis is execution of 
a \emph{hybrid parallel application} in a \emph{heterogeneous high performance computing system}.
The primary components of a \emph{hybrid parallel application} are \emph{computation} and 
\emph{communication operations}:

\begin{definition}
Let \gls{compset} denote the set of all possible \underline{computation operations}, where
$\gls{comp} \in \gls{compset}$ is a \underline{computation operation}.
\end{definition}

\begin{definition}
Let \gls{commset} denote the set of all possible \underline{communication operations}, where
$\gls{comm} \in \gls{commset}$ is a \underline{communication operation}.
\end{definition}

\noindent A \emph{computation operation} utilizes a certain computing device in order to perform
computations, while a \emph{communication operation} utilizes a certain network link to
perform communication between two processes.
\emph{Operations} are characterized by \emph{operation parameters}, for example
computational complexity of a \emph{computation operation} or \emph{data size} for
a \emph{communication operation}.

The \emph{operations} are executed within \emph{processes}:

\begin{definition}
Let $\gls{processes}$ denote a set of all possible \underline{processes}, where \underline{parallel process}
$\gls{process}^j \in \gls{processes}$ is a sequence of \emph{computation} and \emph{communication operations}
($\gls{process}^j_{1}, \gls{process}^j_{2}, ..., \gls{process}^j_{|\gls{process}^j|}$) where
$\gls{process}^j_i \in \gls{compset} \cup \gls{commset}$.
\end{definition}

\noindent Parallelism of a \emph{hybrid parallel application} can be taken into account by the proposed model
on two levels:
first, there might be many processes in an application which are executed simultaneously on different hardware.
Secondly, the \emph{process} itself may execute the computations in parallel.
For example, an application consisting of of many \gls{mpi} processes which execute parallel
codes on \gls{gpu}s is parallel at both levels. 

A \emph{hybrid parallel application} consists of one or more \emph{process implementations}, which 
define the sequence of operations in each \emph{process} of the application:

\begin{definition}
Let $\gls{impl} = \{ \gls{process}^1, \gls{process}^2, \ldots, \gls{process}^{|\gls{impl}|} \} \subset \gls{processes}$ denote a set of \underline{process implementations} provided by a \emph{hybrid parallel application} \gls{app}.
\end{definition}

\noindent A \emph{process implementation} results from the code of the application, but can be 
also represented as an arbitrary schedule of operations, computer program or log from
a completed application. The operation sequences of two \emph{processes} may also depend on
each other if at some moment of execution there occurs communication between these two processes. 

An application has certain \emph{process requirements}:

\begin{definition}
Let vectors $\gls{minrequirements} = [\gls{minrequirement}^j]^{|\gls{impl}|}, \gls{maxrequirements} = [\gls{maxrequirement}^j]^{|\gls{impl}|}$ denote \underline{process requirements} of an application, where $\gls{minrequirement}^j, \gls{maxrequirement}^j$ are minimal and maximal numbers of instances of
\emph{process} $\gls{process}^j \in \gls{impl}$ that are required by an application to be executed.
\end{definition}

\noindent The process requirements may for example depend on the parallel paradigm of the application. For instance, a master/slave application would require exactly one master ($\gls{minrequirement}^{master} = \gls{maxrequirement}^{master} = 1$). It would also require at least one slave ($\gls{minrequirement}^{slave} = 1$) and the maximal number of required slaves could be a value depending on data partitioning limitations or none (infinity).

Summarizing, a \emph{hybrid parallel application} is defined as follows:

\begin{definition}
Let $\gls{app} = \langle \gls{impl}, \gls{minrequirements}, \gls{maxrequirements} \rangle$ denote a \underline{hybrid parallel application}, where $\gls{impl} \subset \gls{processes}$
is a set of process implementations of the application and $\gls{minrequirements}$ and
$\gls{maxrequirements}$ are minimal and maximal process requirements of the application.
\end{definition}

It should be noted that a set of independent \emph{hybrid parallel applications} can be modeled, where processes of each application never communicate with processes of another application. Thus, we can still treat such a set of applications as one \emph{hybrid parallel application}.

\emph{Hybrid parallel applications} are executed on \emph{heterogeneous \gls{hpc} systems}, also called heterogeneous computing systems (\gls{hcs}):

\begin{definition}
Let a graph $\gls{system}(\gls{devices}, \gls{networkset})$ denote \underline{heterogeneous \gls{hpc} system}, where $\gls{devices}$ is a set of \underline{computing devices} $(\gls{devices} = \{\gls{device}_1, \gls{device}_2, ..., \gls{device}_{|\gls{devices}|} \})$
and $\gls{networkset}$ is a set of \underline{network links} between \emph{computing devices} $(\gls{networklink}_{i, j}$ - network link between devices
$\gls{device}_i$ and $\gls{device}_j$).
\end{definition}

\noindent The \emph{computing devices} represented by $\gls{device}_i$ are hardware devices capable of performing computations.
Depending on the considered granularity they might
represent computing cores, processing units, computer nodes, groups of nodes, high performance clusters, voluntary computing systems etc.
Analogously, a \emph{network link} can represent a system bus, local area network (\gls{lan}), wide area network (\gls{wan}) etc.
\emph{Computing devices} and \emph{network links} are characterized by \emph{hardware parameters}, for example
computational performance of a \emph{computing device} or \emph{bandwidth} of a \emph{network link}.
Utilizing hardware with different \emph{hardware parameters} makes a system \emph{heterogeneous}.

Each \emph{computing device} has certain \emph{hardware capabilities}, which resemble technical possibility to execute certain processes 
(depending on software stack possibilities, available computing capabilities etc.):

\begin{definition}
Let function $\gls{hardwarecapabilities}(\gls{device}, \gls{process}) \in N$ denote \underline{hardware capabilities} defining how many instances of process $\gls{process}$ can be executed on device $\gls{device}$. 
\end{definition}

Execution of an \emph{application} depends also on \emph{application execution parameters}:

\begin{definition}
Let $\gls{feasibleexecutionparameters}$ denote the \underline{space of application execution parameters} feasible for a 
\emph{hybrid parallel application} \gls{app} and a \emph{heterogeneous HPC system} \gls{system}.
\end{definition}

\noindent The \emph{space of application execution parameters} can potentially be highly dimensional.
Multiple examples of execution parameters are described in Chapter \ref{chp:applicationstheory}.
The \emph{application execution parameters} consist of two groups:

\begin{itemize}
\item \emph{application parameters} - related to the algorithms used in the application, parallel paradigm, data partitioning,
assumed problem constraints, buffer sizes, etc.;
\item \emph{execution parameters} - related to possible configurations of the hardware used at multiple parallelization levels,
including numbers of threads, thread affinity, \gls{gpu} grid configurations, \gls{dvfs} modes etc.
\end{itemize}

For a specific execution of a \emph{hybrid parallel application} in a \emph{heterogeneous \gls{hpc} system},
specific values of the \emph{application execution parameters} have to be set:

\begin{definition}
Let $\gls{executionparameters} \in \gls{feasibleexecutionparameters}$ denote a vector of \underline{application execution parameters} of a \emph{hybrid parallel application}.
\end{definition}

\noindent Also, the \emph{processes} have to be \emph{mapped} to specific \emph{computing devices}:

\begin{definition}
Let function $\gls{mapping}(\gls{device}, \gls{process})$ denote \underline{process mapping} of a \emph{hybrid parallel application}
to a \emph{heterogeneous \gls{hpc} system}, where $\gls{mapping}(\gls{device}, \gls{process})$ defines how many instances of process
$\gls{process}$ should be run on a \emph{computing device} $\gls{device}$ during a certain application execution.
\end{definition}

\noindent It should be noted that given a \emph{hybrid parallel application}
\gls{app} and a \emph{heterogeneous HPC system} \gls{system}, a feasible \emph{process mapping} function $\gls{mapping}$ belongs to 
the following \underline{feasible set of process} \underline{mapping functions}
$\gls{mappingset}$:
\begin{equation*}
\gls{mappingset} = \big\{\gls{mapping}(\gls{device}, \gls{process}) |
\forall_{\gls{process}^j \in \gls{impl}}
\forall_{\gls{device} \in \gls{devices}}
\big( \gls{mapping}(\gls{device}, \gls{process}^j) <= \gls{hardwarecapabilities}(\gls{device}, \gls{process}^j)\big) \wedge
\forall_{\gls{process}^j \in \gls{impl}}
\big( \gls{minrequirement}^j <= \sum_{i}{\gls{mapping}(\gls{device}_i, \gls{process}^j)} <= \gls{maxrequirement}^j \big) \big\}.
\end{equation*}


The primary outcomes of executing a \emph{hybrid parallel application} are the results that depend on the purpose of the
application. The optimization problem considered in this thesis assumes that these primary results
are correct and focuses on the secondary results of the execution, related to \emph{execution time}
and \emph{power consumption}:

\begin{definition}\label{def:executiontime}
Let function $\gls{executiontime}(\gls{app}, \gls{system}, \gls{mapping}, \gls{executionparameters})$ denote 
\underline{execution time} of a \emph{hybrid parallel application} \gls{app} on a \emph{heterogeneous \gls{hpc} system}
\gls{system} with \emph{process mapping} function \gls{mapping} and vector of \emph{application execution parameters} \gls{executionparameters}. Specifically, the \emph{execution time} is defined as the time from the start of the 
application execution (which involves starting all process instances in the application)
until finishing of all \emph{process} instances on the devices assigned by the \emph{process mapping} function:
$\gls{executiontime}(\gls{app}, \gls{system}, \gls{mapping}, \gls{executionparameters})
= \max_{\gls{process} \in \gls{impl}}
\big(\max_{\gls{device} \in \gls{devices}:\gls{mapping}(\gls{device}, \gls{process}) > 0}
\gls{processexecutiontime}(\gls{process}, \gls{device})
\big)$
where \gls{processexecutiontime}(\gls{process}, \gls{device}) is execution time of a \emph{parallel process}
\gls{process} on \emph{device} \gls{device}.
\end{definition}

\begin{definition}
Let function $\gls{powerconsumption}(\gls{app}, \gls{system}, \gls{mapping}, \gls{executionparameters})$ denote 
\underline{average power consumption} of all \emph{computing devices} of a
\emph{heterogeneous \gls{hpc} system} \gls{system} during the execution a \emph{hybrid parallel application} \gls{app} with \emph{process mapping}
function \gls{mapping} and vector of \emph{application execution parameters} \gls{executionparameters}.
Specifically, the \emph{average power consumption} is defined as the total energy consumption (including while idle)
of all \emph{computing devices} in the system throughout the application execution divided by the execution time:
$\gls{powerconsumption}(\gls{app}, \gls{system}, \gls{mapping}, \gls{executionparameters}) = 
\frac{
\sum\limits_{\gls{device} \in \gls{devices}}{ 
\gls{deviceenergyconsumption}(\gls{device}, \gls{app})
}
}{\gls{executiontime}(\gls{app}, \gls{system}, \gls{mapping}, \gls{executionparameters})}$
where \gls{deviceenergyconsumption}(\gls{device}, \gls{app}) is the total energy consumption of
a computing device \gls{device} during the execution of a \emph{hybrid parallel application} \gls{app}.
\end{definition}

\begin{definition}
Let function $\gls{maxpowerconsumption}(\gls{app}, \gls{system}, \gls{mapping}, \gls{executionparameters})$ denote 
\underline{maximum power consumption} of all \emph{computing devices} of a
\emph{heterogeneous \gls{hpc} system} \gls{system} during the execution a \emph{hybrid parallel application} \gls{app} with \emph{process mapping}
function \gls{mapping} and vector of \emph{application execution parameters} \gls{executionparameters}:
$\gls{maxpowerconsumption}(\gls{app}, \gls{system}, \gls{mapping}, \gls{executionparameters}) = 
\max\limits_{t} 
\sum\limits_{\gls{device} \in \gls{devices}}{ 
\gls{powercons}(\gls{device}, t)
}
$
where \gls{powercons}(\gls{device}, $t$) is the power consumption of
a computing device \gls{device} at time $t$.
\end{definition}

The following multi-objective optimization problem,
is solved within this thesis, with the objective space consisting of
\emph{execution time} \gls{executiontime} and \emph{average power consumption}~\gls{powerconsumption}.
Given a \emph{hybrid parallel application} \gls{app} 
and a heterogeneous high performance computing system~\gls{system}:

\begin{equation}\label{eqn:problem}
\begin{aligned}
& \underset{\gls{mapping}, \gls{executionparameters}}{\min}
& & \gls{paretofunctionvector}(\gls{app}, \gls{system}, \gls{mapping}, \gls{executionparameters}) 
= [\mathrm{\gls{executiontime}}(\gls{app}, \gls{system}, \gls{mapping}, \gls{executionparameters}), 
\mathrm{\gls{powerconsumption}}(\gls{app}, \gls{system}, \gls{mapping}, \gls{executionparameters})] \\
& \text{subject to}
& & \gls{mapping} \in \gls{mappingset}, \\
&&& \gls{executionparameters} \in \gls{feasibleexecutionparameters}.
\end{aligned}
\end{equation}

Pareto method is used for multi-objective optimization where the
expected solution to the optimization problem is a set of Pareto-optimal points from the search space.
The search space consists of the \emph{feasible set of process mapping functions}~$\gls{mappingset}$
and the space of \emph{feasible application execution parameters} $\gls{feasibleexecutionparameters}$.
A point is Pareto-optimal if every other point in the search space 
results in higher execution time or higher average power consumption than this point.
The Pareto method is described in more detail in Section \ref{sec:pareto}.

In this thesis we assume that the \emph{\gls{hpc} system} \gls{system} is given and cannot be modified,
although optimization that considers introducing new, currently unavailable hardware to the system
can be considered as future work. Changes to the \emph{processes} $\gls{process} \in \gls{impl}$
of the \emph{application} \gls{app} are possible, but limited. The computational goals of the
application have to be maintained, so the optimization of the processes can be only performed 
manually by a specialist. Such optimizations are included in the execution steps
proposed in this thesis as an optional first step of preliminary process optimization, after which the application \gls{app}
cannot be modified. The remaining parameters of the execution can be optimized automatically
and define the search space of the considered process mapping and parameter tuning problem.

The following related problems are also considered in the thesis:

\begin{itemize}
\item{Optimization of execution time under power consumption constraints - given a \emph{hybrid parallel application} \gls{app}, heterogeneous high performance computing system \gls{system} and \underline{power consumption limit}~\gls{powerlimit}:
\begin{equation}\label{eqn:problemlimit}
\begin{aligned}
& \underset{\gls{mapping}, \gls{executionparameters}}{\min}
& & \mathrm{\gls{executiontime}}(\gls{app}, \gls{system}, \gls{mapping}, \gls{executionparameters}) \\
& \text{subject to}
& & \gls{mapping} \in \gls{mappingset}, \\
&&& \gls{executionparameters} \in \gls{feasibleexecutionparameters}, \\
&&& \mathrm{\gls{maxpowerconsumption}}(\gls{app}, \gls{system}, \gls{mapping}, \gls{executionparameters}) <= \gls{powerlimit}.
\end{aligned}
\end{equation}}
\item{Optimization of execution time - given a hybrid parallel application \gls{app} and heterogeneous high performance computing system \gls{system}:
\begin{equation}\label{eqn:problemtime}
\begin{aligned}
& \underset{\gls{mapping}, \gls{executionparameters}}{\min}
& & \mathrm{\gls{executiontime}}(\gls{app}, \gls{system}, \gls{mapping}, \gls{executionparameters}) \\
& \text{subject to}
& & \gls{mapping} \in \gls{mappingset}, \\
&&& \gls{executionparameters} \in \gls{feasibleexecutionparameters}.
\end{aligned}
\end{equation}}
\end{itemize}

The model of \emph{hybrid parallel application} execution in a \emph{heterogeneous \gls{hpc} system} proposed in this Section is referenced by Claim \ref{clm:1} of this dissertation.

\newpage
\section{Scope of the Dissertation}

The work described in this thesis is related to six main areas:

\begin{itemize}
\item specific \emph{hybrid parallel applications} and their implementations;
\item solutions for executing \emph{hybrid parallel applications} in \emph{heterogeneous \gls{hpc} systems};
\item modeling and simulation of parallel applications in \emph{heterogeneous \gls{hpc} systems};
\item multi-objective optimization of parallel applications;
\item energy-aware resource management in \emph{heterogeneous \gls{hpc} systems};
\item parameter auto-tuning in parallel applications.
\end{itemize}

\begin{figure}[ht!]
\begin{center}
\includegraphics[scale=0.7]{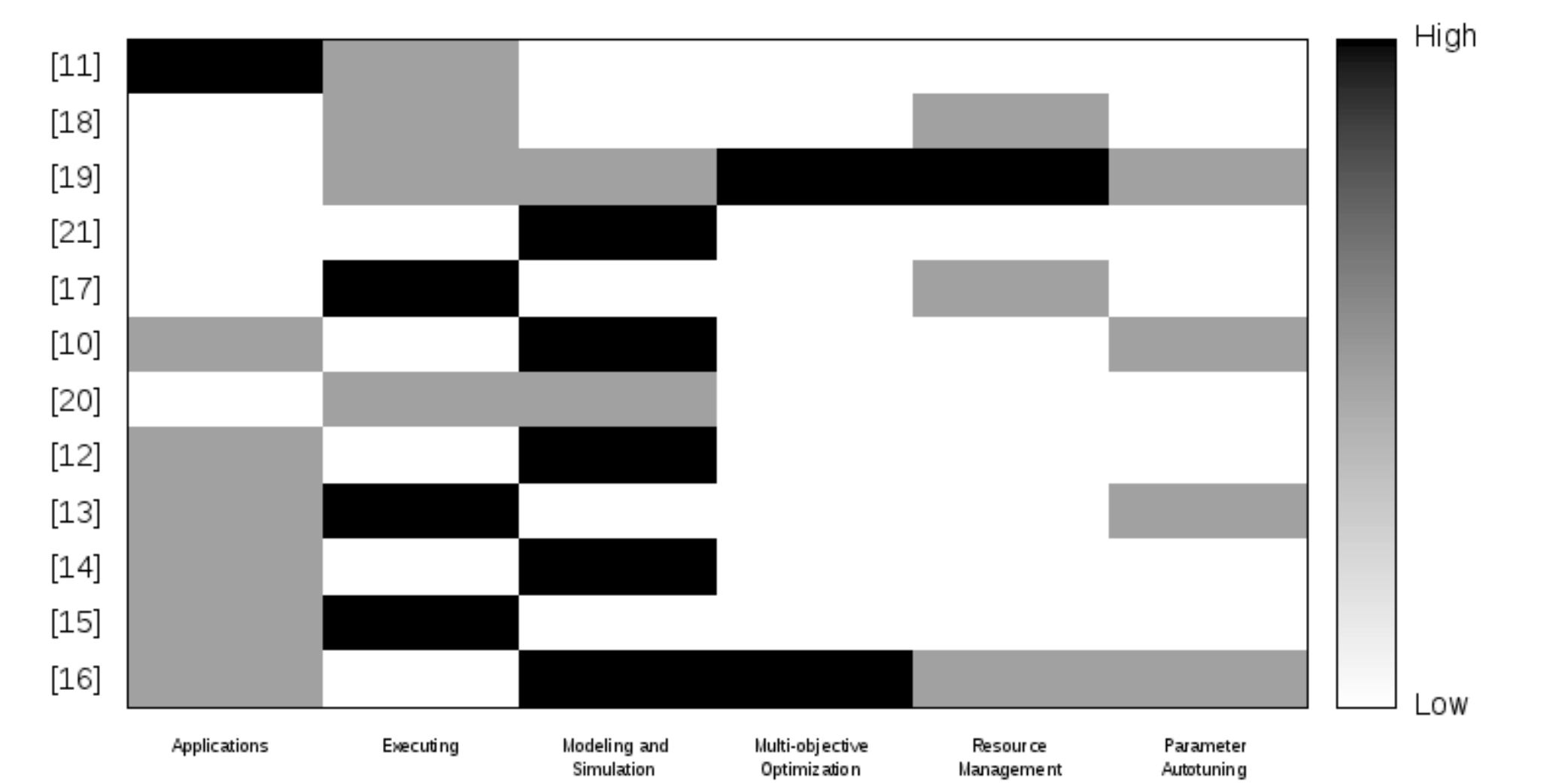}
\end{center}
\caption{Map of contributed papers and their relevant fields\label{fig:fieldmap}}
\end{figure}

Figure \ref{fig:fieldmap} presents to what extent the contributed papers related to this
thesis are relevant to each of these main areas.

In the area of specific \emph{hybrid parallel applications}, the scope of this thesis includes
an overview of existing applications in Section \ref{sec:appexamples} and the applications described
in Section \ref{sec:applications}, used in the experiments. In the most application-oriented
paper \cite{rosciszewski_regular_2014}
we proposed a regular expression matching application with configurable data intensity for
testing \emph{heterogeneous \gls{hpc} systems}. Certain descriptions and analyses of the corresponding
considered applications have also been provided in papers
\cite{czarnul_simulation_2015, czarnul_modeling_2016, rosciszewski_kernelhive:_2016,
czarnul_merpsys:_2017, rosciszewski_minimizing_2017, rosciszewski_modeling_2017}.

Chosen existing approaches to executing parallel applications in \emph{heterogeneous \gls{hpc} systems}
have been described in Section \ref{sec:executing}. The framework for multi-level high
performance computing using clusters and workstations with \gls{cpu}s and \gls{gpu}s introduced in
\cite{rosciszewski_kernelhive:_2016} is proposed in this thesis as a software solution for
executing the proposed optimization methodology. Execution of a parallel deep neural network
training application on a cluster with multiple \gls{gpu}s is described in \cite{rosciszewski_minimizing_2017}.
An approach to using distributed databases for data management in \emph{heterogeneous \gls{hpc} systems} is proposed
in \cite{rosciszewski_network-aware_2014}. Certain aspects regarding execution of the considered
applications have been also discussed in papers \cite{rosciszewski_regular_2014, balicki_runtime_2014, hutchison_optimization_2014,
rosciszewski_executing_2016}.

An important contribution of this thesis is the method for modeling and simulation of
\emph{hybrid parallel application} execution on \emph{heterogeneous \gls{hpc} systems} described in 
Section \ref{sec:simulation}. Section~\ref{sec:modeling} provides an overview of existing
approaches to simulation and modeling in this field. In~\cite{rosciszewski_simulation_2014}
we discussed the existing simulation systems and provided motivations for a new discrete-event
simulation environment introduced in \cite{czarnul_merpsys:_2017}. An example of
execution time modeling using this environment has been described in \cite{czarnul_simulation_2015},
and an example of modeling energy consumption in~\cite{czarnul_modeling_2016}. In \cite{rosciszewski_modeling_2017}
we proposed an approach, described also in Section \ref{sec:iccs}, to exploring power/time trade-off of a deep neural
network training application using this simulation environment. Additionally, simulation
has been used in the approach to optimization of execution time under power consumption constraints
proposed in \cite{hutchison_optimization_2014} and described in Section \ref{sec:icdcn} to determine
optimal data partitioning. In \cite{rosciszewski_executing_2016}, a method of configuring process labels
in the system model has been proposed that simplifies executing simulations with fine-grained granularity.

The problem solved within this thesis, defined in Section \ref{sec:problemformulation}, is a multi-objective
optimization problem. In Section \ref{sec:multiobjective} we discuss approaches to multi-objective optimization
of parallel applications and choose the Pareto method for the approach proposed in this thesis.
In \cite{hutchison_optimization_2014} we focused on the problem of execution time optimization under
power consumption constraints defined in Equation \ref{eqn:problemlimit}. Pareto method has been used
for exploring the power/time trade-off of a parallel deep neural network training application
described in \cite{rosciszewski_modeling_2017} and in Section \ref{sec:iccs}.

The \emph{task mapping} aspect of the optimization problem solved within this thesis,
connected with finding the optimal \emph{process mapping} function \gls{mapping}
lies in the field of resource management. The approach proposed in this thesis
is compared to chosen approaches to energy-aware resource management
in \emph{heterogeneous \gls{hpc} systems} in Section \ref{sec:scheduling}. The task mapping
problem has been considered in the context of optimization of execution time under power
consumption constraints in \cite{hutchison_optimization_2014}. Resource management
has been also considered in \cite{rosciszewski_network-aware_2014} in the context of
network-aware scheduling and in \cite{balicki_runtime_2014} in the context of computing
system monitoring. Finding the optimal number of used computing devices in
\cite{rosciszewski_modeling_2017} is also connected with resource management.

The aspect of the considered optimization problem connected with finding the optimal
set of \emph{application execution variables} \gls{executionparameters} lies in 
the field of \emph{parameter auto-tuning}. Chosen approaches to parameter auto-tuning
in parallel applications are described
in Section \ref{sec:autotuning}. Auto-tuning of execution variables such as
thread numbers, \gls{gpu} grid configurations and data partitioning is an important part of the
optimization methodology proposed in \cite{rosciszewski_kernelhive:_2016} and the generalized
version in Chapter \ref{chp:solution}. Data partitioning has been also tuned
in \cite{hutchison_optimization_2014} and \cite{czarnul_simulation_2015}. 
In \cite{rosciszewski_modeling_2017} the tuned variable is connected with resource
mapping, namely the number of devices used for computations.

The problem solved in this thesis combines the following problems from
the fields of multi-objective optimization, resource management and parameter auto-tuning:

\begin{itemize}
\item multi-objective optimization of execution time and power consumption problem;
\item suboptimal-approximate energy-aware global static task mapping problem;
\item offline auto-tuning of system parameters problem
\end{itemize}

\noindent into one bi-objective Pareto optimization problem of execution time and power
consumption.

\section{Main Contributions of the Dissertation}

This work contains the following original contributions made by the author:

\begin{itemize}
\item proposition of a new model of \emph{hybrid parallel application} execution in
\emph{heterogeneous \gls{hpc} systems} described in Section \ref{sec:problemformulation}
that focuses on execution time and power consumption of \emph{processes} consisting
of \emph{computation and communication operations} and considers \emph{process mapping}
and \emph{application execution parameters};
\item implementations connected with specific \emph{hybrid parallel applications}, namely:
heterogeneous OpenCL implementation of the regular expression matching application described
in Section \ref{sec:regex};
multi-level heterogeneous OpenCL + MPI implementation of the application described in
Section \ref{sec:idw} and integration of the application with the KernelHive framework
(this air pollution interpolation application was used in the SmartCity system developed
for the local government of the city of Gdańsk, Poland);
implementation of the model of the large vector similarity measure computation application
described in Section \ref{sec:big_data_sim}
in the MERPSYS environment; extension of the existing deep neural network training
application described in Section \ref{sec:kaldi_training}
by MPI message passing for multi-level execution and implementation of the model of this
application in the MERPSYS environment 
(the application was used for acoustic model
development by VoiceLab.ai, a company based in Gdańsk, Poland);
\item proposition of specific execution steps for \emph{hybrid parallel applications} in
\emph{heterogeneous \gls{hpc} systems} described in Section \ref{sec:methodology}
consisting of preliminary process optimization, process mapping, parameter tuning and actual execution,
and implementation of these steps in the KernelHive framework;
\item co-design of the simulation method of \emph{hybrid parallel application} execution
in \emph{heterogeneous \gls{hpc} systems} described in Section \ref{sec:simulation} and
proposition of using this method for fast evaluation of \emph{process mappings} and \emph{application
execution parameter} values in the multi-objective execution time and power consumption optimization.
Co-implementation of this method within the MERPSYS simulation environment. Specifically, implementation
of the scheduler mechanism, framework for executing multiple parallel simulations, power consumption
computation and multiple improvements of the simulator; 
\item demonstration of the proposed execution steps on the example of multi-level task
farming applications described in Section \ref{sec:task_farming_case} including 
computation and communication overlapping, network-aware and power constrained scheduling,
tuning of grid configurations and data partitioning and execution in heterogeneous \gls{hpc} systems
with \gls{cpu}s and \gls{gpu}s using the KernelHive framework;
\item demonstration of the proposed optimization methodology as a whole including the execution steps and simulation method 
on the example of a deep neural network training application described in Section \ref{sec:training_case}
including overlapping of training and data preprocessing, power-aware device selection and 
execution on a professional cluster of workstations with \gls{gpu}s using the contributed
multi-level implementation; 
\end{itemize}

\section{Claims of the Dissertation}\label{sec:claims}
\begin{enumerate}
\item \textbf{The execution steps specific in the context of the proposed model, including preliminary process optimization,
process mapping, parameter tuning and actual execution allow
to optimize execution time of hybrid parallel applications in
heterogeneous high performance computing systems.} Empirical proofs 
for this claim are provided in Chapter \ref{chp:experiments}. Specifically,
proofs related to preliminary process optimization are presented in Sections
\ref{sec:prefetching} and \ref{sec:hpcs}, process mapping in Sections
\ref{sec:icdcn} and \ref{sec:iccs}, parameter tuning in Sections
\ref{sec:khtuning} and \ref{sec:iccs}, and actual execution
in Sections \ref{sec:khexecution} and \ref{sec:kaldimpi}.
\label{clm:1}

\item \textbf{The proposed modeling and simulation method allows
for fast and accurate identification of the set of Pareto-optimal solutions to
the problem of multi-objective execution time and power consumption optimization
of hybrid parallel applications in heterogeneous high performance computing systems.}
Empirical proofs for this claim are provided in Section \ref{sec:iccs}.
\label{clm:2}
\end{enumerate}

\section{Overview of the Dissertation}

The remainder of this thesis is organized as follows.
A state of the art review is provided in Chapters
\ref{chp:applicationstheory} and \ref{chp:optimizationtheory}.
Chapter \ref{chp:applicationstheory} discusses \emph{hybrid parallel applications},
examples of applications from various fields, software solutions for executing
them in heterogeneous \gls{hpc} systems and approaches to modeling and simulation of their execution.
Chapter \ref{chp:optimizationtheory} covers approaches to optimization of parallel
application execution in the contexts of multi-objective execution time and power consumption
optimization, energy-aware resource management and parameter auto-tuning.

In Chapter \ref{chp:applicationsandsystems} specific applications and systems are described,
that have been used in the experiments within the thesis. The main contribution of this thesis is the optimization
methodology proposed in Chapter \ref{chp:solution}. An empirical evaluation of the methodology,
described in Chapter \ref{chp:experiments}, is based on case studies involving optimization
of specific executions of \emph{hybrid parallel applications} in \emph{heterogeneous \gls{hpc} systems}.
Finally, conclusions and discussion about the possible future work are provided in Chapter \ref{chp:summary}.

\leavevmode\thispagestyle{empty}\newpage
\leavevmode\thispagestyle{empty}\newpage
\chapter{Execution, Modeling and Simulation of Hybrid Parallel Applications}\label{chp:applicationstheory}

In Section \ref{sec:hybridapplications} we provide examples of \emph{hybrid parallel applications}
and describe the systems, APIs and frameworks used by these applications
as \emph{heterogeneous \gls{hpc} systems} in Section \ref{sec:executing}.
An important part of the optimization methodology proposed in this thesis is a method for modeling 
applications and systems and simulating application execution. In Section \ref{sec:modeling}
we provide an overview of existing approaches to modeling parallel applications and systems, as
well as simulating the execution. 

\section{Hybrid Parallel Applications}\label{sec:hybridapplications}

We characterize the common denominator shared by parallel applications which
could benefit from the methodology proposed in this thesis by naming
them \emph{hybrid parallel applications}.
In Section \ref{sec:hybrid} we explain our interpretation of the term \emph{hybrid} 
in the context of parallel applications by comparing it to existing works.
In Section \ref{sec:appexamples} we provide examples of specific applications
in the fields similar to the applications considered in our experiments.
Possible \emph{application parameters} are emphasized.

\subsection{Hybridity of the Applications}\label{sec:hybrid}

The term \emph{hybrid} in the context of parallel applications appears in two main
meanings. First, mixing different types of computing devices
\cite{wyrzykowski_parallelization_2014, liang_enabling_2012, czarnul_benchmarking_2016, lee_hybrid_2015},
for example \gls{cpu} + \gls{gpu}.
Second, mixing programming APIs on different parallelism levels 
\cite{li_hybrid_2010, danner_hybrid_2012, johnson_billion-scale_2017},
for example \gls{mpi} + OpenMP.
In both cases the aim is utilizing more computing devices in order to achieve better performance
or performance/power consumption ratio. 
The first meaning implies the heterogeneity of the utilized \gls{hpc} system, and the second meaning 
implies that the system is multi-level. It is worth noting that, in essence, both
these meanings are related to the properties of the \gls{hpc} system, namely that it is heterogeneous
and multi-level. Examples of such systems are discussed in Section \ref{sec:hetero}. 


Having said that, the term \emph{hybrid} may seem redundant in the context of parallel applications themselves.
However, we decided to emphasize this word due to another meaning. A crucial factor of the model of parallel application
execution proposed in this thesis is the set of parameters which influence its execution.
In a heterogeneous multi-level system these parameters may
include various aspects of execution, often related to certain types of computing devices or parallelization
levels. The term \emph{hybrid} is gaining importance in the cases where changing one of these application
parameters influences the optimal value of another.

For example, the term \emph{hybrid} has been used in the context of a \emph{hybrid} Xeon/Xeon Phi system in \cite{czarnul_benchmarking_2016}.
Within the paper, a parallel application for computing similarity measures between large vectors
is optimized for scalability in a system consisting of Intel Xeon \gls{cpu}s along with
Intel Xeon Phi coprocessors. Proposed optimizations include load balancing, loop tiling,
overlapping and thread affinity setting. The system is \emph{hybrid} in the sense of utilizing
different types of computing devices, which makes it a heterogeneous computing system.
What is more significant, executing computations on the Intel Xeon Phi influences
the optimal number of cores used on the host processor for computations: one core
should be left free, so it can be efficiently used for handling the accelerator.

Situations when the optimal values of execution parameters depend on each other often occur
in practical approaches.
In the above example, changing one \emph{execution parameter} (whether or not to use the coprocessor)
influences the optimal value of another \emph{execution parameter} (number of used cores of the host processor used for the computations).
Such relationships concern also application-specific parameters. 
For instance, when the authors of \cite{tabatabaee_parallel_2005} fixed one of
three \emph{application parameters} of a GS2 physics application and plotted the performance as
a function of two other \emph{application parameters}, it turned out that the optimization surface
was not smooth and contained multiple local minima.

The mutually dependent parameters include also meta-parameters resulting from
code optimizations.
One example would be an approach to optimizing parallel application execution
using a combination of compiler and runtime techniques proposed in \cite{jordan_multi-objective_2012}.
In this approach, regions of the tuned applications are subject to source-to-source transformations. 
Parameters of these transformations belong to the set of tuned variables. The authors observe that
optimal transformation parameter values or even distinct transformation sequences depend on another
\emph{execution parameter}, namely the number of threads. This observation is a motivation for multi-versioning
approach proposed in the paper, where a set of optimal solutions is encoded by the compiler into one executable and the
runtime system dynamically chooses between the versions depending on changing circumstances and
objectives.

Another example is simultaneous loop tiling and unrolling program transformations for minimizing
execution time of selected numeric benchmarks in \cite{kisuki_combined_2000}. The authors claim that
combining the best tiling transformation with the best unrolling factor does not necessarily give 
the best overall transformation. What is more, it is shown that small deviations from optimal tile
sizes and unroll factors can cause such an increase in execution time, so that it is even higher
than in the original program. Previously used static cost models which attempt to give an analytical
expression for the execution time were vulnerable to such varying optimization spaces. Instead, an
approach called \emph{iterative compilation} is proposed, where many versions of the program are
generated and executed for determining optimal values of the optimization parameters.

Although the two latter approaches concern programs that are neither multi-level nor executed
on a heterogeneous platform, they illustrate non-trivial dependencies between \emph{application execution parameters},
which cause a need for specific optimization algorithms. In contemporary multi-level \gls{hpc} applications
executed in heterogeneous \gls{hpc} systems, dependencies of this type are increasingly likely, and it is
getting harder and harder to describe them using analytical models.

Summarizing, in the sense of this work, a parallel application is \emph{hybrid} if
it is multi-level or executed in a heterogeneous \gls{hpc} system, but more importantly when
there are non-trivial dependencies between the \emph{application execution parameters}
and the optimization objectives, such as execution time and power consumption.
This deeper meaning of the term \emph{hybrid} is essential in the model and optimization approach proposed
in this thesis, because it means that instead of finding optimal parameter values separately,
in many cases there is a need to take into account the whole space of decision parameters.

\subsection{Examples of Hybrid Parallel Applications in Selected Fields}\label{sec:appexamples}

In order to provide examples of
what the discussed class of applications can be useful for, in this section we describe
chosen applications from fields similar those developed and used in the experiments within this thesis,
described in Chapter \ref{sec:applications}.

The sample application described in Section \ref{sec:md5} is distributed
MD5 hash breaking using a cluster with \gls{cpu}s and \gls{gpu}s. A similar approach to 
distributed password cracking has been proposed in \cite{kim_distributed_2012},
where a cluster of up to 88 \gls{gpu}s has been used for password recovery through
brute-force attack, achieving good scalability thanks to the proposed efficient
password distribution scheme. The solution is hybrid in the multi-level sense, but
the computing system is homogeneous (\gls{cpu}s are not used for computations). In the field
of cryptography, parallel applications that are hybrid in the sense of device heterogeneity
are also used. For example in \cite{niewiadomska-szynkiewicz_hybrid_2012}, a cluster system
integrating \gls{cpu} and \gls{gpu} devices from various vendors is used for efficient encryption and decryption
of large amounts of data. \emph{Application parameters} in password cracking
may be related to the method of dividing workload across the used workers, namely
the number of passwords to be checked by each worker in an iteration and expected
minimum/maximum password length which influence the average time of recovering
the password.

The regular expression matching application described in Section \ref{sec:regex}
belongs to the field of text pattern matching. Pattern matching algorithms
are widely used in signature-based network intrusion detection systems (NIDS)
\cite{mukherjee_network_1994}.
The objective of such systems is to examine if incoming network packet payloads
contain malicious content defined as "signatures" or "patterns" and generate
alert messages for system administrators. Examples of \emph{application parameters} 
in this problem are maximum length of a signature and size of slices into
which the traffic is partitioned.
Hybrid \gls{cpu}/\gls{gpu} pattern matching
for deep packet inspection has been proposed in \cite{lee_hybrid_2015},
where the incoming packets are pre-filtered using \gls{cpu} and suspicious packets
are sent to the \gls{gpu} for complete matching. The \gls{gpu} workload is reduced thanks
to the \gls{cpu} pre-filtering stage.

Geostatistical interpolation applications are a fundamental task in geographic
information science and are used for prediction of environmental phenomena at
non-observed locations. Computational cost of the used algorithms grows with
the number of data points from the observed locations and the number of locations
for which the interpolated values are needed. The contemporary interpolation workloads
are critical, for example in weather forecast systems, and efficient implementations of geostatistical
interpolation algorithms are needed. The same algorithm that the one described in Section
\ref{sec:idw} has been adapted to massively parallel computing environments in
\cite{hennebohl_spatial_2011} and \cite{mei_evaluating_2014}. An efficient \gls{gpu}
implementation of another popular geostatistical interpolation method called 
kriging has been proposed in \cite{srinivasan_efficient_2010}.
An exemplary \emph{application parameter} in the geostatistical interpolation problem
could be related to the used strategy of data point partitioning.
For example, quad trees have been used in the approach to another geoscientific problem:
constructing digital elevation models from high resolution point clouds acquired using
LIDAR technology \cite{danner_hybrid_2012}.
A hybrid \gls{mpi}/\gls{gpu} implementation for solving this problem has been proposed.
The application is hybrid in the multi-level sense, with multiple \gls{gpu}-equipped
hosts independently interpolating a portion of data and assembling the final
model from partial results with balancing of I/O, computation and communication in mind.

The application described in Section \ref{sec:big_data_sim} concerns large-vector similarity measure
computation. This task is a crucial part of clustering which means grouping a set of objects into 
classes of similar objects. Algorithms using pattern similarity have been successfully applied to 
large data sets in DNA microarray analysis, e-commerce applications, such as collaborative filtering
\cite{wang_clustering_2002} as well as real-time searching for similar short messages for the purposes
of a social networking service with a dataset of over billion messages \cite{sundaram_streaming_2013}.
\emph{Faiss} \cite{johnson_billion-scale_2017} is a recent open-source library for efficient similarity search
and clustering of dense vectors. A key problem addressed by this approach is to, given a large database
of objects, construct a k-NN graph - a directed graph whose nodes represent objects from the
database and edges connect them to k nearest neighbors, according to one of the supported distance functions.
The solution is capable of using multiple \gls{gpu}s on one server for constructing a high accuracy k-NN graph.
For example, construction of a graph connecting 1 billion vectors in less
than 12 hours on 4 NVIDIA Maxwell Titan X \gls{gpu}s has been reported. The presented application of the
library is constructing a k-NN graph for a database of 95 million images and finding a path in this 
graph, resulting in a sequence of smooth transitions from a given first to a given last image.
\emph{Application parameters} of the large-vector similarity measure computation problem include
the maximum dimensionality of an object and size of packages in which the objects are transferred
to the computing devices.

Finally, there are multiple applications of parallel deep neural network training, the field
of the application described in Section \ref{sec:kaldi_training}. Many of them are hybrid in
the multi-level sense, because often multiple computing devices are used for training.
For example, a neural network for classifying positions in the \emph{Go} game according to archival expert moves
was trained using asynchronous stochastic gradient descent on 50 \gls{gpu}s in \cite{akram_dvfs_2016}.
The training took around three weeks, because 340 million training steps were needed to contribute
to the achievement of winning 99.8\% games against other \emph{Go} programs and defeating
the human European \emph{Go} champion by 5 games to 0. There have also been hybrid \gls{cpu}/\gls{gpu}
approaches to deep neural network training. A version of the popular deep learning framework Caffe
proposed in \cite{hadjis_caffe_2015} allows using both \gls{cpu}s and \gls{gpu}s for training a deep neural network,
which on a single convolutional layer achieves 20\% higher throughput than only on a \gls{gpu}. 
A hybrid \gls{cpu}/\gls{gpu} implementation \cite{babaeizadeh_reinforcement_2016} has been also proposed for A3C,
a parallel method for reinforcement learning, which can for example learn to successfully play an
Atari game only from raw screen inputs. The proposed hybrid implementation generates and consumes
training data for large deep neural networks up to 45 times faster than its \gls{cpu} counterpart.
An \emph{application parameter} in the field of parallel deep neural network training 
can for example be the frequency of model synchronization between many training workers.

\newpage
\section{Executing Parallel Applications in Heterogeneous HPC Systems}\label{sec:executing}

Depending on the hardware utilized by a parallel application, various software solutions are
used for its execution. In this section we describe selected tools, frameworks and APIs used
for executing parallel applications. We put particular emphasis on related \emph{execution parameters},
which may belong to a set of decision variables in the optimization problem solved in this thesis.
Section \ref{sec:shared} is devoted to systems, where multiple threads
can run in parallel and communicate and synchronize through \emph{shared memory}.
\emph{Distributed memory systems}, where each process has its own private memory and 
some form of interconnection is needed for communication 
are described in Section \ref{sec:distributed}.
Finally, Section \ref{sec:hetero} focuses on systems that allow executing applications that
are hybrid in two meanings described in Section \ref{sec:hybrid} -- with multiple levels
of parallelization and with heterogeneous computing devices.

\subsection{Shared memory systems}\label{sec:shared}

From the viewpoint of a parallel application programmer, a program running in a \emph{shared memory system}
typically consists
of one or more threads - sequences of programmed instructions, which are executed concurrently.
POSIX threads \cite{butenhof_programming_1997} is a popular parallel execution model that defines
C-language functions for thread management and synchronization, which implementations are available for many
operating systems. Analogous mechanisms are available for many popular programming languages, for example
\emph{threading} library in Python, Java \emph{threads}~\cite{oaks_java_2004} etc. 
If the utilized runtime supports it, some of the concurrent threads may be executed in parallel, resulting
in reduction of application execution time.

Parallel applications that utilize the threading execution model are very often executed on
\gls{cpu}s with multiple cores (multi-core). An evident \emph{execution parameter} of such applications is the number of used threads.
The optimal number
of threads should allow to efficiently utilize available \gls{cpu} cores, which does not
necessarily mean that the optimal number is equal to the number of \gls{cpu} cores.
The capacity of utilizing many cores in parallel depends on the algorithms
used by the application. Additionally, 
modern \gls{cpu}s support hardware multithreading techniques such as simultaneous multithreading (SMT), which allow   
multiple threads to be executed simultaneously on one core. Although the threads are completely separated
from each other, running them on one core influences the computation performance.
Apart from the proper number of used threads, in such non-uniform computing architectures,
efficient utilization of the multi-core computing devices requires taking into
account how the application threads are mapped to the available cores. This can be achieved
by configuring thread affinity, which allows binding and unbinding certain threads to certain
\gls{cpu} cores and is another example of a parallel application \emph{execution parameter}.

Modern high performance computing accelerators are equipped with dozens of physical computing
cores. For example, the \emph{Knights Landing} architecture used by the second generation of Intel MIC
accelerators is built from up to 72 cores with possibility to run four threads per core.
Shared-memory multiprocessing APIs such as OpenMP are often used to develop
parallel programs for such multi-core architectures. Multithreading can be achieved
using OpenMP by extending sequential C/C++ and Fortran codes with compiler directives
that take care of thread creation, data sharing, synchronization, scheduling etc.
For example, a C/OpenMP implementation of a parallel large vector similarity measure computation
application was used in \cite{czarnul_benchmarking_2016} to test various thread affinities, 
but also allocating memory in large pages for improved data transfer rate.
The latter is an example of an \emph{execution parameter} specific for the used device - in this case the
Intel Xeon Phi coprocessor.

Arguably, the most popular computing accelerators recently are \gls{gpu}s (graphics processing units).
They consist of several streaming multiprocessors, which in turn consist of multiple processing
elements known as CUDA cores. The name is connected with the CUDA parallel computing platform and API
created by NVIDIA. In this execution model, the code written in a form of a \emph{kernel}
is executed by multiple threads at once. 
For example, the NVIDIA Tesla V100 data center \gls{gpu} is equipped with 5120 CUDA cores and
allows to run 163840 threads simultaneously.
The threads are logically aligned in a hierarchy called
\emph{grid} that consists of a number of \emph{blocks} constructed from a number of threads.
Numbers of blocks in a grid and threads in a block can be arranged by the programmer in up to three dimensions.
This setting is called \emph{grid configuration} and its optimal values may depend on the
\gls{gpu} device model, but also on the application, its computation to communication ratio,
code branches etc. Finding the optimal values often requires tuning through testing the application
performance for multiple combinations of the grid size parameters. The \gls{gpu} grid configuration is
another example of a parallel application \emph{execution parameter}.

A similar hierarchical application structure
can be found in OpenCL framework \cite{stone_opencl:_2010},
which allows to write programs that can be executed across
heterogeneous platforms, including \gls{gpu}s. Here, the equivalent of \emph{grid} is called
\emph{NDRange}, consists of \emph{work groups}, which in turn consist of \emph{work items}.
The actual mapping of this structure to the computing device architecture depends on
the chosen \emph{installable client driver} (ICD). Multiple OpenCL implementations offered
by different vendors support various computing devices. Different \emph{NDRange} configurations
can be optimal depending on the OpenCL implementation and the utilized computing device,
which makes the selection of appropriate \emph{execution parameters} even harder if 
the code is executed in a heterogeneous system.

Heterogeneous systems can be also programmed using OpenACC, a programming standard aimed
for simplification of programming of \gls{cpu}/\gls{gpu} systems. Similarly to OpenMP, C/C++ and Fortran
codes can be annotated using compiler directives responsible for parallelization, with a particular
emphasis on parallelizing loops. The API defines also runtime functions responsible for
device and memory management. An important aspect of programming with OpenACC is optimizing
data locality by providing the compiler with additional information about the data location.
This allows to reuse data on the same device and minimize data transfer, which can be particularly beneficial on 
systems where used devices have separate memories. Selecting the appropriate strategies of
data creation, copying and address management is another example of a parallel application \emph{execution parameter}.

An \emph{execution parameter} crucial in terms of power efficiency is setting the "gear" of a computing device
using the Dynamic Voltage and Frequency Scaling (\gls{dvfs}) technique. Often the most efficient setting in terms
of compute capacity is not the most power efficient, because modern computing devices often have asymmetric power 
characteristics. For example Volta, the latest NVIDIA \gls{gpu} architecture
at the time of this writing, is claimed by its vendor to achieve up to 80\% of the peak performance at half
the power consumption.  Performance vs energy trade-offs can be found on both modern \gls{cpu}s and \gls{gpu}s \cite{calore_evaluation_2017}.
The effects of scaling core voltage and frequency depend not only on the computing device architecture,
but also on the application characteristics \cite{mei_survey_2017}, which makes finding the optimal setting
non-trivial.

Summarizing, the optimal values of \emph{execution parameters} of parallel applications executed in
shared memory systems are hard to find, because
they often depend on each other, on algorithms used in the application, its input/output characteristics
as well as utilized hardware. Finding the appropriate value often comes down to empirical verification
and tuning.

\subsection{Distributed memory systems}\label{sec:distributed}

Computing systems in which each computing device has its own private memory are 
called distributed memory systems. Programming applications for such systems
requires taking care of not only process synchronization, but also data transmission
between the processes. Message Passing Interface (MPI) \cite{forum_mpi:_1994} is the de facto standard
communication protocol for point-to-point and collective communication, with several
well-tested and efficient implementations. The MPI interface provides the essential
virtual topology, synchronization and communication functionality between a set of processes.
In order to achieve the best possible performance of an MPI-based application, certain runtime
parameters of the used MPI implementation should be optimized for the target platform,
including the cross-over point for point-to-point operations between the eager and the rendezvous protocol,
network specific values such as internal buffer sizes and algorithms to be used for collective operations.
OpenMPI, a popular MPI implementation has been equipped with the Open Tool for Parameter Optimization (OTPO)
\cite{chaarawi_tool_2008} which systematically tests large numbers of combinations of OpenMPI runtime
tunable parameters to determine the best set for the given platform, based on selected benchmarks.

Message passing interfaces such as \gls{mpi} allow to implement applications employing many well known
parallel processing paradigms \cite{buyya_high_1999}. The \emph{task-farming} paradigm, also known as \emph{master/slave}
consists of two kinds of processes. The first one, \emph{master}, is responsible for decomposing the problem
into small tasks and iteratively distributing them across a farm of processes of the second kind, called \emph{slaves}.
The \emph{slave} processes perform a simple cycle: get a message with the task, process the task, send the
results back to the \emph{master}. Paradigm-related \emph{application parameters} in this case include the number of
\emph{slaves} used at each iteration, task allocation and load balancing strategies as well as granularity of problem
decomposition which may influence message sizes.

In the \emph{Single-Program Multiple-Data} (\gls{spmd}) paradigm, data required by the application is split among 
available processors and each process executes the same code on a different part of the data.
This paradigm is especially useful for problems with a geometric structure, where communication between
the processes is spatially limited, for example physical problems and simulations. Efficient execution of 
\gls{spmd} programs in multi-core environments often requires managing communication heterogenities that cause
unbalanced workload. For example, the methodology evaluated on a \gls{mpi} \gls{spmd} application in \cite{muresano_methodology_2010}
allowed for 43\% efficiency improvement through scheduling policies that allow to determine the number 
of tasks to be assigned to each core, which allows computation and edge communication overlapping and optimal
load balancing. The number of tasks per each core is another example of an \emph{application parameter}.

Another, more fine-grained paradigm which also reflects data dependencies of the application
is \emph{data pipelining}. Functional decomposition of the application allows to organize a
pipeline of processes, each corresponding to a given stage of the algorithm. The efficiency of
this paradigm depends on the capability of balancing the load across the stages of the pipeline.
Efficient implementation of such pipelines becomes even more challenging in the case of real-time
or near-real-time systems with small messages, due to the trade-off between the system throughput and latency.
Processing the messages directly as they occur results in very frequent, small communication operations
which significantly limit the throughput. Batching techniques can be used to improve the throughput
at the cost of the latency. For example, incurring a small latency to group small messages together
allows improving throughput in Kafka \cite{goodhope_building_2012}, a real-time publish-subscribe system
that can handle more than 10 billion message writes each day. The incurred latency value can be tuned
at the application level and is another example of an \emph{application parameter} which optimal value may
depend on the characteristics of the target system.

\emph{Divide and Conquer} (\gls{dac}) is a well known approach in algorithm development which can be also used
as a parallel processing paradigm. In this approach a problem is decomposed into subproblems which can
be solved independently and their results are combined to give the final result. The problems
can be decomposed recursively, which results in an execution tree that defines a hierarchy of \emph{split},
\emph{compute} and \emph{join} computational operations. Given appropriate complexity proportions between 
the splitting and joining operations compared to the computing operations, performance of the \gls{dac} approach
can benefit from parallel execution.
For example, a multi-threaded framework based on OpenMP proposed in \cite{czarnul_parallelization_2015}
allowed to obtain speed-ups around 90 for an irregular adaptive integration code executed on an Intel Xeon Phi
accelerator. The framework allows developing parallel \gls{dac} applications by coding only basic \gls{dac} constructs
such as data partitioning, computations and result integration, while parallelization is handled
by the contributed runtime layer. The degree of parallelism can be controlled by setting two \emph{application parameters}:
k\_{max} which specifies the maximum depth of the execution tree and max\_thread\_count which is
the total number of threads that can be run at any given time.

Frameworks for automatic parallelization of computations are very much needed, for instance 
in the field of big data processing. 
For example \emph{MapReduce} \cite{dean_mapreduce:_2008} is a programming model similar to \emph{Divide and Conquer}
where the programmer specifies two functions: \emph{map} and \emph{reduce}. Given a single key/value
pair, the \emph{map} function generates a set of intermediate key/value pairs. Intermediate values
with the same key are merged by the \emph{reduce} function. This simple model allows to express many
real world tasks, especially in the field of big data processing. Due to the availability of tools
for automatic parallelization of programs written in the \emph{MapReduce} paradigm such as Hadoop \cite{shvachko_hadoop_2010},
programmers can relatively easily utilize the resources of a large distributed system without any
experience in parallel computing. Although input data partitioning, scheduling, failure and communication management
are handled by the runtime system, certain optional \emph{execution parameters} can be tuned, including maximum number
of used machines and memory limits for each \emph{map} and \emph{reduce} task.

One limitation of the \emph{\gls{dac}}-based computing paradigms is that they support only acyclic
data flow, while in many practical applications a working set of data needs to be reused across
multiple parallel operations. Spark \cite{zaharia_spark:_2010} is an example of a popular
framework that supports these applications by introducing data abstraction called resilient
distributed datasets - read-only collections of objects partitioned across a set of machines.
An operation invoked on such a dataset results in creating a \emph{task} to process each partition
of the dataset and sending these \emph{tasks} to preferred worker nodes using a technique called
\emph{delay scheduling} \cite{zaharia_delay_2010}: when the job that should be scheduled next
according to fairness cannot launch a local task, it waits for a small amount of time, letting other jobs
launch tasks instead. This allows for significant performance improvements for
chosen applications, especially those with iterative and interactive computations. Certain \emph{execution
parameters} can be tuned in Spark, including job scheduling policies, data serialization, memory management
policies, data locality and level of parallelism.

\subsection{Multi-level and Heterogeneous Systems}\label{sec:hetero}

In paper \cite{rabenseifner_hybrid_2003} \emph{hybrid parallel programming}
has been narrowed down to combining distributed memory parallelization
on a node inter-connect with shared memory parallelization inside of each node.
The presented experiments included results from executions of applications
written using message passing (namely \gls{mpi}) on the distributed memory level
and directive-based parallelization (OpenMP) on the shared memory level.
Various schemes of \gls{mpi} + OpenMP programming models have been discussed
and compared based on benchmarks executed on several computing platforms.
The paper underlines the importance of the \emph{communication and computation overlapping}
optimization technique. In our work this method falls into the preliminary process optimization
step, one of the execution steps proposed in Section \ref{sec:methodology}
and was used in the experiments described in Section \ref{sec:icdcn}. 

Authors of \cite{hoefler_mpi+_2013} argue that although using \gls{mpi} for internode communication
and another shared-memory programming model for managing intranode parallelism has become the
dominant approach for utilizing multi-level
systems, the significant downside of this approach is complexity of using two APIs in the same application.
They propose a shared-memory programming model which is incorporated into \gls{mpi} as an extension to its
communication interface. The implementation allows to automatically adjust the shared-memory mapping,
which for several provided use-cases resulted in improved performance.

Utilization of large-scale multi-level computing architectures using a single API
is possible in the partitioned global address space (\gls{pgas}) \cite{de_wael_partitioned_2015} parallel programming model.
In this model, a number of parallel processes jointly execute an algorithm
by communicating with each other via a single shared memory address space. The memory
is conceptually shared (technically realized by several interconnected memories),
which aims to improve programmer productivity. At the same time, the \gls{pgas}
model provides additional abstractions to distinguish between local
and remote data accesses, in order to allow implementations aiming for high performance.
About a dozen languages exists that adhere to the \gls{pgas} model \cite{de_wael_partitioned_2015}.
For example, Java programs can use the PCJ library \cite{nowicki_pcj-java_2014}
that has the ability to work on multi-node multi-core systems and hiding the details
of inter- and intra-node communication. Experiments involving running a set of
\gls{hpc} applications show good performance scalability of the PCJ library
compared to native implementations of the same algorithms in C++ with \gls{mpi}
and Java 8 with parallel streams mechanism, while maintaining the usability
typical for the \gls{pgas} model implementations.

Many solutions for executing applications utilizing a single parallelization API on multi-level computing
systems are also aimed at utilizing heterogeneous devices
\cite{barak_package_2010, aoki_hybrid_2011, kegel_dopencl:_2012, ozaydin_opencl_2012, diop_distcl:_2013, grasso_libwater:_2013}.
For example Many \gls{gpu}s Package (MGP)\cite{barak_package_2010}
provides C++ and OpenMP APIs that allow to extend existing applications executed on one hosting-node, so that
they can transparently utilize cluster-wide devices. What is more, the package includes also an
implementation of OpenCL specifications that allows executing OpenCL code on a cluster with many \gls{cpu} and/or \gls{gpu}
devices without any modifications of the code. This reduces the complexity of programming and running parallel
applications on cluster, especially since MGP provides an API for scatter-gather computations and takes
care of dependencies between the split sub-tasks, queuing and scheduling them as well as managing buffers.

Frameworks for multi-level parallel computing on heterogeneous systems are also developed 
with specific type of application in mind. For example, TensorFlow \cite{abadi_tensorflow:_2016}
is an interface for expressing machine learning algorithms and an implementation for their
execution. TensorFlow applications are constructed from \emph{operations} which constitute
a dataflow graph, where each node has zero or more inputs and zero or more outputs. Many implementations
can be provided for each \emph{operation} that can be run on a particular type of device. This
way, various device types can be used by one application in order to efficiently utilize
a heterogeneous system. Various kinds of parallelism can be expressed through replication
and parallelization of the graph, including model parallelism and data parallelism.
Additionally, an \emph{operation} implementation can be parallel
itself, using for example the parallel capabilities of a \gls{gpu}, which means that TensorFlow
applications can be multi-level.

There are solutions for executing parallel applications in hybrid multi-level systems that take
into account energy efficiency. For example the authors of \cite{duato_rcuda:_2010} argue
that high-end \gls{gpu}-based accelerators feature a considerable energy consumption, so a solution
is needed that would enable each node in a cluster to run efficient computations on a \gls{gpu} while
avoiding attaching a \gls{gpu} to every node of the cluster. The proposed rCUDA framework for remote
\gls{gpu} acceleration allows this, introducing only a small overhead for chosen sample
applications from CUDA SDK.

Energy efficiency has been considered also in the context of hybrid \gls{cpu}/\gls{gpu} architectures in
\cite{ma_greengpu:_2012} where GreenGPU, a framework for \gls{gpu}/\gls{cpu} heterogeneous architectures
was proposed. The solution provides load balancing between \gls{cpu} and \gls{gpu} by dynamic splitting and
distributing workloads based on its characteristics, so that idle energy consumption is minimized.
Additionally, \gls{gpu} core and memory frequency as well as \gls{cpu} frequency and voltage are dynamically
throttled based on their utilization. The holistic approach that includes workload division and
frequency scaling achieves an average energy saving of 21.04\% and only 1.7\% longer execution
for the \emph{kmeans} and \emph{hotspot} workloads from the Rodinia benchmark \cite{che_rodinia:_2009}.

\newpage
\section{Modeling and Simulation of Parallel Applications in Heterogeneous HPC Systems}\label{sec:modeling}

Formulating the optimization problems in Chapter \ref{chp:introduction}
and describing the proposed optimization methodology in Chapter \ref{chp:solution}
required defining a formal model of hybrid parallel application execution in a heterogeneous \gls{hpc} system.
The model is also necessary for performing simulation
which is proposed as a method for fast evaluation of process mappings and application execution parameters 
in Section \ref{sec:simulation}.

The proposed model assumes that the a priori knowledge about the behavior of
processes in the system is limited and precise analytical models of execution time and power
consumption of the application are hard to formulate. Instead, a general execution model of hybrid
parallel applications in heterogeneous \gls{hpc} systems is proposed, that allows to express many
practical applications.
In this section we describe chosen existing approaches to modeling and simulation
of parallel application execution in heterogeneous \gls{hpc} systems. In Section \ref{sec:modelingsystems}
we focus on the system models. We divide the considerations of application modeling
to execution time models described in Section \ref{sec:modelingapps} and power
consumption models described in Section \ref{sec:modelingpower}. Finally, in Section \ref{sec:executionsimulation}
we describe chosen simulation approaches.

\subsection{Modeling Heterogeneous HPC Systems}\label{sec:modelingsystems}

A heterogeneous \gls{hpc} system is basically a set of computing devices of various types connected by a network.
In numerous approaches to optimization of parallel application execution, the system is modeled
just as a set of computing devices with certain characteristics. 
For example in~\cite{sajid_energy-efficient_2016}, a~heterogeneous computing system is
modeled as a set of heterogeneous processors with certain characteristics, including voltage and frequency.
The heterogeneity of the system is accounted for by including these hardware characteristics
in the considered objective functions such as performance and power consumption.
The formulas used to compute the objective functions
are usually defined within the system model, aimed for a certain optimization problem setup.
Similarly, the model considered in this thesis assumes that \textbf{execution time functions
are defined for all computation and communication operations} and \textbf{idle and stress power consumption functions
and core numbers are defined for all computing devices}. The model allows for different specifications of these functions
and hardware characteristics which makes expressing various specific problem formulations within the proposed framework.

Treating a \gls{hpc} system simply as a set of computing devices does not take into account
the network which connects the devices and its
topology, however it is sufficient for many optimization problems.
In the cases when network topology is considered, it is usually modeled as a graph.
For example in \cite{wu_hierarchical_2015}, the topology of the computing platform
is denoted as \emph{host graph}, where each vertex denotes a processor or a switch
and each edge represents a direct link or cost of communication.
Although we do not put particular emphasis on network topology in our work,
\textbf{we decided to adopt a graph-based system model}, in order to emphasize
the importance of network in \gls{hpc} systems and express network-aware optimizations
described in Section~\ref{sec:icdcn}.

\subsection{Modeling Execution Time of Parallel Applications}\label{sec:modelingapps}

According to the nomenclature introduced in \cite{norman_models_1993}, there are three basic
approaches to modeling an application understood as a set of modules that make up the computation.
In the first one, the modules execute independently and there is no communication between them.
This is not the case in our work, because \textbf{we assume possible communication between the processes}.

The second approach is the \emph{task-based} model, where the modules are called tasks and are
arranged in a \emph{directed acyclic graph} (\gls{dag}). The model is commonly used by the researchers interested in scheduling problems, 
for example \cite{vivekanandarajah_task_2008, baskiyar_energy_2010, kessaci_parallel_2011, chowdhury_efficient_2015,
sun_re-stream:_2015, wu_hierarchical_2015}.
The application is modeled as an acyclic
graph with nodes representing tasks and edges representing the precedence relationship
between them. Weights are associated with nodes, representing task execution time, as well as with
edges, denoting communication time between the connected tasks. 
There are various forms of this model.
For example, in the \emph{macro-dataflow} model \cite{beaumont_static_2002}, 
there is a limited number of computing resources to execute the tasks. Given
that task $T$ is a predecessor of task $T^\prime$, if both tasks are assigned
to the same processor, no communication overhead is paid. On the contrary,
if $T$ and $T^\prime$ are assigned to different processors, a communication
delay is paid, which depends upon both tasks and both processors.
Another example is the \emph{data stream graph} (DSG) model~\cite{sun_re-stream:_2015},
which emphasizes data streams, modeled as directed edges in the task precedence
graph. Each edge is characterized by a tuple containing an identifier and
communication cost of this edge.
In our work we do not employ the \emph{task-based} model, because \textbf{we assume no a priori knowledge about the precedence relationships
between the \emph{operations}}. 

The third approach is the \emph{process-based} model, where processes are arranged in 
undirected graphs where an edge represents volume of communication between the processes.
Despite the coincidence of names, the \emph{processes} considered in this work
are different. In this work, the equivalent of such \emph{processes} are \emph{operations}. The
sequences of these \emph{operations} are not known a priori and depend, in turn, on the \emph{processes}
defined in this work. It should be
noted all these three approaches to modeling parallel applications 
can be expressed using the model proposed in this thesis by proper implementation of the computation and communication sequences.

Computation and communication operations have been distinguished in the model used 
in~\cite{zhou_task_2014}. The considered problem is mapping a set of independent applications
onto a set of heterogeneous processors including \gls{cpu}s and \gls{gpu}s. Each application consists of several computational kernels,
which are chained together according to some computation logic. However, unlike in our approach,
these dependencies between kernels are known and modeled as edges in the computation graph.
The execution time of each kernel on each processor is estimated a priori using the
sampling functionality of the StarPU \cite{augonnet_starpu:_2011} task scheduling platform.

One apparent way to create a model of an existing parallel application in the
framework proposed in this thesis is to analyze its source code and indicate which
fragments can be simplified to \emph{computation and communication operations}.
Then, simulation can be performed in order to establish the exact sequence of
operations resulting from the code logic and communication between the processes.
The researcher responsible for the modeling process has to decide how detailed
the model should be, bearing in mind that the more detailed the model, the more computationally
costly it is to perform simulation of the application. It should be noted that distinguishing
the appropriate code regions could be done automatically. For example in the PATUS framework~\cite{christen_patus:_2011}
in order to distinguish parts of the
code (stencil computations) to be optimized through code generation, a \emph{strategy}
mechanism is used to separate them from the parts related to parallelization and
bandwidth optimization.

Apart from the dependencies between certain parts of an application, the application model should
define execution time of each part. The general model proposed in this thesis in Section \ref{sec:problemformulation}
delegates this task to the \emph{computation time} function which can adopt different forms depending
on the specific model. For example, the \emph{computation time} functions for models of applications
described in Section \ref{sec:big_data_sim} and Section \ref{sec:kaldi_training} are defined
as functions of the number of floating point operations of the \emph{computation operation}
(which in turn depends on the input data size) and the floating point operation performance of the used processor.

Many optimization approaches are based
on an assumption that a matrix is given with estimated times to compute (\gls{etc}) each 
task on each processor (see also Section \ref{sec:schedulingapriori}).
Some application modeling approaches propose how to construct this
matrix based on the task and processor characteristics. For example 
in \cite{sajid_energy-efficient_2016} the application is modeled as BoT (Batch of Tasks or Bag of Tasks \cite{li_energy-efficient_2014}), 
meaning a set of independent tasks that belong to different applications and can be run
simultaneously. The \gls{etc} matrix is derived from the number of cycles
of a task and frequency of the processor for each task-processor pair. A similar
derivation can be used within the approach proposed in this thesis, through proper
implementation of the \gls{comptime} modeling function (see Section \ref{sec:simulation}).
On the other hand, the approach proposed in this thesis is different from the BoT approach,
because there are communication dependencies between the operations (which
are the equivalent of tasks). 


\subsection{Modeling Power Consumption of Parallel Applications}\label{sec:modelingpower}

According to \cite{li_energy-efficient_2014}, among memory, disks, networks, fans, cooling system and other components
of a heterogeneous computing system, significant portion of energy is consumed by its processors. 
Like in the paper, in this thesis \textbf{only energy consumption of the computing devices
available in the system is considered}. 

Many scheduling optimization approaches focus on \gls{dvfs}-enabled \gls{hpc} systems
\cite{li_energy-efficient_2014, oxley_makespan_2015, sun_re-stream:_2015, zhang_solving_2015, awan_energy-aware_2016},
where processor cores can operate in discrete performance states. This capability
could be included in the system model. For example, a performance predictor
for managed multithreaded applications proposed in \cite{akram_dvfs_2016} can accurately predict
the impact of scaling voltage and frequency of a processor. The authors of \cite{de_langen_trade-offs_2007}
claim that even dynamic processor shutdown should be considered as an option for
reducing power consumption of embedded multiprocessors, due to the trend of significantly increasing
static power consumption. 

Analogously to the \gls{etc} matrix used for execution time modeling, some optimization solutions
assume that an \emph{average power consumption} (\gls{apc}) matrix is given that defines 
average power consumption of a certain type of task on a certain type of machine. The authors
of \cite{tarplee_energy_2016} indicate that these matrices are usually obtained from historical
data in real environments. Some approaches are based on assumptions strictly connected with
the nature of their considerations, like for example in the topology-aware task mapping
approach \cite{wu_hierarchical_2015}, where energy required for data transmission between
processors depends directly on the total traffic load of the interconnect network.
In \cite{sajid_energy-efficient_2016} energy consumption of executing a certain task on a certain processor
depends on two components: dissipation power and dynamic power. The first one is a static
property of the processor. The second one depends on another static property, namely
physical capacitance, as well as two values decided at runtime: supply voltage
and frequency of operation. This approach resembles the one adopted in this thesis,
described in Section \ref{sec:simulation}, 
where idle power consumption is the static component and the dynamic component is
derived from current runtime parameters.

An important decision that must be made in modeling energy consumption of parallel
applications is if the idle energy consumption of the computing devices should be included in
the total energy consumption. In some approaches the idle energy consumption
is not taken into account. For example in the approach to energy-efficient scheduling
of batch-of-tasks applications on heterogeneous computing systems proposed in \cite{sajid_energy-efficient_2016}
the overall energy consumption is computed as a sum of all task executions, assuming that due to static scheduling
the idle slots are negligible.
In the experiments presented in \cite{tarplee_energy_2016} all the systems have zero idle power
consumption, but setting power consumption to 10\% of the mean power for each machine type
is used to model the case when the server is powered off but a management controller is 
still responsive to remote power on signals.

Some approaches to multi-objective time, power and energy optimization
of \gls{hpc} applications focus on finding a set of Pareto-optimal solutions,
in order to give the system administrator or programmer the chance to choose
the most appropriate solution according to their needs.
The authors of \cite{jarvis_multi_2014}
propose to take into account a measurement error margin, which enlarges
the set of optimal solutions so that no potentially important
solution is overlooked.

Analogously to execution time of \emph{computation and communication operations},
the simulation method proposed in this thesis in Section \ref{sec:simulation} delegates the task
of power consumption modeling to the \emph{idle and stress power consumption} functions
which can adopt different forms depending on the specific model.
Although these aspects were not considered in the experiments
within this thesis, the proposed model allows to include the \gls{dvfs} states and processor shutdown 
into the \emph{energy consumption} function and passing the chosen state from the
\emph{application model} as a parameter. The same applies to using an \gls{apc} matrix,
focusing only on the network traffic or considering physical parameters of the processors.

\textbf{The model used in this thesis considers power consumption in idle state of the used
hardware as well as additional power consumption under stress.}
In this approach the idle power consumption can be configured to a certain value or a percentage
of the stress power consumption. In particular, experiments that assume no idle power consumption
can be configured by setting this value to zero.
 
\subsection{Simulation of Parallel Application Execution}\label{sec:executionsimulation}

In computer simulation, the role of a simulator is to compute consecutive states
of a model which is a simplified representation of a certain real object. Models used
for simulation of parallel application execution in heterogeneous \gls{hpc} systems have
been discussed in Sections \ref{sec:modelingsystems}, \ref{sec:modelingapps} and \ref{sec:modelingpower}.
In this section we provide examples in what way such models can be used by a simulator.

A crucial aspect of computer simulation is to choose succeeding points in time for which
the simulator should calculate consecutive states of the modeled system. Majority of the approaches
to simulation of parallel application execution are based on discrete-event simulation
\cite{czarnul_merpsys:_2017},
which means that the consecutive states are computed for consecutive events that occur
in a discrete sequence in time. An event might represent the end of processing of a certain
task by a computing device or transmission of a certain message between two devices. The sequence
of events depends on the used application model. The approach used for simulation in this
thesis is also based on \textbf{discrete-event simulation}.

SimJava \cite{kreutzer_simjavaframework_1997},
a layered discrete-event simulation framework based on the Java programming language
is a foundation for many simulation environments aimed for specific aspects of computing systems.
For example GridSim \cite{buyya_gridsim:_2002} is a toolkit for modeling and simulation of distributed resource
management and scheduling focusing on components of grid systems. It allows for modeling
and simulation of heterogeneous grid resources, users and application models and provides
primitives for creating application tasks and mapping tasks to resources.
Another simulator based on SimJava is OMNeT++ \cite{wehrle_modeling_2010}, a C++ simulation
library and framework focusing on computer
networks and other distributed systems. It is mainly used for building network simulators,
but the MARS \cite{denzel_framework_2010} framework based on it contains also modules modeling
computing components. In that case, the events in simulation depend on replaying traces of \gls{mpi} calls. 

The approaches to simulation based on traces of previous application executions are
called \emph{trace-driven} and can combine different types of traces with
different application models. 
For example SST/macro \cite{hsieh_sst+_2012} uses
an application model based on \gls{mpi} along with \gls{mpi} trace files in two formats:
Open Trace Format and DUMPI. A trace-driven approach has been also proposed 
for assessing the performance of hierarchical dataflow applications in RADA \cite{mazumdar_analysing_2017}.
The simulation environment used for simulations in 
this thesis is \textbf{not trace-driven but tuning of the model to the results of real application executions}
is suggested. There are approaches to simulated evaluation of parallel application execution where
application model is tuned to some existing combinations of execution parameters.
For example, it is the case in \cite{sarood_maximizing_2014}, where curve fitting is used to obtain
model parameters for the purpose of evaluating different scheduling policies.

The authors of \cite{mazumdar_analysing_2017} in further work \cite{scionti_efficient_2017} point out that the performance
of simulations is often limited, because synchronization of a large number of processes/threads
is required. To overcome the limited scalability, they proposed to use virtualization based on \emph{Linux Containers}.
In the environment used for simulations in this thesis, \textbf{the number of
simulation threads is unlikely to become prohibitively large} for one simulation, because instances of processes
with the same \emph{process implementation} running on the same \emph{computing device} are aggregated and
handled by one simulation thread. The simulator is also \textbf{designed for performing multiple concurrent simulations}
by utilizing threads and distributed processes connected to a queue of simulation instances, which
allows for high scalability.

Machine learning along with traces of previous application executions can be used for evaluating application execution time instead of an application model.
For example, a formulation-based predictor is used for evaluating Hadoop MapReduce application execution time
in order to find approximately optimal solutions by tuning the application parameters in \cite{chen_machine_2015}.
The predictor fits to logs of previous Hadoop MapReduce jobs and uses a 2-level machine learning ensemble
model consisting of random forest, extra trees and gradient boosting regressors to predict the median,
standard deviation and wave of the application execution time. 

Established toolkits such as SimGrid \cite{casanova_simgrid:_2001} for studying the operations of large-scale
distributed computing systems, provide good basis for implementation and simulation of a wide range of algorithms.
However, the authors of \cite{bak_gssim_2011} point out that in most cases the experiments have to be developed
from scratch, using just the basic functionality of the toolkit, and the experiments are rarely 
useful for other researchers. To address this issue, the authors propose the Grid Scheduling Simulator (GSSIM),
a comprehensive simulation tool for distributed computing problems. GSSIM includes an advanced web portal for
remote management and execution of simulation experiments which allows to share workloads, algorithms and 
results by many researchers. The MERPSYS environment co-developed by the author of this thesis and described
in Section \ref{sec:simulation} also allows sharing application and system models as well as simulation results
by the users, thanks to provided simulation database and Web application.

\leavevmode\thispagestyle{empty}\newpage
\leavevmode\thispagestyle{empty}\newpage
\chapter{Optimization of Parallel Applications Considering Execution Time and Power Consumption}\label{chp:optimizationtheory}

The most obvious goal of developing parallel programs can be
found in the sole \gls{hpc} (High Performance Computing) keyword. Parallel
solutions to existing, as well as new problems are always developed
with performance in mind, no matter if measured as execution time
of an application or throughput of a real-time system. For this
reason, optimization of parallel applications is often connected
with only this one objective: performance.
However, experience in running \gls{hpc} applications shows that 
in many real life scenarios, other, often contradicting
objectives need to be taken into account.

For example, running an application in parallel can significantly influence
its reliability. In~\cite{dongarra_bi-objective_2007} an 
exponential probability of failure is assumed for each
processor in a heterogeneous system. Task scheduling algorithms
are proposed, which optimize both makespan (performance) and
reliability, letting the user choose a trade-off between 
reliability maximization and makespan minimization. A
scheme for scheduling independent tasks on uniform processors
proposed in \cite{jeannot_bi-objective_2008} allows generating
an approximate Pareto set of the bi-objective makespan 
an reliability optimization problem.  

Another optimization objective that is increasingly important 
is power consumption. In this Chapter we focus on existing
work in the field of multi-objective optimization of parallel
applications with respect to execution time and power consumption.
In Section \ref{sec:multiobjective} we describe the problem of
multi-objective optimization and provide examples of existing
approaches in \gls{hpc}.

One of the execution steps proposed in this thesis is task mapping, and more specifically
proper assignment of individual parts of the computations of an application to the available computing
devices. In Section \ref{sec:scheduling}	we situate our approach in the
area of resource management with special emphasis on energy-awareness and heterogeneous computing systems.
Another proposed execution step is parameter tuning. We describe chosen
solutions in the field of parameter auto-tuning in parallel applications in Section \ref{sec:autotuning}.

\section{Multi-objective Optimization of Parallel Applications}\label{sec:multiobjective}

The outline of this section is as follows:
In Section \ref{sec:pareto} we provide formulation of the multi-objective
optimization problem, discuss the possible approaches to taking many objectives
into account in the optimization process and choose the Pareto method
for consideration in this work.
In Section \ref{sec:powerconstraints} we describe chosen
existing solutions to the problem of optimization of execution time under power consumption constraints,
which is investigated in Section \ref{sec:icdcn} of this thesis.
In Section \ref{sec:tradeoff} we discuss examples of existing work
which show that this trade-off exists for execution time and power consumption
in various hardware configurations and executed applications.
Finally, in Section \ref{sec:paretooptim} we provide examples of existing solutions considering
Pareto optimization in the field of parallel computing.

\subsection{Multi-objective Optimization Problem Formulation}\label{sec:pareto}

In general, the goal of solving an optimization problem is to find a point in a given decision space,
for which the value of a given objective function is minimal. 

\begin{definition}
Let \gls{paretofunction} denote an \underline{objective function} of an optimization problem.
\end{definition}

For example, in our work the objectives are application execution time and 
power consumption. 
Usually there are certain constraints which have to be fulfilled in order for the 
decision point to be a feasible solution to the problem.
In this work the constraints may be related to numbers of available computing devices,
possible sizes of data chunks, numbers of processes required by the application etc.
The constraints are often expressed as mathematical functions, however a more general
statement would be that the decision point should belong to a feasible set, which results from the constraints.

\begin{definition}
Let \gls{feasibleset} denote the \underline{feasible set} of an optimization problem.
\end{definition}

Having defined the objective function $\gls{paretofunction}(\gls{optimparameters})$ and the feasible set
$\gls{feasibleset}$, a mono-objective optimization problem can be stated as follows:

\begin{equation*}
\begin{aligned}
& \underset{\gls{optimparameters}}{\min}
& & \gls{paretofunction}(\gls{optimparameters})\\
& \text{subject to}
& & \gls{optimparameters} \in \gls{feasibleset}
\end{aligned}
\end{equation*}

The convention in optimization problems assumes minimization of a function,
however the definition can be easily used for maximizing (for instance real-time
application throughput), by using a function
$\gls{paretofunction}^{\prime}(\gls{optimparameters}) = - \gls{paretofunction}(\gls{optimparameters})$
as an objective.
Because there is only one objective, the definition of optimum is straightforward:

\begin{definition}\label{def:optimum}
A point $\gls{optimum} \in \gls{feasibleset}$ is an \underline{optimum} of
an optimization problem if and only if:
\begin{equation}
\forall_{\gls{optimparameters} \in \gls{feasibleset}}
\gls{paretofunction}(\gls{optimum}) \leq \gls{paretofunction}(\gls{optimparameters}).
\end{equation}
\end{definition}

In the case of multi-objective optimization problems, defining the optimal solution is less straightforward,
because the objective consists of \gls{paretoobjectives} objective functions:

\begin{definition}
Let vector $\gls{paretofunctionvector}(\gls{optimparameters}) = [\gls{paretofunction}_1(\gls{optimparameters}), \ldots, \gls{paretofunction}_{\gls{paretoobjectives}}(\gls{optimparameters})]$ denote a \underline{vector of objective functions}, where \gls{paretoobjectives} is the \underline{number of} simultaneously optimized \underline{objectives}.
\end{definition}

\noindent Then, following the train of thought proposed in \cite{ehrgott_multicriteria_2005}, the optimization
problem can be written as:

\begin{equation*}
\begin{aligned}
& \underset{\gls{optimparameters}}{"\min"}
& & \gls{paretofunctionvector}(\gls{optimparameters}) = [\gls{paretofunction}_1(\gls{optimparameters}), \ldots, \gls{paretofunction}_{\gls{paretoobjectives}}(\gls{optimparameters})]\\
& \text{subject to}
& & \gls{optimparameters} \in \gls{feasibleset}
\end{aligned}
\end{equation*}

The quotation marks in the \emph{"min"} notation are used for a reason. In the case of multiple
objectives, different interpretations can be associated with the minimization operator.
Various interpretations have been discussed in \cite{ehrgott_multicriteria_2005}.
In the field of energy aware scheduling, a taxonomy of optimization approaches has been presented in 
\cite{kessaci_parallel_2011}. The multi-objective approaches are divided into three classes:
aggregation, lexicographic and Pareto. 

The traditional lexicographic approach assumes some preference order of the objectives.
A set of mono-objective problems is solved sequentially for the consecutive objectives
until a unique solution is found.

In the case of aggregation methods, a function is used to transform the multi-objective
optimization problem into a mono-objective one by combining the objective functions
into a single in one aggregation function.
For example, the authors of \cite{sajid_energy-efficient_2016}
propose algorithms which
optimize two objectives (minimization of makespan and energy consumption) at the same time.
The objectives are combined into one weighted aggregation cost function. A parameter
$\alpha$ denoting the weight of each of the objectives is set to the value 0.5, so
the both objectives are equally prioritized.

In this work we focus on the third approach: the Pareto method. In this case, instead of
one point, the solution to the optimization problem is a set of Pareto-optimal points:

\begin{definition}
We say that \underline{$\gls{paretofunctionvector}(\gls{optimparameters}_1) \leq \gls{paretofunctionvector}(\gls{optimparameters}_2)$} if $\forall_{i=1,\cdots,\gls{paretoobjectives}} \gls{paretofunction}_i(\gls{optimparameters}_1)
\leq \gls{paretofunction}_i(\gls{optimparameters}_2)$. Point $\gls{optimparameters}_1 \in \gls{feasibleset}$
is \underline{Pareto-optimal} if there is no such $\gls{optimparameters}_2 \in \gls{feasibleset}$ that
$\gls{paretofunctionvector}(\gls{optimparameters}_2) \leq \gls{paretofunctionvector}(\gls{optimparameters}_1)$. Let set of points
$\gls{paretoset} \in \gls{feasibleset}$ denote the \underline{Pareto set}.
\end{definition}

In other words, the solution to the optimization problem is the set \gls{paretoset} of all such points
$\gls{optimum} \in \gls{paretoset} \in \gls{feasibleset}$, that for all other points there is at least one
objective function which value is lower for the point \gls{optimum}.
This means that in the case of two objectives considered in this work, if a point
\gls{optimum} is in the Pareto set \gls{paretoset}, then for every other point in the decision space 
the point \gls{optimum} has at least lower execution time or power consumption.

Additionally, it is important to distinguish the notion of Pareto front: 

\begin{definition}
Let set of objective function values of all Pareto-optimal points 
$\gls{paretofront} = \{ \gls{paretofunctionvector}(\gls{optimparameters}) : 
\gls{optimparameters} \in \gls{paretoset} \}$ denote the \underline{Pareto front}.
\end{definition}

As indicated in \cite{jarvis_multi_2014}, the Pareto front contains significantly
richer information that one obtains from single-objective formulations. The solution provides
not only solutions globally optimal in terms of the consecutive objectives, but also fully 
illustrates the trade-off between the objectives. The authors notice that a multi-objective formulation
is needed for the auto-tuning problem. In the context of trade-offs between time, energy and
resource usage, authors of \cite{gschwandtner_multi-objective_2014}
state that the multi-criteria scenario "requires a further development of auto-tuners,
which must be able to capture these trade-offs and offer the user either the
whole Pareto set or a solution within it".

Indeed, authors of \cite{durillo_single-multi-objective_2014} state that multi-objective
approaches have not been extensively researched in the past and 
stress the advantages of generating the whole Pareto front for the problem of multi-objective
auto-tuning of programs. Firstly, the user can visually explore the Pareto set and select the solution
which fits his interest best. Secondly, some kind of aforementioned 
aggregation method can be used for automatic selection of an optimal solution. Having access to the
whole Pareto front before constructing an aggregation function allows for normalizing the function
in order to avoid drawbacks resulting from different value ranges of the objective functions.
Finally, the authors notice that computing the Pareto front does not necessarily require
additional computational effort comparing to a single-objective approach. 

Summing up, in this work \textbf{we focus on the multi-objective Pareto optimization problem} of execution
time and power consumption defined in Equation
\ref{eqn:problem} and two related \textbf{mono-objective optimization problems} defined
in Equations \ref{eqn:problemlimit} and \ref{eqn:problemtime}. 
The difference between the two latter lies in the feasible set: the first one is constrained by the
power consumption limit. 

\subsection{Optimization of Execution Time Under Power Consumption Constraints}\label{sec:powerconstraints}

A particular type of multi-objective optimization problem is
when there are strict constraints imposed on one of the objectives.
In the case of the bi-objective problem formulation considered in our work this would
mean a strict deadline for the execution or limit of power
consumption of the computing system utilized by an application. For example, multi-objective optimization
using Particle Swarm Optimization (PSO)
algorithm with energy-aware cost function and task deadlines has
been proposed in \cite{saad_energy-aware_2012} for
partitioning tasks on heterogeneous multiprocessor platforms.
In both cases of deadlines (energy or performance), although there are
two objectives taken into account, a solution to the problem can
be a single point like in the single-objective problem in Definition
\ref{def:optimum}, provided that the constraint on the other objective
is fulfilled.

Optimization of throughput of single power-constrained \gls{gpu}s has been
investigated in \cite{lee_improving_2011}. The authors notice that although
throughput is proportional to the product of the number of operating cores
and their frequency, because of limited parallelism of some applications,
it might be beneficial to scale these parameters. A technique has been proposed
for dynamic
scaling of the number of operating cores, the voltage/frequency of cores
and bandwidth of on-chip interconnects/caches and off-chip memory 
depending on application characteristics. 
Experimental results of executing 15 CUDA applications from
GPGPU-Sim \cite{bakhoda_analyzing_2009}, Rodinia \cite{che_rodinia:_2009}
and ERCBench~\cite{chang_ercbench:_2010} benchmark suites
show that the proposed technique can provide
on average 20\% higher throughput than the baseline \gls{gpu} under the same power
constraint. Although the benefits from applying the proposed technique have been
presented by extensive exploring of arbitrarily chosen combinations in the search
space, the work leaves open questions in the field of optimization - how to automatically
find the optimal values using an optimization algorithm and simulations or
real executions.

Optimization of a \gls{hpc} system throughput with power constraints has been also
considered on a larger scale of parallelization in data centers \cite{sarood_maximizing_2014}.
The problem of maximizing throughput of \gls{hpc} data centers under a strict power
budget has been formulated as an Integer Linear
Programming (\gls{ilp}) problem. The proposed online resource manager uses
overprovisioning \cite{patki_exploring_2013}, power capping through RAPL
interface \cite{rountree_beyond_2012} and moldable/malleable job scheduling
to achieve high job throughput of power-constrained data centers.
Both real experiments on a 38-node Dell PowerEdge R260 cluster and simulations
of large scale executions show improvements in job throughput compared
to the well-established power-unaware SLURM \cite{yoo_slurm:_2003} scheduling policy.

Because taking into account all variables for start and end time
of jobs would make the \gls{ilp} problem "computationally very intensive and thus impractical in
many online scheduling cases", the authors proposed a greedy objective function
that maximizes the sum of so called "power-aware speedup" for all jobs that
are ready for execution. This "power-aware speedup" is a value resulting
from the Power Aware Strong Scaling (\gls{pass}) model contributed in the paper,
which is a power-aware extension of the model for speedup of parallel programs proposed
in \cite{downey_model_1997}. Using the single objective function made the \gls{ilp}
problem computationally adequate for online scheduling.

A strict constraint on execution time has been imposed on multi-threaded
applications optimized in terms of energy consumption due to off-chip memory accesses
in \cite{chen_memory_2010}. In the proposed \gls{dvfs}-based algorithm, the
throughput-constrained energy minimization problem
has been formulated as multiple-choice knapsack problem (MCKP).
A cost function is defined for binary variables which denote
if a certain frequency level should be assigned or not to a given process.
The algorithm uses a performance model based on regression of data
points reported by hardware performance counters and a power model that focuses
on numbers of floating point instructions, branch instructions, L1 and L2 data cache
references and L2 cache misses.
Similarly to the
scheduling algorithms described in Section \ref{sec:schedulingapriori}, the proposed
optimization algorithm assumes strong a priori knowledge about the application behavior.

However, in order to relax this assumption, authors propose also P-DVFS, a predictive online
version of the algorithm which, similarly to the scheduling algorithms described in Section~\ref{sec:schedulinglimited},
does not require a priori knowledge about the application.
The proposed prediction technique relies on the similarity between present and future
\gls{mpi} distributions. Experimental results from 11 chosen benchmark
applications from SPEC2000 \cite{henning_spec_2000} and ALPBench \cite{li_alpbench_2005}
benchmark suites executed on a Pentium Dual Core processor
proved around 10\% average power reduction as compared to the most advanced related work.  

Dynamic Core and Frequency Scaling (\gls{dcfs}) \cite{imamura_optimizing_2012},
a technique for optimizing the power/performance trade-off for multi-threaded
applications is an extension to \gls{dvfs} which apart from \gls{cpu}
frequency, adjusts also core counts (core throttling). The adjustments are made dynamically
during the application execution, in order to optimize performance under
a certain power consumption constraint. During the training phase, the
application is executed for a short period of time for chosen combinations
of core numbers and \gls{cpu} frequencies. The optimal configuration is chosen
based on measured IPS (instructions per second). The proposed technique
dynamically reacts to the changes in application behavior: the IPS values are
measured during the execution phase and if the value changes by a given
threshold, the application is switched to the training phase for a certain
period.

The proposed technique has been evaluated on ten benchmark
applications from PARSEC \cite{bienia_parsec_2008} executed on AMD Opteron and Intel
Xeon processors. Although the average performance improvement of multiple benchmark applications
is 6\%, the majority of this score depends on one, poorly scalable application from the
domain of enterprise storage
for which the performance improvement is 35\%. There is barely any performance
improvement in the cases of other applications, even for two poorly scalable ones,
because they are the most memory-bounded. These results mean that the proposed
technique is suitable only for a specific kind of applications. One of the
reasons of low performance is the overhead of the training phase.
The authors consider various lengths of the training phase period, as well
as IPS change thresholds. The \gls{dcfs} method could benefit from
a computationally cheaper way to adjust the execution parameters, for
example a simplified application model and/or simulation scheme.

\subsection{Energy/Time and Power/Time Trade-offs in Parallel Applications}\label{sec:tradeoff}

A multi-objective approach to optimization would not be needed 
if the considered objectives were not contradicting in some cases. 
Existence of a trade-off between energy consumption and execution time of
an application may seem counter-intuitive. One could argue that the shorter a program
runs, the less energy it consumes. This was indeed the case for a discrete
Fourier transforms application executed on Intel Pentium M microprocessor
investigated in \cite{telgarsky_spiral:_2006}, as long as the application
was optimized only via an algorithm selection software technique. As expected, the authors state that
for a given voltage-frequency setting of the microprocessor,
the fastest algorithm was also the lowest energy algorithm. However, in
cases when voltage-frequency scaling was also considered, trade-offs
between energy consumption and execution time have been reported.
Namely, for some problem configurations it was possible decrease overall energy consumption
at the cost of increasing the execution time
by executing the program on a lower microprocessor gear.

Energy/time trade-offs have been also shown in \cite{freeh_exploring_2005}
on the example of executing programs from the Numerical Aerodynamic Simulation benchmark \cite{bailey_nas_1991}
on a power-scalable cluster using \gls{mpi}. The hardware in the cluster allowed energy saving by scaling down the \gls{cpu}s.
The decision space concerned choosing among available gears at which the processors were operating,
as well as the number of processors used for execution.
In the cases when the number of processors utilized by the application
was fixed, there were usually single or few Pareto-optimal
points in the decision space. However, when the number of used processors ranging from one
to nine was a degree of freedom in the optimization,
energy/time trade-offs were reported for several applications. The trade-offs vastly depended on the
scalability of the programs. In some cases energy and time could be saved by executing a program
on more nodes at a slower gear rather than on fewer nodes at the fastest gear. This shows
that non-trivial trade-offs between energy consumption and execution time can appear
during parallel execution.

These trade-offs may be even more complicated if the computing devices
available in high performance computing systems are heterogeneous,
the application consists of multiple tasks and execution time of a
particular task can differ depending on the assigned device. Such a model
has been studied in \cite{friese_analyzing_2012}, where the Variation (COV)
method \cite{ali_representing_2000} has been used to model a heterogeneous set
of machine types and task types. Different resource allocations resulted in
different makespan and energy consumption values, giving a wide set of
Pareto-optimal options to choose from. Examples of Pareto fronts for a system modeled in such a way have been also
presented in \cite{tarplee_energy_2016} for various numbers of computing machine types.

The authors of \cite{friese_analyzing_2012} focused on discovering the Pareto
front using the NSGA-II evolutionary algorithm, but also investigated in detail
how solutions in the Pareto front differ from one another. For this purpose,
the authors analyzed the individual finishing times and energy consumptions
for chosen points in the Pareto front. Apparently, the location of a point
in the Pareto front depends on balancing of the individual tasks both in
terms of execution time and energy consumption. In the case of the points
with extremely low execution time, the finishing times of individual jobs
were fairly balanced and relatively low, while the energy consumption values
were uneven with the highest values on a relatively high level. On the other
hand in the case of points with extremely low energy consumption, the task completion
times were uneven, while the energy consumption levels were balanced.
This observation helps to understand that in the case of highly heterogeneous systems,
the trade-off between execution time and energy consumption can be partly
explained as a trade-off between fair balancing of the completion
times and energy consumption values of individual tasks of the application.

The relationships between execution time, power consumption and energy consumption of \gls{hpc}
codes have been studied in \cite{jarvis_multi_2014}. Pareto fronts for 
energy/time and power/time trade-offs are shown for a set of chosen \gls{hpc} applications
executed on chosen parallel architectures. The results include measurements from real
executions of:

\begin{itemize}
\item finite-difference time domain \cite{balaprakash_spapt:_2011}, sparse matrix multiplication \cite{davis_direct_2006} and quick sort \cite{kaiser_torch_2010} on an Intel Xeon Phi coprocessor;
\item finite-difference time domain, bi-conjugate gradient and Jacobi computation \cite{balaprakash_spapt:_2011} on an Intel Xeon E5530 processor; 
\item fine element mini-application miniFE \cite{heroux_improving_2009} on a Vesta IBM Blue Gene/Q 10-petaflop supercomputer. 
\end{itemize}

In all presented power/time charts the Pareto front consisted of several points,
which proved that the power/time trade-off exists for the selected problem configurations.
However, energy/time trade-off was reported only for the miniFE application,
due to considering the parameter of the number of used nodes. In other experiments there
was only one point in the energy/time Pareto front. This means that in these experiments
there was no energy/time trade-off, because there was one solution that was optimal
for both objectives.

The authors notice that since power corresponds to a rate of energy,
the problem of bi-objective optimization of execution time and power consumption is clearly related to the problem
of bi-objective optimization of execution time and energy consumption. Moreover, the authors prove that all points
on the energy/time Pareto front have a corresponding point on the power/time Pareto front. 
Specifically, denoting the time, power and energy objectives by T, P and E respectively,
the authors prove that $\gls{paretoset}^E \subseteq \gls{paretoset}^P$ where
$\gls{paretoset}^P \subseteq \gls{feasibleset}$ is the set of Pareto-optimal
points for the vector of objective functions $\gls{paretofunctionvector} = [T, P]$
and 
$\gls{paretoset}^E \subseteq \gls{feasibleset}$ is the set of Pareto-optimal
points for the vector of objective functions $\gls{paretofunctionvector} = [T, E]$.

The conclusion from these findings is that exploring the power/time trade-off gives
richer information about the potentially favorable execution configurations, including
those in the energy/time Pareto set. For this reason \textbf{in this work we focus on
bi-objective optimization of execution time and power consumption}.

\subsection{Pareto Optimization of Parallel Application Execution}\label{sec:paretooptim}

One method of finding optimal values of decision variables in an optimization problem
is exhaustive search \cite{benkner_automatic_2014}, which means evaluating the values of objective functions for
all possible combinations of decision variable values. This approach may be infeasible
in many cases of optimization of execution time and power consumption in \gls{hpc}.
First, precise evaluation of the objective functions may involve actual execution
of the application, which is often extremely costly. In this work we propose using
the simulation method described in Section \ref{sec:simulation} to evaluate the objective functions
at low cost.

Secondly, the decision space of the optimization problem may be high-dimensional
and, thus, exhaustive search may require vast numbers of evaluations. Even in the case
of using a low cost model or simulator, the number of evaluations often makes the
exhaustive search method infeasible. One approach to solve this problem is to use
derivative based optimization techniques, such as gradient descent. However, this
approach requires the objective functions to be differentiable. Some models
of parallel application execution provide differentiable formulas for execution
time and power consumption, but it is rarely the case considering complexity of
the contemporary parallel applications and systems.

In other cases derivative-free
approaches are needed, such as genetic algorithms, particle swarm optimization etc.
Their assumption is that the explicit mathematical formula behind the objective
functions is unknown and evaluations are possible for certain points in the decision
space, however the number of evaluations is treated as the main computational cost of the algorithm.
Such a simulator for evaluating the objective functions, as the one proposed in
this work can be used as an evaluation function in the derivative-free algorithms.

For example, the framework for multi-objective auto-tuning proposed in 
\cite{jordan_multi-objective_2012} (described more broadly in Section \ref{sec:combinatorial})
allows to compute the Pareto set for the trade-off
between execution time and percentage of hardware utilization of parallel codes. The optimizer
uses a combination of compiler and runtime techniques and the decision parameters
include tile sizes in loop tiling, thread counts and choosing between code versions.
The authors claim that in the considered testbed, the Pareto set is prohibitively
large, making exhaustive search impossible. A simple genetic algorithm is proposed,
but still the number of required steps is too large to represent a viable option.
Finally, the differential evolution \gls{gde3} \cite{kukkonen_gde3:_2005} algorithm is used
in the optimization phase. At runtime, the system can choose an optimal configuration
from the computed Pareto set using weights provided for each optimization goal.  

Pareto fronts have also been explored to trade-off energy and performance in 
a heterogeneous \gls{hpc} system where multiple optimal solutions resulted from different
task assignments~\cite{fidanova_efficient_2015}. The \emph{estimated times to complete} of each task on each
machine were assumed to be given in an \gls{etc} matrix (generated randomly for the experiments).
The finishing times and energy
consumption were modeled by strict mathematical formulas, so evaluating one solution
was relatively cheap computationally. Still, in the testbed with 1100 tasks of 30
types and 36 machines of 9 types, exhaustive search was infeasible for the
scheduling process, which needs to be fast in order not to add a significant overhead
to the overall processing time.

For this reason, as an alternative to the exhaustive search solution, NSGA-II algorithm~\cite{deb_fast_2002}
has been used, which is a popular adaptation of the genetic algorithm
optimized to find the Pareto front for a multi-objective optimization problem. The algorithm modifies
the fitness function
of the genetic algorithm to work well for discovering the Pareto front. An important
phase of a genetic algorithm is seeding the initial population. Employing the basic seeding method 
using the optimal energy solution, suboptimal minimum makespan solution and a random
initial population, the algorithm needed hours of computations to discover a reasonable
approximation of the Pareto front in \cite{fidanova_efficient_2015}. The authors proposed
a different seeding strategy for generating configurations
with full allocations. This modification allowed the optimization algorithm to achieve
significantly closer approximations of the Pareto front in just dozens of seconds.

It should be stressed that in the cases where defining a strict mathematical formula
for the optimization objectives is possible, there is often no need to use a simulator
and search for the optimal solutions either by exhaustive search or evolutionary algorithms.
For example, the performance of the NSGA-II algorithm has been significantly outperformed
by a linear programming solution in \cite{tarplee_energy_2016} for multi-objective optimization
of energy and makespan of Bag of Tasks applications. However, using this technique
requires defining linear objective functions for the considered objectives. 
As discussed in Section \ref{sec:modeling}, in this work \textbf{we focus on cases
where exact formulas for the optimization objectives are unknown} and, thus,
we consider the evolutionary algorithms for potential utilization in the
proposed optimization methodology.

\newpage
\section{Energy-aware Resource Management in Heterogeneous HPC Systems}\label{sec:scheduling}

Mapping processes to computing devices is an important part of the optimization
methodology proposed in this thesis. 
In this section we provide background for our work in
the field of resource management. In Section \ref{sec:schedulingproblem} we introduce
the global static task mapping problem, which is an important part of the problem
formulation proposed in Chapter \ref{chp:introduction}. We discuss
chosen scheduling optimization solutions focusing on network topology in Section \ref{sec:networkscheduling}.
The remaining related work is divided into solutions with strong a priori knowledge about
the optimized application in Section \ref{sec:schedulingapriori} and with limited knowledge
in Section \ref{sec:schedulinglimited}.

\subsection{Global Static Task Mapping Problem}\label{sec:schedulingproblem}

Resource management has been a crucial topic in distributed computing for many years.
Numerous different problem formulations have been stated, also in other fields, such as control theory,
operations research and production management. Comparing the approaches has become
hard because of their vast number and essential differences resulting from particular
setups and applications. In order to achieve categories of comparable approaches,
a taxonomy of distributed scheduling approaches has been proposed in \cite{casavant_taxonomy_1988}.

According to this taxonomy, the approach proposed in this thesis is \emph{global},
because it considers where to execute a process 
and assumes that \emph{local scheduling} is the task of the operating system of the computing device.
This \emph{local scheduling} is connected with assigning processor time to processes, as well as optimizing
the utilization of device internals. These tasks are solved by the operating system and increasingly often 
by internal schedulers in the devices. This work does not explore the details of these tasks,
focusing on the problem of \emph{global scheduling}. 

Global scheduling problems are divided in the taxonomy to \emph{static} and \emph{dynamic scheduling} problems.
In the case of \emph{static scheduling}, information regarding the total mix of possible processes in the
system is available before the application execution. In this sense, the approach proposed 
in this thesis considers a \emph{static scheduling} problem, because a static schedule in the
meaning of assignment of processes to computing devices is fixed at the beginning of the execution.

Global static scheduling problems are divided in the taxonomy to optimal and suboptimal problems.
As indicated in \cite{casavant_taxonomy_1988}: "In the case that all information regarding
the state of the system as well as the resource needs of a process are known, an optimal
assignment can be made based on some criterion function. Examples of optimization measures 
are minimizing total process completion time, maximizing utilization of resources in the system,
or maximizing system throughput. In the event that these 
problems are computationally infeasible, suboptimal solution may be tried". In this sense,
the approach proposed in this thesis is suboptimal, because not all information about the
processes is known a priori. For example, there is no estimation
of execution time or resource needs of each process. We argue that for many applications
it is hard or impossible to
prepare a feasible criterion function, which could be used for preparing an optimal assignment.

Global static suboptimal scheduling problems are divided in the taxonomy to heuristic and approximate
problems. Heuristic algorithms make use of special parameters which are correlated to system performance
in an indirect way, and such alternate parameters are much simpler to monitor or calculate.
Such heuristic algorithms make the most realistic assumptions about a priori knowledge concerning
process and system loading characteristics. The assumptions can be made in approaches to optimization
of specific applications in specific systems. In this thesis we propose a general approach
which cannot include such assumptions. Hence, it is an approximate approach,
in which "instead of searching the entire solution space for an optimal solution,
we are satisfied when we find a 'good' one" \cite{casavant_taxonomy_1988}, based on certain evaluation metrics.

According to \cite{casavant_taxonomy_1988}, important factors that determine
if the suboptimal-approximate approach is worthy of pursuit include availability of a function
to evaluate a solution and the time required to evaluate it. Results from real execution of a parallel application
can be such a function. In cases when the factor of evaluation time is prohibitively large, we propose using functions
based on modeling and simulation, as described in Section \ref{sec:simulation}.

To sum up, the approach proposed in this thesis is giving a \textbf{suboptimal-approximate solution to the global static scheduling problem}, also known as \emph{task scheduling}, \emph{task mapping} or \emph{task allocation}. 
The problem of task partitioning among heterogeneous multiprocessors has been proven NP-hard
in \cite{baruah_task_2004}.

In \cite{balzuweit_local_2016}, the resource management process was divided into three stages:
\begin{itemize}
\item scheduling -- deciding when each job should run;
\item allocation -- determining which nodes take part in the computation;
\item task mapping -- matching the job to individual computational elements (nodes/cores). 
\end{itemize}

The authors assumed that the first two stages are usually done at the system level and focused
on improving the task mapping stage, which in that case meant mapping tasks to \gls{mpi} ranks.

In this nomenclature, the approach proposed in this thesis \textbf{focuses on the task mapping stage}
which includes allocation, because assigning no processes to a device means not allocating the device.
In this sense, scheduling of \emph{operations} in the proposed approach depends on the process implementations given in the application model.
According to \cite{dongarra_bi-objective_2007},
the problem of mapping each task of an application onto the available heterogeneous resources
in order to minimize the application runtime is known to be NP-hard.   

\subsection{Network-aware Scheduling}\label{sec:networkscheduling}

One of the important factors that influence execution of parallel applications in \gls{hpc} systems
is network topology.
In \cite{beaumont_static_2002}, the authors point out that the traditional macro-dataflow
model of application execution was inconvenient, because it assumed unlimited network resources, allowing
simultaneous communications on a given link. They propose a communication-aware and one port model in order
to take into account the influence of network topologies on the scheduling algorithms.
For example, the Data Intensive and Network Aware (\gls{diana}) scheduling technique
proposed in \cite{mcclatchey_data_2007} takes into account not only data and computation
power, but also network characteristics such as bandwidth, latencies, packet loss,
jitter and anomalies of network links. 

Network-aware scheduling is still an active branch of scheduling studies.
For example, the objective for scheduling
optimization in \cite{wu_hierarchical_2015} is defined as minimization of numbers of hops in shortest paths
between devices. This problem is called \emph{topology-aware task mapping} or just \emph{topology mapping}.

Resource scheduling in data centers with two-tiered and three-tiered network architectures
has been studied in \cite{shuja_data_2014}. The authors propose a topology-aware scheduling
heuristic and demonstrate its performance using the GreenCloud \cite{kliazovich_greencloud:_2012}
packet-level simulator.

In \cite{sun_re-stream:_2015} the response time of a real-time stream computing environment is optimized
by minimizing the latency of a critical path in a \gls{dag} representing the application. The proposed Re-Stream
solution has been verified in a simulation environment based on the Storm \cite{aniello_adaptive_2013} platform.

A simulator for evaluating the fitness of the intermediate solutions in an optimization algorithm
was used for example in \cite{balzuweit_local_2016}. 
The authors propose a local search algorithm which tries swapping pairs of tasks in order to minimize
the application execution time by reducing the number of network hops.
Fitness of the solutions is measured by a simulator \cite{bunde_premas:_2014} and a swap is preserved if it
decreases the number of hops. The proposed solution, aimed for applications with stencil communication
patterns, has been proven useful on the example of a shock physics model application executed
on a Cray XE6 system.

In the model proposed in this thesis, network topology can be taken into account in the system
model through proper implementation of the \gls{commtime} function (see Section \ref{sec:simulation}).
Network topology plays also an important role in experiments with real application execution regarding
network-aware optimizations described in Sections \ref{sec:prefetching} and \ref{sec:icdcn}.

\subsection{Heuristics Based on Strong a Priori Assumptions}\label{sec:schedulingapriori}

There are numerous approaches to task scheduling which assume that the (\gls{etc}) matrix
is given \cite{wall_genetic_1996, siegel_techniques_2000, kessaci_parallel_2011, zhou_task_2014, zhang_solving_2015, awan_energy-aware_2016, sajid_energy-efficient_2016}.
The matrix contains the expected execution times of each task on each processor.
Variations of this assumption are sometimes used, for example a computation cost matrix \cite{sun_heuristics_2011}.
A similar approach is often used towards power consumption, for example authors of \cite{tarplee_energy_2016}
assume that the \gls{apc} matrix is given.

In such problem frameworks, meta-heuristics are often used, including genetic algorithms \cite{wall_genetic_1996, siegel_techniques_2000,
kessaci_parallel_2011}, tabu search \cite{siegel_techniques_2000}, simulated annealing \cite{siegel_techniques_2000}, A* \cite{siegel_techniques_2000} and shuffled frog-leaping~\cite{zhang_solving_2015}.
The objective is to obtain an optimal schedule, namely the assignment of tasks to processors as well as
determining order of execution within each processor.
A thorough review of traditional and energy-aware
algorithms based on \gls{etc} matrix has been provided in \cite{sajid_energy-efficient_2016}.
The authors also propose
two new scheduling algorithms which introduce a task migration phase for minimizing the makespan and
energy consumption of the application.

The task mapping problem has been also considered in the topology-aware context in~\cite{wu_hierarchical_2015}.
The authors propose two graph-theoretic
mapping algorithms: a generic
one with inter-node and intra-node mapping and a recursive bipartitioning one for
\emph{torus} network topology, which takes into account compute node coordinates. 

Focusing on the aspect of heterogeneous \gls{cpu}s, paper \cite{garg_exploiting_2009} proposes Heterogeneity Aware
Meta-scheduling Algorithm (HAMA), claimed to reduce between 23 and 50\% of energy consumption.
The grid meta-scheduler described in the paper, collects information about the grid
infrastructure and users, and periodically passes it to the HAMA algorithm. Based on parameters
such as average cooling system efficiency, \gls{cpu} power, frequency and computation time
slots, the algorithm first selects the most energy efficient resources. What is more, if possible,
it utilizes the Dynamic Voltage Scaling capabilities of the \gls{cpu}s.

\subsection{Approaches With Limited a Priori Knowledge}\label{sec:schedulinglimited}

The model proposed in this thesis is especially useful for modeling
applications for which, at a given granularity level, it is hard or impossible to estimate
the exact graph of individual tasks and communication dependencies between them, before actual
execution of the application. 
Chosen approaches that also assume limited a priori knowledge about the application
and its processes are described in this section. For example authors of \cite{oxley_makespan_2015} notice that task execution times
in large heterogeneous computing systems may vary due to factors which are hard to incorporate into the
model, like cache misses or data dependence of the execution times. They propose a stochastic measure
for minimizing the probability of violating the makespan and energy constraints. This robust measure
is used as the objective function for various heuristic algorithms, including tabu search and genetic algorithm with
local search. 

The application model in \cite{li_energy-efficient_2014} is a Bag of Tasks, which could have different
execution time for different inputs. Because of that, the authors interpret task execution
times as random variables and consider \emph{stochastic task scheduling}. They propose algorithms with
an objective to improve the weighted probability of meeting both deadline and energy consumption budget constraints.
The proposed algorithms are performing significantly better than the traditional heuristics in an experimental
setting with \gls{dvfs}-enabled \gls{hcs}, for both randomly generated BoT applications and real-world multimedia applications.

Authors of \cite{yarkhan_experiments_2002} propose a simulated annealing approach to optimizing
task allocation in a grid environment with respect to execution time. Comparison to an ad-hoc greedy scheduler shows
that in certain cases the simulated annealing approach allows to avoid local minima in the optimization.
During the optimization process, the solutions are evaluated using a hand-crafted performance model.
The approach is verified on a simplistic testbed consisting of 15 machines, running a parallel
numerical solver application. The authors emphasize, that the usefulness of their approach depends
vastly on the accuracy of the performance model, for which the simulation method proposed in this thesis
might be a convenient replacement.

The uncertainty about the processes has been also considered in the field of cloud computing.
The authors of \cite{chowdhury_efficient_2015} notice that the existing efficient schedulers
require a priori information about the processes and ignore cluster dynamics like pipelining,
task failures and speculative execution. They propose a new scheduling algorithm for
minimizing average coflow completion time (\gls{cct}) in data centers by prioritizing the processes (coflows)
across a small number of priority queues.
The coflows are separated into the queues based on their past activity in the cluster.
The solution is proven efficient by experiments run on
100-machine EC2 clusters.

Lack of knowledge about the task processing times has been also studied in the context of game theoretic
approach to \emph{distributed scheduling} \cite{durr_non-clairvoyant_2009}. The considered problem is a
\emph{scheduling game}, where each player owns a job and chooses a machine to execute it. Even if there
exists an equilibrium in this game, the global cost (makespan) might be significantly larger than in the optimal scenario.
There exist policies that reduce the price of anarchy, but typically they have access to the announced execution
times of all tasks in the system. Policies studied in the paper are \emph{non-clairvoyant}, which means that they
assume that the task processing times are private for the players and, hence, not available a priori.  

\newpage
\section{Parameter Auto-tuning in Parallel Applications}\label{sec:autotuning}

A significant part of the parallel application optimization methodology contributed in this
thesis can be described in terms of parameter auto-tuning. In this section we provide background
for our work in this field. In Section \ref{sec:blackbox}, the problem solved within
this thesis is classified as a problem of offline auto-tuning of system parameters.
Then, chosen approaches to parallel application auto-tuning are described, divided
into those involving exhaustive search of the optimization search space in Section 
\ref{sec:exhaustive} and those involving combinatorial search in Section \ref{sec:combinatorial}.

\subsection{Offline Auto-tuning of System Parameters Problem}\label{sec:blackbox}

According to the proceedings of a recent seminar in the field of automatic application tuning
for \gls{hpc} architectures \cite{benkner_automatic_2014}, approaches to application auto-tuning
can be divided into black-box and white-box. The search process in white-box
algorithms can be guided, because there is some a priori understanding of the underlying problem.
Notable examples of white-box auto-tuning approaches are ATLAS \cite{whaley_automated_2001} for
automatic tuning of linear algebra applications and Spiral \cite{telgarsky_spiral:_2006} for automatic
generation of linear transform implementations. In the ELASTIC \cite{martinez_elastic:_2014} environment
for large scale dynamic tuning of parallel \gls{mpi} programs, the knowledge required to guide the auto-tuning process is integrated
as plugins which implement an API for modeling performance and abstraction models of the application.

Many auto-tuning solutions focus on optimizing parallel programs by choosing between 
multiple alternatives of semantically equivalent but syntactically different versions
of a program. Two ways of source code adaptation are distinguished in \cite{whaley_automated_2001}.
The first one is to supply various hand-tuned implementations and allow the optimization algorithm
to choose between them. The second method is automatically generating the code by using manual
transformations or compiler options.

For example, an auto-tuning framework introduced in \cite{kamil_auto-tuning_2010} is able
to parse Fortran 95 codes in order to extract Abstract Syntax Tree (\gls{ast}) representations
of stencil computations. The framework generates multiple versions of optimized stencil codes
by multiple transformations of the \gls{ast} code representation. The results of the transformations
depend on a number of serial and parallel optimization parameters. In order to achieve a 
feasible subset of parameter spaces, architecture-aware strategy engines are used. Then, an
auto-tuner performs exhaustive search on the limited parameter space. Additionally, the
framework allows migrating existing Fortran codes to emerging parallel APIs such as CUDA. 
Focus on stencil computation has been also put in the PATUS framework \cite{christen_patus:_2011}, which
allows generating code of stencil computation kernels from initial codes called \emph{specifications}.
The framework automatically distinguishes regions of the code which contain so-called \emph{operations}
responsible for the stencil computations and generates different versions of these regions.

In contrast to white-box approaches, in black-box approaches
there is an assumption that the only knowledge about the the optimized
application can be obtained through evaluating an instance of parameter set,
and not through analysis of its code. In this sense, the execution steps proposed
in this thesis in Section \ref{sec:methodology} consist of both white-box and black-box steps. The first
step, \emph{process optimization} is a white-box optimization step, where
an analysis of \emph{processes} in the application can be made in terms of the underlying
\emph{operations} and modifications of the \emph{operation} sequences can be made.
After this step has finished, the succeeding steps solve a black-box optimization problem, because no further
modifications of the processes are allowed.

Similarly to scheduling approaches described in Section \ref{sec:networkscheduling}, there are
also tuning approaches that stress the importance of network interconnect. Authors of \cite{laros_iii_energy_2012}
argue that optimization of application execution in next generation large-scale platforms,
especially for energy efficient performance, should not only use \gls{cpu} frequency scaling, but
could also benefit from tuning other platform components, for example network bandwidth scaling.
The paper exploits power measurement capabilities of Cray XT architecture and proposes 
a static tuning approach which demonstrates energy savings of up to 39\%. It is noted
that a dynamic approach is also an important area of investigation, though it is challenging due to reliability issues and overhead of frequency state transitions.

According to \cite{durillo_single-multi-objective_2014}, execution parameters
such as numbers of threads, their affinity, processing frequency and work-group/grid sizes of \gls{gpu} applications
are equally important tuning parameters, called \emph{system parameters}. Even the mapping of threads
onto physical cores (discussed in more detail in Section \ref{sec:scheduling})
can be considered as a part of the parallel application auto-tuning process.
Two classes of auto-tuning approaches are distinguished in the paper: offline and online. In the offline
version the program is tuned before running it in production mode. The online version has
challenging aspects, because while being executed during the application execution
it implies performance overhead and makes the execution more exposed to possible poor performing
parameter configurations. The optimization approach proposed in this thesis is based on evaluating
multiple execution configurations before the actual execution, thus in the sense of this classification
it \textbf{focuses on the problem of offline auto-tuning of \emph{system parameters}}, to which
we refer to as \emph{application execution parameters} and \emph{process mappings}. Using simulation
for re-evaluating certain \emph{application execution parameters} during the actual application
execution can potentially be beneficial for specific types of application and is an interesting
direction for future work.

\subsection{Auto-tuning Approaches with Exhaustive Search}\label{sec:exhaustive}

A plugin-based approach has been used in the European AutoTune project to extend the
PERISCOPE \cite{muller_periscope:_2010} performance analysis tool by a number of tuning plugins,
producing the Periscope Tuning Framework (\gls{ptf}) \cite{miceli_autotune:_2012}.
The plugins may employ expert knowledge or machine learning to perform multi-aspect application tuning
with regard to energy consumption, inter-process communication, load balancing, data locality, memory
access and single core performance. The tuning process starts with preprocessing C/C++ or Fortran source
code files using \gls{mpi} or OpenMP in order to distinguish code regions and parameters that may influence
their performance. For each code region, tuning scenarios defined by the plugins perform search
strategies in the parameter search space in order to minimize a tuning objective, defined as a function
which may take into account measurements like execution time and energy consumption.

The applicability of the \gls{ptf} framework was presented in \cite{miceli_autotune:_2012} on the examples of the following plugins:

\begin{itemize}
\item maximizing throughput of high-level pipeline patterns written in C/C++ with OpenMP pragma-annotated
while loops, executed on single-node heterogeneous manycore architectures using StarPU \cite{augonnet_starpu:_2011}
for execution on \gls{cpu}s and \gls{gpu}s. The main tuning parameters were stage replication factors and buffer sizes;
\item minimizing execution time of HMPP codelets - computational units written in C or Fortran, annotated with
directives which allow the special CAPS compiler to translate them to hardware-specific languages such as CUDA
and OpenCL. Considered tuning parameters were connected with the codelet internals such as unrolling factors,
grid sizes, loop permutations and also target-specific variables and callbacks available at runtime;
\item minimizing energy consumption of applications executed on shared memory processors with \gls{cpu} frequency
scaling. The tuned parameters were energy efficiency policies and used \gls{cpu} frequencies;
\item minimizing execution time of \gls{mpi} \gls{spmd} programs by tuning \gls{mpi} annotated code variants and
environment parameters including numbers of executed tasks, task affinity, communication buffer sizes and message
size limits;
\item reducing execution time of sequential programs by tuning the selection of compiler flags. 
\end{itemize} 

The PTF framework has been also used in \cite{liu_automatic_2014} to minimize execution time of
\gls{mpi} programs by tuning the parameters of MPI-IO communication interface. The proposed PTF plugin
aimed for automatically optimizing the values of selected MPI-IO hints and \gls{mpi} parameters, which
are normally optimized by programmers who have a deep understanding of the application behavior
on the target system. The authors state that because of high dimensions, the space of tuning
parameters still needs to be restricted using expert knowledge. Exhaustive search is used
to find the optimal parameter values. Exploring more elaborate search algorithms as well as
parallel application models is listed as future work.

The importance of application auto-tuning has been stressed in the context
of code portability in \cite{tsai_performance-portable_2016}, where an OpenCL implementation
of convolutional layers for deep neural networks is proposed. Codes in OpenCL
can be executed without changes on various hardware by compiling
them using local compilers dedicated to certain computing architectures.
However, usually due to the differences between architectures, in order to develop
a highly efficient implementation, one needs to take into account specific
coding practices and low-level details. The authors propose to implement the
kernel in a tunable way, accepting size of the input images, filters and computing
thread work-groups for each layer of the optimized neural network as inputs.
The approach achieves full portability of the kernels
without the need to develop multiple specific implementations, while
maintaining good performance.
For the auto-tuning problem, the optimization space is searched exhaustively, however automatic space pruning is
done, so that only nearly 20\% of the configurations are tested. Auto-tuning
a single layer out of the five layers of the neural network takes about one hour, which
is claimed not to be a significant overhead compared to the entire network training
time that could take weeks of repeatedly running the same set of kernels.

Often the search space of an auto-tuning problem is high dimensional and
prohibitively large to perform exhaustive search. One approach to perform auto-tuning
in such situations is to explicitly prune the search space. For example,
a search space reduction procedure has been proposed for auto-tuning of a parallel
implementation of the 2D MPDATA EULAG algorithm, executed in a hybrid \gls{cpu}-\gls{gpu} architecture
\cite{wyrzykowski_parallelization_2014}.
The algorithm consists of 16 stages linked with non-trivial data dependencies and the
implementation consists of 16 kernels. The parameters that constitute the search space
in the auto-tuning problem have been divided into two groups. The parameters in the first
group create a local search space for each kernel individually, and include work-group 
sizes and sizes of vectors for vectorization. The second group consists of specific
parameters related to the entire algorithm (this division of parameters is similar to
the one proposed in this thesis, with the local parameters resembling execution parameters
and algorithm-related parameters resembling application parameters).
The size of the global
search space defined by ranges of all applicable parameters is above 524 million combinations,
which makes testing all the configurations unacceptably expensive. The authors provide a group
of methods which allow to radically reduce the search space by applying certain domain-specific
constraints. This allows to prune the search space to over 379 thousand and 965 thousand 
combinations for ATI Radeon and NVIDIA \gls{gpu} respectively. Then, the auto-tuning mechanism evaluates
all configurations in the search space to select the best configuration corresponding to the
shortest execution time.

\subsection{Auto-tuning Approaches with Combinatorial Search}\label{sec:combinatorial}

In many cases when search space pruning is infeasible, approaches alternative to exhaustive search
are used, where only chosen combinations of the parameters are evaluated.
This section discusses chosen approaches that use such methods, that are called
\emph{combinatorial search} methods.

The \emph{Insieme Compiler and Runtime infrastructure}\footnote{http://insieme-compiler.org} has
been used in \cite{jordan_multi-objective_2012} as a test platform for tuning loop tiling in
cache-sensitive parallel programs. A combination of compiler and runtime techniques allows
tuning parameters of code regions, such as tile sizes, loop ordering and unrolling factors.
An optimizer is proposed, which generates multiple application configurations and evaluates
them by running the programs on the targeted platform in order to find optimal solutions.
However, because the solutions are evaluated by real program executions and the parameter
space is large, it is impossible to perform an exhaustive search evaluating all the parameter
combinations. To address this problem, a RS-GDE3 search algorithm based on Differential Evolution
and rough set theory is proposed.

Approximate optimal values of the application parameters are found by the 
Generalized Differential Evolution (\gls{gde3}) algorithm, which allows to decrease the search
time by evaluating each of the configurations from a population in parallel. A search space
reduction using the rough sets method proposed in \cite{santana-quintero_demors:_2010} 
is used to reduce the search space in every iteration of the search algorithm.
The solution is evaluated on a case study of a nested loop matrix multiplication
application on a target platform employing 10-core Xeon E7-4870 processors.
The proposed search algorithm finds similar solutions as an exhaustive brute-force search, but
uses from 90\% to 99\% fewer evaluations.

The Insieme infrastructure and the RS-GDE3 search algorithm have been used for
multi-objective optimization of parallel applications also with regard to energy
consumption \cite{gschwandtner_multi-objective_2014}. The solution was tested
on matrix multiplication and linear algebra applications, stencil codes and n-body
simulation executed in a shared-memory system with 8-core Intel Xeon E5-4650 Sandy
Bridge EP processor. The proposed algorithm outperformed chosen general-purpose
multi\-/objective optimization algorithms such as hierarchical and random
search and NSGA-II \cite{deb_fast_2002}. 

The 8-core Intel Xeon E5-4650 processors were also used for a comparison of the
RS-GDE3 search algorithm with mono-objective auto-tuners using local search, simulated annealing,
genetic algorithm and NSGA-II based on a n-body simulation application \cite{durillo_single-multi-objective_2014}.
Loop tile sizes, thread 
numbers and processor clock frequencies were the tunable parameters. These experiments
confirmed superiority of RS-GDE3 in the multi-objective setup.

Application-specific parameters have been tuned in \cite{tabatabaee_parallel_2005} to
optimize performance of a GS2 physics application for studying low frequency turbulence
in magnetized plasma. Unlike in our work, the considered problem is online tuning, focused
on performance variability. The tuned parameters can be changed during the program execution
and optimal parameter values can change during the runtime. The PRO (Parallel Rank Ordering)
algorithm is proposed as an alternative to the traditional Simplex algorithm, which is claimed
to have unpredictable performance in the case of tuning more than one parameter. The proposed
algorithm belongs to a class of direct search algorithms known as GSS methods and is resilient
to performance variability.

Execution time of large-scale dataset processing applications with Apache Hadoop is optimized 
by parameter tuning in \cite{chen_machine_2015}. Out of more than 130 configuration parameters
of the Hadoop MapReduce system, 20 most affecting the system performance have been chosen arbitrarily.
These parameters specify the way how the data should be processed in each phase of the MapReduce job
execution with regard to parallelism, memory capacity, job flow and data compression. The impact
of these features on the application execution time is identified using the random forest feature importance method.
Five most influential parameters are chosen for optimization in the experiments.

In the optimization process, the application configuration parameters are repeatedly modified
by an optimizer and evaluated by a predictor, a model that predicts median, standard deviation
and wave of the application execution time, as described in Section \ref{sec:executionsimulation}.
The optimizer uses the predicted values as input to machine learning ensemble regression methods.
The authors assume adequate efficiency of the predictor and solve the black-box optimization problem
using the RHC method (combination of Random Sampling and Hill-Climbing). A high dimensional space
is used as the feasible set of parameters. 
In the exploration phase,
the parameter space is examined by random sampling to find areas with high
probability of approximately optimal parameters. In the exploitation phase, the hill-climbing
algorithm is used to search these areas more deeply.

The solution is evaluated on original MapReduce jobs such as TeraSort and WordCount, text processing
and hive-aggregation, executed on an 8-node cluster with Intel i7 \gls{cpu}s. The approximately optimal
parameter settings found by the proposed algorithm enable up to 8.8-fold improvement of execution times
as compared to the default values of the parameters. The authors claim that this automatic tuning
approach is useful in real life setups because it is hard for the system administrators to set the
multiple parameters by hand.

\addtocontents{toc}{\protect\newpage}
\leavevmode\thispagestyle{empty}\newpage
\leavevmode\thispagestyle{empty}\newpage
\chapter{Investigated Applications and Systems}\label{chp:applicationsandsystems}

Apart from the literature review in Chapters \ref{chp:applicationstheory} and
\ref{chp:optimizationtheory}, the contributions of this thesis are driven
by the needs revealed by empirical
experiences from developing, executing, simulating and optimizing multiple parallel applications
in various \gls{hpc} systems. Chosen applications that directly served for the experiments and examples
provided in this thesis are described in Section \ref{sec:applications}. Chosen
\gls{hpc} systems used for execution of these applications are described in Section \ref{sec:systems}.

\section{Investigated Hybrid Parallel Applications}\label{sec:applications}

In this thesis, big emphasis is put on the practical use of the proposed
contributions. For this reason, we aimed to base our experiments on diverse applications
that are useful in real life. In this section we describe the applications developed for the sake of this
thesis, including MD5 hash breaking in Section \ref{sec:md5},
regular expression matching in Section \ref{sec:regex}, geostatistical interpolation
of air pollution values in Section \ref{sec:idw}, large vector similarity measure computation
in Section \ref{sec:big_data_sim} and training deep neural networks for automatic
speech recognition in Section \ref{sec:kaldi_training}. A description of each application
is given, including the specification in what way the application is hybrid.

\subsection{Multi-level Heterogeneous CPU/GPU MD5 Hash Breaking}\label{sec:md5}

The MD5 hash breaking application has been used as a verifying application for the
framework for automatic parallelization of computations in heterogeneous HPC systems
co-developed by the author of this thesis and described in \cite{rosciszewski_kernelhive:_2016}. The purpose
of the application is to retrieve a lost password, given its MD5
hash. The brute-force attack method is used, which means performing the encryption procedure for
all passwords from a given range and comparing their hashes to the given one. 
The application implements the task farming parallel programming paradigm, 
where the master
orders the slaves to search specific subranges of the feasible password space.

In order to ensure
fair comparison of execution time for various data partitionings, the application checks all passwords
in a given range, whether or not the appropriate password is found. Thus, depending on the
assumptions about the possible password length and character set, quite different problem sizes
can be achieved, which makes the application useful for execution experiments in systems with
various computing powers. For example, employing both \gls{cpu} and \gls{gpu} of one "des" workstation
from the department laboratory described in Section \ref{sec:lab527} it took around 30 seconds
to search all passwords up to six characters
long, while for passwords up to eight characters long the execution time was much larger - around 4.5 hours.
In the context of recovering passwords which often consist of multiple characters, reducing the application execution
time would be a crucial improvement of the application.
 
The application is hybrid in two meanings. First, using the functionality of \emph{KernelHive}, the application
can be executed on multiple clusters consisting of multiple nodes equipped with multiple computing devices,
which in turn can be parallel on a lower level. This way, the application is hybrid in the multi-level sense.
Secondly, due to the implementation in OpenCL and parallelization capabilities of \emph{KernelHive}, the 
application can be executed on heterogeneous computing devices if they provide an OpenCL runtime. This makes
the application hybrid in the sense of computing device heterogeneity.

\subsection{Heterogeneous Regular Expression Matching with Configurable Data Intensity}\label{sec:regex}

Optimal configuration for efficient execution of a parallel application strongly depends on the application
profile, namely
whether it is computationally intensive, requires frequent inter-process communication or
if it mostly depends on the input data and how efficiently it can be delivered to the computing device.
The idea behind the regular expression matching application proposed by the author of this thesis in
\cite{rosciszewski_regular_2014}
was to develop one application that has a different ratio of computational intensity to data intensity
depending on the given input data.

The goal of the application is to find all matches of a given regular expression in a given
text file. The regular expressions consist of characters which have to appear in the text in
the given order and a special character "*" which is a wildcard that matches one or more occurrences
of any character. This way, a complex signature of the sought text can be defined. As the application
searches the whole given text file regardless of line endings, the computational cost of the search
strongly depends on the assumed maximal number of characters matched by the wildcard character, as
well as the number of wildcard characters in the signature. Because of this, various application profiles
can be achieved, ranging from extremely compute intensive (\textasciitilde431s for searching a 1MiB file) to
extremely data intensive (\textasciitilde3s for searching a 512MiB file).

The application is hybrid both in the multi-level and heterogeneous sense, because it is implemented
in OpenCL and integrated with the \emph{KernelHive} framework. This allows for testing the influence of
the computational/data intensity ratio on the application execution on various computing devices. 

\subsection{Multi-level Heterogeneous IDW Interpolation for Air Pollution Analysis}\label{sec:idw}

The geostatistical interpolation application investigated within this thesis has been implemented
within the master thesis by the author of this dissertation \cite{rosciszewski_smart_2015}
as a part of a module for the SmartCity system
designed to support the local government of the city of Gdańsk, Poland.
The goal of the module is to provide the user with visualizations of particulate matter air
pollution. Preparing the visualizations requires estimating the air pollution level at non-observed
locations based on real measurements from ten regional monitoring stations, taken in hourly
schedule.

The used interpolation method, inverse distance weighting (\gls{idw}), derives the interpolated
value for each point on the requested area from the real measurement values, normalized proportionally
to the distance between the interpolated point and the measured point. To perform the interpolation
for one point in time, a basic Python implementation running on one core of the Intel Core i7-4770
processor needed around 36 minutes. Given the hourly measurement schedule, this
performance would be enough to render visualizations in real time, however a practical use case
was to render visualizations for multiple sets of historical data. Rendering the visualizations
for measurements from one year would require around seven months of computations.

For this reason, a massively parallel implementation has been proposed and tested
on a single computing device in \cite{rosciszewski_smart_2015}.
In \cite{rosciszewski_kernelhive:_2016} the author of this thesis contributed 
reducing the execution time of the application time by scaling it to a multi-level setup
using the proposed KernelHive framework.
Being implemented in OpenCL and integrated with the \emph{KernelHive} framework, the application
is hybrid both in the multi-level and heterogeneous sense. 
Additionally, a hybrid multi-level version using \gls{mpi} + OpenCL has been contributed by the author
of this thesis as a test application in \cite{rosciszewski_kernelhive:_2016}.

\subsection{Multi-level Large Vector Similarity Measure Computation}\label{sec:big_data_sim}

The application first proposed in \cite{czarnul_simulation_2015} for verification of the proposed simulation method 
is large vector similarity measure computation for big data analysis in a parallel environment.
The goal of the application is to compute a similarity matrix for a large set of points in a 
multidimensional space, assuming that the size of the processed data does not fully fit into the
memory. The implementation in C/\gls{mpi} uses the master-slave parallel programming paradigm, where the
master partitions the input data into chunks and distributes them to slaves which compute Euclidean
distances between the points in a given chunk. 

The implementation that uses \gls{mpi} allows to spawn slave processes across multiple nodes in a cluster 
where each process utilizes a single \gls{cpu} core. Thus, in the case of running multiple processes per node,
the application is hybrid in the multi-level sense, because it is parallelized across many nodes
in a cluster and many cores within each node. An important parameter of the application
that influences its execution time is the number of points in each data chunk. Finding the optimal
value of this parameter through comparing times of many real executions could be prohibitively time consuming.
For example, computing similarity measures for 2000 points in a space of 200000 dimensions
on all 128 virtual cores of 16 "des" workstations (see Section \ref{sec:lab527}) takes 
around 8 hours.

The main contribution of the author of this thesis in the paper was developing a model of the application
and conducting the described experiments with simulation of the application.
In order to estimate relatively fast the execution times of the application for different problem sizes, application
parameters and utilized hardware, a simulation model was proposed. Execution times of 
the most significant computation and communication operations have been modeled as functions of the
data chunk size. The model allows to find the optimal number of points in data chunk. What is more,
it can be used to predict execution times while using different, currently unavailable computing resources.

\subsection{Multi-level Deep Neural Network Training for Automatic Speech Recognition}\label{sec:kaldi_training}

The deep neural network training application optimized and modeled in the case study described
in Section \ref{sec:training_case} of this thesis is a part of the automatic speech recognition (\gls{asr})
system developed at the VoiceLab.ai company based in Gdańsk, Poland.
One of the crucial elements
of the recipe based on the Kaldi toolkit \cite{povey_kaldi_2011} is the acoustic model implemented
as an artificial neural network. The goal of the acoustic model is to classify each audio frame
in a given recording to one of the possible speech units called phonemes. The input of the network
for each frame is a set of its features, in this case 13 mel-frequency cepstral coefficients (MFCC).

The specific application considered in this thesis is parallel neural network training 
with natural gradient and parameter averaging \cite{povey_parallel_2014} using chosen 100 hours
from the internal VoiceLab.ai corpora consisting of over 4200 hours of Polish speech from about 5700 speakers.
The model is a recurrent neural network constructed from 4 layers of long short-term memory (\gls{lstm}) cells.
The training consists of iterations of backpropagation algorithm performed by multiple \gls{gpu}s on separate copies
of the model on different training data chunks and averaging the model weights at the end of each iteration.
One training epoch consists of such a number of iterations that all training examples are used. In a usual
training procedure, 15 training epochs are executed, which on two workstations, cuda5 and cuda6 from the
VoiceLab.ai cluster (see Section \ref{sec:voicelab_cluster}) takes over 11 hours. Developing an efficient
acoustic model requires testing many neural network architectures trained with multiple values of training parameters
and thus, running many instances of the training. Execution time reduction would be a crucial improvement of the application.

The training can utilize \gls{gpu}s from multiple computing nodes in a cluster,
which makes the application multi-level. It is also heterogeneous, because the backpropagation
algorithm is executed on \gls{gpu}s, while data preprocessing and model averaging are executed on a \gls{cpu}.
What is more, the capability of the application to efficiently utilize different \gls{gpu} models
is often a practical requirement. Clusters used by companies are regularly upgraded
with new, more powerful computing devices and proper load balancing is required to make the
most of the whole cluster without inefficiencies resulting from lower speed of the older devices.

\newpage
\section{Investigated HPC Systems}\label{sec:systems}

Similarly to the applications described in Section \ref{sec:applications},
we aimed to base our experiments on many diverse high performance
computing systems. In this section we describe chosen utilized
systems, including a collection of laboratory workstations with \gls{gpu}s in 
Section \ref{sec:lab527}, a collection of servers with computing accelerators in
Section \ref{sec:apl}, a cluster with 864 \gls{cpu} cores in Section
\ref{sec:k2}, a pilot laboratory for massively parallel systems in Section \ref{sec:miclab}
and a professional cluster of workstations with \gls{gpu}s in Section \ref{sec:voicelab_cluster}.

\subsection{Des - Department Laboratory Workstations}\label{sec:lab527}

"Des" workstations are machines available in the high performance computing and artificial intelligence laboratory 
located at the Department of Computer Architecture - home department of the author of this thesis.
Although the main purpose of the laboratory is engaging students in classes
concerning parallel algorithms, high performance computing systems and massively parallel processing,
in certain reserved time windows it can also be used for experiments involving the heterogeneous
hardware resources of the laboratory.
In particular, the 18 laboratory machines called "des01-18", equipped with an Intel i7-2600K \gls{cpu} with 8 logical cores 
and 8GB of RAM each
can be utilized as a heterogeneous computing cluster, because all nodes have also NVIDIA GeForce GTS 450
\gls{gpu}s with 192 CUDA cores installed, except for one node with NVIDIA GeForce GTX 480 with 480 CUDA cores.

\subsection{Apl - Department Computing Accelerator Servers}\label{sec:apl}

The computing resources available at the Department of Computer Architecture include also four
servers called "apl09-12" with high performance \gls{cpu}s
and computing accelerators. Apl09 and apl10 are equipped with Intel Xeon W3540 \gls{cpu} with 8 logical cores,
12 GB of RAM and \gls{gpu}s: NVIDIA Tesla C2050 and NVIDIA GeForce GTX 560 Ti, both with 448 CUDA cores. Apl09 is
also equipped with a NVIDIA Quadro FX 3800 \gls{gpu} with 192 CUDA cores. Apl11 is a server with two Intel Xeon 
E5-2680 v2 \gls{cpu}s with 12 logical cores each, 64GB of RAM and two Tesla K20m \gls{gpu}s with 2496 CUDA cores each.
Apl12 is a server with two Intel Xeon E5-2680 v2 \gls{cpu}s with 20 logical cores each and two Intel Xeon Phi 5100
accelerators with 240 logical cores each. Despite fast aging of equipment, especially the \gls{gpu}s, regular upgrades
of the apl high performance computing server infrastructure makes it a good experiment environment,
particularly in the context of efficient utilization of heterogeneous computing infrastructure by a single application.

\subsection{K2 - Department High Performance Computing Cluster}\label{sec:k2}

The Department of Computer Architecture maintains also K2, a high performance computing cluster
consisting of 3 racks of 36 nodes each. Each node is equipped with two
Intel Xeon E5345 4-core \gls{cpu}s and 8GB of RAM, giving a total of 864 \gls{cpu} cores.
The nodes are connected with an InfiniBand interconnect supported by
Mellanox Technologies MT25204 network cards.
The cluster is particularly useful for applications that can scale to hundreds of
cores, such as the \gls{idw} interpolation application described in Section \ref{sec:idw}.

\subsection{MICLAB - Pilot Laboratory for Massively Parallel Systems}\label{sec:miclab}

MICLAB is a laboratory at the Institute of Computer and Information Sciences of
the Technical University of Częstochowa, built within the project
"Pilot laboratory for massively parallel systems". The aim of the project
is to create a virtual laboratory, where the nationwide scientific community
can investigate the usage possibilities and define application directions of
contemporary massively parallel computing architectures in leading fields of science.

The computing infrastructure of the laboratory consists of 10 high performance computing
nodes. Eight of them are equipped with two Intel Xeon E5-2699 v3 \gls{cpu}s and 256 GB of RAM each.
The processors have 36 logical cores each, but the significant computing power lies also
in the Intel Xeon Phi 7120P coprocessors with 244 logical cores each. Two such coprocessors
are installed in the eight latter nodes, and also in two other nodes, each equipped with
two Intel Xeon E5-2695 v2 \gls{cpu} with 24 logical cores each and 128 GB of RAM. The two remaining
nodes are also equipped with two Intel Xeon E5-2695 v2 \gls{cpu}s and 128 GB of RAM, but in their
case the installed computing accelerators are two NVIDIA Tesla K80 \gls{gpu}s with 4992 CUDA cores each.

\subsection{Cuda567 - Professional Workstations With GPUs}\label{sec:voicelab_cluster}

The professional high performance computing infrastructure at the VoiceLab.ai company
is dedicated to deep neural network training applications, such as the one described
in Section \ref{sec:kaldi_training}. The computing power for deep learning is based mostly on \gls{gpu}s, which
are commonly used accelerators for this purpose. Cuda567 is a subset of the infrastructure,
consisting of three nodes, each with 4 NVIDIA GeForce GTX Titan X \gls{gpu}s with 3072 CUDA
cores. The first node, cuda5 is equipped with two Intel Xeon E5-2620 v3
\gls{cpu}s with 12 logical cores each and 128GB of RAM. The two other nodes, cuda6 and cuda7
are also equipped with 128GB of RAM, but stronger \gls{cpu}s: two Intel Xeon E5-4650 v2 \gls{cpu}s
with 20 logical cores each.

\chapter{Proposed Optimization Methodology}\label{chp:solution}

The main contributions of this thesis result from the analysis of applications,
systems and optimizations described in Chapters \ref{chp:applicationstheory} and
\ref{chp:optimizationtheory}, as well as multiple experiments involving
real executions of hybrid parallel applications on heterogeneous \gls{hpc} systems.
Selected applications and systems, described in Chapter \ref{chp:applicationsandsystems}
were used for the experiments described in Chapter \ref{chp:experiments} and motivations provided
in Chapter \ref{chp:introduction}. Based on the literature analysis and experiments,
a general optimization methodology for executing hybrid parallel applications in heterogeneous
\gls{hpc} systems has been developed, that focuses on execution time and power consumption of the applications.
The proposed optimization methodology consists of execution steps described 
in Section \ref{sec:methodology} and a simulation procedure proposed
in Section \ref{sec:simulation} as a method for accomplishing two of the execution steps.

\section{Execution Steps}\label{sec:methodology}

The following steps related to Claim \ref{clm:1} of this thesis are proposed to optimize the
execution of a \emph{hybrid parallel application}:

\begin{enumerate}
\item preliminary process optimization - if possible, modification
of the implementations of \emph{parallel processes} $\gls{process} \in \gls{impl}$,
in such a way that the result of the application remains valid,
but the sequence of \emph{operations} is changed in order to 
reduce the process execution time;
\item application execution optimization:
\begin{enumerate}[label=(\alph*)]
\item process mapping - finding the \emph{process mapping} function
$\gls{mapping} \in \gls{mappingset}$;
\item parameter tuning - finding the vector of \emph{application execution parameters} 
$\gls{executionparameters}~\in~\gls{feasibleexecutionparameters}$;
\end{enumerate}
\item actual execution.
\end{enumerate}

The first step has been included in the proposed execution steps in order
to stress the importance of profiling and performance analysis of the application.
In many practical situations, the most significant reduction of application execution time can
be achieved through relatively straightforward modifications of the parallel algorithm
that result for example in overlapping of certain operations or more efficient sequence of operations
in terms of memory performance (e.g. loop tiling \cite{li_automatic_2004} or avoiding false sharing \cite{bolosky_false_1993}).
An analysis should be performed to determine the most time consuming parts of the application
and identify the corresponding utilized hardware resources. If two consecutive operations
require different hardware elements (for example a computing device and a network device),
often an overlapping technique allows to perform these operations simultaneously. This step
is particularly useful in cases of \emph{hybrid parallel applications} where different
hardware elements can be specialized for certain types of tasks, for example a \gls{cpu} for handling
I/O operations and a computing accelerator for massively parallel computations.

Step 2a is connected with the global static
task mapping problem described in Section \ref{sec:scheduling}. Step 2b is connected with the
offline auto-tuning of system parameters problem described in Section \ref{sec:autotuning}.

Steps 1, 2a or 2b may be omitted. For certain applications introducing \emph{preliminary process optimization} may
be infeasible. Similarly, there might be only one feasible \emph{process mapping} 
or value of certain \emph{application execution parameters}. The appropriate choice
of performed execution steps may differ throughout applications and systems.
If at all, the \emph{preliminary process optimization} step should be performed as the first one,
because the optimal \emph{process mappings} and \emph{application execution parameters}
depend on the exact process implementations. They might also depend on each other,
hence Steps 2a and 2b could be performed repeatedly in turns or performed simultaneously.

Although the last of the proposed steps, actual execution, may seem
obvious and straightforward, performing it might require significant technical effort,
especially in multi-level and heterogeneous systems. Dozens of software frameworks and programming
interfaces are used for execution of parallel applications, depending on the target system,
application characteristics and field, used programming language etc. 
In a multi-level heterogeneous system, a software solution is required that
allows execution of the application on various types of computing devices available
in the system, as well as communication through a hierarchical network infrastructure.
Chosen software solutions for executing parallel
applications in multi-level heterogeneous \gls{hpc} systems have been described in
Section \ref{sec:hetero}. Many of them are mixing different APIs in one solution,
which requires know-how and specialized programming effort.

\begin{figure}[ht!]
\begin{center}
\includegraphics[scale=0.35]{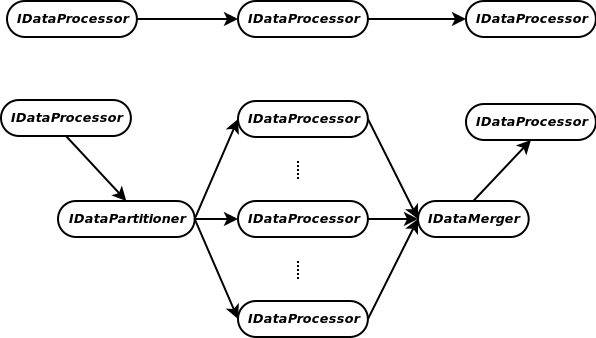}
\end{center}
\caption{Illustration of the \emph{unrollable node} mechanism in KernelHive \cite{rosciszewski_kernelhive:_2016} \label{fig:khnodeunrolling}}
\end{figure}

In order to perform all proposed steps using one, easy to 
use software environment, we propose using \emph{KernelHive}, a framework for parallelization of 
computations in multi-level heterogeneous \gls{hpc} systems first introduced in the master
thesis \cite{rosciszewski_system_2012} by the author of this dissertation, available as free
software\footnote{https://github.com/roscisz/KernelHive}. The system allows parallelization of applications among clusters and
workstations with \gls{cpu}s and \gls{gpu}s. \emph{KernelHive} applications are
developed using a graphical tool \emph{hive-gui} by constructing a dataflow graph and implementing
computational kernels assigned to the graph nodes. Custom graph nodes can be developed
by implementing the \emph{IDataProcessor} kernel interface, and a library of sample implementations
and templates is provided. 
Automatic parallelization is possible through an \emph{unrollable node} mechanism illustrated by Figure \ref{fig:khnodeunrolling},
where apart from
a data \emph{IDataProcessor} kernel, two other kernels are defined for the node: \emph{IDataPartitioner} responsible
for dividing the problem into a given number of subproblems and \emph{IDataMerger} responsible for merging
the results of multiple tasks solving these subproblems.
This way, any application that implements the \emph{IDataPartitioner}, \emph{IDataProcessor}
and \emph{IDataMerger} kernels can be automatically parallelized to a given number of computing devices.

\begin{figure}[ht!]
\begin{center}
\includegraphics[scale=0.33]{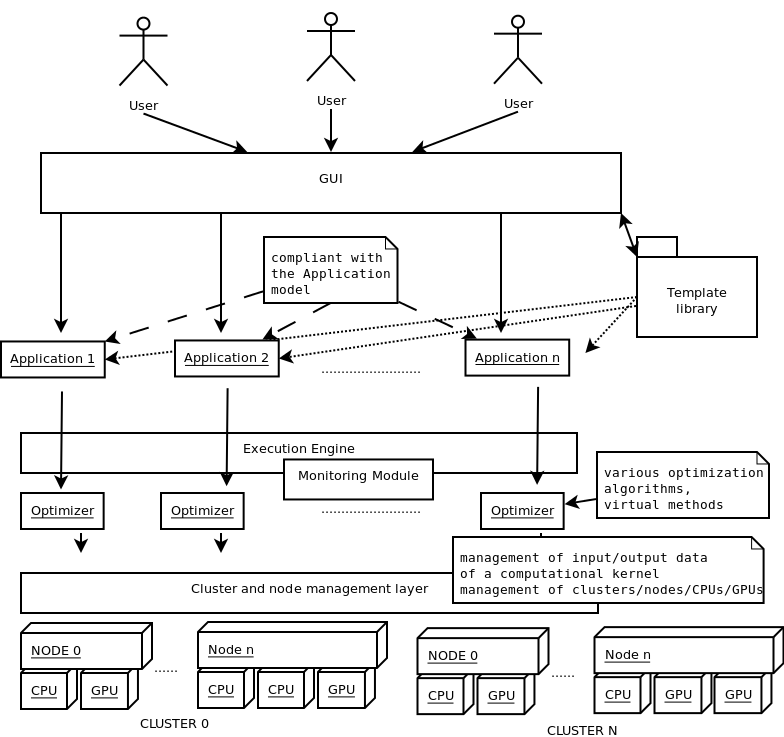}
\end{center}
\caption{Architecture of the KernelHive framework \cite{rosciszewski_kernelhive:_2016} \label{fig:kharchitecture}}
\end{figure}

Task mapping and allocation, data transfer and automatic parallelization
through the \emph{unrollable nodes}
is performed under the hood, allowing programmers
to benefit from parallel execution while focusing only on the application
rather than the complicated parallelization internals. Details about the
available infrastructure along with application progress can be monitored
in a graphical tool \cite{balicki_runtime_2014}. Overview of the architecture
of the KernelHive system is presented in Figure~\ref{fig:kharchitecture}.

The components of the framework are arranged in
a hierarchical structure corresponding to the used computing system.
An example of such a hierarchy can be seen in Figure \ref{fig:powerlimitarchitecture}.
The central component is the \emph{Engine} which interacts via a Simple Object Access Protocol (SOAP)
interface with the computing
nodes through instances of the \emph{Cluster} component, part of the Cluster and node management layer implemented
as Java system daemons installed in access nodes to particular clusters available in the system.
The \emph{Cluster} instances interact via a Transmission Control Protocol (TCP) interface 
with the underlying \emph{Unit} component instances, running as C++ system
daemons in each available computing node. Using this hierarchical architecture,
the framework is able to discover available computing devices in all connected nodes.
Data about the currently available computing devices, their hierarchy and state
is gathered in an object-oriented data structure in the central \emph{Engine} subsystem.

The proposed hierarchical framework architecture allows to take various
characteristics of the system into account during optimization of the execution and,
what is more, setting and auto-tuning various \emph{application execution parameters} concerning
different levels of the computing system. Technically, the optimization is done using a
mechanism of interchangeable \emph{Optimizers} focused on different goals, such
as optimization of execution time, power consumption or application reliability.
The user of the framework can develop and plug in a new \emph{Optimizer} corresponding
to their needs, but choosing from the available implementations or mixing them is also
possible.

The essential processing in the \emph{KernelHive} framework is performed
by tasks implemented as OpenCL kernels. The process
responsible for running consecutive tasks is \emph{unit}, a subsystem of \emph{KernelHive}
running as a system daemon on a given node. 
Tasks in \emph{KernelHive} are orchestrated by the \emph{engine} subsystem, the central module
of \emph{KernelHive}, which supports submitting multiple applications for execution. 
The applications are represented as \gls{dag} which nodes
represent computing tasks and edges represent data flow. The \emph{Engine} keeps
track of the current state of each application, in particular which tasks have
already been completed and for which the input data is already gathered so they
are ready for execution.
An interchangeable optimizer
interface allows plugging in different scheduling implementations
that periodically analyze
the set of jobs ready for execution and decide which ones should be executed
next and on which available computing devices, according to a given optimization strategy.

An improved and tested version of \emph{KernelHive} has been described in 
\cite{rosciszewski_kernelhive:_2016} along with a specific execution
methodology consisting of the following steps:
\begin{itemize}
\item selection of computing devices;
\item determination of best grid configurations for particular compute devices;
\item determination of the preferred data partitioning and granularity;
\item actual execution.
\end{itemize}

The execution steps proposed in this thesis is a more general version
of the latter, extended with Step 1, \emph{preliminary process optimization}.
Step 2a, \emph{process mapping} is a broader term for selection of computing devices
while Step 2b, \emph{parameter tuning} includes determination of both grid
configurations (one of the possible \emph{execution parameters}) and data
partitioning (one of the possible \emph{application parameters}). The \emph{actual execution}
step remains unchanged.

\newpage
\section{Modeling and Simulation for Fast Evaluation of Execution Configurations}\label{sec:simulation}			

As noted in Section \ref{sec:methodology}, steps 2a (process mapping) and 2b (parameter tuning) 
could be performed simultaneously. In fact, these two steps represent
solving the optimization problem defined in Equation \ref{eqn:problem} in Section \ref{sec:problemformulation} when the process 
implementations \gls{impl} are already optimized and will not be changed any more.
Chosen solutions to similar optimization problems are discussed in Chapter \ref{chp:optimizationtheory}.
Different methods can be suitable for this task depending on the size of the search space and cost of evaluating each
solution. If both the search space and the cost of solution evaluation are small, exhaustive search
can be used, which guarantees finding a global optimum, because it consists of systematic evaluation
of every possible alternative. In all other cases we propose using simulation method for fast solution evaluation.

In the cases of
small search space but high cost of solution evaluation or small cost of solution evaluation but 
big search space without possibility of space pruning, combinatorial search methods could be used,
such as local search, simulated annealing or evolutionary algorithms which do not guarantee finding
neither a local nor global optimum. Availability of an accurate enough simulation method would allow
to still perform exhaustive search. In extreme cases of prohibitively big search spaces and
high costs of solution evaluation, such a simulation method could also be useful as
a fast evaluation method for combinatorial search algorithms. In this Section we propose a simulation
for these purposes, related to Claim \ref{clm:2} of this Thesis.

Searching for a suitable simulation tool, in \cite{rosciszewski_simulation_2014} we reviewed
chosen existing parallel application simulators and provided motivations for developing a new
discrete-event simulator of parallel application execution on large-scale distributed systems.
MERPSYS, the simulation environment proposed in \cite{czarnul_merpsys:_2017} allows
to accurately predict execution time and power consumption of parallel applications
and analyze the power/time trade-off by performing the following steps:

\begin{enumerate}
\item preparing the \emph{application model} $\gls{app} = \langle \gls{impl},
\gls{minrequirements}, \gls{maxrequirements} \rangle $
by defining:
\begin{itemize}
\item the \emph{process implementations} $\gls{impl}$ using the \emph{Editor} graphical tool for 
writing code in a Java-based meta-language which provides
API for modeling various types of \emph{computation and communication operations}.
This requires identifying the crucial \emph{operations} and
thus deciding on the granularity of the model by analyzing code of an existing
application or providing them from scratch. The granularity level should allow to define
the modeling functions described in point 3;
\item \emph{process requirements} $\gls{minrequirements}, \gls{maxrequirements}$ by inserting their values into
a form;
\end{itemize}
\item preparing the \emph{system model} $\gls{system}(\gls{devices}, \gls{networkset})$ by
using the Editor graphical tool for building the \emph{hardware} graph from \emph{computing devices} and \emph{network links} available in a database;
\item defining \emph{hardware capabilities} $\gls{hardwarecapabilities}$ by inserting values for each \emph{process} into a form
available after selecting a certain device in the \emph{Editor};
\item defining certain modeling functions using a Web application for
filling in JavaScript snippets that have access to \emph{computing device}
characteristics and \emph{operation parameters}. The functions may be based
on analytical performance and power consumption models. If possible, we suggest tuning
these functions using results of real application executions. The following modeling
functions are required by the proposed simulator:
\begin{itemize}
\item $\gls{comptime}(\gls{comp}, \gls{device})$ - execution time of a \emph{computation operation} \gls{comp} using a \emph{computing device} \gls{device};
\item $\gls{commtime}(\gls{comm}, \gls{networklink})$ - execution time of a \emph{communication operation} \gls{comm} using a \emph{network link} \gls{networklink};
\item $\gls{pcidle}(\gls{device})$ - idle power consumption of a \emph{computing device} \gls{device};
\item $\gls{pcpeak}(\gls{device})$ - peak power consumption of a \emph{computing device} \gls{device}.
\end{itemize}
\noindent Additionally, a \emph{hardware parameter} $\gls{ncores}(\gls{device})$ is required, which
denotes the number of cores of a \emph{computing device} \gls{device}.
\item simulating the application execution and analysis of the resultant values of \emph{execution time} $\gls{executiontime}(\gls{app}, \gls{system}, \gls{mapping}, \gls{executionparameters})$ and \emph{average power consumption} $\gls{powerconsumption}(\gls{app}, \gls{system}, \gls{mapping}, \gls{executionparameters})$ through:
\begin{itemize}
\item providing a scheduling mechanism which defines the \emph{process mapping} function \gls{mapping}
by choosing or writing an implementation of a \emph{Scheduler} programming interface;
\item using the \emph{Editor} graphical tool for choosing specific values of
\emph{application execution parameters}~\gls{executionparameters}, enqueuing a single \emph{simulation instance} and analyzing its results;
\item running one or more instances of the \emph{Simulator} program which would execute
in parallel all \emph{simulation instances} enqueued in the \emph{simulation queue};
\item using the \emph{Web} interface for enqueuing an \emph{optimizer suite}
- an automatically populated
set of \emph{simulation instances} based on the previously executed single \emph{simulation instance}, with
a range of varying values of certain \emph{application execution parameters} in \gls{executionparameters} and,
thus, defining the \emph{space of application execution parameters} \gls{feasibleexecutionparameters}; 
\item using a \emph{ParetoVisualizer} tool for viewing a chart of results for all
\emph{simulation instances} in a \emph{suite} with \emph{execution time} and \emph{power consumption}
as axes, indicated set of Pareto-optimal solutions and values of the varying \emph{application execution parameters} accessible by hovering over a data point.
\end{itemize}
\end{enumerate} 

The proposed simulation environment performs a discrete-event simulation that runs
the \emph{application model} codes of all defined \emph{processes} and increases appropriate
execution time and energy consumption counters for all \emph{computation and communication operations}.
It should be noted that, in the case of \emph{communication operations}, the simulator 
ensures proper synchronization between the processes, so that for each process,
possible waiting for another process is included in the execution time of the operation.
The \emph{execution time} of a process is modeled as the sum of the execution times of all
\emph{computation and communication operations}:

\begin{equation}
\gls{processexecutiontime}(\gls{process}, \gls{device}) = \sum_{i: \gls{process}_i \in \gls{compset}}
{\gls{comptime}(\gls{process}_i, \gls{device})}
+ \sum_{i: \gls{process}_i \in \gls{commset}}
{\gls{commtime}(\gls{process}_i, \gls{device})}
\end{equation}

\noindent Hence, considering Definition \ref{def:executiontime}, the \emph{execution time} of the
whole application is modeled as:

\begin{equation}
\gls{executiontime}(\gls{app}, \gls{system}, \gls{mapping}, \gls{executionparameters})
= \max_{\gls{process} \in \gls{impl}}
\Big(\max_{\gls{device} \in \gls{devices}:\gls{mapping}(\gls{device}, \gls{process}) > 0}
\big(\sum_{i: \gls{process}_i \in \gls{compset}}
{\gls{comptime}(\gls{process}_i, \gls{device})}
+ \sum_{i: \gls{process}_i \in \gls{commset}}
{\gls{commtime}(\gls{process}_i, \gls{device})}
\big)\Big)
\end{equation}

The \emph{average power consumption} of the application is computed as sum of idle power consumptions
of all devices in the system plus all additional energy consumption caused by the \emph{computation operations}
of the application divided by the total application \emph{execution time}, computed as the
\emph{execution time} of the operation multiplied by the power consumption of the used
device at the time $t_{\gls{process}_i}$ of execution of the operation $\gls{process}_i$
($\gls{poperation}(\gls{device}, t_{\gls{process}_i})$):

\begin{equation}
\gls{powerconsumption}(\gls{app}, \gls{system}, \gls{mapping}, \gls{executionparameters}) = 
\sum\limits_{\gls{device} \in \gls{devices}}{
\gls{pcidle}(\gls{device})
}
+
\frac{
\sum\limits_{\gls{device} \in \gls{devices}}{
\Big( 
\sum\limits_{\gls{process} \in \gls{impl}: \gls{mapping}(\gls{device}, \gls{process}) > 0}{
\big(
\sum\limits_{i:\gls{process}_i \in \gls{compset}}{
\gls{comptime}(\gls{process}_i, \gls{device}) \cdot
\gls{poperation}(\gls{device}, t_{\gls{process}_i})
}
}
\big)
}\Big)
}{\gls{executiontime}(\gls{app}, \gls{system}, \gls{mapping}, \gls{executionparameters})}
\end{equation}

The simulator keeps track of the number of operations running on each \emph{computing device}
at each time throughout the application execution ($\gls{activeoperations}(\gls{device}, t)$).
This allows to simulate the power consumption of the device at a given moment of execution 
as a value between the idle level $\gls{pcidle}$ and the peak level $\gls{pcpeak}$, proportionally
to the number of cores used by the operations, but not exceeding the number of available cores:

\begin{equation}
\gls{poperation}(\gls{device}, t) = \min\Big(
\frac{\gls{activeoperations}(\gls{device}, t)}{\gls{ncores}(\gls{device})} 
, 1
\Big) \cdot \Big(\gls{pcpeak}(\gls{device}) - \gls{pcidle}(\gls{device})\Big)
\end{equation}

\noindent This way, the power consumption estimation takes into account that the
\emph{computing device} can be utilized by multiple \emph{processes}.
It should be noted that the simulator allows also to implement a non-linear increase of
power consumption until a saturation point, as proposed in \cite{wyrzykowski_performance_2016}.

The novelty of the proposed simulation method is that it combines mathematical modeling
(at the granularity level for which it is feasible) with discrete-event simulation at the level of
parallel processes for which operation sequences and synchronization with
other processes are difficult to model analytically. Additionally, the contributions
of the author of this thesis include experiments with modeling execution of various applications.

In \cite{czarnul_simulation_2015} we provided an example of 
developing and tuning an application model in MERPSYS based on the large vector similarity
measure computation application described in Section \ref{sec:big_data_sim}. The proposed
simulator allowed to predict the shapes of time curves beyond the area where empirical
results could be obtained.
In \cite{czarnul_modeling_2016} we provided examples of energy consumption modeling
based on two types of parallel applications: geometric SPMD and \gls{dac}. Experiments
of running up to 512 and 1024 processes of the two applications respectively,
on a large cluster from Academic Computer Center in Gdańsk, demonstrated
high degree of accuracy between simulated and measured results.

\begin{figure}[ht!]
\begin{center}
\includegraphics[scale=0.32]{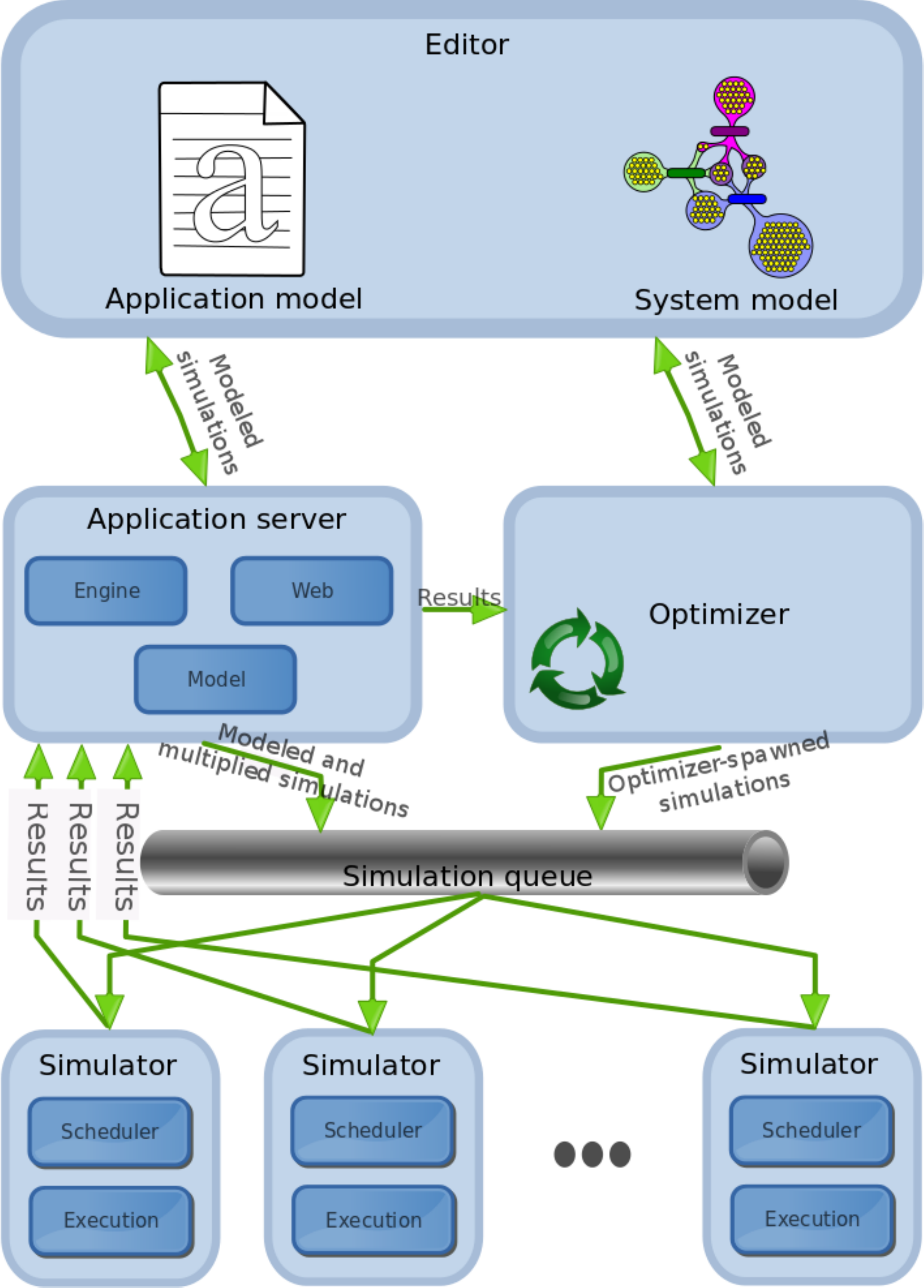}
\end{center}
\caption{Architecture of the distributed simulation framework \cite{rosciszewski_executing_2016} in MERPSYS \label{fig:merpsysarchitecture}}
\end{figure}

The web application included in the MERPSYS simulation environment allows to define
suites of simulations through providing ranges of selected parameters, so that multiple
simulation instances with different values of these parameters are created automatically.
A distributed simulation framework contributed by the author of this thesis in \cite{rosciszewski_executing_2016}
whose architecture is presented in in Figure \ref{fig:merpsysarchitecture}
provides two ways of handling such simulation suites. In the first one, all simulation
instances in the suite are enqueued to a simulation queue and processed by distributed
simulators. The results of the simulations can be browsed in the web application afterwards.
This way, exhaustive search of parameter space can be performed. 

The second way of handling the simulation suites involves an Optimizer component.
The MERPSYS environment provides a \emph{SimulationOptimizer} programming interface
whose implementations have access to the parameter search space and can adopt various
strategies of evaluating simulation instances in the suite. The input for such an optimizer
is a \emph{SimulationTask} that includes the application model, system model, feasible combinations
of \emph{application execution variables} and \emph{process mapping} constraints. 
This allows the Optimizer implementations to enqueue selected groups of simulation instances
in iterations, where results of the instances selected in the previous iteration can be used
to select new candidates, until the optimizer algorithm decides that a suboptimal approximate
solution has been found. This way, combinatorial search methods can be used for application optimization.
The architecture allows also to develop custom optimization tools, for example the ParetoVisualizer contributed
by the author of this thesis in \cite{rosciszewski_modeling_2017} which, instead of iterative evaluation, performs
exhaustive search and provides visualization of the found set of optimal solutions.

\leavevmode\thispagestyle{empty}\newpage
\leavevmode\thispagestyle{empty}\newpage
\chapter{Empirical Evaluation of the Proposed Methodology}\label{chp:experiments}

In this Chapter we describe case studies of using the methodology
proposed in Chapter~\ref{chp:solution} to optimize execution of specific
\emph{hybrid parallel applications} in specific \emph{heterogeneous \gls{hpc} systems}.
In Section \ref{sec:task_farming_case} we describe examples of applying the
individual execution steps to independent cases of multi-level task
farming applications.
Section \ref{sec:training_case} describes a case study of applying the
proposed optimization methodology as a whole to optimize one deep neural
network training application
executed on a professional cluster of workstations with \gls{gpu}s.
The aim of these case studies is empirical evaluation of the proposed methodology.
Table \ref{fig:casestudytable} lists the specific
actions that were performed within the proposed execution steps.

\begin{table}[ht!]
\begin{center}
\begin{tabular}{ | p{3.5cm} | p{5cm} | p{5cm} |}
\hline
\centering Step & \centering Individual steps & \centering Full Case Study \tabularnewline \hline \hline
1. Preliminary process optimization & Computation and communication overlapping & \gls{gpu} training and \gls{cpu} data preprocessing overlapping \\ \hline
2a. Process mapping & Network-aware and Power Constrained Scheduling & Exploring the front of Pareto-optimal solutions using the proposed \\ \cline{1-2}
2b. Parameter tuning  & Tuning data partitioning and grid configurations &  simulation environment \\ \hline
3. Actual execution & Execution in KernelHive & Execution using \gls{mpi} and Kaldi \\ \hline
\end{tabular}
\end{center}
\caption{Action selection for particular execution steps for the considered case studies}\label{fig:casestudytable}
\end{table}

As indicated in the table, method of simulation for evaluating \emph{process mappings} and 
\emph{application execution parameters} proposed in Section \ref{sec:simulation} has been
used in the full case study within the combined \emph{process mapping} and \emph{parameter tuning}
steps.

\section{Multi-level Task Farming on Heterogeneous Systems with CPUs and GPUs}\label{sec:task_farming_case}

On the example of multi-level task farming applications, the case studies described in this section
show how all execution steps proposed in this thesis can be performed using the KernelHive framework
and address Claim \ref{clm:1} of this thesis by proving the significance of these steps.
 In the \emph{preliminary process optimization} step,
a data prefetching optimization is proposed that results in overlapping
of computations and communications, as described in Section \ref{sec:prefetching}.
Example of the \emph{process mapping} step, namely
network-aware and power constrained scheduling 
are described in Section \ref{sec:icdcn}.
Tuning of \gls{gpu} grid configurations and data partitioning within the \emph{parameter tuning}
step is described in Section \ref{sec:khtuning}. Finally, the step of actual application execution
using the KernelHive framework is described in Section \ref{sec:khexecution}.

\subsection{Preliminary Process Optimization - Computation and Communication Overlapping}\label{sec:prefetching}

In the first of the execution steps proposed in this thesis, \emph{preliminary process optimization},
overlapping of certain operations can be implemented.
This requires analyzing the \emph{operations} of the \emph{processes} in the
application at a chosen granularity level and searching for the possibility of
overlapping certain operations that use different hardware.
In this section we provide such an analysis for the regular expression matching application described
in Section \ref{sec:regex}, implemented in the \emph{KernelHive} framework 
and we describe the proposed modification that allows overlapping of certain communication
and computation operations.

In the basic version of the application, the default \emph{Optimizer} implementation is used
that, in each iteration of application workflow processing, schedules the jobs ready for execution
to devices available in the system in a round-robin fashion.
In terms of processes and operations, this means that the engine process $\gls{process}^{engine} \in \gls{impl}$
that runs on the central node $\gls{device}_{engine}$
consists of iterations of the following communication operations:
ordering the appropriate \emph{unit} subsystem to execute a next available task
and receiving the ID of the computation result package afterwards.

For each task, the sequence of operations of the \emph{unit} process
resulting from its \emph{process implementation} $\gls{process}^{unit} \in \gls{impl}$ is as follows:

\begin{enumerate}
\item \emph{communication} operation that downloads input data from a given database;
\item \emph{computation} operation that is execution of a given task kernel on a given target device;
\item \emph{communication} operation that uploads the resulting output data to a given
database;
\item \emph{communication} operation that reports the ID assigned to the output data to the \emph{engine}.
\end{enumerate}

Depending on the computationally intensive, data intensive or communication intensive 
profile of the specific application, different operations of the \emph{unit} process
can have significant execution time.
Profiling of a moderately data intensive configuration of the regular expression
matching application described in Section \ref{sec:regex} executed in a multi-level 
\gls{hpc} system revealed that the most costly operations are the two first ones.
Since they require different hardware elements (a chosen computing
device and a network interconnect), the operations could be potentially overlapped
if only in the application there were multiple tasks without mutual data dependencies.

\begin{figure}[ht!]
\begin{center}
\includegraphics[scale=0.7]{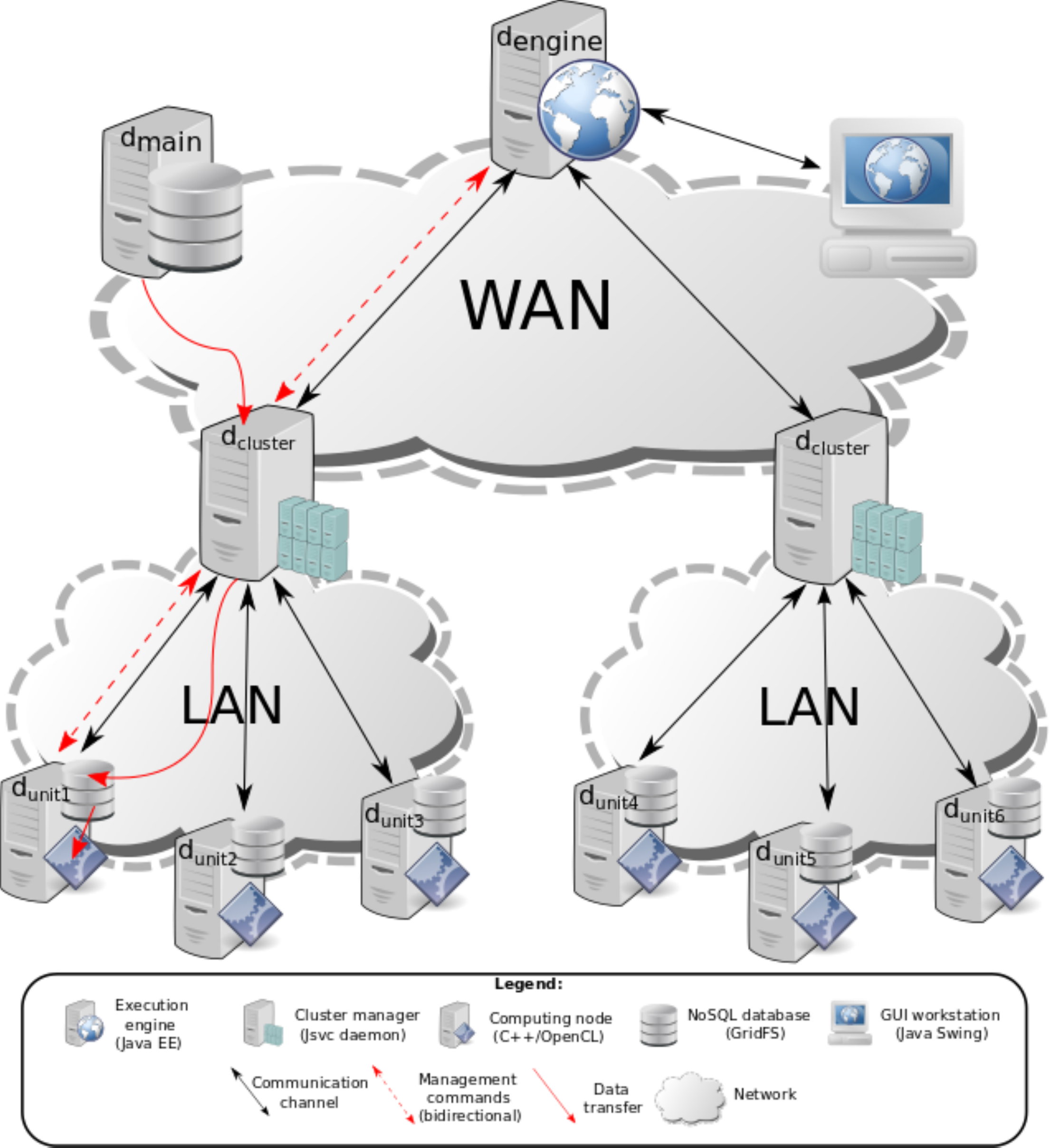}
\end{center}
\caption{Architecture of the proposed data prefetching optimization \label{fig:prefetchingarchitecture}}
\end{figure}

This is the case in the considered application, which gives the motivation for the 
overlapping optimization contributed by the author of this thesis in \cite{rosciszewski_dynamic_2014}.
The optimization benefits from the feature of \emph{KernelHive}, that
multiple databases can be used for storing the input, output and partial
data packages. In terms of the notation proposed in Section \ref{sec:problemformulation},
the optimization modifies the \emph{process implementations} belonging to the application
$\gls{impl}$ by introducing a modified implementation of the $\gls{process}^{engine}$ 
process - $\gls{process}^{engine\prime} \in \gls{impl}$ and a new supporting process  $\gls{process}^{prefetching} \in \gls{impl}$.
The architecture of the proposed prefetching scheme is shown in Figure \ref{fig:prefetchingarchitecture}.
Apart from the main database, accessible by the \emph{unit}
processes through a network interconnect, one locally installed database
per each \emph{unit} has been used. The \emph{unit} process remained unchanged,
but an additional prefetching process has been defined, executed by another
subsystem, \emph{cluster} placed on the cluster access node $\gls{device}_{cluster}$, which is an entry point
for accessing the \emph{units} that are in a local network, possibly not 
directly visible by the \emph{engine}. The $\gls{process}^{prefetching}$ process executes
the following \emph{communication operations}: 

\begin{enumerate}
\item downloading input data from the main database;
\item uploading the data to a given local database;
\item reporting the ID of the prefetched data package to the \emph{engine}.
\end{enumerate}

In order to utilize this prefetching mechanism, a new \emph{Optimizer}
has been developed for the \emph{engine}, which changes the sequence of 
\emph{communication operations}.
For each selected computing device, at the beginning of the sequence
resulting from the modified process $\gls{process}^{engine\prime} \in \gls{impl}$,
there is an operation of ordering the appropriate \emph{cluster} to execute a prefetching
job related to the next available task. Then, the following operations are performed in each iteration:

\begin{enumerate}
\item receiving the ID of the prefetched data package from the appropriate \emph{cluster};
\item ordering the appropriate \emph{cluster} to execute a prefetching job for a next available task;
\item ordering the appropriate \emph{unit} to execute the task for which data has been prefetched; 
\item receiving the ID of the computation result package.
\end{enumerate}

As a result, the first operation in the \emph{unit} process downloads the input
data directly from the local database, which reduces the operation execution time.
Assuming that the main database is installed on device $\gls{device}_{main} \in \gls{devices}$
and the \emph{unit} subsystems with their local databases are running on devices $\gls{device}_{unit_{i}} \in \gls{devices}$,
apart from data intensity of the application, the benefit from the proposed optimization depends on the
properties of the network link $\gls{networklink}_{main, unit} \in \gls{networkset}$ between these devices, which depends
on the \emph{heterogeneous HPC system} graph $\gls{system}(\gls{devices}, \gls{networkset})$.
The gain from the proposed optimization depends also on the number of tasks in the application
and thus the number of iterations.

The experiments described in \cite{rosciszewski_dynamic_2014} involved executing the application on a
heterogeneous cluster consisting
of apl09 and apl10 servers described in Section \ref{sec:apl} and a remote machine was
used as the device $\gls{device}_{main}$.
In the case of one iteration the difference in
execution time was negligible, because in fact no overlapping occurred. However, for
higher iteration numbers the gain from the optimization ranged from around 11\%
to 16\%.
These results prove that the \emph{preliminary process optimization} step included in Claim
\ref{clm:1} of this dissertation allows to significantly reduce execution time of the optimized application.

\subsection{Process Mapping - Network-aware and Power Constrained Scheduling}\label{sec:icdcn}

Apart from \emph{preliminary process optimization} described in the previous Section, 
the topography of the \gls{hpc} system resulting from the \emph{heterogeneous HPC system}
graph $\gls{system}(\gls{devices}, \gls{networkset})$ can be also taken into
account in the \emph{process mapping} execution step.
For example, a network-aware scheduling scheme for a \emph{KernelHive optimizer}
proposed in \cite{rosciszewski_network-aware_2014} schedules the tasks equally
between the clusters, understood as sets of nodes connected by a local area network. 
Reported simulation results show that such an approach can significantly reduce
total execution time of the application, provided that only a subset of computing
devices available in the system are used for computations.

Utilizing only a subset of devices available in a \gls{hpc} system is a common case.
This is often caused by limited scalability of the application, when there is
an optimal number of subtasks into the computations could be divided and further
division results in reduced efficiency. 
Another reason for using only a subset of available devices
is an imposed power consumption limit. Solving the problem of execution time optimization
under power consumption constraints, defined in Equation \ref{eqn:problemlimit} in Section
\ref{sec:problemformulation}
requires optimal selection of the computing devices subset within the \emph{process mapping}
$\gls{mapping} \in \gls{mappingset}$, so that their total maximum power consumption does not exceed
a given limit~$\gls{powerlimit}$.

\begin{figure}[ht!]
\begin{center}
\includegraphics[scale=0.7]{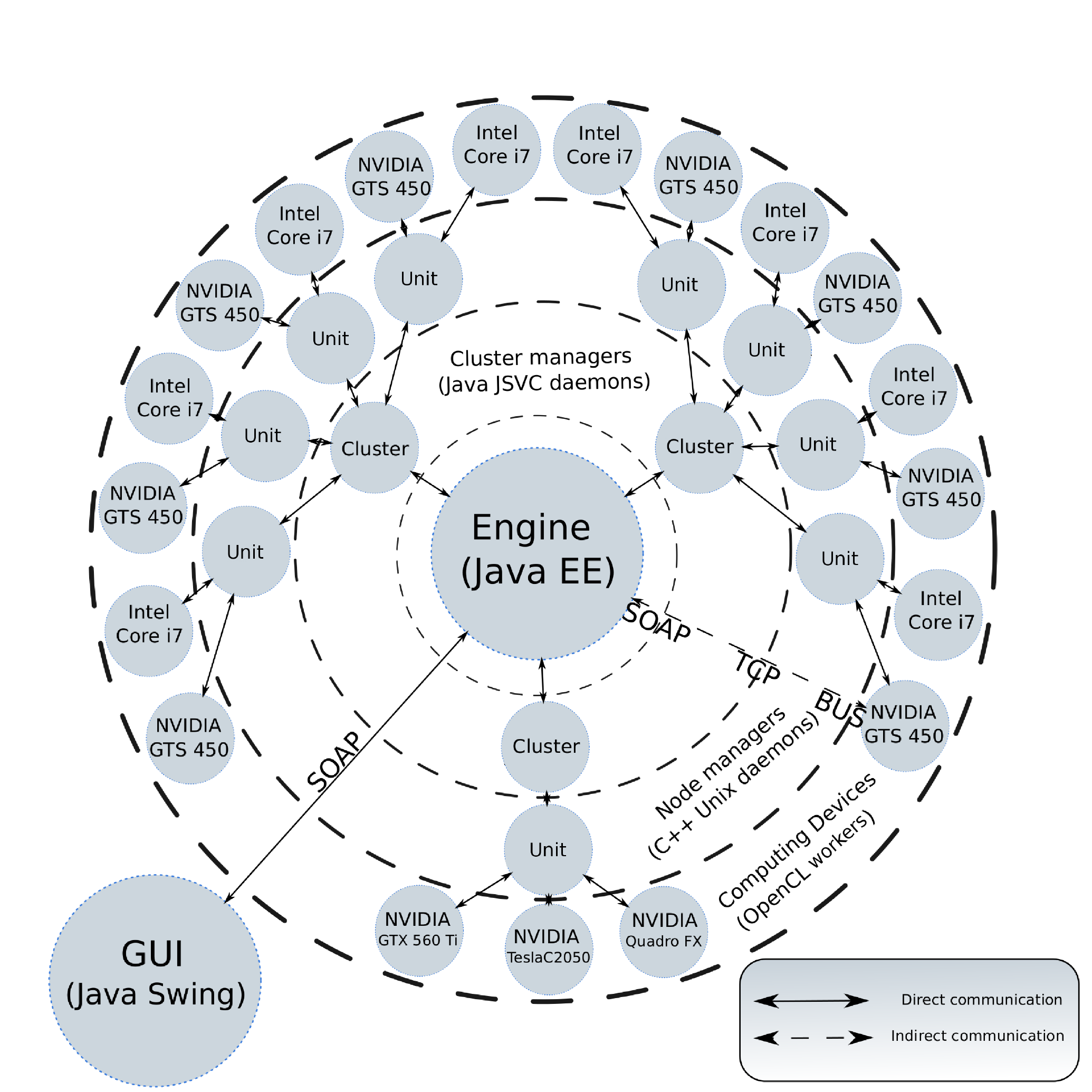}
\end{center}
\caption{Heterogeneous computing system used in the experiments with imposed power limits \label{fig:powerlimitarchitecture}}
\end{figure}

In \cite{hutchison_optimization_2014} we formulated this selection problem as a 0/1
knapsack problem (given power consumption levels and performance of $|\gls{devices}|$ devices, put
these items in a knapsack of capacity $\gls{powerlimit}$).
The aim of the experiments was to minimize the execution
time of the task farming application for password breaking described in Section \ref{sec:md5}
under different values of imposed power limit in a heterogeneous system combined from the
computing infrastructure described in Sections \ref{sec:lab527} and \ref{sec:apl}, presented
in Figure \ref{fig:powerlimitarchitecture}.
The paper shows how a greedy solution to the knapsack problem can be deployed
in a real framework able to parallelize computations in a multi-level heterogeneous
computing infrastructure.

One of the challenges of this approach was the ability to dynamically divide the computations
into a number of subtasks which could change at application runtime. In order to meet this
challenge, the \emph{unrollable node} functionality of \emph{KernelHive} has been used. 
The application was represented  as a dataflow graph with a single node implementing
the \emph{unrollable node} mechanism.
The \emph{partitioner} kernel implemented division of the range of examined passwords
and the \emph{merger} implemented checking if the password has been broken in any task.
Our approach is similar to the "moldable jobs" concept discussed in \cite{sarood_maximizing_2014},
where "the user specifies the range of nodes (the minimum and the maximum number of nodes) on which the job
can run. The job scheduler decides the number of nodes within the specified range to be
allocated to the job".
It should be noted that apart from the \emph{process mapping} function $\gls{mapping}$,
during the process of computing device selection, also an \emph{execution parameter}
is modified, namely the number of running tasks. This is an example how in some situations
the \emph{process mapping} and \emph{parameter tuning} steps have to be performed
simultaneously.

\begin{figure}[ht!]
\begin{center}
\includegraphics[scale=0.8]{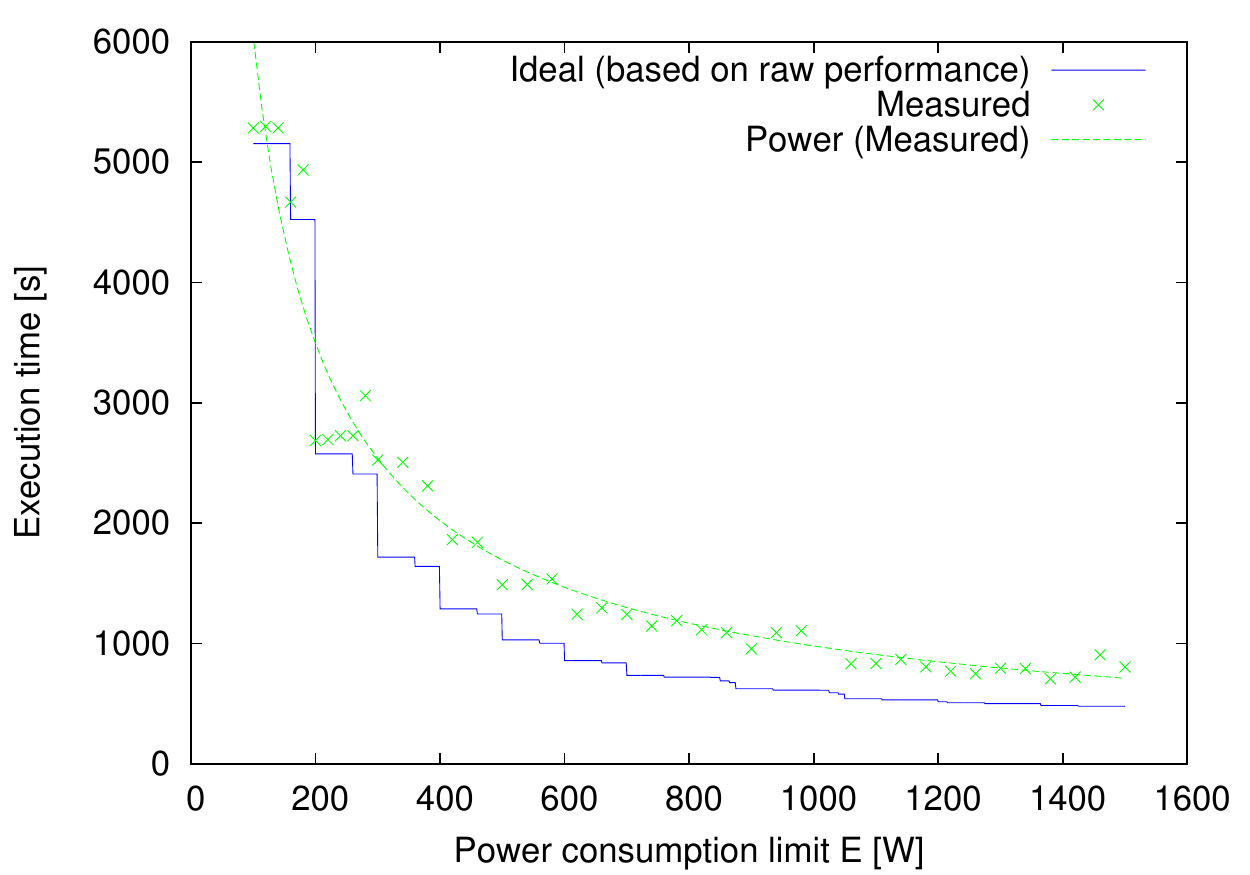}
\end{center}
\caption{Comparison of execution times under power limits with theoretical ideal values \label{fig:powerlimittime}}
\end{figure}

For the sake of the experiments described in \cite{hutchison_optimization_2014},
maximum power consumption of each used device model when running the application has
been measured using a hardware power meter. The application has been executed
on a heterogeneous computing system consisting of eight "des" nodes described
in Section \ref{sec:lab527} (seven with a GTS450 \gls{gpu} and one with a GT4X80 \gls{gpu})
and "apl09" server described in Section \ref{sec:apl}.
Real execution times of the application under varying power consumption limits 
have been compared 
to the theoretical ideal execution times based on relative performance of the used devices.
The comparison of execution times is shown in Figure \ref{fig:powerlimittime}.

\begin{figure}[ht]
\begin{center}
\includegraphics[scale=0.8]{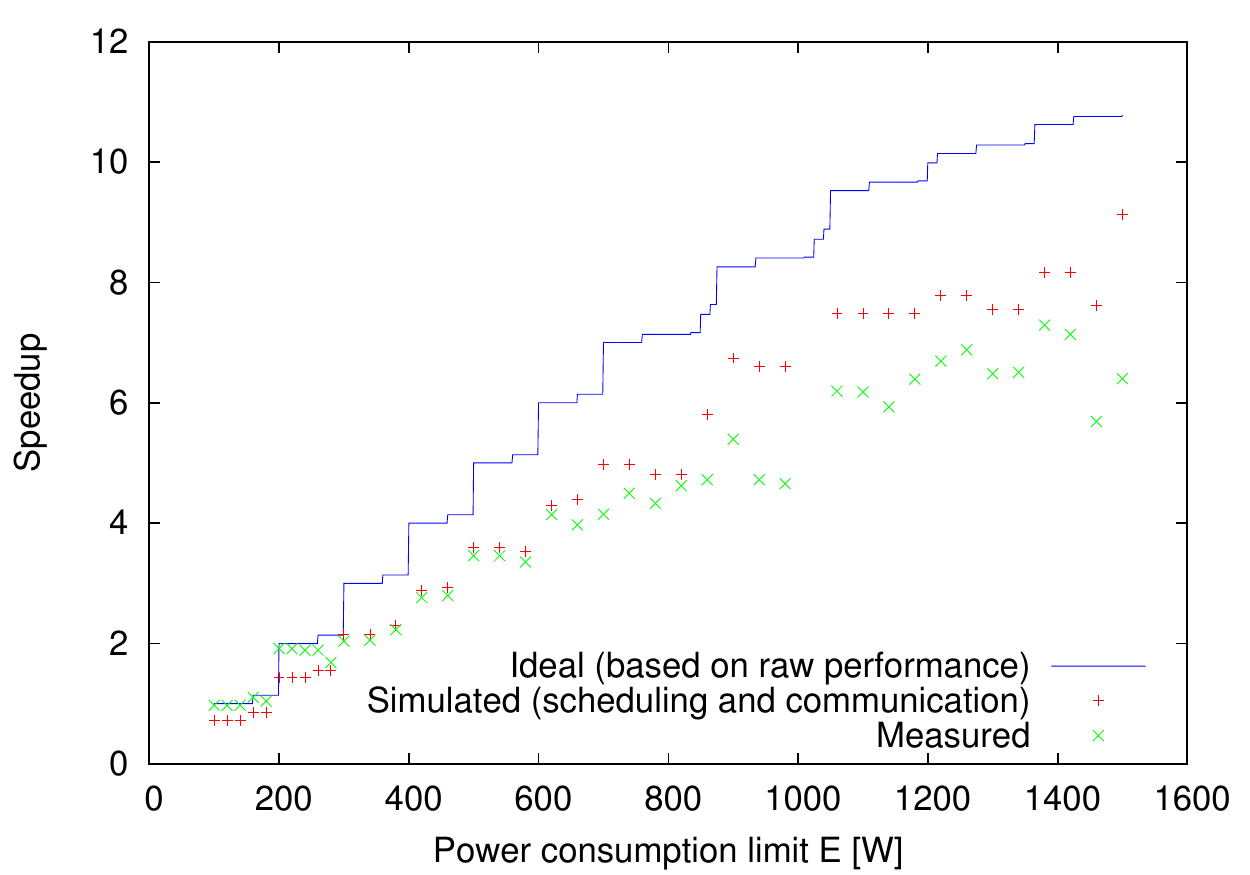}
\end{center}
\caption{Speedup comparison of real executions under power limits with theoretical and simulated values \label{fig:powerlimitspeedup}}
\end{figure}

Although the shapes of the real and ideal curves are similar, the higher power
limit, the bigger the difference between them. The reason for this is that the
ideal curve does not consider communication and scheduling overheads which
occur in real executions. In order to better explain the real execution results,
the speedups were compared also with the values resulting from simulations
that reflected the key dependencies by taking into account the scheduling
and communication overheads. The speedup comparison in Figure \ref{fig:powerlimitspeedup}
shows that the proposed \emph{process mapping} method allows to achieve
nearly ideal speedups of the application under a strict power consumption constraint.
This proves that the \emph{process mapping} step included in Claim \ref{clm:1} of
this dissertation allows to optimize the execution time of the considered application.

\subsection{Parameter Tuning - Grid Configurations and Data Partitioning}\label{sec:khtuning}

One of the advantages of using the \emph{KernelHive} framework for executing
parallel applications is the possibility to dynamically tune the values of certain
\emph{application execution parameters} depending on the specific executed computational
kernel or the specific device used for the computations. 
For example, the \gls{gpu} grid configuration \emph{execution parameters} described
in Section~\ref{sec:shared} including the total number of threads
$v_{threadstotal}$ and number of threads per block $v_{threadsblock}$
can be fixed using an appropriate form in the \emph{hive-gui}
user interface or programmed in the \emph{engine} subsystem to take into account
the used device model and determine the best configuration by probing chosen values.
The latter approach has been used in \cite{rosciszewski_kernelhive:_2016}
for the password breaking application described in Section \ref{sec:md5}.
In terms of the notation proposed in Section \ref{sec:problemformulation},
the problem solved in this case study can be expressed as follows:

\begin{equation}
\begin{aligned}
& \underset{\gls{executionparameters}}{\min}
& & \mathrm{\gls{executiontime}}(\gls{app}, \gls{system}, \gls{mapping}, \gls{executionparameters}) \\
& \text{subject to}
& & \gls{executionparameters} = [v_{threadstotal} \in T, v_{threadsblock} \in B].
\end{aligned}
\end{equation}

\noindent where T is the set of feasible total numbers of threads and B is the set of feasible block sizes.

\begin{figure}[ht!]
\begin{center}
\includegraphics[scale=0.8]{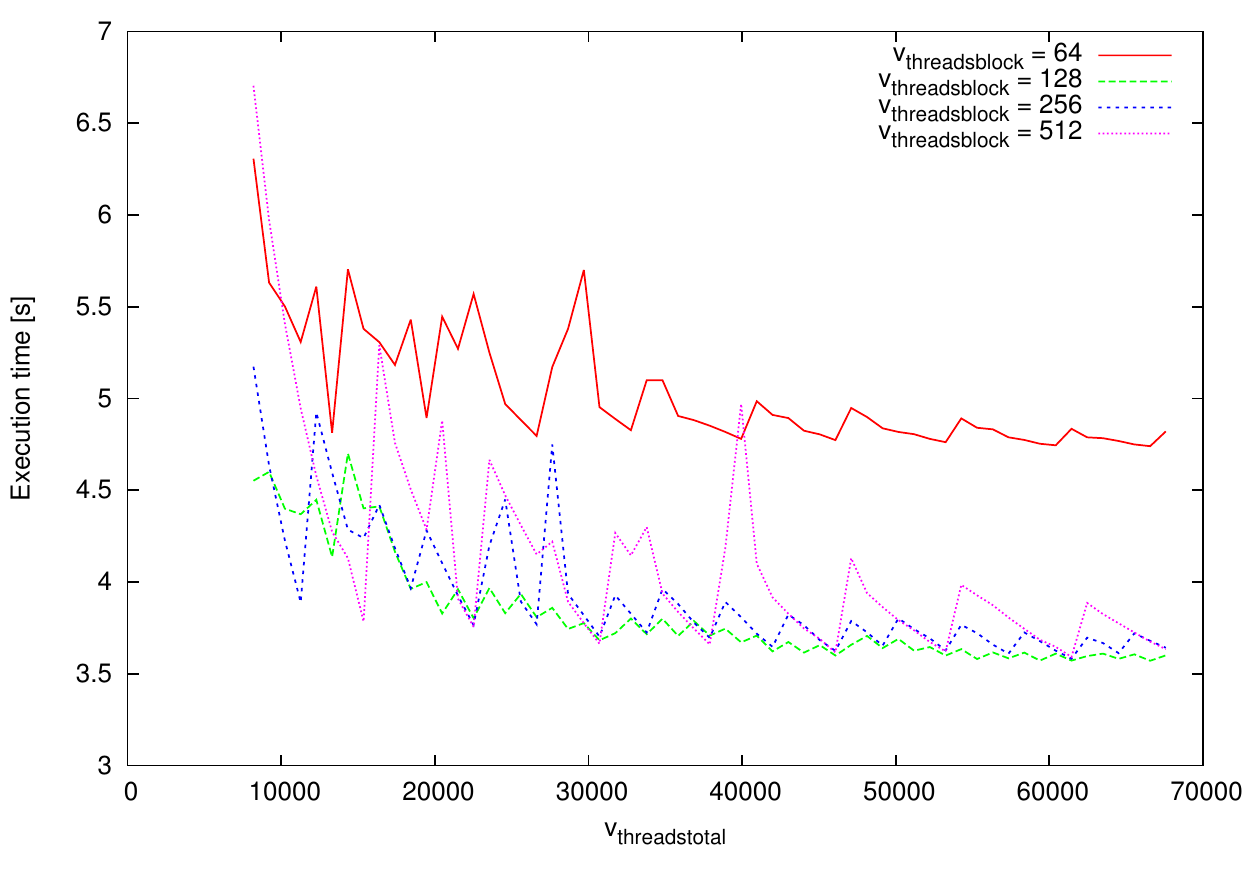}
\end{center}
\caption{Execution time depending on grid configurations for GeForce GTX 480\label{fig:gridgtx}}
\end{figure}

\begin{figure}[ht!]
\begin{center}
\includegraphics[scale=0.8]{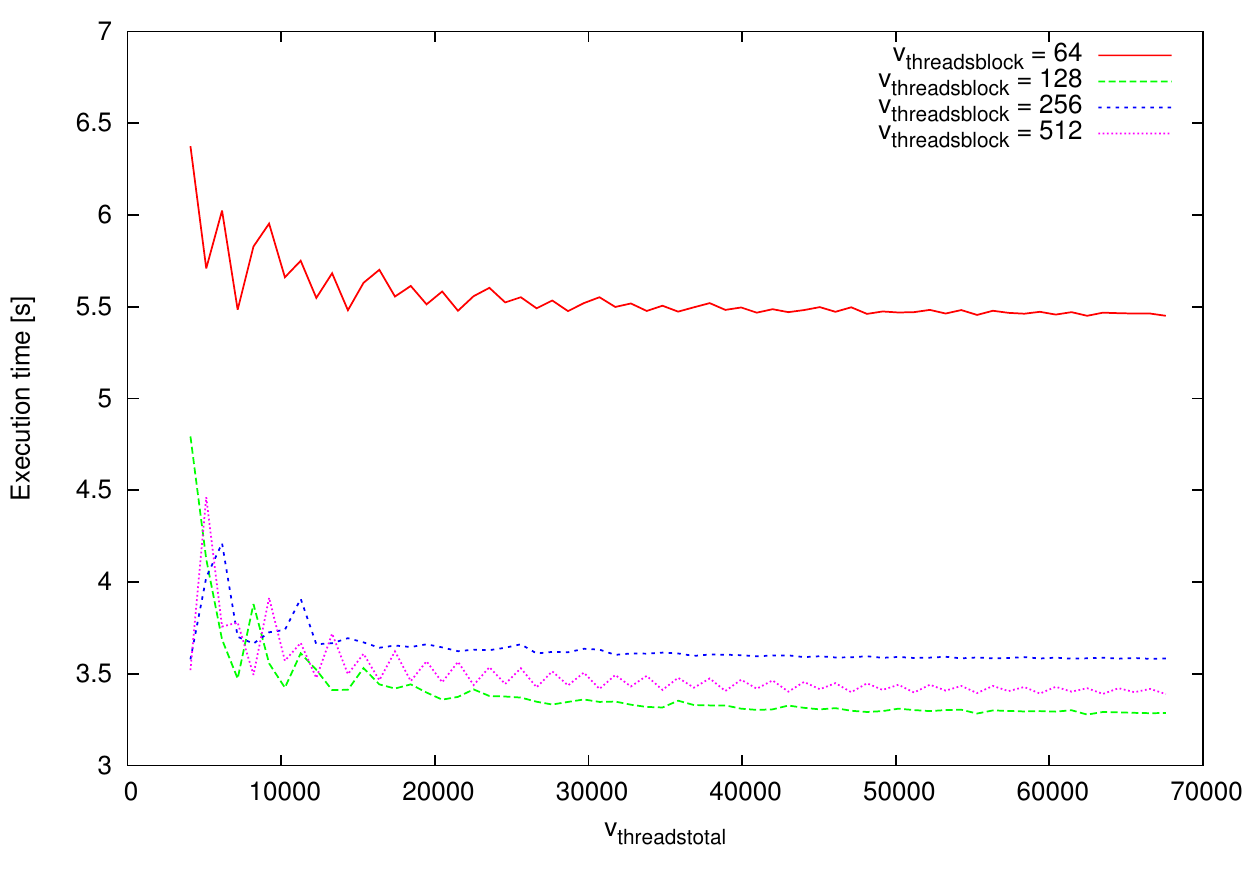}
\end{center}
\caption{Execution time depending on grid configurations for GeForce GTS 450\label{fig:gridgts}}
\end{figure}

\begin{figure}[ht!]
\begin{center}
\includegraphics[scale=0.8]{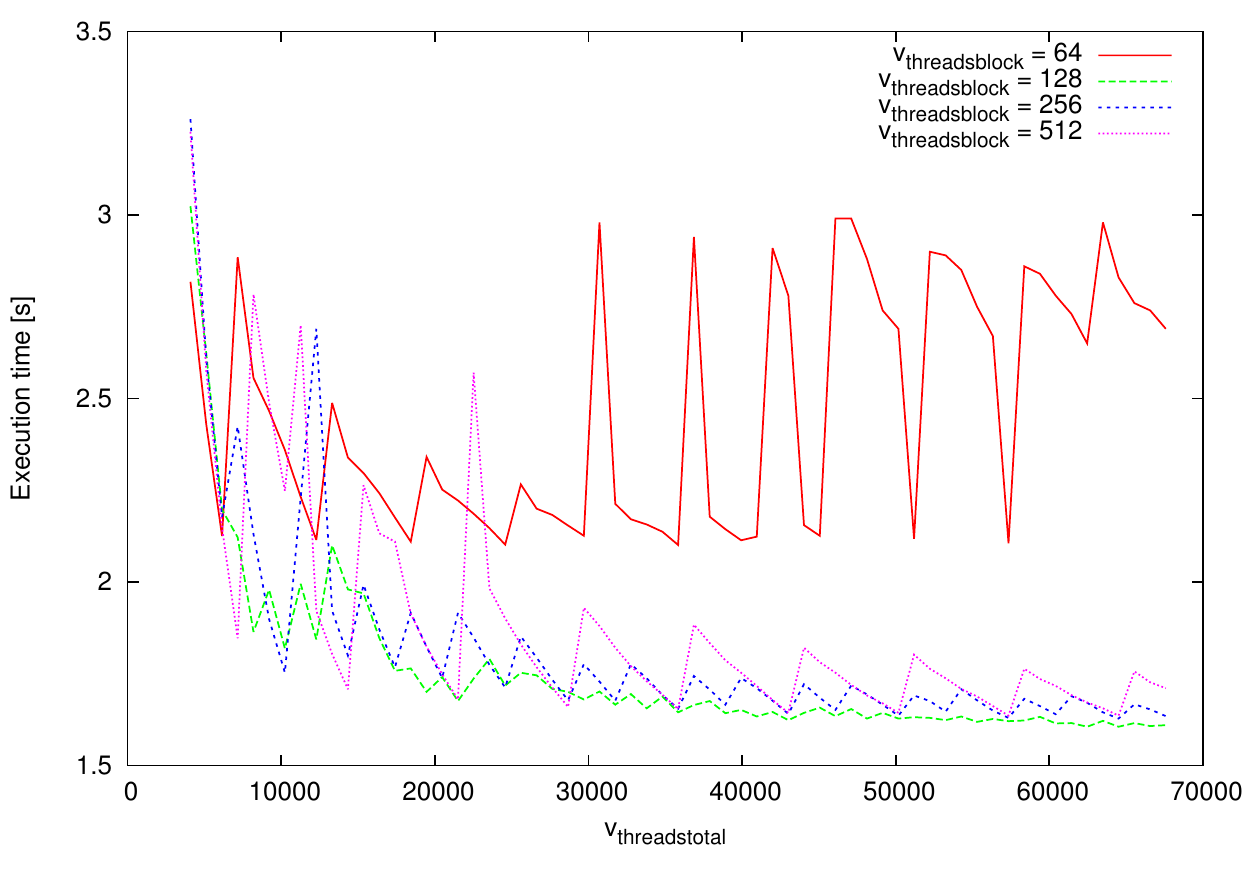}
\end{center}
\caption{Execution time depending on grid configurations for Tesla C2050\label{fig:gridtesla}}
\end{figure}

Comparison of execution times of searching through $10^9$ hashes depending
on \emph{grid configurations} is presented in Figure \ref{fig:gridgtx}
for GeForce GTX 480 GPU, Figure \ref{fig:gridgts} for GeForce GTS 450 GPU
and Figure \ref{fig:gridtesla} for Tesla C2050 GPU.

The comparison shows that setting the appropriate values of \emph{execution parameters} can
have a great impact on the application execution time and what is more, optimal values
of one parameter can depend on the selected value of another parameter.
Setting too low number of $v_{threadsblock}$ (64) results in significantly longer execution time for
all analyzed computing devices. For example, for the GeForce GTS 450 \gls{gpu} it results in
execution time of roughly 5.5s compared to roughly 3.5s in the case of optimal setting.
However, increasing this number above a certain
value also increases the execution time. The value of $v_{threadsblock}$ chosen in the
\emph{parameter tuning} step
for the analyzed application on each analyzed device is 128. Regarding
the $v_{threadstotal}$ parameter, increasing it
to a certain value results in a significant
speedup. While GTS 450 is able to maintain a constant performance with more threads,
GTX 480 and Tesla C2050 have difficulties with scheduling the threads, resulting
in peaks in execution time. To avoid such problems, the value of $v_{threadstotal}$
chosen in the \emph{parameter tuning} step for all devices was 32 768.

Apart from \emph{execution parameters}, also certain \emph{application parameters}
can be tuned using \emph{KernelHive}. Let the \emph{application execution parameter}
$v_{nnodes} \in \mathbb{N}$ denote the number of nodes used by the application.
In task farming applications executed in heterogeneous computing systems, there
often occurs a load imbalance resulting from differing speeds of the used computing
devices. For example, if the computational problem is divided into a number of subproblems
equal to the number of nodes $v_{nnodes}$, the time of computing the resulting single
task on the slowest used device determines the total application execution time.
All the faster devices are idle while waiting for the slowest device to finish,
which makes the execution inefficient.

A common way to address this problem
is to divide the computational problem to a higher number of smaller subproblems,
so that the execution time on the slowest device is reduced, and 
the faster devices can be dynamically loaded with consecutive tasks.
The higher performance differences across the devices in $\gls{devices}$,
the more tasks are needed to ensure proper load balancing. On the other hand,
increasing the number of tasks comes at a cost of additional communication
overheads for distributing the tasks. In the \emph{parameter tuning} step,
the optimal number of tasks should be determined for the specific application
and utilized system.
In \emph{KernelHive}, the number of tasks can be dynamically set thanks to
the \emph{unrollable node} mechanism described in Section \ref{sec:icdcn}.
For the considered task farming applications,
the number of subproblems is derived from the number of utilized devices multiplied
by a \emph{application parameter} $v_{dpm} \in \mathbb{N}$
called \emph{data package multiplier}. In terms of the notation proposed in
Section \ref{sec:problemformulation}, finding the optimal number of used \emph{computing devices}
and value of the \emph{data package multiplier} can be expressed as follows:

\begin{equation}
\begin{aligned}
& \underset{\gls{executionparameters}}{\min}
& & \mathrm{\gls{executiontime}}(\gls{app}, \gls{system}, \gls{mapping}, \gls{executionparameters}) \\
& \text{subject to}
& & \gls{executionparameters} = [v_{nnodes} \in \mathbb{N}, v_{dpm} \in \mathbb{N}].
\end{aligned}
\end{equation}

\begin{figure}[ht!]
\begin{center}
\includegraphics[scale=0.9]{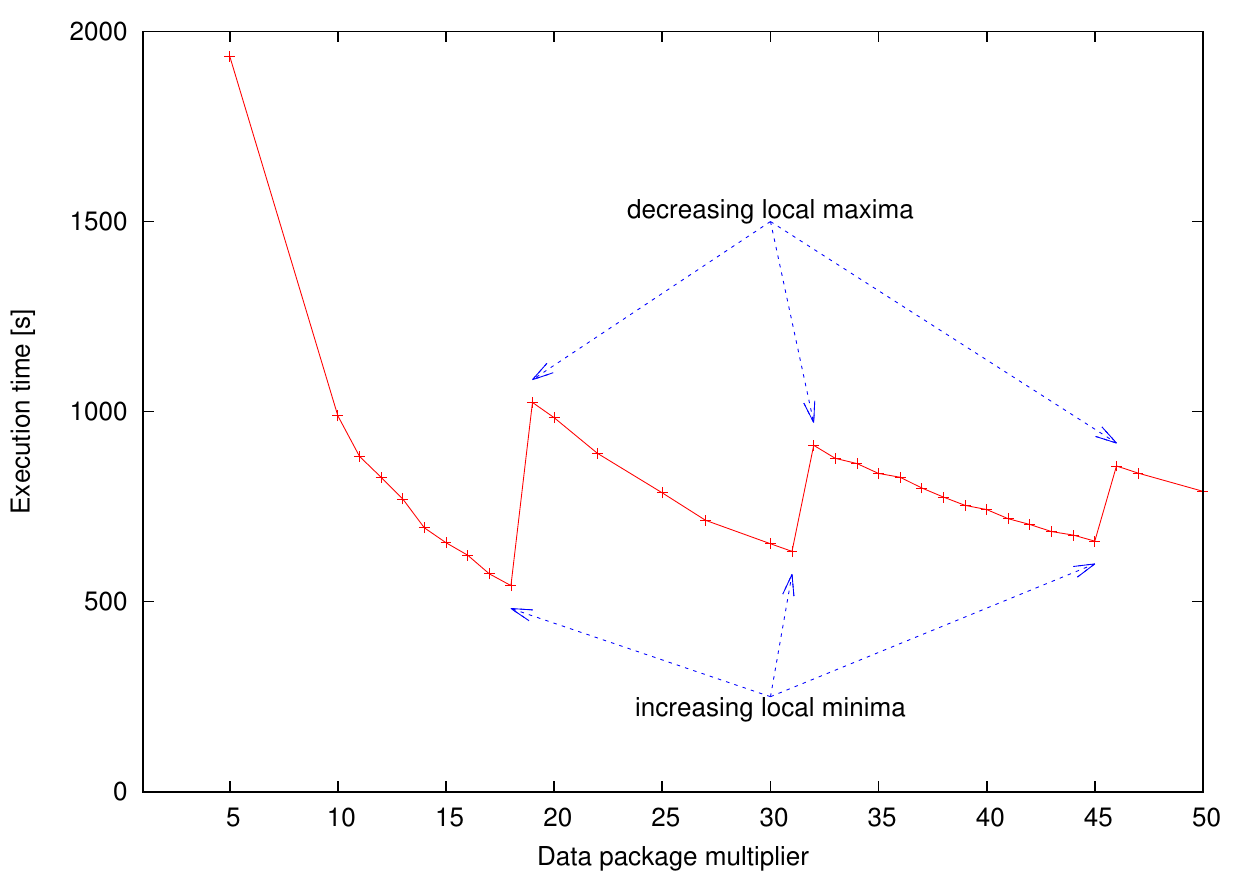}
\end{center}
\caption{Influence of the package number multiplier $v_{dpm}$ on the real application execution time \label{fig:pna}}
\end{figure}

The influence of the $v_{dpm}$ parameter
on the execution time of the password breaking application described in Section \ref{sec:md5},
searching through $5 * 10^{12} hashes$ is shown in Figure \ref{fig:pna}. The application has been executed on a 
system consisting of $v_{nnodes}$ = 16 "des" nodes described in Section \ref{sec:lab527},
each consisting of an Intel i7-2600K \gls{cpu} and a NVIDIA GeForce GTS 450 \gls{gpu}.

\begin{figure}[ht!]
\begin{center}
\includegraphics[scale=0.9]{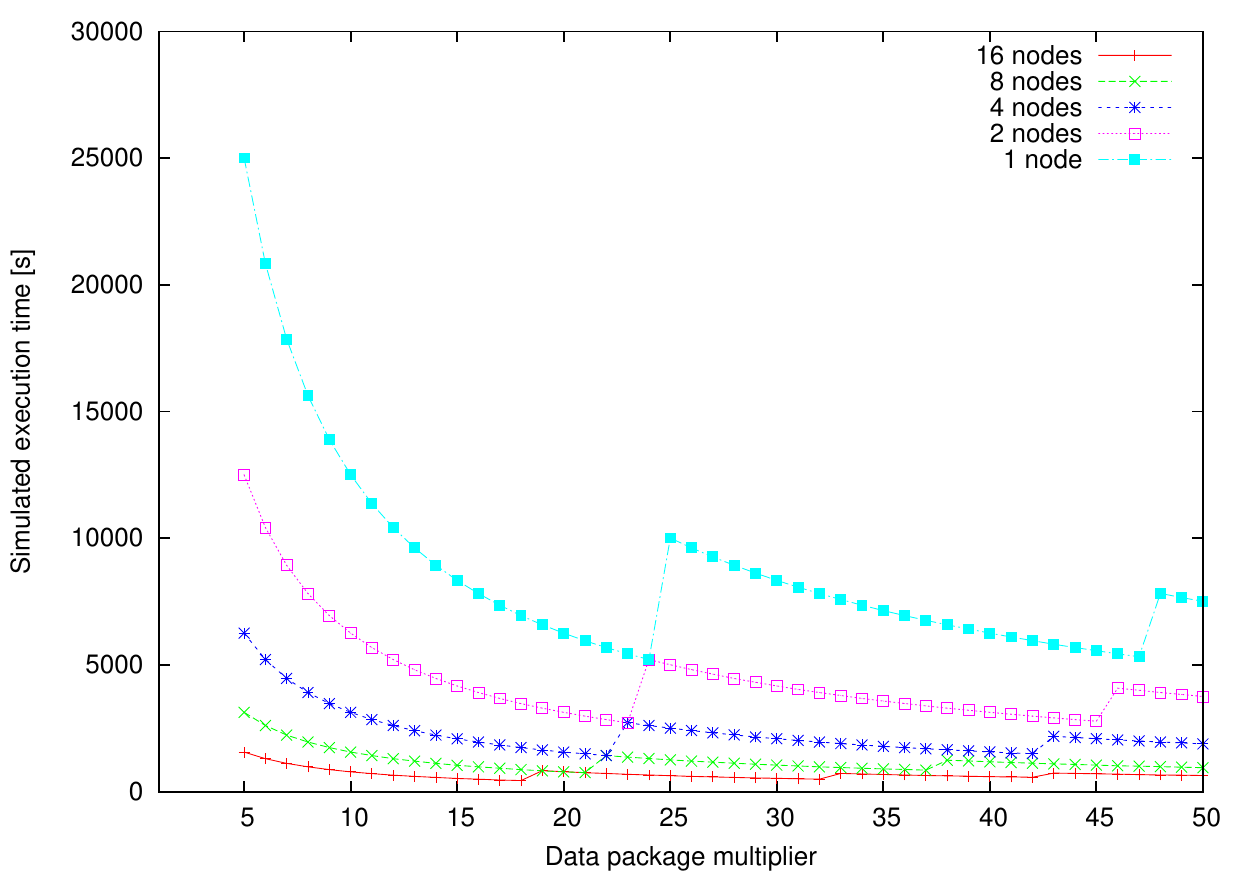}
\end{center}
\caption{Influence of the package number multiplier $v_{dpm}$ and node number $v_{nnodes}$ on simulated application execution times \label{fig:multipliers}}
\end{figure}

The figure shows how significant the influence of a single \emph{application parameter}
can be on the application execution time. While there is a point of optimal load balancing with
$v_{dpm}$ equal to 18, increasing it only to 19 ruins the load balancing
and results in nearly two-fold increase of the execution time. Such "steps" in the chart
are periodic and result from imbalance of processing times between \gls{cpu}s and \gls{gpu}s. Because
for this application the \gls{cpu}s are less efficient than the \gls{gpu}s, for some numbers of packets
the \gls{gpu}s are idle waiting for the \gls{cpu}s to finish previously ordered data packets. The height
of the consecutive "steps" in the chart is decreasing, because higher number of data packets
also means that they are smaller, hence smaller differences between processing times of \gls{cpu}s and \gls{gpu}s.
On the other hand, the local minima before the "steps" are increasing, because high numbers of
data packets introduce additional communication overheads of packet transfers. Based on results
of simulations that take into account these overheads, for the same application,
Figure \ref{fig:multipliers} shows how the shape of the "steps" changes depending
on the number of used nodes $v_{nnodes}$.

\begin{figure}[ht!]
\begin{center}
\includegraphics[scale=0.8]{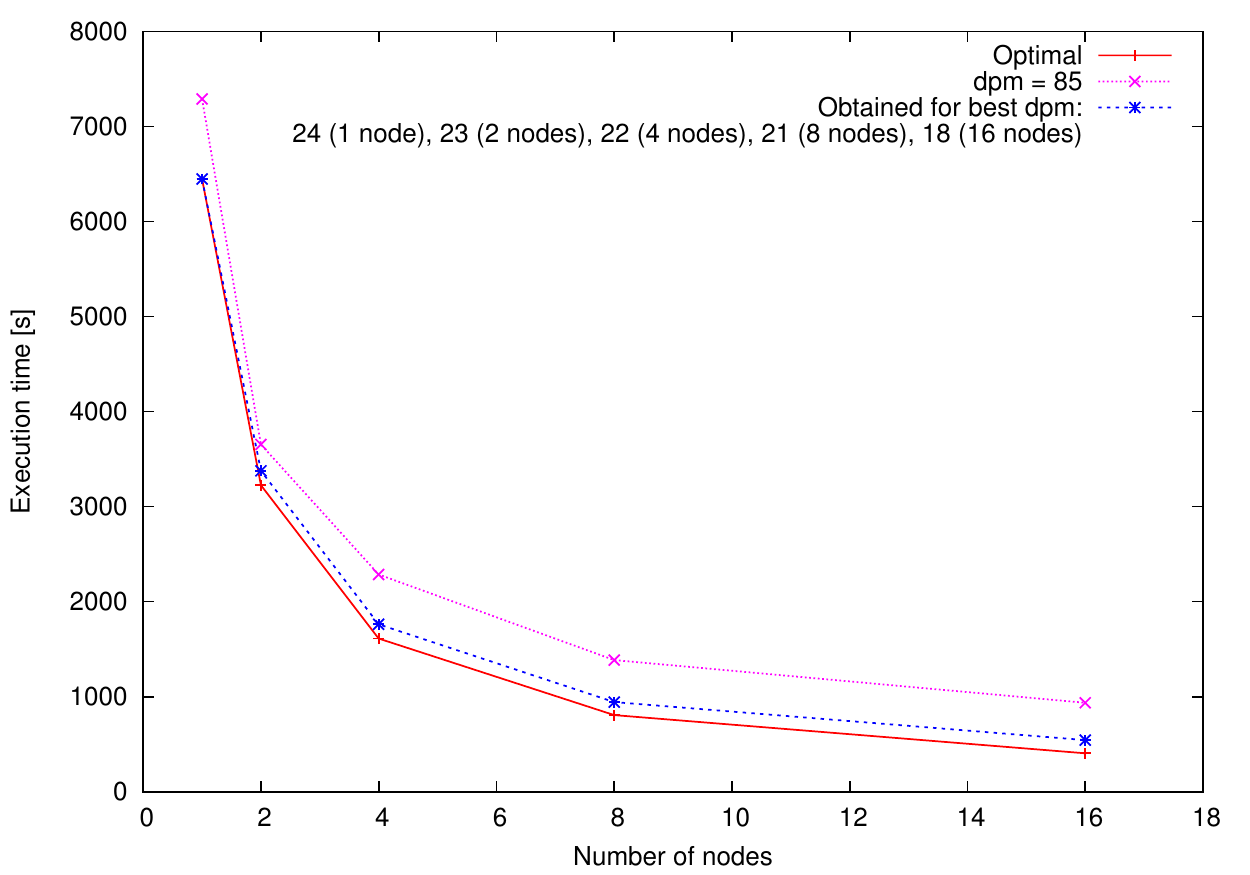}
\end{center}
\caption{Execution time of the password breaking application depending on $v_{nnodes}$ and $v_{dpm}$\label{fig:khtimemultipliers}}
\end{figure}

\begin{figure}[ht!]
\begin{center}
\includegraphics[scale=0.8]{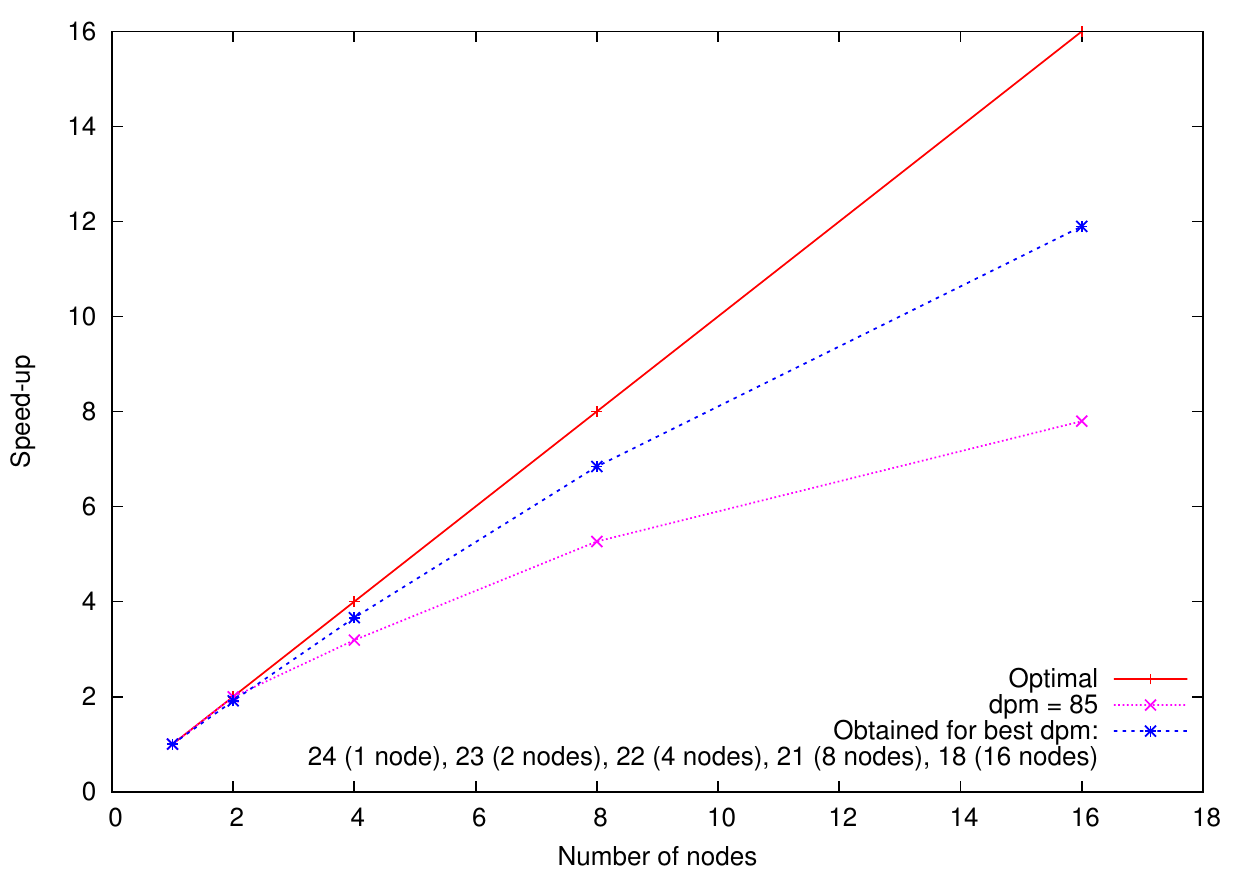}
\end{center}
\caption{Speedup the password breaking application depending on $v_{nnodes}$ and $v_{dpm}$\label{fig:khspeedupmultipliers}}
\end{figure}

The execution time is significantly higher for lower $v_{nnodes}$, but the
Figure shows another important fact: the optimal value of the \emph{application parameter}
$v_{dpm}$ depends on the value of an \emph{execution parameter} $v_{nnodes}$.
In particular, a local execution time minimum for a certain value of $v_{nnodes}$, 
can be a local maximum for another $v_{nnodes}$. For example, if only one node is computing
the application with $v_{dpm}$ set to the optimal value of 24, adding a second computing
node without changing $v_{dpm}$ does not improve the execution time due to load imbalance.
Because the local minima are increasing and local maxima are decreasing, a constant high value
of $v_{dpm}$ can be used to avoid extreme execution time values, regardless of the $v_{nnodes}$
value. However, this comes at a cost of unnecessary communication overheads.  
In order to execute the application efficiently, both $v_{dpm}$ and number of used nodes
have to be tuned together. Execution times (Figure \ref{fig:khtimemultipliers}) and speedups
(Figure \ref{fig:khspeedupmultipliers}) are compared depending on $v_{nnodes}$ in two cases:
using a constant high $v_{dpm} = 85$
and using an optimal $v_{dpm}$ value for each number of used nodes.
Although both approaches achieve good scalability, the second one results in significantly
lower execution times, close to theoretically ideal ones.

The \emph{parameter tuning} examples discussed in this section show
that proper tuning of both \emph{application parameters} and \emph{execution parameters}
has a significant influence on the application execution time.
Tuning this parameters is crucial for solving the problem of execution time optimization, defined in Equation \ref{eqn:problemtime} in Section \ref{sec:problemformulation}.  
This proves that the \emph{parameter tuning} step included in Claim \ref{clm:1} of
this dissertation allows to optimize the execution time of the considered application.
Simulation can be used to establish optimal parameter configuration in cases when the
dependencies between the parameters are non-trivial and an analytical formula 
for the optimal parameter values is unavailable.

\subsection{Actual Execution - KernelHive}\label{sec:khexecution}

In this section we describe how the execution configurations established by
aforementioned optimization steps are used in practice during \emph{actual execution} of 
a \emph{hybrid parallel application}.
Provided the application defined by the user as a \gls{dag}
with OpenCL kernel implementations corresponding to the graph nodes,
\emph{KernelHive} can perform the step of \emph{actual execution}
through distributing the appropriate kernels across the chosen nodes, dynamically compiling
them for the target platform and handling input and output data transfers.
Such execution can be achieved using a mix of existing APIs, for example
OpenCL + \gls{mpi}, but such approach would require programming expertise in these
APIs. Using \emph{KernelHive} simplifies
application development and provides automatic parallelization of the application
and efficient utilization of the available \gls{hpc} system.

\begin{figure}[ht!]
    \centering
    \begin{subfigure}[b]{0.35\textwidth}
        \centering
                \includegraphics[scale=0.5]{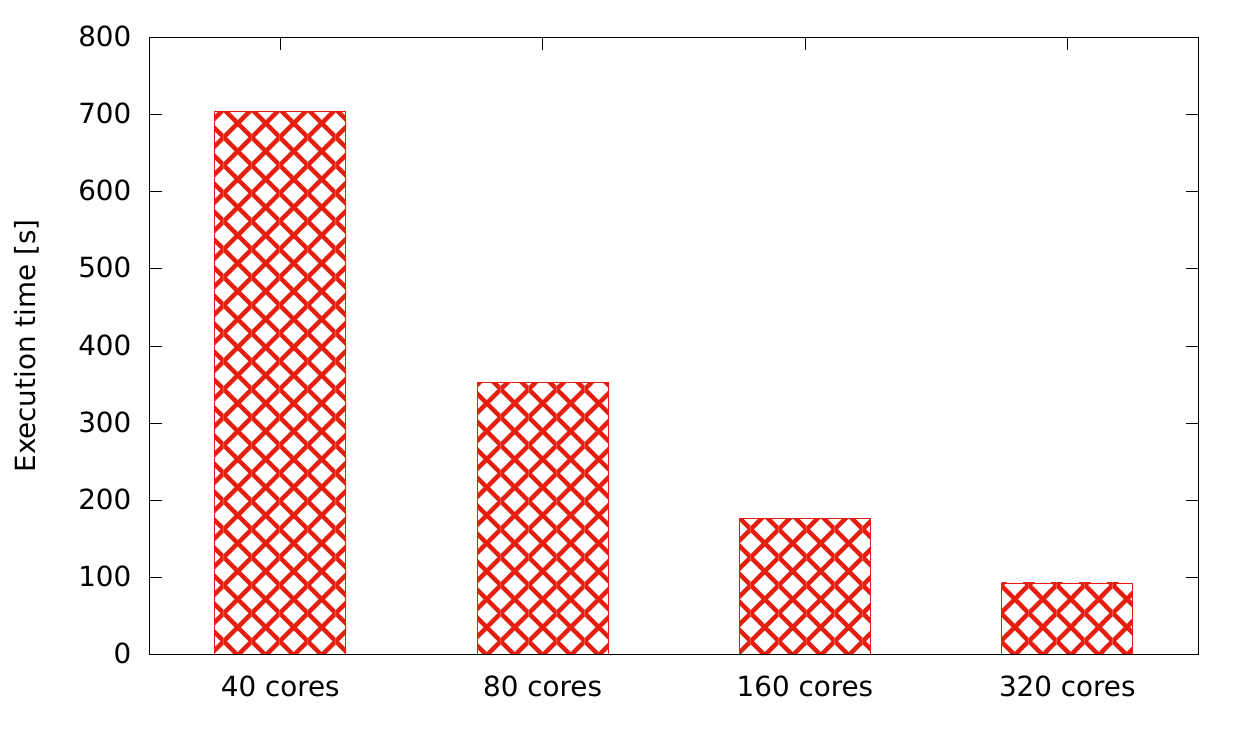}
                \caption{Geostatistical interpolation in a cluster with \gls{cpu}s (K2) \label{fig:khscalability}}
    \end{subfigure}
    ~
    \begin{subfigure}[b]{0.6\textwidth}
        \centering
                \includegraphics[scale=0.56]{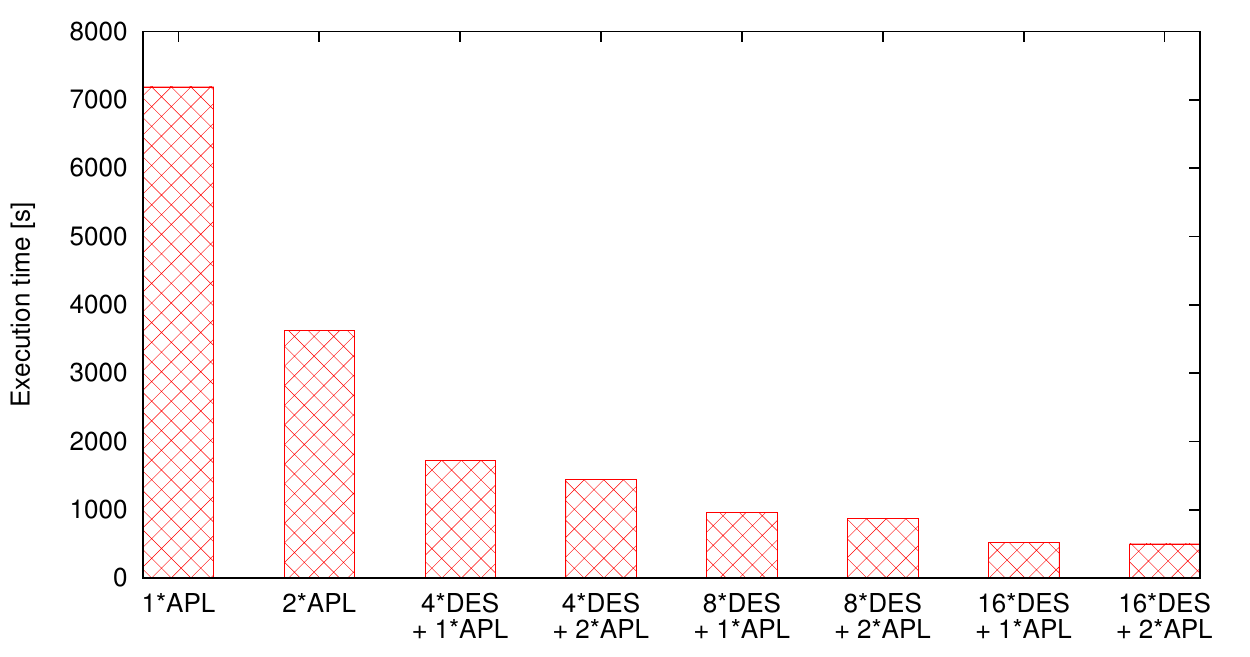}
        \caption{Password breaking in a heterogeneous system with \\ \gls{cpu}s and \gls{gpu}s (des + apl workstations) \label{fig:khheterogeneity}}
    \end{subfigure}
    \caption{Scalability of chosen task farming applications executed using \emph{KernelHive}}\label{fig:khscal}
\end{figure}

Figure \ref{fig:khscal} presents scalability of two chosen task farming applications
executed using \emph{KernelHive} on two different systems. Figure \ref{fig:khscalability}
shows good scalability of the geostatistical interpolation application described in Section
\ref{sec:idw} executed in the K2 system described in Section \ref{sec:k2}
with the number of used cores exponentially increasing up to 320 (40 nodes with two 4-core \gls{cpu}s each).
Improvement of execution time of the task farming applications executed using \emph{KernelHive}
is also possible in the case of utilizing heterogeneous \gls{hpc} systems where 
significant differences in performance of the available computing devices are present. For example,
Figure~\ref{fig:khheterogeneity} shows execution times of the password breaking application
described in Section \ref{sec:md5}, searching through $5 * 10^{12}$ hashes, 
depending on the selection of used "des" and "apl" workstations described in
Sections \ref{sec:lab527} and \ref{sec:apl}, equipped with both \gls{cpu}s and \gls{gpu}s. 
Even though for this application
the \gls{cpu}s are significantly slower than the \gls{gpu}s \cite{rosciszewski_kernelhive:_2016},
using additional resources, up to 284 heterogeneous compute units in the largest case,
improves the execution time.

These results show that proper implementation of the \emph{actual execution} of \emph{hybrid parallel applications}
allows to reduce their execution time through adding more and more utilized computing devices.
This applies to applications that are parallel in both hybridity meanings: executed in multi-level systems combining
intra-node and inter-node parallelism, as well as in heterogeneous systems combining different models
of computing devices (in this case CPUs and GPUs). This proves the importance of the \emph{actual execution} step
included in Claim \ref{clm:1} of this dissertation.
 
\newpage
\section{Deep Neural Network Training on a Professional Cluster of Workstations With GPUs}\label{sec:training_case}

The case study described in this section concerns the deep neural network training
application described in Section \ref{sec:kaldi_training} executed in a professional
cluster of workstations with \gls{gpu}s, described in Section \ref{sec:voicelab_cluster}. 
The case study concerns performing all proposed execution steps on one application, and
the provided results support Claim \ref{clm:1} of this thesis.
As indicated in Table~\ref{fig:casestudytable}, the \emph{preliminary process optimization} 
execution step in this case involves
minimization of distribution overheads and overlapping of \gls{cpu} and \gls{gpu}
computations described in Section \ref{sec:hpcs}. Section \ref{sec:iccs} describes
how the \emph{process mapping} and \emph{parameter tuning} steps were performed simultaneously
using the simulation method proposed in Section \ref{sec:simulation} and its results support
Claim \ref{clm:2} of this thesis. Finally, actual execution
of the application using a hybrid Kaldi + \gls{mpi} solution is described in Section \ref{sec:kaldimpi}.

\subsection{Preliminary Process Optimization - Training and Data Preprocessing Overlapping}\label{sec:hpcs}

The considered application is based on Kaldi \cite{povey_kaldi_2011}, a free, 
open-source toolkit for speech recognition research. A crucial part of a speech
recognition solution is the acoustic model, trained on a corpus of transcribed
speech data. The aim of the training process is to achieve the best efficiency
of the acoustic model which requires running several training epochs.
An epoch consists of such a number of iterations that every example from the training
set is used.
In each iteration the backpropagation algorithm \cite{rumelhart_learning_1986}
is used to train the model using a certain number of audio frames,
which is an \emph{application parameter} $v_{fpi}$ (frames per iteration).
Due to the used method of parallel training with model averaging \cite{povey_parallel_2014},
$v_{fpi}$ influences both final acoustic model efficiency and training time.
In this method, multiple copies of the model are trained in parallel on different 
parts of the training set and at the end of each iteration the parameters of the
model copies are averaged. Each iteration of the parallel training using Kaldi consists
of the following steps:

\begin{enumerate}
\item distributing and launching the training jobs;
\item reading the model from a network filesystem by each training job;
\item reading and preprocessing on CPU the training data from the local filesystem by each training job;
\item performing the training on a \gls{gpu} by each training job on the current data;
\item writing the partial model by each training job to the network filesystem;
\item launching the averaging job;
\item reading model from the network filesystem by the averaging job;
\item averaging the models on a \gls{cpu};
\item saving the averaged model to the network filesystem.
\end{enumerate}

For the experiments described in this section, the training data was a set
of randomly selected 100 hours (around 90k utterances) 
from the internal VoiceLab.ai corpora of Polish speech data.
The used \gls{hpc} system consisted of "cuda5" and "cuda6" workstations
described in Section \ref{sec:voicelab_cluster} and the number of
used \gls{gpu}s was gradually increased from 4 to 8 throughout the training.

In order to perform the first of the proposed execution steps,
\emph{preliminary process optimization}, the baseline version of the application
has been profiled and three most
time consuming \emph{operations} have been extracted. First,
the \emph{communication operation} of distributing and launching the
training programs, because at each time it required using
Sun Grid Engine for running a binary on the target machine and allocating a \gls{gpu}.
Secondly, the \emph{computation operation} of preprocessing (on CPU) the training data from the local filesystem,
because it included copying, shuffling and merging the training examples.
Finally, the \emph{computation operation} of actual training using the backpropagation algorithm (on GPU).
The other steps were negligible in terms of execution time. The models
weighed 27MB each, so reading and saving them did not introduce significant overhead.
Neither did launching and executing the averaging program, because it required only
a few weight scaling and adding operations on the \gls{cpu}.

The training program in Kaldi consists of a set of Bash scripts, which perform consecutive
steps by launching appropriate binaries and ensuring data flow between them.
In the cases when the binaries need to be distributed across available hardware,
Sun Grid Engine (\gls{sge}) \cite{gentzsch_sun_2001} is used, which introduces overhead of launching, managing and
queuing of the distributed job binaries. Additionally, each training job introduces
overhead of allocating the \gls{gpu} device. In order to minimize these overheads,
a modified, \gls{mpi}-based training program has been contributed by the author of this thesis
in \cite{rosciszewski_minimizing_2017}.
The application consists of two \emph{processes}: one \emph{master} process and
a number of \emph{slave} processes equal to the final, maximum number of used \gls{gpu}s.
The \emph{communication operation} of job distribution and
\gls{gpu} allocation is performed only once, at the beginning of the training.
Between the consecutive iterations, only sending the model back and forth
between the \emph{master} and the \emph{slaves} is done, which is a significantly
less time consuming \emph{communication operation}.

Since the two \emph{computation operations} in the baseline version of the application
(training data preprocessing and actual training) are executed on different devices
in the heterogeneous computing system (\gls{cpu} and \gls{gpu} respectively), there is a potential
for overlapping these \emph{operations}. The \gls{mpi} version of the training program proposed
in \cite{rosciszewski_minimizing_2017} has been extended with such overlapping.
In this version, the following steps are performed at the beginning of the training:

\begin{enumerate}
\item \emph{master} and \emph{slave} jobs are distributed and launched using \gls{mpi};
\item \emph{master} distributes the archive numbers for the first iteration through message passing across the \emph{slaves};
\item data reading and preprocessing is executed by each \emph{slave} in a separate \gls{cpu} thread.
\end{enumerate}

\noindent Then, each iteration consists of the following steps:

\begin{enumerate}
\item \emph{master} distributes the model and training parameters for the current iteration and archive numbers for the next iteration through message passing across the \emph{slaves};
\item \emph{slaves} perform in parallel:
\begin{itemize}
\item training on a \gls{gpu} on the previously loaded data;
\item reading and preprocessing the next archive of training data from the local filesystem in a separate \gls{cpu} thread;
\end{itemize}
\item sending the model back to the \emph{master} by each \emph{slave} through message passing;
\item averaging the models by the \emph{master};
\item saving the averaged model to the network filesystem.
\end{enumerate}

\begin{figure}[ht!]
\begin{center}
\includegraphics[scale=0.7]{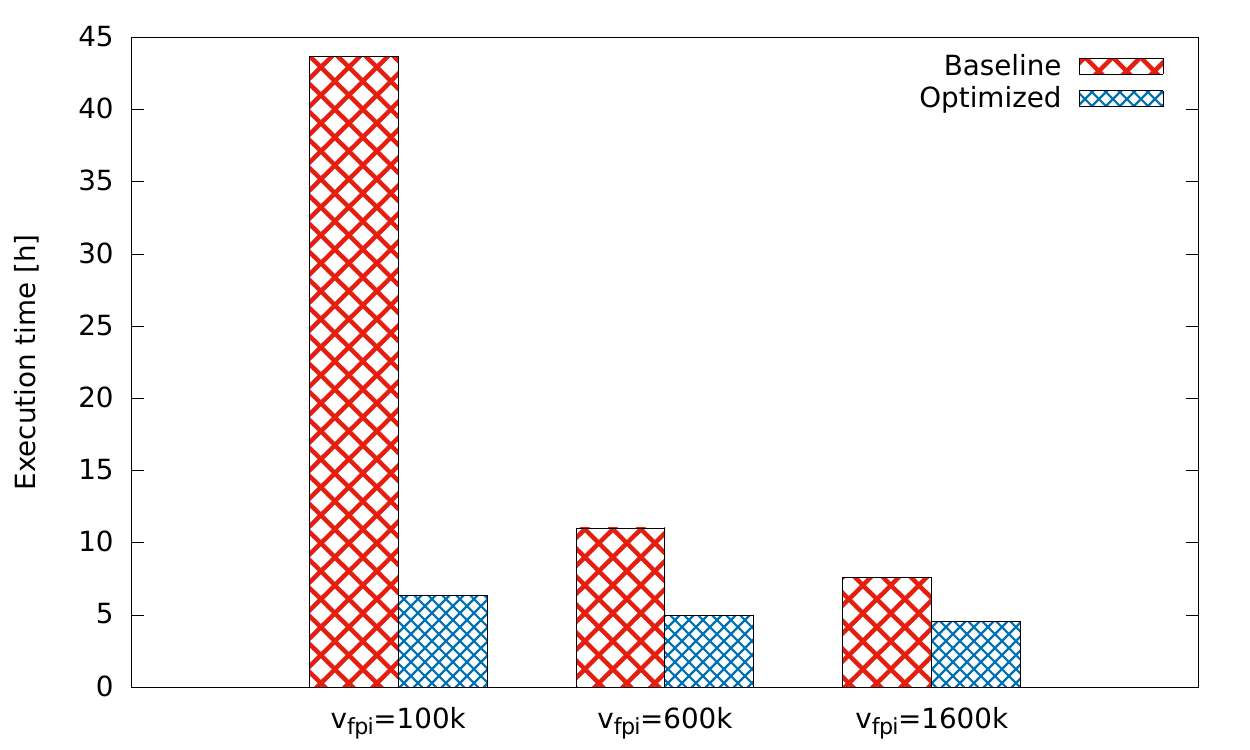}
\end{center}
\caption{15 training epochs execution time comparison depending on the used version of the training program
for three chosen values of $v_{fpi}$\label{fig:sgempiperf}}
\end{figure}

Figure \ref{fig:sgempiperf} presents a comparison of execution times of 15 training epochs
depending on three chosen values of $v_{fpi}$, for the baseline and the optimized version
of the application.

The following facts can be concluded from the comparison.
First, the lower value of $v_{fpi}$, the higher the execution time. This is
due to the communication overheads. Low value of $v_{fpi}$ means that a small 
number of audio frames is processed per each iteration, so the model averaging
is performed very often, which increases the communication overheads. Analogously,
the lower value of $v_{fpi}$, the higher the benefits of the proposed optimizations:
comparing to the baseline version, the optimized version reduces the execution
time by 85.5\% for $v_{fpi} = 100k$, by 55\% for $v_{fpi} = 600k$ and by 40\% for $v_{fpi} = 1600k$.

If the $v_{fpi}$ parameter had influence only on the training performance, the most
desirable value would be the highest possible, so that the model averaging would be
done only once. In such case, the benefits from the proposed optimization would be negligible.
However, model averaging is an important part of the parallel training algorithm
which influences the performance of the final model measured by the word error rate (\gls{wer})
metric which is the edit distance between a reference word sequence and its automatic transcription.
The influence of $v_{fpi}$ on the efficiency of the trained model has been examined by the author of this thesis
in \cite{rosciszewski_minimizing_2017}, and out of several considered values, 600k turned
out to be the best. In machine learning applications the efficiency of the trained 
model is of highest priority and cannot be traded off for computation efficiency. This
is the motivation for the proposed optimizations. The execution times for 
$v_{fpi}$ values of 100k and 1600k are presented as extreme values illustrating
the influence of the proposed optimizations but the $v_{fpi}$ parameter is not
subject to the \emph{parameter tuning} step, and the true outcome of the proposed
optimizations is execution time reduction by 55\%. It is a significant improvement,
because in practice it allows to try out twice as many machine learning hypotheses in
the same time period. In the case of running a single training, reducing the execution time
from over 11 hours to 4 hours and 58 minutes allows to obtain the
trained model and draw conclusions during the same working day. It should be noted
that the main influence (47.7\%) on this significant execution time reduction was contributed by 
the elimination of redundant initialization operations, which was necessary to implement the
overlapping. The overlapping itself resulted in relative execution time reduction by 
13.6\% \cite{rosciszewski_minimizing_2017} which is still a significant improvement.

\subsection{Process Mapping and Parameter Tuning - MERPSYS Simulation}\label{sec:iccs}

In the deep neural network training case study we used the simulation approach
proposed in Section \ref{sec:simulation} to simultaneously perform the 
\emph{process mapping} and \emph{parameter tuning} execution steps 
proposed in Section \ref{sec:methodology}. Evaluating each execution configuration
through actual execution of the application would take from 10 to 45 hours depending on the configuration,
so testing multiple configurations would be prohibitively long.
The proposed modeling and simulation approach allowed exploring power/time trade-off of the application
and was the main contribution of the author of this thesis in \cite{rosciszewski_modeling_2017}.
The considered optimization problem was multi-objective optimization of execution time and power
consumption, defined in Section \ref{sec:problemformulation} in Equation \ref{eqn:problem}.
The decision space of the \emph{parameter tuning} step consisted of the number of used \gls{gpu}
devices and thus, it was directly connected with \emph{process mapping}.
From the practical viewpoint, finding the set of Pareto-optimal solutions in this multi-objective
optimization problem helps engineers responsible for running the training to choose the trade-off between performance
and power consumption. Having access to a set of computing devices, they can decide
if in the given circumstances the computations should finish as soon as possible or if it
is acceptable that they run longer but consume less resources, which could be used more
efficiently for other applications or stay idle in order to save energy. 

In order to perform the simulations, an \emph{application model} 
$\gls{app} = \langle \gls{impl}, \gls{minrequirements}, \gls{maxrequirements} \rangle$
has been developed in MERPSYS (see Section \ref{sec:simulation}).
The modeled application was the one described in Section \ref{sec:kaldi_training} in the
optimized master/slave version described in Section \ref{sec:hpcs}.
A set of two \emph{process implementations} $\gls{impl} = \{\gls{process}^{master}, \gls{process}^{slave} \}$
has been implemented in the proposed Java based meta-language.
The $\gls{process}^{master}$ process is responsible for orchestrating the training through
distributing and averaging the neural network models and ordering data archive numbers,
and the $\gls{process}^{slave}$ process is responsible for performing the model optimization.
The modeled \emph{process implementations} reflect the code of the real application where
chosen fragments are replaced by API calls representing \emph{operations}.
Three \emph{operations} are used in the application model: 
\begin{itemize}
\item $\gls{comp}_{train}$ - a \emph{computation operation} representing an iteration of deep neural network training on a \gls{gpu};
\item $\gls{comm}_{send}$ - a \emph{communication operation} representing sending the neural network model;
\item $\gls{comm}_{recv}$ - a \emph{communication operation} representing receiving the neural network model. 
\end{itemize}

The $\gls{process}^{master}$ process runs multiple iterations of $\gls{comm}_{send}$ and
$\gls{comm}_{recv}$ \emph{operations} for each instance of the $\gls{process}^{slave}$ process. 
The $\gls{process}^{slave}$ process consists of a loop that consists of thee \emph{operations}:
$\gls{comm}_{recv}$ and $\gls{comm}_{send}$ \emph{communication operations} and a $\gls{comp}_{train}$
operation between them.
 
In the experiments discussed in this section, the \emph{process requirements} for 
$\gls{process}^{master}$ were set to $\gls{minrequirement}^{master} = \gls{maxrequirement}^{master} = 1$,
because there is always exactly one $\gls{process}^{master}$ process in the modeled application.
The \emph{process requirements} for $\gls{process}^{slave}$ were set to
$\gls{minrequirement}^{slave}~=~1, \gls{maxrequirement}^{slave} = 8$, so that at least
one $\gls{process}^{slave}$ process could be executed, but maximally 8 (the number
of \gls{gpu}s available in the system). The number of actually used \gls{gpu}s was correlated
with the $v_{nslaves}$ \emph{execution parameter},
which in this experiment was the only degree of freedom in the \emph{space of application
execution parameters} \gls{feasibleexecutionparameters} \cite{rosciszewski_modeling_2017}.

\begin{figure}[ht!]
\begin{center}
\includegraphics[scale=0.22]{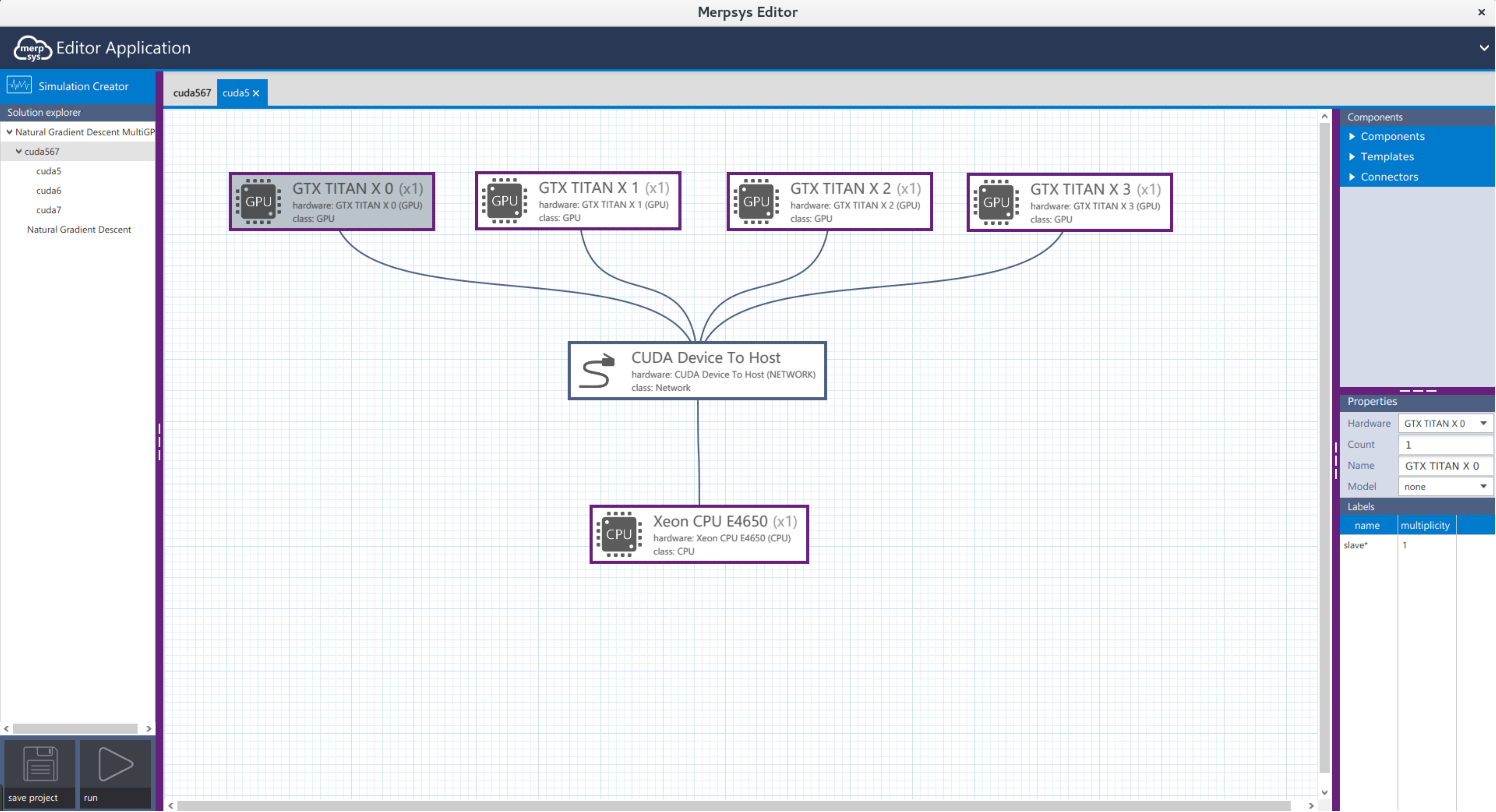}
\end{center}
\caption{Screenshot of the System Model in the MERPSYS Simulation Environment\label{fig:msscreen1}}
\end{figure}

As shown in Figure \ref{fig:msscreen1}, the graphical \emph{Editor} tool from the MERPSYS environment has been used to prepare the
\emph{system model} resembling the real hardware configuration used at the VoiceLab.ai company.
The \emph{heterogeneous HPC system} $\gls{system}(\gls{devices}, \gls{networkset})$ consisted of 2 workstations connected
through a Gigabit Ethernet interconnect, each containing of 4 GeForce GTX TITAN X \gls{gpu} \emph{computing devices}
connected with a Xeon \gls{cpu} through an artificial interconnect called
"CUDA Device To Host", representing the CPU-GPU interconnect.
The \emph{hardware capabilities} function allowed to run a \emph{master} process on one \gls{cpu} device
\big($\gls{hardwarecapabilities}(\gls{device}_{\gls{cpu}\_0}, \gls{process}^{master}) = 1$\big)
and one \emph{slave} process on each \gls{gpu} device:
\big($\forall_{i} \gls{hardwarecapabilities}(\gls{device}_{\gls{gpu}\_i}, \gls{process}^{slave}) = 1$\big).

The function modeling \emph{computation time} for the $\gls{comp}_{train}$ \emph{computation operation}
on each $\gls{device}_{\gls{gpu}\_i}$ device is a linear function
$\gls{comptime}(\gls{comp}_{train}, \gls{device}) = \varphi \cdot
\frac{dataSize(\gls{comp}_{train})}{performance(\gls{device})} + \psi$ where $\varphi$
and $\psi$ are tunable constants, $dataSize(\gls{comp})$ is an
\emph{operation parameter} denoting size in bytes of the processed data package and $performance(\gls{device})$
is a \emph{hardware parameter} denoting performance of the used \emph{computing device}.
The \emph{communication time} function modeling the $\gls{comm}_{send}$ and $\gls{comm}_{receive}$ operations,
is a linear function $\gls{comptime}(\gls{comm}, \gls{networklink}) = tstartup(\gls{networklink}) + \frac{dataSize(\gls{comm})}{bandwidth(\gls{networklink})}$,
where $tstartup(\gls{networklink})$ and $bandwidth(\gls{networklink})$ are \emph{hardware parameters}
denoting startup time and bandwidth of a network link, and $dataSize(\gls{comm})$ is \emph{operation parameter}
denoting the size of the transferred data package.
For the \gls{gpu}s utilized by the \emph{computation operation}, power consumption was set
to approximate values measured for real GeForce GTX TITAN X \gls{gpu}s through the nvidia-smi tool:
$\forall_{i} \big( \gls{pcidle}(\gls{device}_{\gls{gpu}\_i}) = 70W, \gls{pcpeak}(\gls{device}_{\gls{gpu}\_i}) = 140W\big)$.
Analogously to the real application, the model assumes that a \gls{gpu} is either fully utilized or idle,
so the power consumption scaling according to number of threads is not used in this experiment and
$\forall_{i} \big(\gls{ncores}(\gls{device}_{\gls{gpu}_i}) = 1\big)$.

\begin{figure}[ht!]
    \centering
    \begin{subfigure}[b]{0.46\textwidth}
        \includegraphics[width=\textwidth]{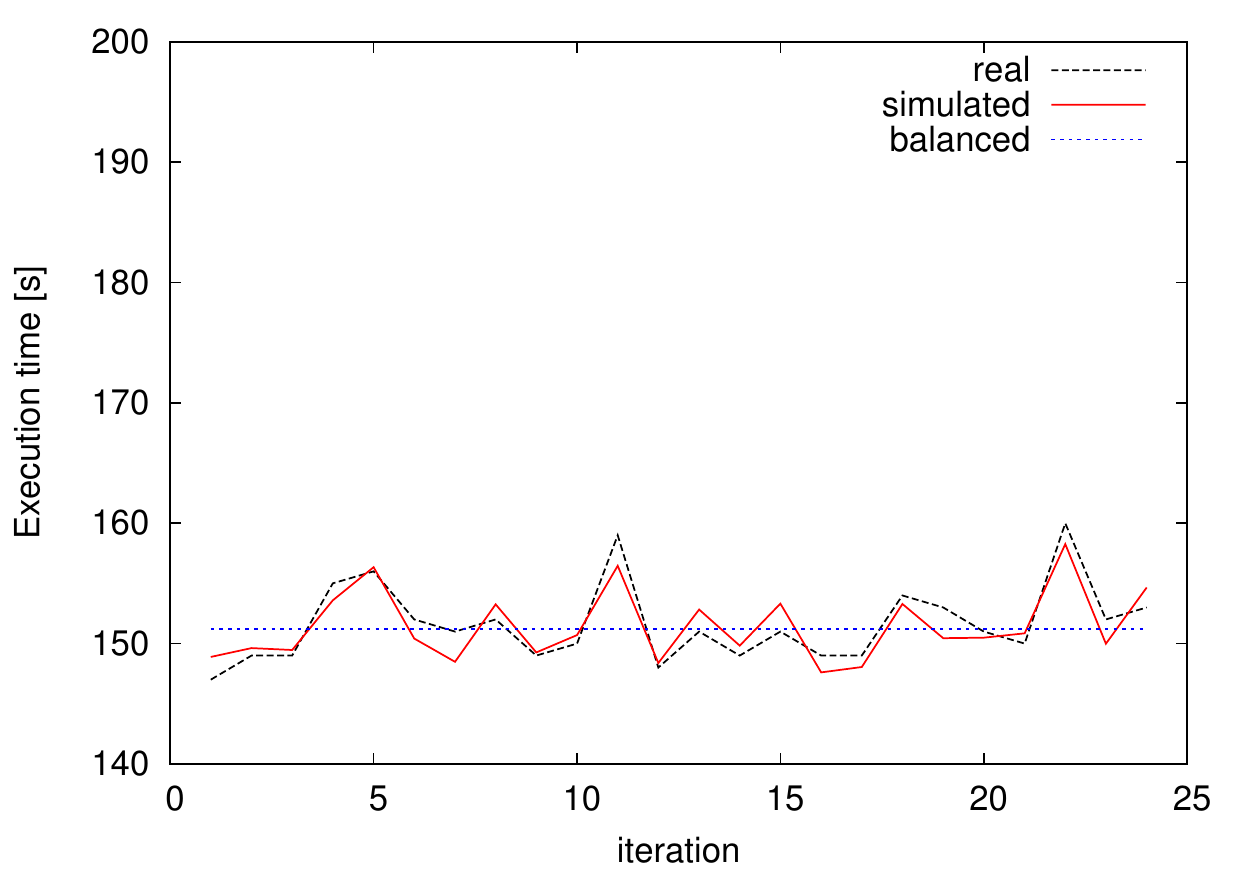}
        \caption{1 slave}
        \label{fig:t1}
    \end{subfigure}
    ~ 
    \begin{subfigure}[b]{0.46\textwidth}
        \includegraphics[width=\textwidth]{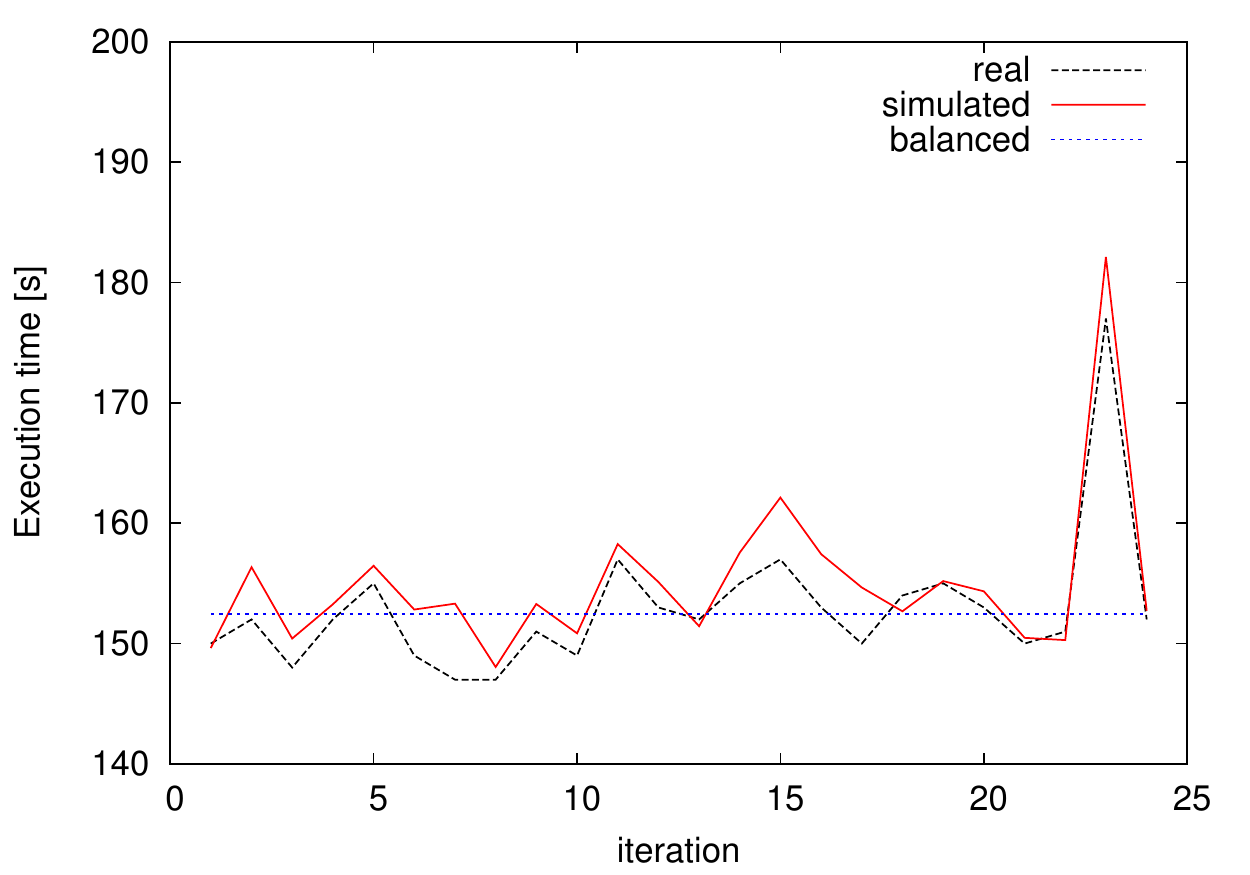}
        \caption{2 slaves}
        \label{fig:t2}
    \end{subfigure}

    \begin{subfigure}[b]{0.46\textwidth}
        \includegraphics[width=\textwidth]{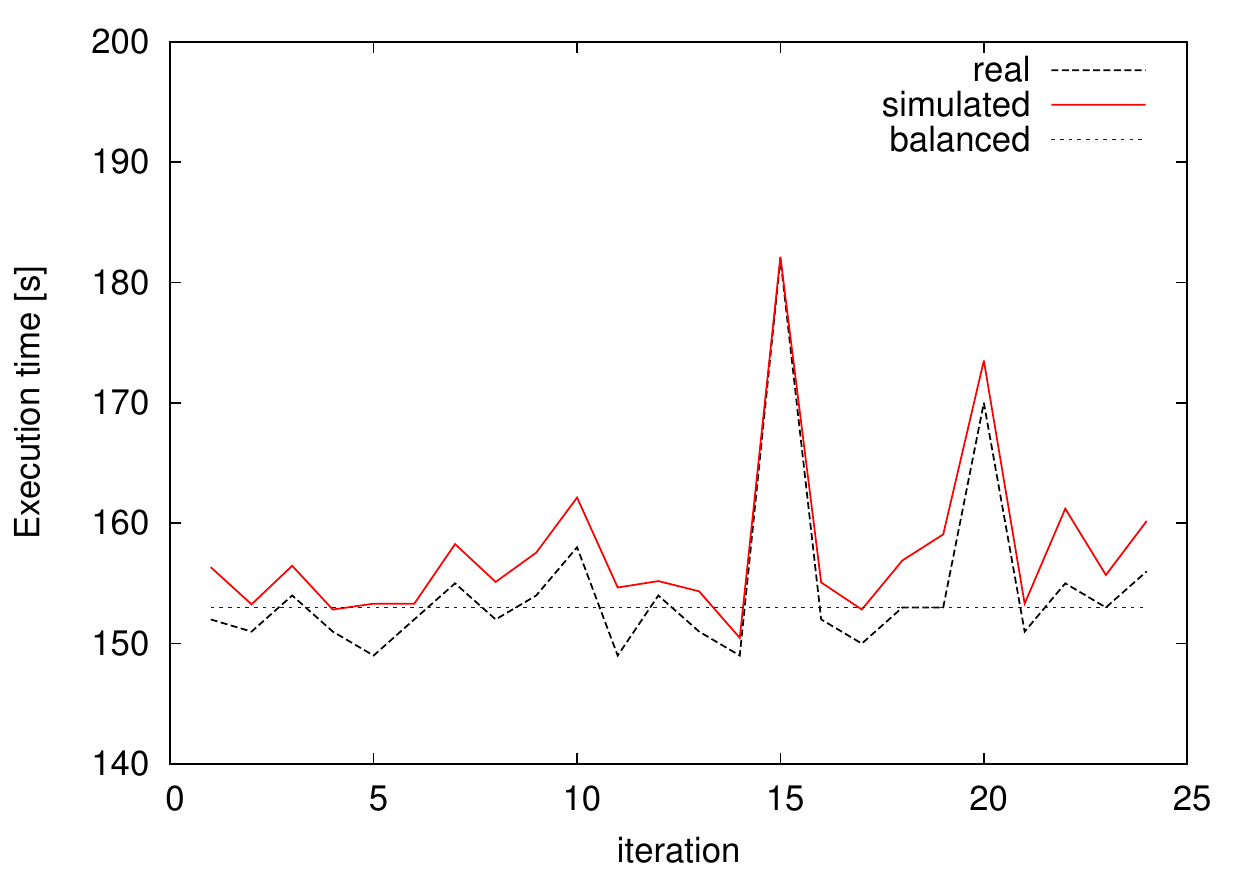}
        \caption{3 slaves}
        \label{fig:t3}
    \end{subfigure}
    \begin{subfigure}[b]{0.46\textwidth}
    \includegraphics[width=\textwidth]{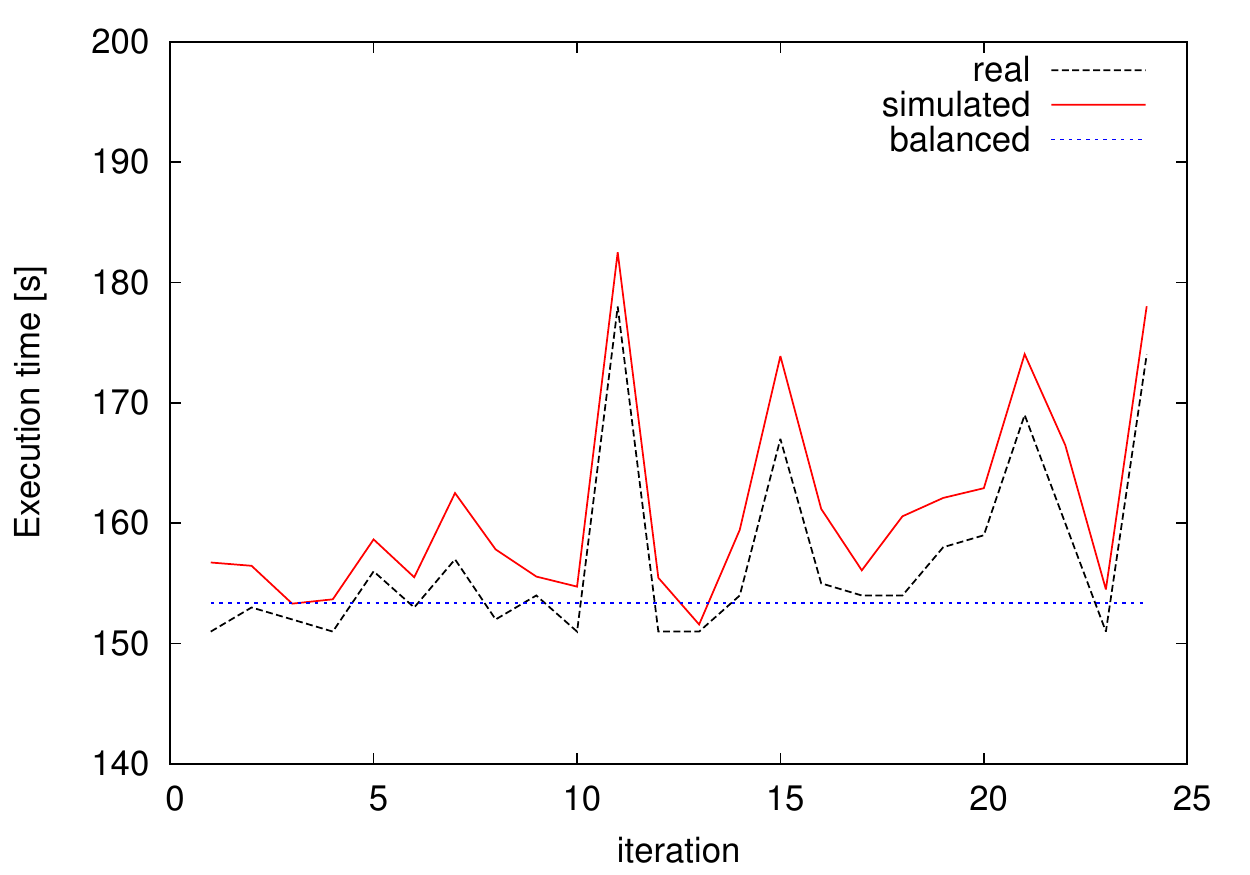}
    \caption{4 slaves}
    \label{fig:t4}
    \end{subfigure}
    
    \begin{subfigure}[b]{0.46\textwidth}
        \includegraphics[width=\textwidth]{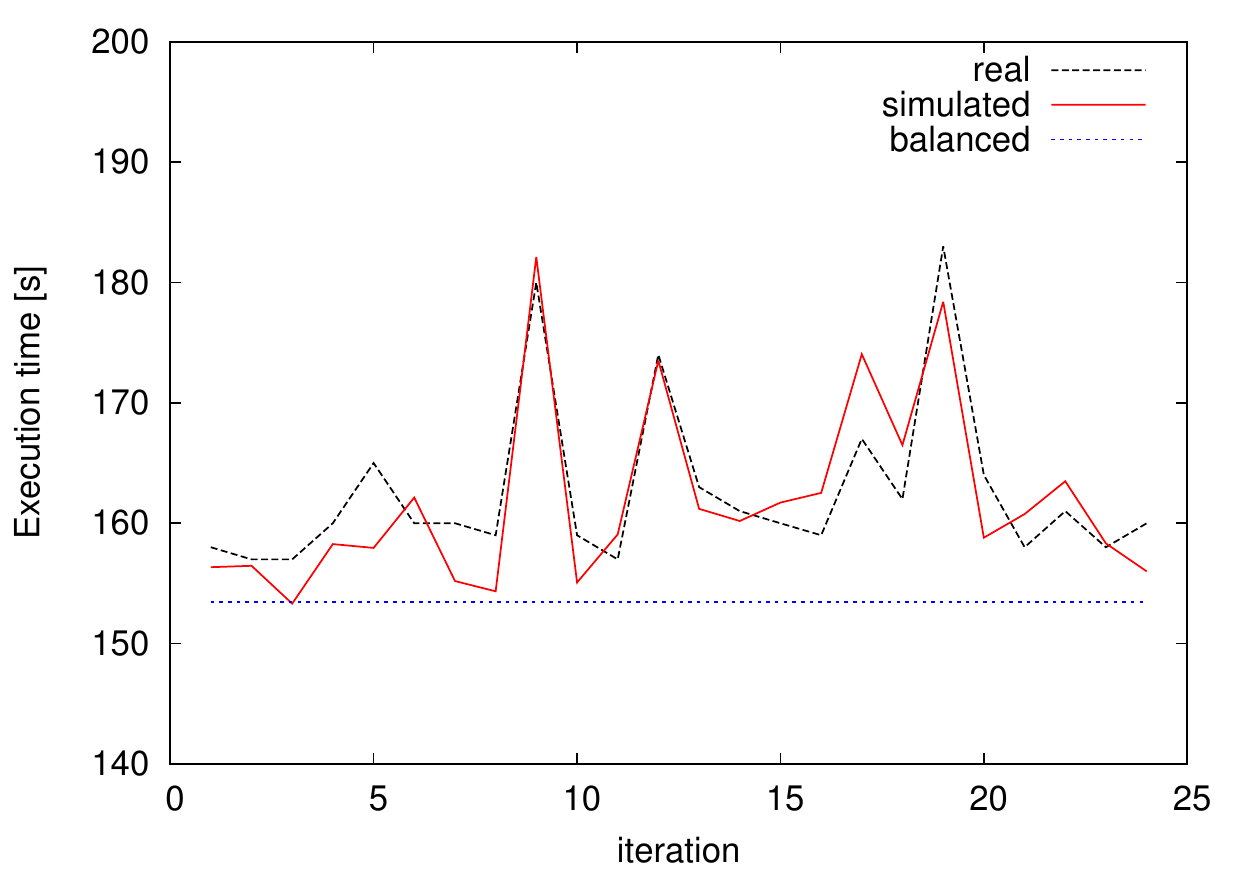}
        \caption{5 slaves}
        \label{fig:t5}
    \end{subfigure}
    ~ 
    \begin{subfigure}[b]{0.46\textwidth}
        \includegraphics[width=\textwidth]{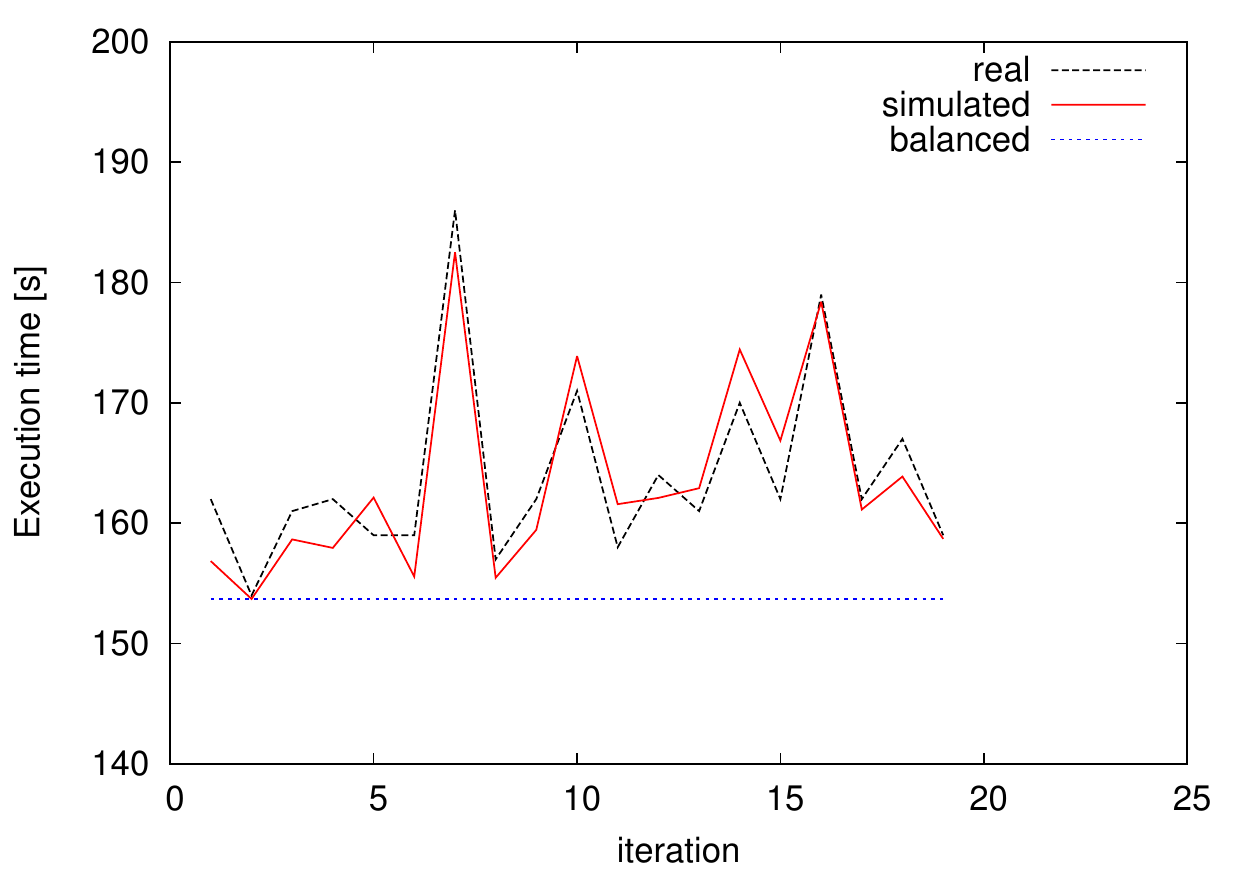}
        \caption{6 slaves}
        \label{fig:t6}
    \end{subfigure}

    \begin{subfigure}[b]{0.46\textwidth}
        \includegraphics[width=\textwidth]{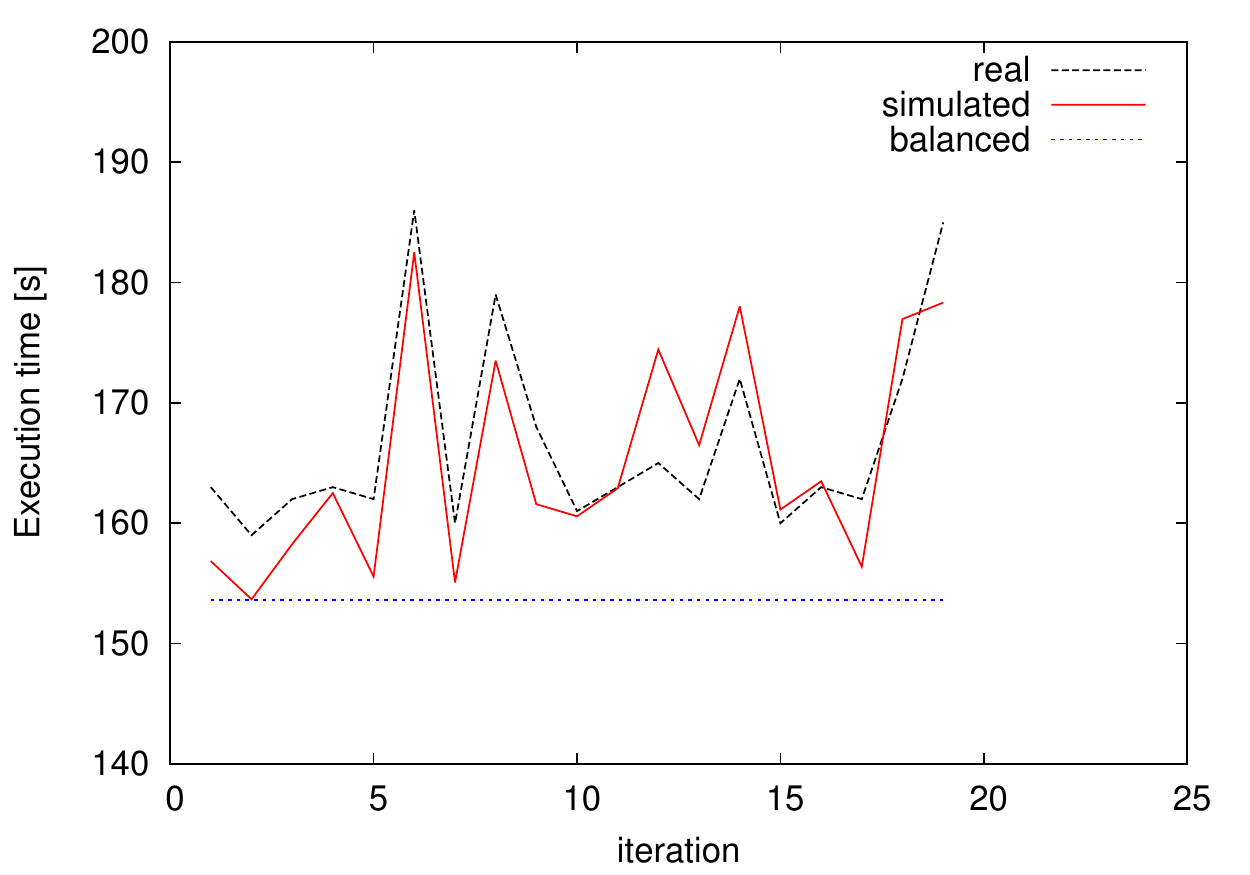}
        \caption{7 slaves}
        \label{fig:t7}
    \end{subfigure}
    \begin{subfigure}[b]{0.46\textwidth}
    \includegraphics[width=\textwidth]{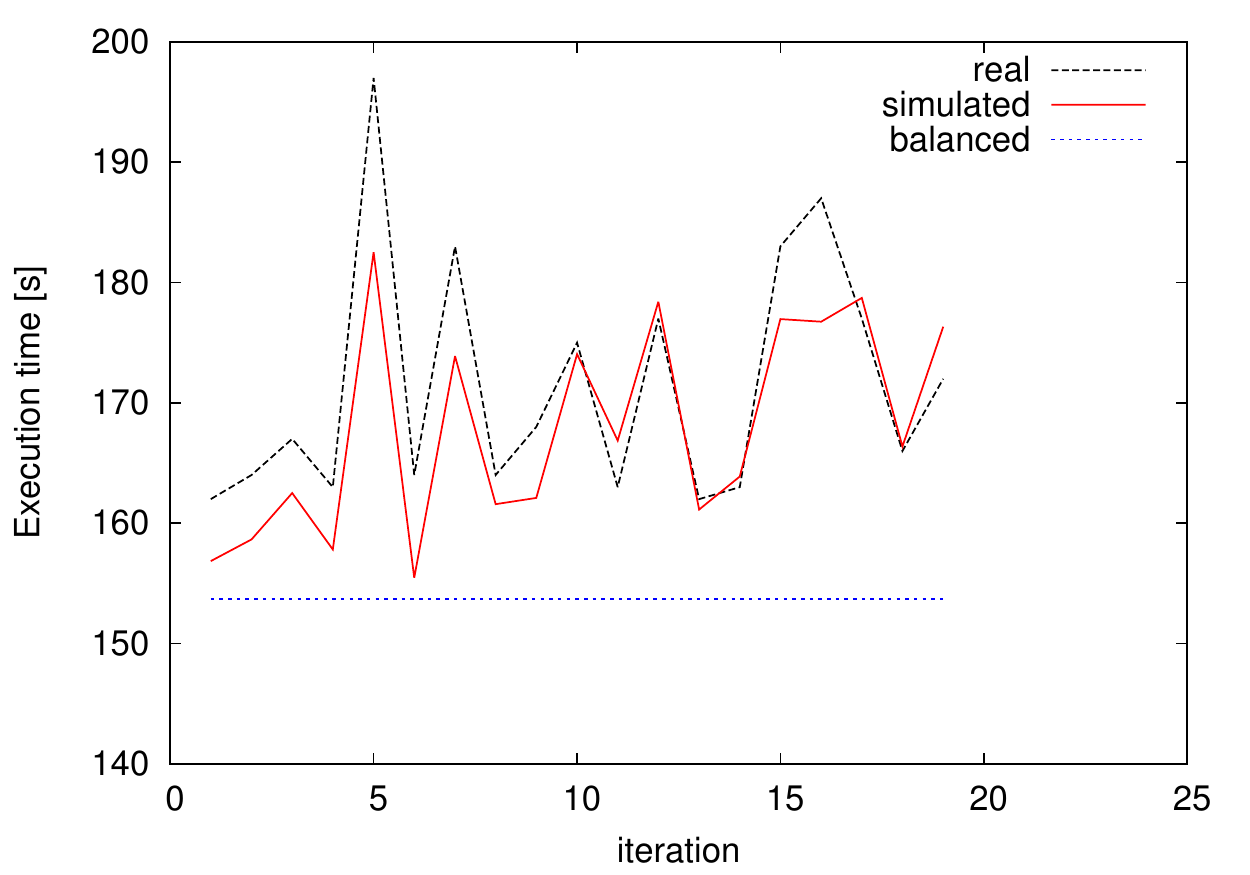}
    \caption{8 slaves}
    \label{fig:t8}
    \end{subfigure}    
    \caption{Execution times of consecutive iterations depending on $v_{nslaves}$ (number of $\gls{process}^{slave}$ process instances) in the deep neural network training application.}\label{fig:tuningperf}
\end{figure}

The coefficients $\psi$ and $\varphi$ in the \emph{computation time} function
have been found using ordinary least squares regression to results from real
executions of the training application on one \gls{gpu}. Execution times of the
training \emph{computation operation} have been measured and averaged from four
runs, for 25 consecutive iterations. As shown in Figure \ref{fig:tuningperf}, the
model tuned on the results from executions with $v_{nslaves} = 1$
is accurate also in the case of multiple instances of $\gls{process}^{slave}$, with
mean percentage error up to 2.7\%. Consequently, the simulation results are reliable
and can be used for evaluating execution times of whole training epochs that consist
of hundreds of training iterations.

It should be noted that in the first attempt to tuning the \emph{computation time}
function, a different \emph{operation parameter} was used, namely the number of training
examples from the given archive used in the neural network training algorithm. However, in
that case the regression results did not show correlation between this \emph{operation parameter}
and execution time of an iteration, while in the case of data size in bytes, the correlation
was present. Although the number of training examples used by the training algorithm was the same
across the archives, it appeared that the data preprocessing step  which shuffled all data in
the archive was also significant in terms of execution time. This way, the modeling process allowed
the author of this thesis to find a data imbalance bottleneck that the engineers using the
application were not aware of. Getting rid of this data imbalance bottleneck would require
significant engineering effort, so an estimate of the influence of this bottleneck on the
performance of the whole application was needed in order to make the decision if engineering
resources should be assigned to this task.

The simulations were repeated for a hypothetical case when the input data would be equally distributed across
the input packages. The execution times for this case is denoted in Figure \ref{fig:tuningperf} by dotted lines.  
In the case of $v_{nslaves} = 1$ the execution times with balanced data packages are equal to the mean of
execution times of the original application. However, the higher value of $v_{nslaves}$, the higher would
be the benefit from fixing the data package imbalance, because the more \gls{gpu}s have to sit idle until
the largest data package is processed.

\begin{figure}[ht!]
\begin{center}
\includegraphics[scale=0.4]{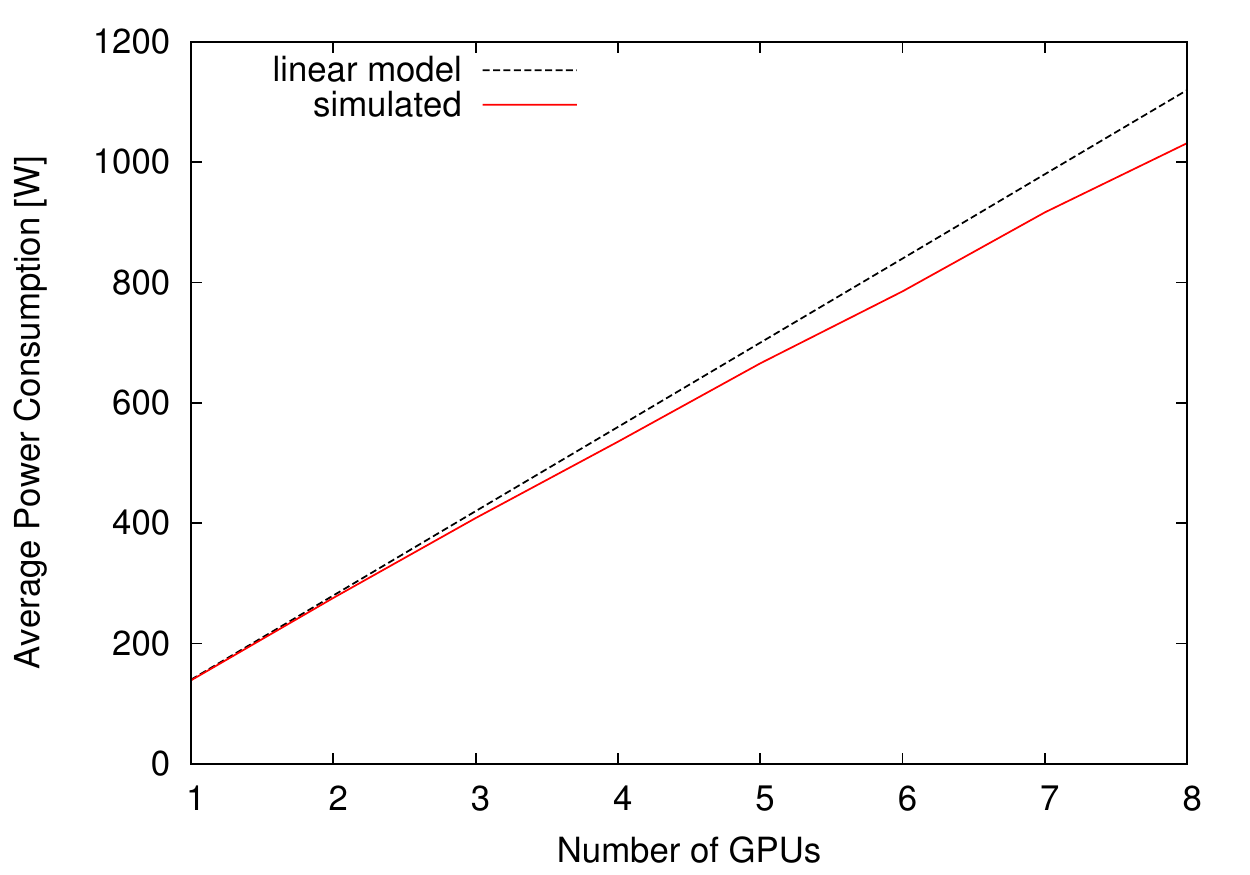}
\end{center}
\caption{Verification of the Simulated Average Power Consumption Estimations with a Linear Model\label{fig:tuningpower}}
\end{figure}

To verify the average power consumption estimations returned by the simulator, their values were
compared to a linear model that multiplies the number of used GPUs by the peak power consumption of a single
GPU, as shown in Figure \ref{fig:tuningpower}.
The simulation results are close to the linear model estimations. It should be noted that the simulated
values are getting lower than the ones based on the linear model with the increase of the number of 
utilized GPUs. This is because the simulator takes into account that the GPUs do not consume additional
energy during the idle time resulting from communication, which significance increases with the number of
simultaneously utilized devices.

Execution model tuned this way was used to explore the power/time trade-off of the application,
which, in terms of the notation proposed in Section \ref{sec:problemformulation}, means
using the Pareto method to solving the following optimization problem:

\begin{equation}
\begin{aligned}
& \underset{\gls{mapping}, \gls{executionparameters}}{\min}
& & \gls{paretofunctionvector}(\gls{app}, \gls{system}, \gls{mapping}, \gls{executionparameters}) 
= [\mathrm{\gls{executiontime}}(\gls{app}, \gls{system}, \gls{mapping}, \gls{executionparameters}), 
\mathrm{\gls{powerconsumption}}(\gls{app}, \gls{system}, \gls{mapping}, \gls{executionparameters})] \\
& \text{subject to}
& & \gls{mapping} \in \gls{mappingset} = \big\{\gls{mapping}(\gls{device}, \gls{process}) | \\
&&& \big( \gls{mapping}(\gls{device}_{CPU\_0}, \gls{process}^{master}) = 1
\big) \\
&&& \wedge \big( \forall_{i} \gls{mapping}(\gls{device}_{GPU\_i}, \gls{process}^{slave})< = 1
\big) \\
&&& \wedge \big( 1 <= (\sum_{i}{\gls{mapping}(\gls{device}_{GPU\_i}, \gls{process}^{slave})} = v_{nslaves}) <= 8 \big) \big\},
\\
&&& \gls{executionparameters} = [v_{nslaves} \in \mathbb{N}^{[1, 8]}].
\end{aligned}
\end{equation}

\begin{figure}[ht!]
\begin{center}
\includegraphics[scale=0.33]{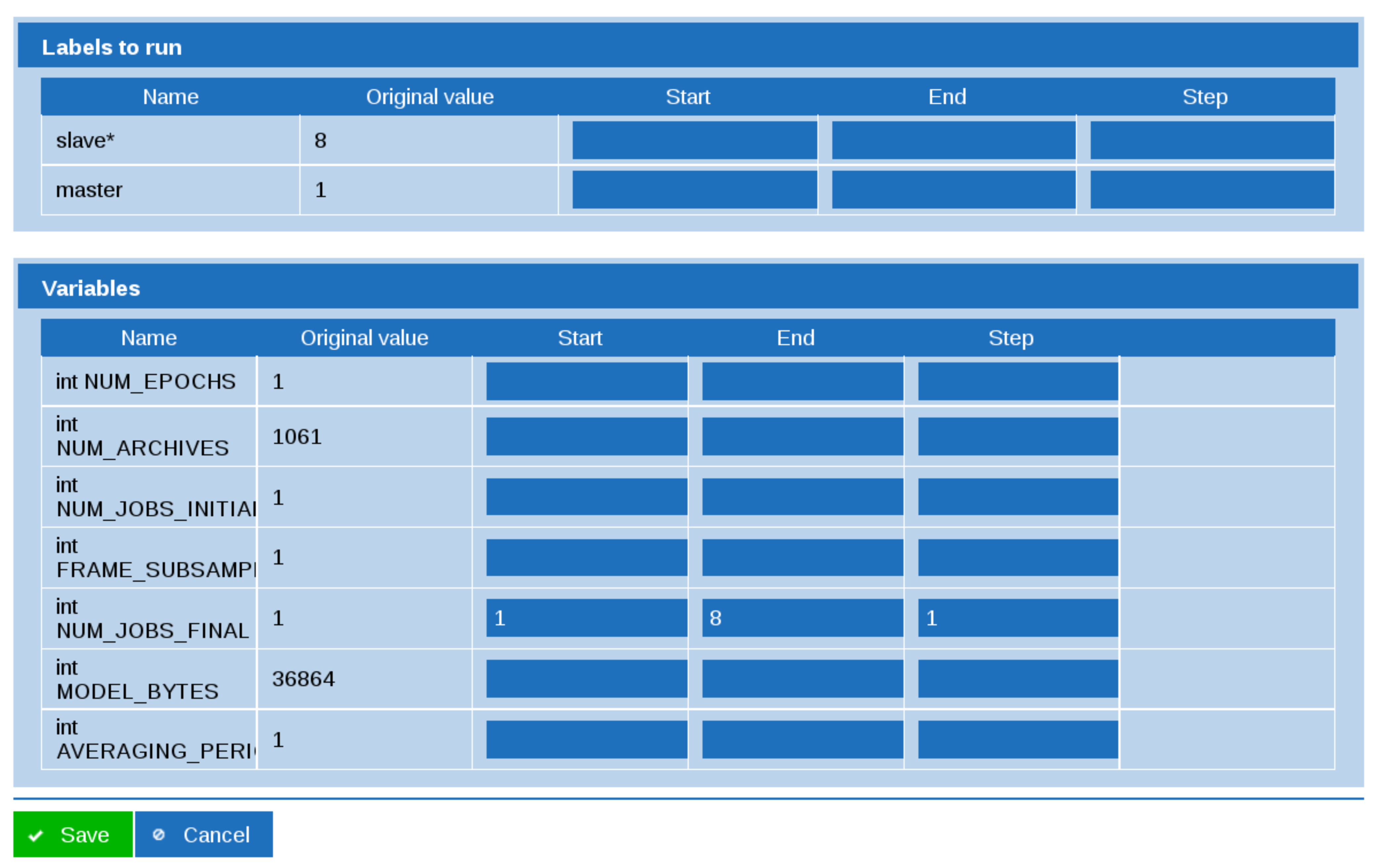}
\end{center}
\caption{Screenshot of the Optimizer Suite Definition Interface in the MERPSYS Simulation Environment\label{fig:msscreen2}}
\end{figure}

In order to do this, we first executed a simulation 
of one training epoch consisting of 1061 iterations, where only one \gls{gpu} took part in the computations. Then, as presented
in Figure \ref{fig:msscreen2}, the MERPSYS
Web interface was used to order an \emph{optimizer suite} based on the resultant single \emph{simulation instance},
allowing the $v_{nslaves}$ to vary from 1 to 8 with a step of 1.

We defined two such suites - "Suite 1" resembling the real application and "Suite 2" resembling the aforementioned version with equal data
distribution. We started the MERPSYS \emph{Optimizer}
component configured to run the \emph{ParetoVisualizer} module.
The \emph{Optimizer} enqueued all 8 feasible \emph{simulation instances} for each of
the suites and passed the results to the \emph{ParetoVisualizer}. 

\begin{figure}[ht!]
    \includegraphics[width=0.55\textwidth]{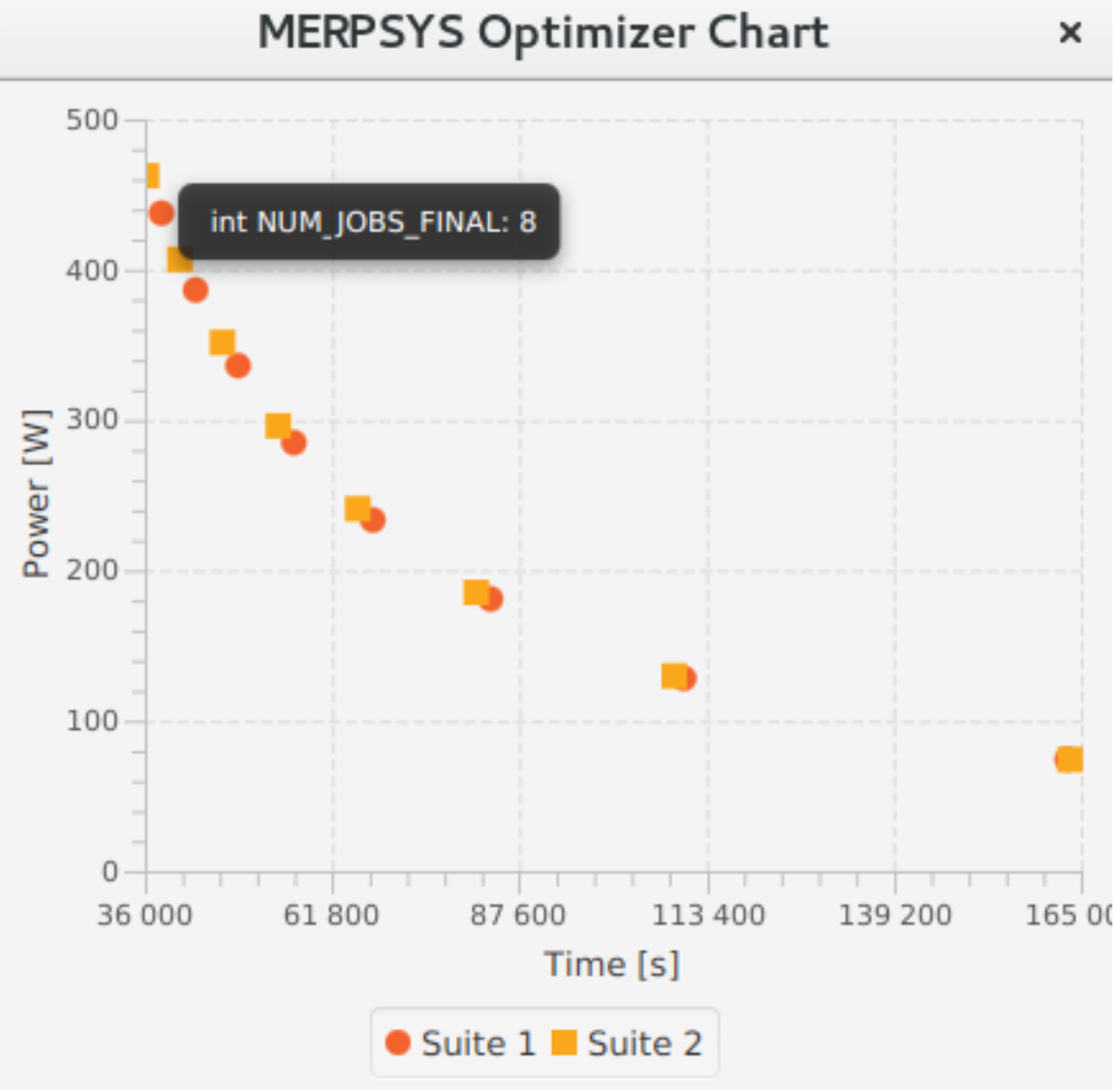}
    \caption{Screenshot from the MERPSYS \emph{ParetoVisualizer} after simulating Suite 1
    with real archive sizes and Suite 2 with hypothetical ideal load balancing.}
    \label{fig:fronts}
\end{figure}

\begin{figure}[ht!]

\begin{tabular}{@{\extracolsep{4pt}}crrrr@{}}
\firsthline
& \multicolumn{2}{c}{Actual} & \multicolumn{2}{c}{Balanced} \\
\cline{2-3} \cline{4-5}
$v_{nslaves}$ & Time (s) & Power (W) & Time (s) & Power (W) \\
\hline
1       & 163 528    & 74.189	  & 163 528 & 74.189    \\
2       & 110 282    & 128.267  & 108 965 & 129.804  \\
3       & 83 646     & 181.006  & 81 685  & 185.332  \\
4       & 67 425     & 233.532  & 65 348  & 240.939 \\
5		& 56 538     & 284.916  & 54 405  & 296.125  \\
6		& 48 846     & 336.171  & 46 699  & 351.602   \\
7		& 42 995     & 386.517  & 40 842  & 406.780  \\
8		& 38 286     & 437.595  & 36 219  & 462.498   \\
\lasthline
\end{tabular}

\caption{Simulated execution time and average power consumption values for the deep neural network training application, depending on $v_{nslaves}$ and presence of load balancing}
\label{fig:frontvals}
\end{figure}

A screenshot of the \emph{ParetoVisualizer} is shown in Figure \ref{fig:fronts}.
It presents a chart, where horizontal axis represents one optimization objective
(execution time) and the vertical axis the second objective (average power consumption).
Each point in the chart represents results for one \emph{simulation instance}.
The results depending on $v_{nslaves}$ are presented in Figure \ref{fig:frontvals}, both
for the actual and hypothetical balanced case.
In both cases, all points in the decision space belong to the Pareto set,
because each number of used \gls{gpu}s results in either lower power consumption
or lower execution time than all others.
Hovering with the cursor over a point, the user can see values of
$v_{nslaves}$ that were used in the given simulation.

It should be noted that total time of the simulations run on an Intel Core i7-4712HQ \gls{cpu}
used to draw the chart in Figure \ref{fig:fronts} was 2 hours and 10 minutes, while
real execution times of the 16 configurations vary from 10 to 45 hours, giving the
total execution time of 335 hours. This shows that the proposed simulation method can
predict the execution parameters in significantly shorter time than the proper application execution. 
This proves Claim \ref{clm:2} of this thesis.
Although preparing the model and performing the simulation requires a few hours of additional work, the model
can be useful for further evaluations in cases of other execution configurations, even currently
unavailable. An example of such an unavailable configuration is the hypothetical version of the application
with balanced load described in this Section, but evaluating different hardware configurations or 
\emph{application execution parameters} could also be a practical use case.

\subsection{Actual Execution - Hybrid Kaldi + MPI}\label{sec:kaldimpi}

The Kaldi framework used by the considered application consists of 
a collection of various speech recognition tools, among which there are
multiple versions of deep neural network training programs for acoustic modeling.
The program used originally by the considered application is called
"nnet3-ctc-train", because it uses the "nnet3" deep neural network setup and the 
Connectionist Temporal Classification \cite{graves_connectionist_2006}, 
loss function. The program utilizes massively parallel \gls{gpu} computations for
deep neural network training through a built-in matrix operation library called CUDA Matrix.
The program itself does not handle multiple computing devices, however the Kaldi framework
includes a collection of scripts that support queuing the training programs
using the \gls{sge} cluster resource management system. Multiple training
programs can be executed simultaneously on potentially heterogeneous computing devices
and a distributed filesystem is required to support data transfers between the programs.
This makes makes the solution a \emph{hybrid parallel application}, both in the sense
of multi-level and heterogeneous. 

Apart from the overhead of
\gls{gpu} initialization by each separate program described in Section \ref{sec:hpcs},
a disadvantage of the original version of the application 
is that it requires installing and configuring both the \gls{sge} cluster management
software and a distributed filesystem. In \cite{rosciszewski_minimizing_2017} the author
of this thesis contributed a modified implementation of the training program called "nnet3-ctc-train-mpi",
which mixes the Kaldi API with MPI and uses message passing for transferring the 
current neural network model data between distributed training programs, coordination
of consecutive training iterations and controlling the $v_{nslaves}$ parameter.
The proposed program is a hybrid Kaldi + MPI multi-level application that supports
overlapping of data preprocessing on \gls{cpu} and neural network training on \gls{gpu}.
Running the training program does not require additional effort of configuring
cluster management software or distributed filesystem. The proposed version
of the training program is available as open-source software
\footnote{https://github.com/roscisz/kaldi-mpi}.

The actual execution step in the case of \emph{hybrid parallel applications}, which are
often multi-level and executed on heterogeneous computing infrastructure, requires advanced,
efficient implementation. 
The proposed hybrid Kaldi + MPI implementation is compared to the original version in terms
of performance in Figure \ref{fig:sgempiperf} in Section \ref{sec:hpcs}.
This implementation is a practical example of using
the execution parameters, established within the optimization steps, during real
execution of the application in a professional cluster of workstations
with \gls{gpu}s. This supports the importance of the \emph{actual execution}
step included in Claim \ref{clm:1} of this dissertation.

\newpage
\chapter{Conclusions and Future Work}\label{chp:summary}

The goal of this thesis was to develop an optimization methodology
for \emph{hybrid parallel application} execution in \emph{heterogeneous
\gls{hpc} systems}. For this purpose, in Chapter \ref{chp:applicationstheory},
existing \emph{hybrid parallel
applications} have been investigated considering the meaning of
their hybridity, methods of their
execution, possible execution parameters and approaches to modeling
their execution on large-scale \gls{hpc} systems with execution time and 
power consumption in mind.
Approaches to optimization of parallel applications have been
discussed in Chapter \ref{chp:optimizationtheory}, with a particular
emphasis on the trade-offs between execution time and power consumption,
computing resource management and parameter auto-tuning.

Multiple experiments have been conducted with real executions involving
\emph{hybrid parallel applications} and \emph{heterogeneous \gls{hpc} systems}
described in Chapter \ref{chp:applicationsandsystems}.
Based on the experiences from this research, an optimization methodology
has been proposed in Chapter \ref{chp:solution}. The methodology has been
verified on a series of case studies described in Chapter \ref{chp:experiments}.
Taking into account the results presented in Chapter \ref{chp:experiments},
the claims of the thesis stated in Section \ref{sec:claims} can be addressed
and proven as follows:

\begin{enumerate}
\item \textbf{The execution steps specific in the context of the proposed model, including preliminary process optimization,
process mapping, parameter tuning and actual execution allow
to optimize execution time of hybrid parallel applications in
heterogeneous high performance computing systems.}

\begin{itemize}
\item The preliminary process optimization step allowed for a significant reduction of execution
time in both relevant case studies described in Chapter \ref{chp:experiments}. Section \ref{sec:prefetching} 
provides results for the case study where the step was implemented
as computation and communication overlapping that reduced execution time
of the considered application up to $11-16\%$. Section \ref{sec:hpcs} describes 
a case study where the step was implemented as overlapping of training and data preprocessing,
resulting in reduction of execution time of the considered application by 13.6\%;
\item The process mapping step implemented as power-aware selection of computing devices
allowed to achieve nearly ideal speedups under a strict power consumption constraint in the case study
described in Section \ref{sec:icdcn};
\item In the case study described in Section \ref{sec:khtuning}, the parameter tuning step
implemented as tuning of \gls{gpu} grid configuration allowed
to efficiently utilize the computing devices. Additionally, implemented
as selection of data partitioning granularity and number of used computing devices,
the step allowed to achieve execution times close to theoretically ideal ones;
\item The combined process mapping and parameter tuning steps, implemented as simulation for
exploring the power/time trade-off of the application, allowed to find a set of Pareto-optimal
execution configurations in the second case study. The results are described in Section \ref{sec:iccs};
\item Section \ref{sec:khexecution} describes a practical example how the \emph{actual execution} of two task farming
applications can be implemented and executed in a large-scale cluster and a heterogeneous system with
\gls{cpu}s and \gls{gpu}s using the proposed KernelHive framework. The implementation allows
to reduce the application execution time through adding utilized computing devices,
both in a homogeneous and a heterogeneous hardware configuration.
Another example of the \emph{actual execution} step considering execution of a parallel deep neural network training application,
in this case in a professional cluster of workstations with \gls{gpu}s, is described in Section \ref{sec:kaldimpi}.
The performance results of this implementation presented in Section \ref{sec:hpcs} show that 
proper execution of a \emph{hybrid parallel application} lead to reduction of 
its execution time by $47.7\%$, because it eliminated redundant initialization operations.
\end{itemize}

\item \textbf{The proposed modeling and simulation method allows
for fast and accurate identification of the set of Pareto-optimal solutions to
the problem of multi-objective execution time and power consumption optimization
of hybrid parallel applications in heterogeneous high performance computing systems.}

\begin{itemize}
\item The proposed modeling and simulation method allowed to identify the
set of Pareto-optimal solutions to the problem of multi-objective time and power consumption
optimization in the case study described in Section \ref{sec:training_case}.
The simulation results were highly accurate and the 
simulation time was multiple times lower than the time of the simulated executions.
These conclusions are based on the results presented in Section \ref{sec:iccs}.
\end{itemize}

\end{enumerate}

Although applying the proposed optimization methodology is indeed useful for multiple practical
applications, it should not be blindly applied to a given new application. There is a multitude of possible 
choices of specific actions that could be performed within the proposed execution steps, as well
as approaches to execution modeling and simulation, deciding on the application model granularity level,
\emph{application execution parameters}, \emph{operation parameters} etc. Applying the methodology
to new hybrid parallel applications, both in terms of performing the execution steps and developing
simulation models, possibly representing various parallel computing paradigms, should be considered as future work.
Regarding the proposed simulation method, a~possible future work direction could be enabling it
for on-line auto-tuning of applications with dynamic re-evaluation of the \emph{application execution parameters}
at application runtime, as well as using the simulation method for supporting decisions about purchasing new
hardware, depending on the expected computational workloads.

Future work could also include approaches to making relaxed assumptions about the applications, for example accepting a
certain simplified problem formulation such as \gls{etc} and \gls{apc} matrices, and generating synthetic problem
instances. Such an approach might allow to develop sophisticated optimization algorithms
aimed for more complex formulations of the optimization problem, in particular considering high-dimensional search spaces.
Such algorithms could become useful for emerging real applications, where increasing complexity of the
application and execution parameters as well as process mapping constraints can be expected.

The author hopes that the presented viewpoint on the current challenges,
selection of discussed literature and proposed practical solutions will be found
a valuable contribution.

\chapter*{Acknowledgements}
\addcontentsline{toc}{chapter}{Acknowledgements}

I would like to express my deep gratitude to my supervisor, Prof. Paweł Czarnul. During the long years of close cooperation,
never once did he refuse to get personally involved in scientific, technical or organizational problems that I have been facing.
He has been the perfect supervisor, since while very demanding towards others, he is always the most demanding of himself.

I would like to thank my auxiliary supervisor, Dr Tomasz Dziubich, who devotes a lot of his efforts to care
about PhD students in our Department and is passionate about practical applications of computer science research.
My sincere thanks also goes to Prof. Henryk Krawczyk for his trust and wisdom as my research supervisor during the first years
of my PhD studies.

Big thanks are also due to my fellow PhD students: Adam Blokus for setting an example that both studies and life are an
exciting journey, Adam Brzeski for his solid faith and reasonable judgement, 
Jan Cychnerski for his perfection in solving the long-term problem of multi-objective optimization
of functioning together at the campus, and Karol Draszawka for inspiring me to get involved
in the arts of machine learning and philosophy of life.

I am also grateful to Prof. Wojciech Jędruch for the most inspiring lectures and conversations which motivated me
to directing my research towards artificial intelligence and its applications. 
I would like to offer my special thanks to Mrs. Izabela Dziedzic for making the Department feel like second home
and dealing with many organizational problems that I know of, as well as probably even more about which I will never even know.

The MERPSYS simulation environment was developed within grant 2012/07/B/ST6/01516 financed by National
Science Center in Poland, entitled "Modeling efficiency, reliability and power consumption of multilevel parallel
HPC systems using CPUs and GPUs". Access to the MICLAB laboratory was possible by courtesy of Technical University
of Częstochowa. I would like to sincerely thank Prof. Cezary Orłowski and Prof. Edward Szczerbicki for inviting me to the
project concerning SmartCity system for the city of Gdańsk, as well as Mr. Jacek Kawalec of VoiceLab.ai for
sharing practical problems related to automatic speech recognition.

Years of trainings, sports camps and skiing competitions helped me during the PhD studies
to maintain balance and develop character. For this I am particularly grateful to
coach Marta Parafiniuk and my teammates from the Gdańsk University of Technology Skiing Team.

I am very grateful to my grandmother Basia, who has always supported and put great emphasis on my education.

Finally, the biggest thanks belong to my fiancée Magda. She was giving me the strength and love every day of
my studies. In this journey that explored many corners of the world, she was always with me and never stopped believing in me. 

\singlespacing
\bibliography{bibliography}

\begin{thebibliography}{164}
\providecommand{\natexlab}[1]{#1}
\providecommand{\url}[1]{\texttt{#1}}
\expandafter\ifx\csname urlstyle\endcsname\relax
  \providecommand{\doi}[1]{doi: #1}\else
  \providecommand{\doi}{doi: \begingroup \urlstyle{rm}\Url}\fi

\bibitem[Villa et~al.(2014)Villa, Johnson, Oconnor, Bolotin, Nellans, Luitjens,
  Sakharnykh, Wang, Micikevicius, Scudiero, Keckler, and
  Dally]{villa_scaling_2014}
O.~Villa, D.~R. Johnson, M.~Oconnor, E.~Bolotin, D.~Nellans, J.~Luitjens,
  N.~Sakharnykh, P.~Wang, P.~Micikevicius, A.~Scudiero, S.~W. Keckler, and
  W.~J. Dally.
\newblock Scaling the {Power} {Wall}: {A} {Path} to {Exascale}.
\newblock In \emph{{SC}14: {International} {Conference} for {High}
  {Performance} {Computing}, {Networking}, {Storage} and {Analysis}}, pages
  830--841, November 2014.
\newblock \doi{10.1109/SC.2014.73}.

\bibitem[Stevens(2017)]{stevens_deep_2017}
Rick Stevens.
\newblock Deep {Learning} in {Cancer} and {Infectious} {Disease}: {Novel}
  {Driver} {Problems} for {Future} {HPC} {Architecture}.
\newblock pages 65--65. ACM Press, 2017.
\newblock ISBN 978-1-4503-4699-3.
\newblock \doi{10.1145/3078597.3091526}.
\newblock URL \url{http://dl.acm.org/citation.cfm?doid=3078597.3091526}.

\bibitem[Abadi et~al.(2016)Abadi, Agarwal, Barham, Brevdo, Chen, Citro,
  Corrado, Davis, Dean, Devin, and {others}]{abadi_tensorflow:_2016}
Martın Abadi, Ashish Agarwal, Paul Barham, Eugene Brevdo, Zhifeng Chen, Craig
  Citro, Greg~S. Corrado, Andy Davis, Jeffrey Dean, Matthieu Devin, and
  {others}.
\newblock Tensorflow: {Large}-scale machine learning on heterogeneous
  distributed systems.
\newblock \emph{arXiv preprint arXiv:1603.04467}, 2016.
\newblock URL \url{http://arxiv.org/abs/1603.04467}.

\bibitem[LeCun et~al.(2015)LeCun, Bengio, and Hinton]{lecun_deep_2015}
Yann LeCun, Yoshua Bengio, and Geoffrey Hinton.
\newblock Deep learning.
\newblock \emph{Nature}, 521\penalty0 (7553):\penalty0 436--444, May 2015.
\newblock ISSN 0028-0836, 1476-4687.
\newblock \doi{10.1038/nature14539}.
\newblock URL \url{http://www.nature.com/doifinder/10.1038/nature14539}.

\bibitem[You et~al.(2017)You, Zhang, Hsieh, Demmel, and
  Keutzer]{you_imagenet_2017}
Yang You, Zhao Zhang, Cho-Jui Hsieh, James Demmel, and Kurt Keutzer.
\newblock {ImageNet} {Training} in {Minutes}.
\newblock \emph{arXiv:1709.05011 [cs]}, September 2017.
\newblock URL \url{http://arxiv.org/abs/1709.05011}.
\newblock arXiv: 1709.05011.

\bibitem[Shalf et~al.(2010)Shalf, Dosanjh, and Morrison]{shalf_exascale_2010}
John Shalf, Sudip Dosanjh, and John Morrison.
\newblock Exascale computing technology challenges.
\newblock In \emph{International {Conference} on {High} {Performance}
  {Computing} for {Computational} {Science}}, pages 1--25. Springer, 2010.

\bibitem[Dongarra et~al.(2011)Dongarra, Beckman, Moore, Aerts, Aloisio, Andre,
  Barkai, Berthou, Boku, Braunschweig, Cappello, Chapman, Chi, Choudhary,
  Dosanjh, Dunning, Fiore, Geist, Gropp, Harrison, Hereld, Heroux, Hoisie,
  Hotta, Jin, Ishikawa, Johnson, Kale, Kenway, Keyes, Kramer, Labarta,
  Lichnewsky, Lippert, Lucas, Maccabe, Matsuoka, Messina, Michielse, Mohr,
  Mueller, Nagel, Nakashima, Papka, Reed, Sato, Seidel, Shalf, Skinner, Snir,
  Sterling, Stevens, Streitz, Sugar, Sumimoto, Tang, Taylor, Thakur, Trefethen,
  Valero, Steen, Vetter, Williams, Wisniewski, and
  Yelick]{dongarra_international_2011}
Jack Dongarra, Pete Beckman, Terry Moore, Patrick Aerts, Giovanni Aloisio,
  Jean-Claude Andre, David Barkai, Jean-Yves Berthou, Taisuke Boku, Bertrand
  Braunschweig, Franck Cappello, Barbara Chapman, Xuebin Chi, Alok Choudhary,
  Sudip Dosanjh, Thom Dunning, Sandro Fiore, Al~Geist, Bill Gropp, Robert
  Harrison, Mark Hereld, Michael Heroux, Adolfy Hoisie, Koh Hotta, Zhong Jin,
  Yutaka Ishikawa, Fred Johnson, Sanjay Kale, Richard Kenway, David Keyes, Bill
  Kramer, Jesus Labarta, Alain Lichnewsky, Thomas Lippert, Bob Lucas, Barney
  Maccabe, Satoshi Matsuoka, Paul Messina, Peter Michielse, Bernd Mohr,
  Matthias~S. Mueller, Wolfgang~E. Nagel, Hiroshi Nakashima, Michael~E. Papka,
  Dan Reed, Mitsuhisa Sato, Ed~Seidel, John Shalf, David Skinner, Marc Snir,
  Thomas Sterling, Rick Stevens, Fred Streitz, Bob Sugar, Shinji Sumimoto,
  William Tang, John Taylor, Rajeev Thakur, Anne Trefethen, Mateo Valero, Aad
  van~der Steen, Jeffrey Vetter, Peg Williams, Robert Wisniewski, and Kathy
  Yelick.
\newblock The {International} {Exascale} {Software} {Project} roadmap.
\newblock \emph{The International Journal of High Performance Computing
  Applications}, 25\penalty0 (1):\penalty0 3--60, 2011.
\newblock \doi{10.1177/1094342010391989}.
\newblock URL \url{https://doi.org/10.1177/1094342010391989}.

\bibitem[Ramapantulu et~al.(2015)Ramapantulu, Loghin, and
  Teo]{ramapantulu_approach_2015}
L.~Ramapantulu, D.~Loghin, and Yong~Meng Teo.
\newblock An {Approach} for {Energy} {Efficient} {Execution} of {Hybrid}
  {Parallel} {Programs}.
\newblock In \emph{Parallel and {Distributed} {Processing} {Symposium}
  ({IPDPS}), 2015 {IEEE} {International}}, pages 1000--1009, 2015.
\newblock \doi{10.1109/IPDPS.2015.71}.

\bibitem[Ma et~al.(2012)Ma, Li, Chen, Zhang, and Wang]{ma_greengpu:_2012}
K.~Ma, X.~Li, W.~Chen, C.~Zhang, and X.~Wang.
\newblock {GreenGPU}: {A} {Holistic} {Approach} to {Energy} {Efficiency} in
  {GPU}-{CPU} {Heterogeneous} {Architectures}.
\newblock In \emph{2012 41st {International} {Conference} on {Parallel}
  {Processing}}, pages 48--57, September 2012.
\newblock \doi{10.1109/ICPP.2012.31}.

\bibitem[Czarnul et~al.(2015)Czarnul, Rościszewski, Matuszek, and
  Szymanski]{czarnul_simulation_2015}
Paweł Czarnul, Paweł Rościszewski, Mariusz Matuszek, and Julian Szymanski.
\newblock Simulation of parallel similarity measure computations for large data
  sets.
\newblock In \emph{2015 {IEEE} 2nd {International} {Conference} on
  {Cybernetics} ({CYBCONF})}, pages 472--477, June 2015.
\newblock \doi{10.1109/CYBConf.2015.7175980}.

\bibitem[Rościszewski et~al.(2014)Rościszewski, Cychnerski, and
  Brzeski]{rosciszewski_regular_2014}
Paweł Rościszewski, Jan Cychnerski, and Adam Brzeski.
\newblock A {Regular} {Expression} {Matching} {Application} with {Configurable}
  {Data} {Intensity} for {Testing} {Heterogeneous} {HPC} {Systems}.
\newblock In \emph{Contemporary {Approaches} to {Design} and {Evaluation} of
  {Information} {Systems}}, pages 39 -- 48. Oficyna Wydawnicza Politechniki
  Wrocławskiej, Wrocław, 2014.

\bibitem[Czarnul et~al.(2016)Czarnul, Kuchta, Rościszewski, and
  Proficz]{czarnul_modeling_2016}
Paweł Czarnul, Jarosław Kuchta, Paweł Rościszewski, and Jerzy Proficz.
\newblock Modeling energy consumption of parallel applications.
\newblock In \emph{Federated {Conference} on {Computer} {Science} and
  {Information} {Systems}}, pages 855--864, October 2016.
\newblock \doi{10.15439/2016F308}.
\newblock URL \url{https://fedcsis.org/proceedings/2016/drp/308.html}.

\bibitem[Rościszewski et~al.(2016)Rościszewski, Czarnul, Lewandowski, and
  Schally-Kacprzak]{rosciszewski_kernelhive:_2016}
Paweł Rościszewski, Paweł Czarnul, Rafał Lewandowski, and Marcel
  Schally-Kacprzak.
\newblock {KernelHive}: a new workflow-based framework for multilevel high
  performance computing using clusters and workstations with {CPUs} and {GPUs}.
\newblock \emph{Concurrency and Computation: Practice and Experience},
  28\penalty0 (9):\penalty0 2586--2607, June 2016.
\newblock ISSN 15320626.
\newblock \doi{10.1002/cpe.3719}.
\newblock URL \url{http://doi.wiley.com/10.1002/cpe.3719}.

\bibitem[Czarnul et~al.(2017)Czarnul, Kuchta, Matuszek, Proficz, Rościszewski,
  Wójcik, and Szymański]{czarnul_merpsys:_2017}
Paweł Czarnul, Jarosław Kuchta, Mariusz Matuszek, Jerzy Proficz, Paweł
  Rościszewski, Michał Wójcik, and Julian Szymański.
\newblock {MERPSYS}: {An} environment for simulation of parallel application
  execution on large scale {HPC} systems.
\newblock \emph{Simulation Modelling Practice and Theory}, 77:\penalty0
  124--140, September 2017.
\newblock ISSN 1569190X.
\newblock \doi{10.1016/j.simpat.2017.05.009}.
\newblock URL
  \url{http://linkinghub.elsevier.com/retrieve/pii/S1569190X17300916}.

\bibitem[Rościszewski and Kaliski(2017)]{rosciszewski_minimizing_2017}
Paweł Rościszewski and Jakub Kaliski.
\newblock Minimizing {Distribution} and {Data} {Loading} {Overheads} in
  {Parallel} {Training} of {DNN} {Acoustic} {Models} with {Frequent}
  {Parameter} {Averaging}.
\newblock pages 560--565. IEEE, July 2017.
\newblock ISBN 978-1-5386-3249-9 978-1-5386-3250-5.
\newblock \doi{10.1109/HPCS.2017.89}.
\newblock URL \url{http://ieeexplore.ieee.org/document/8035128/}.

\bibitem[Rościszewski(2017)]{rosciszewski_modeling_2017}
Paweł Rościszewski.
\newblock Modeling and {Simulation} for {Exploring} {Power}/{Time} {Trade}-off
  of {Parallel} {Deep} {Neural} {Network} {Training}.
\newblock \emph{Procedia Computer Science}, 108:\penalty0 2463--2467, 2017.
\newblock ISSN 18770509.
\newblock \doi{10.1016/j.procs.2017.05.214}.
\newblock URL
  \url{http://linkinghub.elsevier.com/retrieve/pii/S1877050917308074}.

\bibitem[Rościszewski(2014{\natexlab{a}})]{rosciszewski_network-aware_2014}
Paweł Rościszewski.
\newblock Network-{Aware} {Data} {Prefetching} {Optimization} of {Computations}
  in a {Heterogeneous} {HPC} {Framework}.
\newblock \emph{International journal of Computer Networks \& Communications},
  6\penalty0 (5):\penalty0 85--98, September 2014{\natexlab{a}}.
\newblock ISSN 09752293, 09749322.
\newblock \doi{10.5121/ijcnc.2014.6506}.
\newblock URL \url{http://www.airccse.org/journal/cnc/6514cnc06.pdf}.

\bibitem[Bultrowicz et~al.(2014)Bultrowicz, Czarnul, and
  Rościszewski]{balicki_runtime_2014}
Szymon Bultrowicz, Paweł Czarnul, and Paweł Rościszewski.
\newblock Runtime {Visualization} of {Application} {Progress} and {Monitoring}
  of a {GPU}-enabled {Parallel} {Environment}.
\newblock In Jerzy Balicki, {WSEAS (Organization)}, Parallel {and}
  Distributed~Systems International Conference~on Software~Engineering, and
  {International Conference on Bioscience and Bioinformatics}, editors,
  \emph{Applications of information systems in engineering and bioscience:
  proceedings of the 13th international conference on software engineering,
  parallel and distributed systems ({SEPADS} '14) ; proceedings of the 5th
  international conference on bioscience and bioinformatics ({ICBB} '14) :
  {Gdansk}, {Poland}, {May} 15-17, 2014}. 2014.
\newblock ISBN 978-960-474-381-0.
\newblock URL
  \url{http://www.wseas.us/e-library/conferences/2014/Gdansk/SEBIO/SEBIO-07.pdf}.

\bibitem[Czarnul and Rościszewski(2014)]{hutchison_optimization_2014}
Paweł Czarnul and Paweł Rościszewski.
\newblock Optimization of {Execution} {Time} under {Power} {Consumption}
  {Constraints} in a {Heterogeneous} {Parallel} {System} with {GPUs} and
  {CPUs}.
\newblock In David Hutchison, Takeo Kanade, Josef Kittler, Jon~M. Kleinberg,
  Friedemann Mattern, John~C. Mitchell, Moni Naor, Oscar Nierstrasz,
  C.~Pandu~Rangan, Bernhard Steffen, Madhu Sudan, Demetri Terzopoulos, Doug
  Tygar, Moshe~Y. Vardi, Gerhard Weikum, Mainak Chatterjee, Jian-nong Cao,
  Kishore Kothapalli, and Sergio Rajsbaum, editors, \emph{Distributed
  {Computing} and {Networking}}, volume 8314, pages 66--80. Springer Berlin
  Heidelberg, Berlin, Heidelberg, 2014.
\newblock ISBN 978-3-642-45248-2 978-3-642-45249-9.
\newblock URL \url{http://link.springer.com/10.1007/978-3-642-45249-9_5}.

\bibitem[Rościszewski(2016)]{rosciszewski_executing_2016}
Paweł Rościszewski.
\newblock Executing {Multiple} {Simulations} in the {MERPSYS} {Environment}.
\newblock In Paweł Czarnul, editor, \emph{Modeling {Large}-scale {Computing}
  {Systems}: {Practical} {Approaches} in {MERPSYS}}. Gdańsk University of
  Technology, 2016.
\newblock URL \url{https://repository.os.niwa.gda.pl/handle/niwa_item/138}.

\bibitem[Rościszewski and Sidorczak(2014)]{rosciszewski_simulation_2014}
Paweł Rościszewski and Piotr Sidorczak.
\newblock Simulation of {Parallel} {Applications} on {Large}-scale
  {Distributed} {Systems}.
\newblock In Paweł Czarnul, editor, \emph{Modeling {Large}-{Scale} {Computing}
  {Systems}: {Concepts} and {Models}}, pages 135 -- 147. Gdańsk University of
  Technology, Faculty of ETI, Gdańsk, 2014.

\bibitem[Wyrzykowski et~al.(2014)Wyrzykowski, Szustak, and
  Rojek]{wyrzykowski_parallelization_2014}
Roman Wyrzykowski, Łukasz Szustak, and Krzysztof Rojek.
\newblock Parallelization of 2d {MPDATA} {EULAG} algorithm on hybrid
  architectures with {GPU} accelerators.
\newblock \emph{Parallel Computing}, 40\penalty0 (8):\penalty0 425--447, August
  2014.
\newblock ISSN 01678191.
\newblock \doi{10.1016/j.parco.2014.04.009}.
\newblock URL
  \url{http://linkinghub.elsevier.com/retrieve/pii/S0167819114000520}.

\bibitem[Liang et~al.(2012)Liang, Li, and Chiu]{liang_enabling_2012}
T.~Y. Liang, H.~F. Li, and J.~Y. Chiu.
\newblock Enabling {Mixed} {OpenMP}/{MPI} {Programming} on {Hybrid} {CPU}/{GPU}
  {Computing} {Architecture}.
\newblock In \emph{2012 {IEEE} 26th {International} {Parallel} and
  {Distributed} {Processing} {Symposium} {Workshops} {PhD} {Forum}}, pages
  2369--2377, May 2012.
\newblock \doi{10.1109/IPDPSW.2012.294}.

\bibitem[Czarnul(2016)]{czarnul_benchmarking_2016}
Paweł Czarnul.
\newblock Benchmarking {Performance} of a {Hybrid} {Intel} {Xeon}/{Xeon} {Phi}
  {System} for {Parallel} {Computation} of {Similarity} {Measures} {Between}
  {Large} {Vectors}.
\newblock \emph{International Journal of Parallel Programming}, September 2016.
\newblock ISSN 0885-7458, 1573-7640.
\newblock \doi{10.1007/s10766-016-0455-0}.
\newblock URL \url{http://link.springer.com/10.1007/s10766-016-0455-0}.

\bibitem[Lee et~al.(2015)Lee, Lin, and Chen]{lee_hybrid_2015}
Chun-Liang Lee, Yi-Shan Lin, and Yaw-Chung Chen.
\newblock A {Hybrid} {CPU}/{GPU} {Pattern}-{Matching} {Algorithm} for {Deep}
  {Packet} {Inspection}.
\newblock \emph{PLOS ONE}, 10\penalty0 (10):\penalty0 e0139301, October 2015.
\newblock ISSN 1932-6203.
\newblock \doi{10.1371/journal.pone.0139301}.
\newblock URL \url{http://dx.plos.org/10.1371/journal.pone.0139301}.

\bibitem[Li et~al.(2010)Li, de~Supinski, Schulz, Cameron, and
  Nikolopoulos]{li_hybrid_2010}
Dong Li, Bronis~R. de~Supinski, Martin Schulz, Kirk~W. Cameron, and
  Dimitrios~S. Nikolopoulos.
\newblock Hybrid {MPI}/{OpenMP} power-aware computing.
\newblock In \emph{{IPDPS}}, volume~10, pages 1--12, 2010.
\newblock URL \url{http://pure.qub.ac.uk/portal/files/3640017/ipdps10+.pdf}.

\bibitem[Danner et~al.(2012)Danner, Breslow, Baskin, and
  Wilikofsky]{danner_hybrid_2012}
Andrew Danner, Alexander Breslow, Jake Baskin, and David Wilikofsky.
\newblock Hybrid {MPI}/{GPU} interpolation for grid {DEM} construction.
\newblock In \emph{{SIGSPATIAL} 2012 {International} {Conference} on {Advances}
  in {Geographic} {Information} {Systems} (formerly known as {GIS}),
  {SIGSPATIAL}'12, {Redondo} {Beach}, {CA}, {USA}, {November} 7-9, 2012}, pages
  299--308, 2012.
\newblock \doi{10.1145/2424321.2424360}.
\newblock URL \url{http://doi.acm.org/10.1145/2424321.2424360}.

\bibitem[Johnson et~al.(2017)Johnson, Douze, and
  Jégou]{johnson_billion-scale_2017}
Jeff Johnson, Matthijs Douze, and Hervé Jégou.
\newblock Billion-scale similarity search with {GPUs}.
\newblock \emph{arXiv preprint arXiv:1702.08734}, 2017.
\newblock URL \url{https://arxiv.org/abs/1702.08734}.

\bibitem[Tabatabaee et~al.(2005)Tabatabaee, Tiwari, and
  Hollingsworth]{tabatabaee_parallel_2005}
V.~Tabatabaee, A.~Tiwari, and J.K. Hollingsworth.
\newblock Parallel {Parameter} {Tuning} for {Applications} with {Performance}
  {Variability}.
\newblock pages 57--57. IEEE, 2005.
\newblock ISBN 978-1-59593-061-3.
\newblock \doi{10.1109/SC.2005.52}.
\newblock URL
  \url{http://ieeexplore.ieee.org/lpdocs/epic03/wrapper.htm?arnumber=1560009}.

\bibitem[Jordan et~al.(2012)Jordan, Thoman, Durillo, Pellegrini, Gschwandtner,
  Fahringer, and Moritsch]{jordan_multi-objective_2012}
Herbert Jordan, Peter Thoman, Juan~J. Durillo, Sara Pellegrini, Philipp
  Gschwandtner, Thomas Fahringer, and Hans Moritsch.
\newblock A multi-objective auto-tuning framework for parallel codes.
\newblock In \emph{High {Performance} {Computing}, {Networking}, {Storage} and
  {Analysis} ({SC}), 2012 {International} {Conference} for}, pages 1--12. IEEE,
  2012.
\newblock URL
  \url{http://ieeexplore.ieee.org/xpls/abs_all.jsp?arnumber=6468452}.

\bibitem[Kisuki et~al.(2000)Kisuki, Knijnenburg, and
  O'Boyle]{kisuki_combined_2000}
T.~Kisuki, P.~M.~W. Knijnenburg, and M.~F.~P. O'Boyle.
\newblock Combined selection of tile sizes and unroll factors using iterative
  compilation.
\newblock In \emph{International {Conference} on {Parallel} {Architectures} and
  {Compilation} {Techniques}, 2000. {Proceedings}}, pages 237--246, 2000.
\newblock \doi{10.1109/PACT.2000.888348}.

\bibitem[Kim(2012)]{kim_distributed_2012}
Keonwoo Kim.
\newblock Distributed password cracking on {GPU} nodes.
\newblock In \emph{Computing and {Convergence} {Technology} ({ICCCT}), 2012 7th
  {International} {Conference} on}, pages 647--650. IEEE, 2012.
\newblock URL \url{http://ieeexplore.ieee.org/abstract/document/6530414/}.

\bibitem[Niewiadomska-Szynkiewicz et~al.(2012)Niewiadomska-Szynkiewicz, Marks,
  Jantura, and Podbielski]{niewiadomska-szynkiewicz_hybrid_2012}
Ewa Niewiadomska-Szynkiewicz, Michał Marks, Jarosław Jantura, and Mikołaj
  Podbielski.
\newblock A hybrid {CPU}/{GPU} cluster for encryption and decryption of large
  amounts of data.
\newblock \emph{Journal of Telecommunications and Information Technology},
  pages 32--39, 2012.
\newblock URL
  \url{https://yadda.icm.edu.pl/baztech/element/bwmeta1.element.baztech-article-BATA-0017-0004}.

\bibitem[Mukherjee et~al.(1994)Mukherjee, Heberlein, and
  Levitt]{mukherjee_network_1994}
Biswanath Mukherjee, L.~Todd Heberlein, and Karl~N. Levitt.
\newblock Network intrusion detection.
\newblock \emph{IEEE network}, 8\penalty0 (3):\penalty0 26--41, 1994.

\bibitem[Henneböhl et~al.(2011)Henneböhl, Appel, and
  Pebesma]{hennebohl_spatial_2011}
Katharina Henneböhl, Marius Appel, and Edzer Pebesma.
\newblock Spatial interpolation in massively parallel computing environments.
\newblock In \emph{Proceedings of the 14th {AGILE} {International} {Conference}
  on {Geographic} {Information} {Science}}, 2011.

\bibitem[Mei(2014)]{mei_evaluating_2014}
Gang Mei.
\newblock Evaluating the {Power} of {GPU} {Acceleration} for {IDW}
  {Interpolation} {Algorithm}.
\newblock \emph{The Scientific World Journal}, 2014:\penalty0 1--8, 2014.
\newblock ISSN 2356-6140, 1537-744X.
\newblock \doi{10.1155/2014/171574}.
\newblock URL \url{http://www.hindawi.com/journals/tswj/2014/171574/}.

\bibitem[Srinivasan et~al.(2010)Srinivasan, Duraiswami, and
  Murtugudde]{srinivasan_efficient_2010}
Balaji~Vasan Srinivasan, Ramani Duraiswami, and Raghu Murtugudde.
\newblock Efficient kriging for real-time spatio-temporal interpolation.
\newblock In \emph{Proceedings of the 20th {Conference} on {Probability} and
  {Statistics} in the {Atmospheric} {Sciences}}, pages 228--235. American
  Meteorological Society Atlanta GA, 2010.
\newblock URL \url{https://ams.confex.com/ams/pdfpapers/161982.pdf}.

\bibitem[Wang et~al.(2002)Wang, Wang, Yang, and Yu]{wang_clustering_2002}
Haixun Wang, Wei Wang, Jiong Yang, and Philip~S. Yu.
\newblock Clustering by pattern similarity in large data sets.
\newblock In \emph{Proceedings of the 2002 {ACM} {SIGMOD} international
  conference on {Management} of data}, pages 394--405. ACM, 2002.
\newblock URL \url{http://dl.acm.org/citation.cfm?id=564737}.

\bibitem[Sundaram et~al.(2013)Sundaram, Turmukhametova, Satish, Mostak, Indyk,
  Madden, and Dubey]{sundaram_streaming_2013}
Narayanan Sundaram, Aizana Turmukhametova, Nadathur Satish, Todd Mostak, Piotr
  Indyk, Samuel Madden, and Pradeep Dubey.
\newblock Streaming similarity search over one billion tweets using parallel
  locality-sensitive hashing.
\newblock \emph{Proceedings of the VLDB Endowment}, 6\penalty0 (14):\penalty0
  1930--1941, 2013.
\newblock URL \url{http://dl.acm.org/citation.cfm?id=2556574}.

\bibitem[Akram et~al.(2016)Akram, Sartor, and Eeckhout]{akram_dvfs_2016}
Shoaib Akram, Jennifer~B. Sartor, and Lieven Eeckhout.
\newblock {DVFS} performance prediction for managed multithreaded applications.
\newblock In \emph{Performance {Analysis} of {Systems} and {Software}
  ({ISPASS}), 2016 {IEEE} {International} {Symposium} on}, pages 12--23. IEEE,
  2016.
\newblock URL \url{http://ieeexplore.ieee.org/abstract/document/7482070/}.

\bibitem[Hadjis et~al.(2015)Hadjis, Abuzaid, Zhang, and Ré]{hadjis_caffe_2015}
Stefan Hadjis, Firas Abuzaid, Ce~Zhang, and Christopher Ré.
\newblock Caffe con {Troll}: {Shallow} {Ideas} to {Speed} {Up} {Deep}
  {Learning}.
\newblock \emph{arXiv:1504.04343 [cs, stat]}, April 2015.
\newblock URL \url{http://arxiv.org/abs/1504.04343}.
\newblock arXiv: 1504.04343.

\bibitem[Babaeizadeh et~al.(2016)Babaeizadeh, Frosio, Tyree, Clemons, and
  Kautz]{babaeizadeh_reinforcement_2016}
Mohammad Babaeizadeh, Iuri Frosio, Stephen Tyree, Jason Clemons, and Jan Kautz.
\newblock Reinforcement {Learning} through {Asynchronous} {Advantage}
  {Actor}-{Critic} on a {GPU}.
\newblock \emph{arXiv:1611.06256 [cs]}, November 2016.
\newblock URL \url{http://arxiv.org/abs/1611.06256}.
\newblock arXiv: 1611.06256.

\bibitem[Butenhof(1997)]{butenhof_programming_1997}
David~R. Butenhof.
\newblock \emph{Programming with {POSIX} {Threads}}.
\newblock Addison-Wesley Longman Publishing Co., Inc., Boston, MA, USA, 1997.
\newblock ISBN 0-201-63392-2.

\bibitem[Oaks and Wong(2004)]{oaks_java_2004}
Scott Oaks and Henry Wong.
\newblock \emph{Java {Threads}}.
\newblock O'Reilly Media, Inc., 2004.
\newblock ISBN 0-596-00782-5.

\bibitem[Stone et~al.(2010)Stone, Gohara, and Shi]{stone_opencl:_2010}
John~E. Stone, David Gohara, and Guochun Shi.
\newblock {OpenCL}: {A} parallel programming standard for heterogeneous
  computing systems.
\newblock \emph{Computing in science \& engineering}, 12\penalty0 (3):\penalty0
  66--73, 2010.

\bibitem[Calore et~al.(2017)Calore, Gabbana, Schifano, and
  Tripiccione]{calore_evaluation_2017}
Enrico Calore, Alessandro Gabbana, Sebastiano~Fabio Schifano, and Raffaele
  Tripiccione.
\newblock Evaluation of {DVFS} techniques on modern {HPC} processors and
  accelerators for energy-aware applications.
\newblock \emph{Concurrency and Computation: Practice and Experience},
  29\penalty0 (12), 2017.
\newblock URL \url{http://onlinelibrary.wiley.com/doi/10.1002/cpe.4143/full}.

\bibitem[Mei et~al.(2017)Mei, Wang, and Chu]{mei_survey_2017}
Xinxin Mei, Qiang Wang, and Xiaowen Chu.
\newblock A survey and measurement study of {GPU} {DVFS} on energy
  conservation.
\newblock \emph{Digital Communications and Networks}, 3\penalty0 (2):\penalty0
  89--100, 2017.
\newblock URL
  \url{http://www.sciencedirect.com/science/article/pii/S2352864816300736}.

\bibitem[Forum(1994)]{forum_mpi:_1994}
Message~P Forum.
\newblock {MPI}: {A} {Message}-{Passing} {Interface} {Standard}.
\newblock Technical report, University of Tennessee, Knoxville, TN, USA, 1994.

\bibitem[Chaarawi et~al.(2008)Chaarawi, Squyres, Gabriel, and
  Feki]{chaarawi_tool_2008}
Mohamad Chaarawi, Jeffrey~M. Squyres, Edgar Gabriel, and Saber Feki.
\newblock A tool for optimizing runtime parameters of open mpi.
\newblock In \emph{European {Parallel} {Virtual} {Machine}/{Message} {Passing}
  {Interface} {Users}’ {Group} {Meeting}}, pages 210--217. Springer, 2008.
\newblock URL
  \url{http://link.springer.com/chapter/10.1007/978-3-540-87475-1_30}.

\bibitem[Buyya(1999)]{buyya_high_1999}
Rajkumar Buyya.
\newblock \emph{High {Performance} {Cluster} {Computing}: {Programming} and
  {Applications}}.
\newblock Prentice Hall PTR, Upper Saddle River, NJ, USA, 1st edition, 1999.
\newblock ISBN 0-13-013785-5.

\bibitem[Muresano et~al.(2010)Muresano, Rexachs, and
  Luque]{muresano_methodology_2010}
Ronal Muresano, Dolores Rexachs, and Emilio Luque.
\newblock Methodology for {Efficient} {Execution} of {SPMD} {Applications} on
  {Multicore} {Environments}.
\newblock pages 185--195. IEEE, 2010.
\newblock ISBN 978-1-4244-6987-1.
\newblock \doi{10.1109/CCGRID.2010.67}.
\newblock URL \url{http://ieeexplore.ieee.org/document/5493479/}.

\bibitem[Goodhope et~al.(2012)Goodhope, Koshy, Kreps, Narkhede, Park, Rao, and
  Ye]{goodhope_building_2012}
Ken Goodhope, Joel Koshy, Jay Kreps, Neha Narkhede, Richard Park, Jun Rao, and
  Victor~Yang Ye.
\newblock Building {LinkedIn}'s {Real}-time {Activity} {Data} {Pipeline}.
\newblock \emph{IEEE Data Eng. Bull.}, 35\penalty0 (2):\penalty0 33--45, 2012.
\newblock URL
  \url{http://sites.computer.org/debull/A12june/A12JUN-CD.pdf#page=35}.

\bibitem[Czarnul(2015)]{czarnul_parallelization_2015}
Paweł Czarnul.
\newblock Parallelization of {Divide}-and-{Conquer} {Applications} on {Intel}
  {Xeon} {Phi} with an {OpenMP} {Based} {Framework}.
\newblock In Jerzy Swiatek, Leszek Borzemski, Adam Grzech, and Zofia
  Wilimowska, editors, \emph{{ISAT} (3)}, volume 431 of \emph{Advances in
  {Intelligent} {Systems} and {Computing}}, pages 99--111. Springer, 2015.
\newblock ISBN 978-3-319-28562-7.
\newblock URL
  \url{http://dblp.uni-trier.de/db/conf/isat/isat2015-3.html#Czarnul15}.

\bibitem[Dean and Ghemawat(2008)]{dean_mapreduce:_2008}
J.~Dean and S.~Ghemawat.
\newblock {MapReduce}: simplified data processing on large clusters.
\newblock \emph{Commun. ACM}, 51\penalty0 (1):\penalty0 107--113, 2008.
\newblock ISSN 0001-0782.

\bibitem[Shvachko et~al.(2010)Shvachko, Kuang, Radia, and
  Chansler]{shvachko_hadoop_2010}
Konstantin Shvachko, Hairong Kuang, Sanjay Radia, and Robert Chansler.
\newblock The hadoop distributed file system.
\newblock In \emph{Mass storage systems and technologies ({MSST}), 2010 {IEEE}
  26th symposium on}, pages 1--10. IEEE, 2010.
\newblock URL \url{http://ieeexplore.ieee.org/abstract/document/5496972/}.

\bibitem[Zaharia et~al.(2010{\natexlab{a}})Zaharia, Chowdhury, Franklin,
  Shenker, and Stoica]{zaharia_spark:_2010}
Matei Zaharia, Mosharaf Chowdhury, Michael~J. Franklin, Scott Shenker, and Ion
  Stoica.
\newblock Spark: {Cluster} computing with working sets.
\newblock \emph{HotCloud}, 10\penalty0 (10-10):\penalty0 95,
  2010{\natexlab{a}}.
\newblock URL
  \url{http://static.usenix.org/legacy/events/hotcloud10/tech/full_papers/Zaharia.pdf}.

\bibitem[Zaharia et~al.(2010{\natexlab{b}})Zaharia, Borthakur, Sen~Sarma,
  Elmeleegy, Shenker, and Stoica]{zaharia_delay_2010}
Matei Zaharia, Dhruba Borthakur, Joydeep Sen~Sarma, Khaled Elmeleegy, Scott
  Shenker, and Ion Stoica.
\newblock Delay scheduling: a simple technique for achieving locality and
  fairness in cluster scheduling.
\newblock In \emph{Proceedings of the 5th {European} conference on {Computer}
  systems}, pages 265--278. ACM, 2010{\natexlab{b}}.
\newblock URL \url{http://dl.acm.org/citation.cfm?id=1755940}.

\bibitem[Rabenseifner(2003)]{rabenseifner_hybrid_2003}
Rolf Rabenseifner.
\newblock Hybrid parallel programming on {HPC} platforms.
\newblock In \emph{proceedings of the {Fifth} {European} {Workshop} on
  {OpenMP}, {EWOMP}}, volume~3, pages 185--194. Citeseer, 2003.
\newblock URL
  \url{http://citeseerx.ist.psu.edu/viewdoc/download?doi=10.1.1.71.6450&rep=rep1&type=pdf}.

\bibitem[Hoefler et~al.(2013)Hoefler, Dinan, Buntinas, Balaji, Barrett,
  Brightwell, Gropp, Kale, and Thakur]{hoefler_mpi+_2013}
Torsten Hoefler, James Dinan, Darius Buntinas, Pavan Balaji, Brian Barrett, Ron
  Brightwell, William Gropp, Vivek Kale, and Rajeev Thakur.
\newblock {MPI}+ {MPI}: a new hybrid approach to parallel programming with
  {MPI} plus shared memory.
\newblock \emph{Computing}, 95\penalty0 (12):\penalty0 1121--1136, 2013.
\newblock URL \url{http://link.springer.com/article/10.1007/s00607-013-0324-2}.

\bibitem[De~Wael et~al.(2015)De~Wael, Marr, De~Fraine, Van~Cutsem, and
  De~Meuter]{de_wael_partitioned_2015}
Mattias De~Wael, Stefan Marr, Bruno De~Fraine, Tom Van~Cutsem, and Wolfgang
  De~Meuter.
\newblock Partitioned {Global} {Address} {Space} {Languages}.
\newblock \emph{ACM Computing Surveys}, 47\penalty0 (4):\penalty0 62:1--62:27,
  June 2015.
\newblock ISSN 0360-0300.
\newblock \doi{10.1145/2716320}.

\bibitem[Nowicki et~al.(2014)Nowicki, Górski, Grabarczyk, and
  Bała]{nowicki_pcj-java_2014}
Marek Nowicki, Łukasz Górski, Patryk Grabarczyk, and Piotr Bała.
\newblock {PCJ}-{Java} library for high performance computing in {PGAS} model.
\newblock In \emph{High {Performance} {Computing} \& {Simulation} ({HPCS}),
  2014 {International} {Conference} on}, pages 202--209. IEEE, 2014.

\bibitem[Barak et~al.(2010)Barak, Ben-Nun, Levy, and
  Shiloh]{barak_package_2010}
A.~Barak, T.~Ben-Nun, E.~Levy, and A.~Shiloh.
\newblock A package for {OpenCL} based heterogeneous computing on clusters with
  many {GPU} devices.
\newblock In \emph{2010 {IEEE} {International} {Conference} {On} {Cluster}
  {Computing} {Workshops} and {Posters} ({CLUSTER} {WORKSHOPS})}, pages 1--7,
  September 2010.
\newblock \doi{10.1109/CLUSTERWKSP.2010.5613086}.

\bibitem[Aoki et~al.(2011)Aoki, Oikawa, Nakamura, and Miki]{aoki_hybrid_2011}
Ryo Aoki, Shuichi Oikawa, Takashi Nakamura, and Satoshi Miki.
\newblock Hybrid {OpenCL}: {Enhancing} {OpenCL} for {Distributed} {Processing}.
\newblock pages 149--154. IEEE, May 2011.
\newblock ISBN 978-1-4577-0391-1.
\newblock \doi{10.1109/ISPA.2011.28}.
\newblock URL \url{http://ieeexplore.ieee.org/document/5951897/}.

\bibitem[Kegel et~al.(2012)Kegel, Steuwer, and Gorlatch]{kegel_dopencl:_2012}
Philipp Kegel, Michel Steuwer, and Sergei Gorlatch.
\newblock {dOpenCL}: {Towards} a {Uniform} {Programming} {Approach} for
  {Distributed} {Heterogeneous} {Multi}-/{Many}-{Core} {Systems}.
\newblock pages 174--186. IEEE, May 2012.
\newblock ISBN 978-1-4673-0974-5.
\newblock \doi{10.1109/IPDPSW.2012.16}.
\newblock URL \url{http://ieeexplore.ieee.org/document/6270637/}.

\bibitem[Ozaydin and Altilar(2012)]{ozaydin_opencl_2012}
Ridvan Ozaydin and D.~Turgay Altilar.
\newblock {OpenCL} {Remote}: {Extending} {OpenCL} {Platform} {Model} to
  {Network} {Scale}.
\newblock pages 830--835. IEEE, June 2012.
\newblock ISBN 978-1-4673-2164-8 978-0-7695-4749-7.
\newblock \doi{10.1109/HPCC.2012.117}.
\newblock URL \url{http://ieeexplore.ieee.org/document/6332255/}.

\bibitem[Diop et~al.(2013)Diop, Gurfinkel, Anderson, and
  Jerger]{diop_distcl:_2013}
Tahir Diop, Steven Gurfinkel, Jason Anderson, and Natalie~Enright Jerger.
\newblock {DistCL}: {A} {Framework} for the {Distributed} {Execution} of
  {OpenCL} {Kernels}.
\newblock pages 556--566. IEEE, August 2013.
\newblock ISBN 978-0-7695-5102-9.
\newblock \doi{10.1109/MASCOTS.2013.77}.
\newblock URL \url{http://ieeexplore.ieee.org/document/6730812/}.

\bibitem[Grasso et~al.(2013)Grasso, Pellegrini, Cosenza, and
  Fahringer]{grasso_libwater:_2013}
Ivan Grasso, Simone Pellegrini, Biagio Cosenza, and Thomas Fahringer.
\newblock {LibWater}: heterogeneous distributed computing made easy.
\newblock In \emph{Proceedings of the 27th international {ACM} conference on
  {International} conference on supercomputing}, pages 161--172. ACM, 2013.
\newblock URL \url{http://dl.acm.org/citation.cfm?id=2465008}.

\bibitem[Duato et~al.(2010)Duato, Peña, Silla, Mayo, and
  Quintana-Ortí]{duato_rcuda:_2010}
J.~Duato, A.~J. Peña, F.~Silla, R.~Mayo, and E.~S. Quintana-Ortí.
\newblock {rCUDA}: {Reducing} the number of {GPU}-based accelerators in high
  performance clusters.
\newblock In \emph{2010 {International} {Conference} on {High} {Performance}
  {Computing} {Simulation}}, pages 224--231, June 2010.
\newblock \doi{10.1109/HPCS.2010.5547126}.

\bibitem[Che et~al.(2009)Che, Boyer, Meng, Tarjan, Sheaffer, Lee, and
  Skadron]{che_rodinia:_2009}
Shuai Che, Michael Boyer, Jiayuan Meng, David Tarjan, Jeremy~W. Sheaffer,
  Sang-Ha Lee, and Kevin Skadron.
\newblock Rodinia: {A} benchmark suite for heterogeneous computing.
\newblock In \emph{Workload {Characterization}, 2009. {IISWC} 2009. {IEEE}
  {International} {Symposium} on}, pages 44--54. IEEE, 2009.
\newblock URL
  \url{http://ieeexplore.ieee.org/xpls/abs_all.jsp?arnumber=5306797}.

\bibitem[Sajid et~al.(2016)Sajid, Raza, and
  Shahid]{sajid_energy-efficient_2016}
Mohammad Sajid, Zahid Raza, and Mohammad Shahid.
\newblock Energy-efficient scheduling algorithms for batch-of-tasks ({BoT})
  applications on heterogeneous computing systems.
\newblock \emph{Concurrency and Computation: Practice and Experience},
  28\penalty0 (9):\penalty0 2644--2669, 2016.
\newblock ISSN 1532-0634.
\newblock \doi{10.1002/cpe.3728}.
\newblock URL
  \url{http://onlinelibrary.wiley.com/doi/10.1002/cpe.3728/abstract}.

\bibitem[Wu et~al.(2015)Wu, Xiong, and Lan]{wu_hierarchical_2015}
Jingjin Wu, Xuanxing Xiong, and Zhiling Lan.
\newblock Hierarchical task mapping for parallel applications on
  supercomputers.
\newblock \emph{The Journal of Supercomputing}, 71\penalty0 (5):\penalty0
  1776--1802, May 2015.
\newblock ISSN 0920-8542, 1573-0484.
\newblock \doi{10.1007/s11227-014-1324-5}.
\newblock URL \url{http://link.springer.com/10.1007/s11227-014-1324-5}.

\bibitem[Norman and Thanisch(1993)]{norman_models_1993}
Michael~G. Norman and Peter Thanisch.
\newblock Models of machines and computation for mapping in multicomputers.
\newblock \emph{ACM Computing Surveys (CSUR)}, 25\penalty0 (3):\penalty0
  263--302, 1993.
\newblock URL \url{http://dl.acm.org/citation.cfm?id=158908}.

\bibitem[Vivekanandarajah and Pilakkat(2008)]{vivekanandarajah_task_2008}
Kugan Vivekanandarajah and Santhosh~Kumar Pilakkat.
\newblock Task {Mapping} in {Heterogeneous} {MPSoCs} for {System} {Level}
  {Design}.
\newblock pages 56--65. IEEE, March 2008.
\newblock ISBN 978-0-7695-3139-7.
\newblock \doi{10.1109/ICECCS.2008.18}.
\newblock URL
  \url{http://ieeexplore.ieee.org/lpdocs/epic03/wrapper.htm?arnumber=4492879}.

\bibitem[Baskiyar and Abdel-Kader(2010)]{baskiyar_energy_2010}
Sanjeev Baskiyar and Rabab Abdel-Kader.
\newblock Energy aware {DAG} scheduling on heterogeneous systems.
\newblock \emph{Cluster Computing}, 13\penalty0 (4):\penalty0 373--383,
  December 2010.
\newblock ISSN 1386-7857, 1573-7543.
\newblock \doi{10.1007/s10586-009-0119-6}.
\newblock URL \url{http://link.springer.com/10.1007/s10586-009-0119-6}.

\bibitem[Kessaci et~al.(2011)Kessaci, Mezmaz, Melab, Talbi, and
  Tuyttens]{kessaci_parallel_2011}
Yacine Kessaci, Mohand Mezmaz, Nouredine Melab, El-Ghazali Talbi, and Daniel
  Tuyttens.
\newblock Parallel evolutionary algorithms for energy aware scheduling.
\newblock In \emph{Intelligent {Decision} {Systems} in {Large}-{Scale}
  {Distributed} {Environments}}, pages 75--100. Springer, 2011.
\newblock URL
  \url{http://link.springer.com/chapter/10.1007/978-3-642-21271-0_4}.

\bibitem[Chowdhury and Stoica(2015)]{chowdhury_efficient_2015}
Mosharaf Chowdhury and Ion Stoica.
\newblock Efficient {Coflow} {Scheduling} {Without} {Prior} {Knowledge}.
\newblock pages 393--406. ACM Press, 2015.
\newblock ISBN 978-1-4503-3542-3.
\newblock \doi{10.1145/2785956.2787480}.
\newblock URL \url{http://dl.acm.org/citation.cfm?doid=2785956.2787480}.

\bibitem[Sun et~al.(2015)Sun, Zhang, Yang, Zheng, Khan, and
  Li]{sun_re-stream:_2015}
Dawei Sun, Guangyan Zhang, Songlin Yang, Weimin Zheng, Samee~U. Khan, and Keqin
  Li.
\newblock Re-{Stream}: {Real}-time and energy-efficient resource scheduling in
  big data stream computing environments.
\newblock \emph{Information Sciences}, 319:\penalty0 92--112, October 2015.
\newblock ISSN 00200255.
\newblock \doi{10.1016/j.ins.2015.03.027}.
\newblock URL
  \url{http://linkinghub.elsevier.com/retrieve/pii/S0020025515001929}.

\bibitem[Beaumont et~al.(2002)Beaumont, Legrand, Robert, and
  {others}]{beaumont_static_2002}
Olivier Beaumont, Arnaud Legrand, Yves Robert, and {others}.
\newblock Static scheduling strategies for heterogeneous systems.
\newblock In \emph{{ISCIS} {XVII}, {Seventeenth} {International} {Symposium}
  {On} {Computer} and {Information} {Sciences}}, pages 18--22. CRC Press, 2002.
\newblock URL
  \url{https://books.google.com/books?hl=en&lr=&id=QvEljkOMy7gC&oi=fnd&pg=PA18&dq=%22MINIMUM+MAKESPAN%22+%22to+execute+the+tasks.+Communication+delays+are+taken+into+account+as%22+%22T+and+T+%E2%80%B2+are+assigned+to+the+same%22+%22which+of+course+is+not+realistic+as+soon+as+the+processor+number+exceeds+a%22+&ots=oGr7lgr6G_&sig=zXZOY78Ok1x24stdyFqvJRSb-wc}.

\bibitem[Zhou and Liu(2014)]{zhou_task_2014}
Husheng Zhou and Cong Liu.
\newblock Task mapping in heterogeneous embedded systems for fast completion
  time.
\newblock pages 1--10. ACM Press, 2014.
\newblock ISBN 978-1-4503-3052-7.
\newblock \doi{10.1145/2656045.2656074}.
\newblock URL \url{http://dl.acm.org/citation.cfm?doid=2656045.2656074}.

\bibitem[Augonnet et~al.(2011)Augonnet, Thibault, Namyst, and
  Wacrenier]{augonnet_starpu:_2011}
Cédric Augonnet, Samuel Thibault, Raymond Namyst, and Pierre-André Wacrenier.
\newblock {StarPU}: a unified platform for task scheduling on heterogeneous
  multicore architectures.
\newblock \emph{Concurrency and Computation: Practice and Experience},
  23\penalty0 (2):\penalty0 187--198, 2011.
\newblock URL \url{http://onlinelibrary.wiley.com/doi/10.1002/cpe.1631/full}.

\bibitem[Christen et~al.(2011)Christen, Schenk, and
  Burkhart]{christen_patus:_2011}
Matthias Christen, Olaf Schenk, and Helmar Burkhart.
\newblock {PATUS}: {A} {Code} {Generation} and {Autotuning} {Framework} for
  {Parallel} {Iterative} {Stencil} {Computations} on {Modern}
  {Microarchitectures}.
\newblock pages 676--687. IEEE, May 2011.
\newblock ISBN 978-1-61284-372-8.
\newblock \doi{10.1109/IPDPS.2011.70}.
\newblock URL
  \url{http://ieeexplore.ieee.org/lpdocs/epic03/wrapper.htm?arnumber=6012879}.

\bibitem[Li et~al.(2014)Li, Tang, and Li]{li_energy-efficient_2014}
K.~Li, X.~Tang, and K.~Li.
\newblock Energy-{Efficient} {Stochastic} {Task} {Scheduling} on
  {Heterogeneous} {Computing} {Systems}.
\newblock \emph{IEEE Transactions on Parallel and Distributed Systems},
  25\penalty0 (11):\penalty0 2867--2876, 2014.
\newblock ISSN 1045-9219.
\newblock \doi{10.1109/TPDS.2013.270}.

\bibitem[Oxley et~al.(2015)Oxley, Pasricha, Maciejewski, Siegel, Apodaca,
  Young, Briceño, Smith, Bahirat, Khemka, Ramirez, and
  Zou]{oxley_makespan_2015}
M.~A. Oxley, S.~Pasricha, A.~A. Maciejewski, H.~J. Siegel, J.~Apodaca,
  D.~Young, L.~Briceño, J.~Smith, S.~Bahirat, B.~Khemka, A.~Ramirez, and
  Y.~Zou.
\newblock Makespan and {Energy} {Robust} {Stochastic} {Static} {Resource}
  {Allocation} of a {Bag}-of-{Tasks} to a {Heterogeneous} {Computing} {System}.
\newblock \emph{IEEE Transactions on Parallel and Distributed Systems},
  26\penalty0 (10):\penalty0 2791--2805, 2015.
\newblock ISSN 1045-9219.
\newblock \doi{10.1109/TPDS.2014.2362921}.

\bibitem[Zhang et~al.(2015)Zhang, Bai, He, and Cheng]{zhang_solving_2015}
Weizhe Zhang, Enci Bai, Hui He, and Albert~M.K. Cheng.
\newblock Solving {Energy}-{Aware} {Real}-{Time} {Tasks} {Scheduling} {Problem}
  with {Shuffled} {Frog} {Leaping} {Algorithm} on {Heterogeneous} {Platforms}.
\newblock \emph{Sensors (Basel, Switzerland)}, 15\penalty0 (6):\penalty0
  13778--13804, June 2015.
\newblock ISSN 1424-8220.
\newblock \doi{10.3390/s150613778}.
\newblock URL \url{http://www.ncbi.nlm.nih.gov/pmc/articles/PMC4507628/}.

\bibitem[Awan et~al.(2016)Awan, Yomsi, Nelissen, and
  Petters]{awan_energy-aware_2016}
Muhammad~Ali Awan, Patrick~Meumeu Yomsi, Geoffrey Nelissen, and Stefan~M.
  Petters.
\newblock Energy-aware task mapping onto heterogeneous platforms using {DVFS}
  and sleep states.
\newblock \emph{Real-Time Systems}, 52\penalty0 (4):\penalty0 450--485, July
  2016.
\newblock ISSN 0922-6443, 1573-1383.
\newblock \doi{10.1007/s11241-015-9236-x}.
\newblock URL \url{http://link.springer.com/10.1007/s11241-015-9236-x}.

\bibitem[De~Langen and Juurlink(2007)]{de_langen_trade-offs_2007}
Pepijn De~Langen and Ben Juurlink.
\newblock Trade-offs between voltage scaling and processor shutdown for
  low-energy embedded multiprocessors.
\newblock In \emph{International {Workshop} on {Embedded} {Computer}
  {Systems}}, pages 75--85. Springer, 2007.
\newblock URL
  \url{http://link.springer.com/chapter/10.1007/978-3-540-73625-7_10}.

\bibitem[Tarplee et~al.(2016)Tarplee, Friese, Maciejewski, Siegel, and
  Chong]{tarplee_energy_2016}
K.~M. Tarplee, R.~Friese, A.~A. Maciejewski, H.~J. Siegel, and E.~K.~P. Chong.
\newblock Energy and {Makespan} {Tradeoffs} in {Heterogeneous} {Computing}
  {Systems} using {Efficient} {Linear} {Programming} {Techniques}.
\newblock \emph{IEEE Transactions on Parallel and Distributed Systems},
  27\penalty0 (6):\penalty0 1633--1646, June 2016.
\newblock ISSN 1045-9219.
\newblock \doi{10.1109/TPDS.2015.2456020}.

\bibitem[Balaprakash et~al.(2014)Balaprakash, Tiwari, and
  Wild]{jarvis_multi_2014}
Prasanna Balaprakash, Ananta Tiwari, and Stefan~M. Wild.
\newblock Multi {Objective} {Optimization} of {HPC} {Kernels} for
  {Performance}, {Power}, and {Energy}.
\newblock In Stephen~A. Jarvis, Steven~A. Wright, and Simon~D. Hammond,
  editors, \emph{High {Performance} {Computing} {Systems}. {Performance}
  {Modeling}, {Benchmarking} and {Simulation}}, volume 8551, pages 239--260.
  Springer International Publishing, Cham, 2014.
\newblock ISBN 978-3-319-10213-9 978-3-319-10214-6.
\newblock URL \url{http://link.springer.com/10.1007/978-3-319-10214-6_12}.

\bibitem[Kreutzer et~al.(1997)Kreutzer, Hopkins, and
  Van~Mierlo]{kreutzer_simjavaframework_1997}
Wolfgang Kreutzer, Jane Hopkins, and Marcel Van~Mierlo.
\newblock {SimJAVA}—a framework for modeling queueing networks in {Java}.
\newblock In \emph{Proceedings of the 29th conference on {Winter} simulation},
  pages 483--488. IEEE Computer Society, 1997.
\newblock URL \url{http://dl.acm.org/citation.cfm?id=268548}.

\bibitem[Buyya and Murshed(2002)]{buyya_gridsim:_2002}
Rajkumar Buyya and Manzur Murshed.
\newblock Gridsim: {A} toolkit for the modeling and simulation of distributed
  resource management and scheduling for grid computing.
\newblock \emph{Concurrency and computation: practice and experience},
  14\penalty0 (13-15):\penalty0 1175--1220, 2002.
\newblock URL \url{http://onlinelibrary.wiley.com/doi/10.1002/cpe.710/full}.

\bibitem[Wehrle et~al.(2010)Wehrle, Güneş, and Gross]{wehrle_modeling_2010}
Klaus Wehrle, Mesut Güneş, and James Gross, editors.
\newblock \emph{Modeling and {Tools} for {Network} {Simulation}}.
\newblock Springer Berlin Heidelberg, Berlin, Heidelberg, 2010.
\newblock ISBN 978-3-642-12330-6 978-3-642-12331-3.
\newblock \doi{10.1007/978-3-642-12331-3}.
\newblock URL \url{http://link.springer.com/10.1007/978-3-642-12331-3}.

\bibitem[Denzel et~al.(2010)Denzel, {Jian Li}, Walker, and {Yuho
  Jin}]{denzel_framework_2010}
Wolfgang~E. Denzel, {Jian Li}, Peter Walker, and {Yuho Jin}.
\newblock A {Framework} for {End}-to-{End} {Simulation} of {High}-performance
  {Computing} {Systems}.
\newblock \emph{SIMULATION}, 86\penalty0 (5-6):\penalty0 331--350, May 2010.
\newblock ISSN 0037-5497, 1741-3133.
\newblock \doi{10.1177/0037549709340840}.
\newblock URL \url{http://journals.sagepub.com/doi/10.1177/0037549709340840}.

\bibitem[Hsieh et~al.(2012)Hsieh, Pedretti, Meng, Coskun, Levenhagen, and
  Rodrigues]{hsieh_sst+_2012}
Mingyu Hsieh, Kevin Pedretti, Jie Meng, Ayse Coskun, Michael Levenhagen, and
  Arun Rodrigues.
\newblock Sst+ gem5= a scalable simulation infrastructure for high performance
  computing.
\newblock In \emph{Proceedings of the 5th {International} {ICST} {Conference}
  on {Simulation} {Tools} and {Techniques}}, pages 196--201. ICST (Institute
  for Computer Sciences, Social-Informatics and Telecommunications
  Engineering), 2012.
\newblock URL \url{http://dl.acm.org/citation.cfm?id=2263045}.

\bibitem[Mazumdar and Scionti(2017)]{mazumdar_analysing_2017}
Somnath Mazumdar and Alberto Scionti.
\newblock Analysing {Dataflow} {Multi}-{Threaded} {Applications} at {Runtime}.
\newblock In \emph{Advance {Computing} {Conference} ({IACC}), 2017 {IEEE} 7th
  {International}}, pages 744--749. IEEE, 2017.
\newblock URL \url{http://ieeexplore.ieee.org/abstract/document/7976888/}.

\bibitem[Sarood et~al.(2014)Sarood, Langer, Gupta, and
  Kale]{sarood_maximizing_2014}
Osman Sarood, Akhil Langer, Abhishek Gupta, and Laxmikant Kale.
\newblock Maximizing throughput of overprovisioned {HPC} data centers under a
  strict power budget.
\newblock In \emph{Proceedings of the {International} {Conference} for {High}
  {Performance} {Computing}, {Networking}, {Storage} and {Analysis}}, pages
  807--818. IEEE Press, 2014.
\newblock URL \url{http://dl.acm.org/citation.cfm?id=2683682}.

\bibitem[Scionti et~al.(2017)Scionti, Mazumdar, and
  Portero]{scionti_efficient_2017}
Alberto Scionti, Somnath Mazumdar, and Antoni Portero.
\newblock Efficient {Data}-{Driven} {Task} {Allocation} for {Future}
  {Many}-{Cluster} {On}-chip {Systems}.
\newblock pages 503--510. IEEE, July 2017.
\newblock ISBN 978-1-5386-3249-9 978-1-5386-3250-5.
\newblock \doi{10.1109/HPCS.2017.81}.
\newblock URL \url{http://ieeexplore.ieee.org/document/8035120/}.

\bibitem[Chen et~al.(2015)Chen, Zhuo, Yeh, Lin, and Liao]{chen_machine_2015}
Chi-Ou Chen, Ye-Qi Zhuo, Chao-Chun Yeh, Che-Min Lin, and Shih-wei Liao.
\newblock Machine {Learning}-{Based} {Configuration} {Parameter} {Tuning} on
  {Hadoop} {System}.
\newblock In \emph{2015 {IEEE} {International} {Congress} on {Big} {Data}
  ({BigData} {Congress})}, pages 386--392, June 2015.
\newblock \doi{10.1109/BigDataCongress.2015.64}.

\bibitem[Casanova(2001)]{casanova_simgrid:_2001}
H.~Casanova.
\newblock Simgrid: a toolkit for the simulation of application scheduling.
\newblock In \emph{Proceedings {First} {IEEE}/{ACM} {International} {Symposium}
  on {Cluster} {Computing} and the {Grid}}, pages 430--437, 2001.
\newblock \doi{10.1109/CCGRID.2001.923223}.

\bibitem[Bąk et~al.(2011)Bąk, Krystek, Kurowski, Oleksiak, Piątek, and
  Węglarz]{bak_gssim_2011}
Sławomir Bąk, Marcin Krystek, Krzysztof Kurowski, Ariel Oleksiak, Wojciech
  Piątek, and Jan Węglarz.
\newblock {GSSIM} - {A} tool for distributed computing experiments.
\newblock \emph{Scientific Programming}, \penalty0 (4):\penalty0 231--251,
  2011.
\newblock ISSN 1058-9244.
\newblock \doi{10.3233/SPR-2011-0332}.
\newblock URL
  \url{http://www.medra.org/servlet/aliasResolver?alias=iospress&genre=article&issn=1058-9244&volume=19&issue=4&spage=231&doi=10.3233/SPR-2011-0332}.

\bibitem[Dongarra et~al.(2007)Dongarra, Jeannot, Saule, and
  Shi]{dongarra_bi-objective_2007}
Jack~J. Dongarra, Emmanuel Jeannot, Erik Saule, and Zhiao Shi.
\newblock Bi-objective scheduling algorithms for optimizing makespan and
  reliability on heterogeneous systems.
\newblock In \emph{Proceedings of the nineteenth annual {ACM} symposium on
  {Parallel} algorithms and architectures}, pages 280--288. ACM, 2007.
\newblock URL \url{http://dl.acm.org/citation.cfm?id=1248423}.

\bibitem[Jeannot et~al.(2008)Jeannot, Saule, and
  Trystram]{jeannot_bi-objective_2008}
Emmanuel Jeannot, Erik Saule, and Denis Trystram.
\newblock Bi-objective approximation scheme for makespan and reliability
  optimization on uniform parallel machines.
\newblock In \emph{European {Conference} on {Parallel} {Processing}}, pages
  877--886. Springer, 2008.
\newblock URL
  \url{http://link.springer.com/chapter/10.1007/978-3-540-85451-7_94}.

\bibitem[Ehrgott(2005)]{ehrgott_multicriteria_2005}
Matthias Ehrgott.
\newblock \emph{Multicriteria optimization}.
\newblock Springer, Berlin ; New York, 2nd ed edition, 2005.
\newblock ISBN 978-3-540-21398-7.

\bibitem[Gschwandtner et~al.(2014)Gschwandtner, Durillo, and
  Fahringer]{gschwandtner_multi-objective_2014}
Philipp Gschwandtner, Juan~J. Durillo, and Thomas Fahringer.
\newblock Multi-objective auto-tuning with insieme: {Optimization} and
  trade-off analysis for time, energy and resource usage.
\newblock In \emph{European {Conference} on {Parallel} {Processing}}, pages
  87--98. Springer, 2014.
\newblock URL
  \url{http://link.springer.com/chapter/10.1007/978-3-319-09873-9_8}.

\bibitem[Durillo and Fahringer(2014)]{durillo_single-multi-objective_2014}
Juan Durillo and Thomas Fahringer.
\newblock From single-to multi-objective auto-tuning of programs: {Advantages}
  and implications.
\newblock \emph{Scientific Programming}, 22\penalty0 (4):\penalty0 285--297,
  2014.
\newblock URL \url{http://www.hindawi.com/journals/sp/2014/818579/abs/}.

\bibitem[Saad et~al.(2012)Saad, Awadalla, Shalan, and
  Elewi]{saad_energy-aware_2012}
Elsayed Saad, Medhat Awadalla, Mohamed Shalan, and Abdullah Elewi.
\newblock Energy-aware task partitioning on heterogeneous multiprocessor
  platforms.
\newblock \emph{arXiv preprint arXiv:1206.0396}, 2012.
\newblock URL \url{http://arxiv.org/abs/1206.0396}.

\bibitem[Lee et~al.(2011)Lee, Sathisha, Schulte, Compton, and
  Kim]{lee_improving_2011}
Jungseob Lee, Vijay Sathisha, Michael Schulte, Katherine Compton, and Nam~Sung
  Kim.
\newblock Improving {Throughput} of {Power}-{Constrained} {GPUs} {Using}
  {Dynamic} {Voltage}/{Frequency} and {Core} {Scaling}.
\newblock pages 111--120. IEEE, October 2011.
\newblock ISBN 978-1-4577-1794-9 978-0-7695-4566-0.
\newblock \doi{10.1109/PACT.2011.17}.
\newblock URL
  \url{http://ieeexplore.ieee.org/lpdocs/epic03/wrapper.htm?arnumber=6113793}.

\bibitem[Bakhoda et~al.(2009)Bakhoda, Yuan, Fung, Wong, and
  Aamodt]{bakhoda_analyzing_2009}
Ali Bakhoda, George~L. Yuan, Wilson~WL Fung, Henry Wong, and Tor~M. Aamodt.
\newblock Analyzing {CUDA} workloads using a detailed {GPU} simulator.
\newblock In \emph{Performance {Analysis} of {Systems} and {Software}, 2009.
  {ISPASS} 2009. {IEEE} {International} {Symposium} on}, pages 163--174. IEEE,
  2009.
\newblock URL
  \url{http://ieeexplore.ieee.org/xpls/abs_all.jsp?arnumber=4919648}.

\bibitem[Chang et~al.(2010)Chang, Jenkins, Garcia, Gilani, Aguilera, Nagarajan,
  Anderson, Kenny, Bauer, Schulte, and Compton]{chang_ercbench:_2010}
D.~W. Chang, C.~D. Jenkins, P.~C. Garcia, S.~Z. Gilani, P.~Aguilera,
  A.~Nagarajan, M.~J. Anderson, M.~A. Kenny, S.~M. Bauer, M.~J. Schulte, and
  K.~Compton.
\newblock {ERCBench}: {An} {Open}-{Source} {Benchmark} {Suite} for {Embedded}
  and {Reconfigurable} {Computing}.
\newblock In \emph{2010 {International} {Conference} on {Field} {Programmable}
  {Logic} and {Applications}}, pages 408--413, 2010.
\newblock \doi{10.1109/FPL.2010.85}.

\bibitem[Patki et~al.(2013)Patki, Lowenthal, Rountree, Schulz, and
  De~Supinski]{patki_exploring_2013}
Tapasya Patki, David~K. Lowenthal, Barry Rountree, Martin Schulz, and Bronis~R.
  De~Supinski.
\newblock Exploring hardware overprovisioning in power-constrained, high
  performance computing.
\newblock In \emph{Proceedings of the 27th international {ACM} conference on
  {International} conference on supercomputing}, pages 173--182. ACM, 2013.
\newblock URL \url{http://dl.acm.org/citation.cfm?id=2465009}.

\bibitem[Rountree et~al.(2012)Rountree, Ahn, Supinski, Lowenthal, and
  Schulz]{rountree_beyond_2012}
B.~Rountree, D.~H. Ahn, B.~R.~de Supinski, D.~K. Lowenthal, and M.~Schulz.
\newblock Beyond {DVFS}: {A} {First} {Look} at {Performance} under a
  {Hardware}-{Enforced} {Power} {Bound}.
\newblock In \emph{Parallel and {Distributed} {Processing} {Symposium}
  {Workshops} {PhD} {Forum} ({IPDPSW}), 2012 {IEEE} 26th {International}},
  pages 947--953, 2012.
\newblock \doi{10.1109/IPDPSW.2012.116}.

\bibitem[Yoo et~al.(2003)Yoo, Jette, and Grondona]{yoo_slurm:_2003}
Andy~B. Yoo, Morris~A. Jette, and Mark Grondona.
\newblock Slurm: {Simple} linux utility for resource management.
\newblock In \emph{Workshop on {Job} {Scheduling} {Strategies} for {Parallel}
  {Processing}}, pages 44--60. Springer, 2003.
\newblock URL \url{http://link.springer.com/chapter/10.1007/10968987_3}.

\bibitem[Downey(1997)]{downey_model_1997}
Allen~B. Downey.
\newblock \emph{A model for speedup of parallel programs}.
\newblock University of California, Berkeley, Computer Science Division, 1997.
\newblock URL
  \url{http://digitalassets.lib.berkeley.edu/techreports/ucb/text/CSD-97-933.pdf}.

\bibitem[Chen et~al.(2010)Chen, Xu, and Dick]{chen_memory_2010}
X.~Chen, C.~Xu, and R.~P. Dick.
\newblock Memory access aware on-line voltage control for performance and
  energy optimization.
\newblock In \emph{2010 {IEEE}/{ACM} {International} {Conference} on
  {Computer}-{Aided} {Design} ({ICCAD})}, pages 365--372, 2010.
\newblock \doi{10.1109/ICCAD.2010.5653631}.

\bibitem[Henning(2000)]{henning_spec_2000}
John~L. Henning.
\newblock {SPEC} {CPU}2000: {Measuring} {CPU} performance in the new
  millennium.
\newblock \emph{Computer}, 33\penalty0 (7):\penalty0 28--35, 2000.
\newblock URL
  \url{http://ieeexplore.ieee.org/xpls/abs_all.jsp?arnumber=869367}.

\bibitem[Li et~al.(2005)Li, Sasanka, Adve, Chen, and Debes]{li_alpbench_2005}
Man-Lap Li, Ruchira Sasanka, Sarita~V. Adve, Yen-Kuang Chen, and Eric Debes.
\newblock The {ALPBench} benchmark suite for complex multimedia applications.
\newblock In \emph{{IEEE} {International}. 2005 {Proceedings} of the {IEEE}
  {Workload} {Characterization} {Symposium}, 2005.}, pages 34--45. IEEE, 2005.
\newblock URL
  \url{http://ieeexplore.ieee.org/xpls/abs_all.jsp?arnumber=1525999}.

\bibitem[Imamura et~al.(2012)Imamura, Sasaki, Fukumoto, Inoue, and
  Murakami]{imamura_optimizing_2012}
Satoshi Imamura, Hiroshi Sasaki, Naoto Fukumoto, Koji Inoue, and Kazuaki
  Murakami.
\newblock Optimizing power-performance trade-off for parallel applications
  through dynamic core and frequency scaling.
\newblock \emph{Proceedings of the RESoLVE}, 12, 2012.

\bibitem[Bienia et~al.(2008)Bienia, Kumar, Singh, and Li]{bienia_parsec_2008}
Christian Bienia, Sanjeev Kumar, Jaswinder~Pal Singh, and Kai Li.
\newblock The {PARSEC} {Benchmark} {Suite}: {Characterization} and
  {Architectural} {Implications}.
\newblock In \emph{Proceedings of the 17th {International} {Conference} on
  {Parallel} {Architectures} and {Compilation} {Techniques}}, {PACT} '08, pages
  72--81, New York, NY, USA, 2008. ACM.
\newblock ISBN 978-1-60558-282-5.
\newblock \doi{10.1145/1454115.1454128}.
\newblock URL \url{http://doi.acm.org/10.1145/1454115.1454128}.

\bibitem[Telgarsky et~al.(2006)Telgarsky, Hoe, and
  Moura]{telgarsky_spiral:_2006}
Marek Telgarsky, James~C. Hoe, and José~MF Moura.
\newblock {SPIRAL}: {Joint} runtime and energy optimization of linear
  transforms.
\newblock In \emph{Acoustics, {Speech} and {Signal} {Processing}, 2006.
  {ICASSP} 2006 {Proceedings}. 2006 {IEEE} {International} {Conference} on},
  volume~3, pages III--III. IEEE, 2006.
\newblock URL
  \url{http://ieeexplore.ieee.org/xpls/abs_all.jsp?arnumber=1660837}.

\bibitem[Freeh et~al.(2005)Freeh, Pan, Kappiah, Lowenthal, and
  Springer]{freeh_exploring_2005}
Vincent~W. Freeh, Feng Pan, Nandini Kappiah, David~K. Lowenthal, and Rob
  Springer.
\newblock Exploring the energy-time tradeoff in {MPI} programs on a
  power-scalable cluster.
\newblock In \emph{Parallel and {Distributed} {Processing} {Symposium}, 2005.
  {Proceedings}. 19th {IEEE} {International}}, pages 4a--4a. IEEE, 2005.
\newblock URL
  \url{http://ieeexplore.ieee.org/xpls/abs_all.jsp?arnumber=1419817}.

\bibitem[Bailey et~al.(1991)Bailey, Barszcz, Barton, Browning, Carter, Dagum,
  Fatoohi, Frederickson, Lasinski, Schreiber, Simon, Venkatakrishnan, and
  Weeratunga]{bailey_nas_1991}
D.~H. Bailey, E.~Barszcz, J.~T. Barton, D.~S. Browning, R.~L. Carter, L.~Dagum,
  R.~A. Fatoohi, P.~O. Frederickson, T.~A. Lasinski, R.~S. Schreiber, H.~D.
  Simon, V.~Venkatakrishnan, and S.~K. Weeratunga.
\newblock The {NAS} {Parallel} {Benchmarks} - {Summary} and {Preliminary}
  {Results}.
\newblock In \emph{Proceedings of the 1991 {ACM}/{IEEE} {Conference} on
  {Supercomputing}}, Supercomputing '91, pages 158--165, New York, NY, USA,
  1991. ACM.
\newblock ISBN 0-89791-459-7.
\newblock \doi{10.1145/125826.125925}.
\newblock URL \url{http://doi.acm.org/10.1145/125826.125925}.

\bibitem[Friese et~al.(2012)Friese, Brinks, Oliver, Siegel, and
  Maciejewski]{friese_analyzing_2012}
Ryan Friese, Tyler Brinks, Curt Oliver, Howard~Jay Siegel, and Anthony~A.
  Maciejewski.
\newblock Analyzing the trade-offs between minimizing makespan and minimizing
  energy consumption in a heterogeneous resource allocation problem.
\newblock In \emph{{INFOCOMP}, {The} {Second} {International} {Conference} on
  {Advanced} {Communications} and {Computation}}, pages 81--89, 2012.
\newblock URL \url{http://www.engr.colostate.edu/~aam/pdf/conferences/145.pdf}.

\bibitem[Ali et~al.(2000)Ali, Siegel, Maheswaran, Hensgen, and
  Ali]{ali_representing_2000}
Shoukat Ali, Howard~Jay Siegel, Muthucumaru Maheswaran, Debra Hensgen, and
  Sahra Ali.
\newblock Representing task and machine heterogeneities for heterogeneous
  computing systems.
\newblock \emph{Tamkang Journal of Science and Engineering}, 3\penalty0
  (3):\penalty0 195--207, 2000.
\newblock URL
  \url{http://www.airitilibrary.com/Publication/alDetailedMesh?docid=15606686-200009-3-3-195-207-a}.

\bibitem[Balaprakash et~al.(2011)Balaprakash, Wild, and
  Norris]{balaprakash_spapt:_2011}
Prasanna Balaprakash, S.~M. Wild, and Boyana Norris.
\newblock {SPAPT}: {Search} {Problems} in {Automatic} {Performance} {Tuning}.
\newblock April 2011.

\bibitem[Davis(2006)]{davis_direct_2006}
Timothy~A. Davis.
\newblock \emph{Direct {Methods} for {Sparse} {Linear} {Systems}}.
\newblock Society for Industrial and Applied Mathematics, January 2006.
\newblock ISBN 978-0-89871-613-9 978-0-89871-888-1.
\newblock URL \url{http://epubs.siam.org/doi/book/10.1137/1.9780898718881}.

\bibitem[Kaiser et~al.(2010)Kaiser, Williams, Madduri, Ibrahim, Bailey, Demmel,
  and Strohmaier]{kaiser_torch_2010}
Alex Kaiser, Samuel Williams, Kamesh Madduri, Khaled Ibrahim, David Bailey,
  James Demmel, and Erich Strohmaier.
\newblock {TORCH} {Computational} {Reference} {Kernels}: {A} {Testbed} for
  {Computer} {Science} {Research}.
\newblock Technical Report UCB/EECS-2010-144, EECS Department, University of
  California, Berkeley, 2010.
\newblock URL
  \url{http://www2.eecs.berkeley.edu/Pubs/TechRpts/2010/EECS-2010-144.html}.

\bibitem[Heroux et~al.(2009)Heroux, Doerer, Crozier, and
  Willenbring]{heroux_improving_2009}
M.~A. Heroux, D.~W. Doerer, P.~S. Crozier, and J.~M. Willenbring.
\newblock Improving performance via mini-applications.
\newblock Technical Report SAND2009-5574, Sandia National Laboratories, 2009.

\bibitem[Benkner et~al.(2014)Benkner, Franchetti, Gerndt, and
  Hollingsworth]{benkner_automatic_2014}
Siegfried Benkner, Franz Franchetti, Hans~Michael Gerndt, and Jeffrey~K.
  Hollingsworth.
\newblock Automatic {Application} {Tuning} for {HPC} {Architectures}
  ({Dagstuhl} {Seminar} 13401).
\newblock \emph{Dagstuhl Reports}, 3\penalty0 (9), 2014.
\newblock URL \url{http://drops.dagstuhl.de/opus/volltexte/2014/4423/}.

\bibitem[Kukkonen and Lampinen(2005)]{kukkonen_gde3:_2005}
S.~Kukkonen and J.~Lampinen.
\newblock {GDE}3: {The} third {Evolution} {Step} of {Generalized}
  {Differential} {Evolution}.
\newblock volume~1, pages 443--450. IEEE, 2005.
\newblock ISBN 978-0-7803-9363-9.
\newblock \doi{10.1109/CEC.2005.1554717}.
\newblock URL \url{http://ieeexplore.ieee.org/document/1554717/}.

\bibitem[Tarplee et~al.(2015)Tarplee, Friese, Maciejewski, and
  Siegel]{fidanova_efficient_2015}
Kyle~M. Tarplee, Ryan Friese, Anthony~A. Maciejewski, and Howard~Jay Siegel.
\newblock Efficient and {Scalable} {Pareto} {Front} {Generation} for {Energy}
  and {Makespan} in {Heterogeneous} {Computing} {Systems}.
\newblock In Stefka Fidanova, editor, \emph{Recent {Advances} in
  {Computational} {Optimization}}, volume 580, pages 161--180. Springer
  International Publishing, Cham, 2015.
\newblock ISBN 978-3-319-12630-2 978-3-319-12631-9.
\newblock URL \url{http://link.springer.com/10.1007/978-3-319-12631-9_10}.

\bibitem[Deb et~al.(2002)Deb, Pratap, Agarwal, and Meyarivan]{deb_fast_2002}
K.~Deb, A.~Pratap, S.~Agarwal, and T.~Meyarivan.
\newblock A fast and elitist multiobjective genetic algorithm: {NSGA}-{II}.
\newblock \emph{IEEE Transactions on Evolutionary Computation}, 6\penalty0
  (2):\penalty0 182--197, 2002.
\newblock ISSN 1089-778X.
\newblock \doi{10.1109/4235.996017}.

\bibitem[Casavant and Kuhl(1988)]{casavant_taxonomy_1988}
Thomas~L. Casavant and Jon~G. Kuhl.
\newblock A taxonomy of scheduling in general-purpose distributed computing
  systems.
\newblock \emph{Software Engineering, IEEE Transactions on}, 14\penalty0
  (2):\penalty0 141--154, 1988.
\newblock URL \url{http://ieeexplore.ieee.org/xpls/abs_all.jsp?arnumber=4634}.

\bibitem[Baruah(2004)]{baruah_task_2004}
Sanjoy~K. Baruah.
\newblock Task {Partitioning} {Upon} {Heterogeneous} {Multiprocessor}
  {Platforms}.
\newblock In \emph{{IEEE} {Real}-{Time} and {Embedded} {Technology} and
  {Applications} {Symposium}}, pages 536--543. Citeseer, 2004.
\newblock URL
  \url{http://citeseerx.ist.psu.edu/viewdoc/download?doi=10.1.1.86.1151&rep=rep1&type=pdf}.

\bibitem[Balzuweit et~al.(2016)Balzuweit, Bunde, Leung, Finley, and
  Lee]{balzuweit_local_2016}
Evan Balzuweit, David~P. Bunde, Vitus~J. Leung, Austin Finley, and Alan~C.S.
  Lee.
\newblock Local search to improve coordinate-based task mapping.
\newblock \emph{Parallel Computing}, 51:\penalty0 67--78, January 2016.
\newblock ISSN 01678191.
\newblock \doi{10.1016/j.parco.2015.10.012}.
\newblock URL
  \url{http://linkinghub.elsevier.com/retrieve/pii/S0167819115001441}.

\bibitem[McClatchey et~al.(2007)McClatchey, Anjum, Stockinger, Ali, Willers,
  and Thomas]{mcclatchey_data_2007}
Richard McClatchey, Ashiq Anjum, Heinz Stockinger, Arshad Ali, Ian Willers, and
  Michael Thomas.
\newblock Data {Intensive} and {Network} {Aware} ({DIANA}) {Grid} {Scheduling}.
\newblock \emph{Journal of Grid Computing}, 5\penalty0 (1):\penalty0 43--64,
  March 2007.
\newblock ISSN 1570-7873, 1572-9184.
\newblock \doi{10.1007/s10723-006-9059-z}.
\newblock URL
  \url{https://link.springer.com/article/10.1007/s10723-006-9059-z}.

\bibitem[Shuja et~al.(2014)Shuja, Bilal, Madani, and Khan]{shuja_data_2014}
Junaid Shuja, Kashif Bilal, Sajjad~Ahmad Madani, and Samee~U. Khan.
\newblock Data center energy efficient resource scheduling.
\newblock \emph{Cluster Computing}, 17\penalty0 (4):\penalty0 1265--1277,
  December 2014.
\newblock ISSN 1386-7857, 1573-7543.
\newblock \doi{10.1007/s10586-014-0365-0}.
\newblock URL \url{http://link.springer.com/10.1007/s10586-014-0365-0}.

\bibitem[Kliazovich et~al.(2012)Kliazovich, Bouvry, and
  Khan]{kliazovich_greencloud:_2012}
Dzmitry Kliazovich, Pascal Bouvry, and Samee~Ullah Khan.
\newblock {GreenCloud}: a packet-level simulator of energy-aware cloud
  computing data centers.
\newblock \emph{The Journal of Supercomputing}, 62\penalty0 (3):\penalty0
  1263--1283, December 2012.
\newblock ISSN 0920-8542, 1573-0484.
\newblock \doi{10.1007/s11227-010-0504-1}.
\newblock URL \url{http://link.springer.com/10.1007/s11227-010-0504-1}.

\bibitem[Aniello et~al.(2013)Aniello, Baldoni, and
  Querzoni]{aniello_adaptive_2013}
Leonardo Aniello, Roberto Baldoni, and Leonardo Querzoni.
\newblock Adaptive online scheduling in storm.
\newblock In \emph{Proceedings of the 7th {ACM} international conference on
  {Distributed} event-based systems}, pages 207--218. ACM, 2013.
\newblock URL \url{http://dl.acm.org/citation.cfm?id=2488267}.

\bibitem[Bunde and Leung(2014)]{bunde_premas:_2014}
David~P. Bunde and Vitus~J. Leung.
\newblock {PReMAS}: {Simulator} for resource management.
\newblock In \emph{2014 43rd {International} {Conference} on {Parallel}
  {Processing} {Workshops}}, pages 226--234. IEEE, 2014.
\newblock URL
  \url{http://ieeexplore.ieee.org/xpls/abs_all.jsp?arnumber=7103457}.

\bibitem[Wall(1996)]{wall_genetic_1996}
Matthew~Bartschi Wall.
\newblock \emph{A genetic algorithm for resource-constrained scheduling}.
\newblock PhD thesis, Massachusetts Institute of Technology, 1996.
\newblock URL \url{http://lancet.mit.edu/~mwall/phd/thesis/thesis.pdf}.

\bibitem[Siegel and Ali(2000)]{siegel_techniques_2000}
Howard~Jay Siegel and Shoukat Ali.
\newblock Techniques for mapping tasks to machines in heterogeneous computing
  systems.
\newblock \emph{Journal of Systems Architecture}, 46\penalty0 (8):\penalty0
  627--639, 2000.
\newblock URL
  \url{http://www.sciencedirect.com/science/article/pii/S1383762199000338}.

\bibitem[Sun and Sugawara(2011)]{sun_heuristics_2011}
W.~Sun and T.~Sugawara.
\newblock Heuristics and {Evaluations} of {Energy}-{Aware} {Task} {Mapping} on
  {Heterogeneous} {Multiprocessors}.
\newblock In \emph{2011 {IEEE} {International} {Symposium} on {Parallel} and
  {Distributed} {Processing} {Workshops} and {Phd} {Forum} ({IPDPSW})}, pages
  599--607, 2011.
\newblock \doi{10.1109/IPDPS.2011.209}.

\bibitem[Garg and Buyya(2009)]{garg_exploiting_2009}
Saurabh~Kumar Garg and Rajkumar Buyya.
\newblock Exploiting heterogeneity in {Grid} computing for energy-efficient
  resource allocation.
\newblock In \emph{Proceedings of the 17th {International} {Conference} on
  {Advanced} {Computing} and {Communications}}, 2009.
\newblock URL \url{http://cloudbus.org/papers/GreenGrid-ADCOM2009.pdf}.

\bibitem[YarKhan and Dongarra(2002)]{yarkhan_experiments_2002}
Asim YarKhan and Jack~J. Dongarra.
\newblock Experiments with scheduling using simulated annealing in a grid
  environment.
\newblock In \emph{Grid {Computing}—{GRID} 2002}, pages 232--242. Springer,
  2002.
\newblock URL \url{http://link.springer.com/chapter/10.1007/3-540-36133-2_21}.

\bibitem[Dürr and Nguyen(2009)]{durr_non-clairvoyant_2009}
Christoph Dürr and Kim~Thang Nguyen.
\newblock Non-clairvoyant scheduling games.
\newblock In \emph{International {Symposium} on {Algorithmic} {Game} {Theory}},
  pages 135--146. Springer, 2009.
\newblock URL
  \url{http://link.springer.com/chapter/10.1007/978-3-642-04645-2_13}.

\bibitem[Whaley et~al.(2001)Whaley, Petitet, and
  Dongarra]{whaley_automated_2001}
R.~Clint Whaley, Antoine Petitet, and Jack~J. Dongarra.
\newblock Automated empirical optimizations of software and the {ATLAS}
  project.
\newblock \emph{Parallel Computing}, 27\penalty0 (1):\penalty0 3--35, 2001.
\newblock URL
  \url{http://www.sciencedirect.com/science/article/pii/S0167819100000879}.

\bibitem[Martínez et~al.(2014)Martínez, Sikora, César, and
  Sorribes]{martinez_elastic:_2014}
Andrea Martínez, Anna Sikora, Eduardo César, and Joan Sorribes.
\newblock {ELASTIC}: {A} large scale dynamic tuning environment.
\newblock \emph{Scientific Programming}, 22\penalty0 (4):\penalty0 261--271,
  2014.
\newblock URL \url{http://www.hindawi.com/journals/sp/2014/403695/abs/}.

\bibitem[Kamil et~al.(2010)Kamil, Chan, Oliker, Shalf, and
  Williams]{kamil_auto-tuning_2010}
Shoaib Kamil, Cy~Chan, Leonid Oliker, John Shalf, and Samuel Williams.
\newblock An auto-tuning framework for parallel multicore stencil computations.
\newblock In \emph{Parallel \& {Distributed} {Processing} ({IPDPS}), 2010
  {IEEE} {International} {Symposium} on}, pages 1--12. IEEE, 2010.
\newblock URL
  \url{http://ieeexplore.ieee.org/xpls/abs_all.jsp?arnumber=5470421}.

\bibitem[Laros~III et~al.(2012)Laros~III, Pedretti, Kelly, Shu, and
  Vaughan]{laros_iii_energy_2012}
James~H. Laros~III, Kevin~T. Pedretti, Suzanne~M. Kelly, Wei Shu, and
  Courtenay~T. Vaughan.
\newblock Energy based performance tuning for large scale high performance
  computing systems.
\newblock In \emph{Proceedings of the 2012 {Symposium} on {High} {Performance}
  {Computing}}, page~6. Society for Computer Simulation International, 2012.
\newblock URL \url{http://dl.acm.org/citation.cfm?id=2338822}.

\bibitem[Benedict et~al.(2010)Benedict, Petkov, and
  Gerndt]{muller_periscope:_2010}
Shajulin Benedict, Ventsislav Petkov, and Michael Gerndt.
\newblock {PERISCOPE}: {An} {Online}-{Based} {Distributed} {Performance}
  {Analysis} {Tool}.
\newblock In Matthias~S. Müller, Michael~M. Resch, Alexander Schulz, and
  Wolfgang~E. Nagel, editors, \emph{Tools for {High} {Performance} {Computing}
  2009}, pages 1--16. Springer Berlin Heidelberg, Berlin, Heidelberg, 2010.
\newblock ISBN 978-3-642-11260-7 978-3-642-11261-4.
\newblock URL \url{http://link.springer.com/10.1007/978-3-642-11261-4_1}.

\bibitem[Miceli et~al.(2012)Miceli, Civario, Sikora, César, Gerndt, Haitof,
  Navarrete, Benkner, Sandrieser, Morin, and {others}]{miceli_autotune:_2012}
Renato Miceli, Gilles Civario, Anna Sikora, Eduardo César, Michael Gerndt,
  Houssam Haitof, Carmen Navarrete, Siegfried Benkner, Martin Sandrieser,
  Laurent Morin, and {others}.
\newblock Autotune: {A} plugin-driven approach to the automatic tuning of
  parallel applications.
\newblock In \emph{International {Workshop} on {Applied} {Parallel}
  {Computing}}, pages 328--342. Springer, 2012.
\newblock URL
  \url{http://link.springer.com/chapter/10.1007/978-3-642-36803-5_24}.

\bibitem[Liu et~al.(2014)Liu, Ureña, Gerndt, and Gong]{liu_automatic_2014}
W.~Liu, I.~A.~C. Ureña, M.~Gerndt, and B.~Gong.
\newblock Automatic {MPI}-{IO} {Tuning} with the {Periscope} {Tuning}
  {Framework}.
\newblock In \emph{Parallel {Distributed} {Processing} {Symposium} {Workshops}
  ({IPDPSW}), 2014 {IEEE} {International}}, pages 352--360, 2014.
\newblock \doi{10.1109/IPDPSW.2014.46}.

\bibitem[Tsai et~al.(2016)Tsai, Luszczek, Kurzak, and
  Dongarra]{tsai_performance-portable_2016}
Y.~M. Tsai, P.~Luszczek, J.~Kurzak, and J.~Dongarra.
\newblock Performance-{Portable} {Autotuning} of {OpenCL} {Kernels} for
  {Convolutional} {Layers} of {Deep} {Neural} {Networks}.
\newblock In \emph{2016 2nd {Workshop} on {Machine} {Learning} in {HPC}
  {Environments} ({MLHPC})}, pages 9--18, 2016.
\newblock \doi{10.1109/MLHPC.2016.005}.

\bibitem[Santana-Quintero et~al.(2010)Santana-Quintero, Hernández-Díaz,
  Molina, Coello~Coello, and Caballero]{santana-quintero_demors:_2010}
Luis~V. Santana-Quintero, Alfredo~G. Hernández-Díaz, Julián Molina,
  Carlos~A. Coello~Coello, and Rafael Caballero.
\newblock {DEMORS}: {A} hybrid multi-objective optimization algorithm using
  differential evolution and rough set theory for constrained problems.
\newblock \emph{Computers \& Operations Research}, 37\penalty0 (3):\penalty0
  470--480, March 2010.
\newblock ISSN 03050548.
\newblock \doi{10.1016/j.cor.2009.02.006}.
\newblock URL
  \url{http://linkinghub.elsevier.com/retrieve/pii/S0305054809000409}.

\bibitem[Rościszewski(2015)]{rosciszewski_smart_2015}
Paweł Rościszewski.
\newblock Smart {City} data management for analyzing {PM}10 particle pollution.
\newblock Master's thesis, Gdańsk University of Technology, Gdańsk, 2015.

\bibitem[Povey et~al.(2011)Povey, Ghoshal, Boulianne, Burget, Glembek, Goel,
  Hannemann, Motlicek, Qian, Schwarz, and {others}]{povey_kaldi_2011}
Daniel Povey, Arnab Ghoshal, Gilles Boulianne, Lukas Burget, Ondrej Glembek,
  Nagendra Goel, Mirko Hannemann, Petr Motlicek, Yanmin Qian, Petr Schwarz, and
  {others}.
\newblock The {Kaldi} speech recognition toolkit.
\newblock In \emph{{IEEE} 2011 {ASRU} workshop}. IEEE Signal Processing
  Society, 2011.
\newblock URL \url{http://infoscience.epfl.ch/record/192584}.

\bibitem[Povey et~al.(2014)Povey, Zhang, and Khudanpur]{povey_parallel_2014}
Daniel Povey, Xiaohui Zhang, and Sanjeev Khudanpur.
\newblock Parallel training of deep neural networks with natural gradient and
  parameter averaging.
\newblock \emph{CoRR, vol. abs/1410.7455}, 2014.
\newblock URL
  \url{https://www.researchgate.net/profile/Daniel_Povey/publication/267515250_Parallel_training_of_Deep_Neural_Networks_with_Natural_Gradient_and_Parameter_Averaging/links/545536f50cf2bccc490ccb0f.pdf}.

\bibitem[Li and Song(2004)]{li_automatic_2004}
Zhiyuan Li and Yonghong Song.
\newblock Automatic tiling of iterative stencil loops.
\newblock \emph{ACM Transactions on Programming Languages and Systems
  (TOPLAS)}, 26\penalty0 (6):\penalty0 975--1028, 2004.

\bibitem[Bolosky and Scott(1993)]{bolosky_false_1993}
William~J. Bolosky and Michael~L. Scott.
\newblock False sharing and its effect on shared memory performance.
\newblock In \emph{Proceedings of the {Fourth} symposium on {Experiences} with
  distributed and multiprocessor systems}, 1993.

\bibitem[Rościszewski(2012)]{rosciszewski_system_2012}
Paweł Rościszewski.
\newblock A system for parallel reliable and efficient computations on {GPUs}
  and {CPUs}.
\newblock Master's thesis, Gdańsk University of Technology, Gdańsk, 2012.

\bibitem[Proficz and Czarnul(2016)]{wyrzykowski_performance_2016}
Jerzy Proficz and Paweł Czarnul.
\newblock Performance and {Power}-{Aware} {Modeling} of {MPI} {Applications}
  for {Cluster} {Computing}.
\newblock In Roman Wyrzykowski, Ewa Deelman, Jack Dongarra, Konrad Karczewski,
  Jacek Kitowski, and Kazimierz Wiatr, editors, \emph{Parallel {Processing} and
  {Applied} {Mathematics}}, volume 9574, pages 199--209. Springer International
  Publishing, Cham, 2016.
\newblock ISBN 978-3-319-32151-6 978-3-319-32152-3.
\newblock \doi{10.1007/978-3-319-32152-3_19}.
\newblock URL \url{http://link.springer.com/10.1007/978-3-319-32152-3_19}.

\bibitem[Rościszewski(2014{\natexlab{b}})]{rosciszewski_dynamic_2014}
Paweł Rościszewski.
\newblock Dynamic {Data} {Management} {Among} {Multiple} {Databases} for
  {Optimization} of {Parallel} {Computations} in {Heterogeneous} {HPC}
  {Systems}.
\newblock pages 09--19. Academy \& Industry Research Collaboration Center
  (AIRCC), July 2014{\natexlab{b}}.
\newblock ISBN 978-1-921987-08-3.
\newblock \doi{10.5121/csit.2014.4709}.
\newblock URL \url{http://www.airccj.org/CSCP/vol4/csit42502.pdf}.

\bibitem[Rumelhart et~al.(1986)Rumelhart, Hinton, and
  Williams]{rumelhart_learning_1986}
David~E. Rumelhart, Geoffrey~E. Hinton, and Ronald~J. Williams.
\newblock Learning representations by back-propagating errors.
\newblock \emph{Nature}, 323\penalty0 (6088):\penalty0 533--536, October 1986.
\newblock ISSN 0028-0836.
\newblock \doi{10.1038/323533a0}.
\newblock URL \url{http://www.nature.com/doifinder/10.1038/323533a0}.

\bibitem[Gentzsch(2001)]{gentzsch_sun_2001}
W~Gentzsch.
\newblock Sun {Grid} {Engine}: {Towards} {Creating} a {Compute} {Power} {Grid}.
\newblock In \emph{Proceedings of the 1st {International} {Symposium} on
  {Cluster} {Computing} and the {Grid}}, {CCGRID} '01, pages 35--, Washington,
  DC, USA, 2001. IEEE Computer Society.
\newblock ISBN 0-7695-1010-8.
\newblock URL \url{http://dl.acm.org/citation.cfm?id=560889.792378}.

\bibitem[Graves et~al.(2006)Graves, Fernández, Gomez, and
  Schmidhuber]{graves_connectionist_2006}
Alex Graves, Santiago Fernández, Faustino Gomez, and Jürgen Schmidhuber.
\newblock Connectionist temporal classification: labelling unsegmented sequence
  data with recurrent neural networks.
\newblock In \emph{Proceedings of the 23rd international conference on
  {Machine} learning}, pages 369--376. ACM, 2006.
\newblock URL \url{http://dl.acm.org/citation.cfm?id=1143891}.

\end{thebibliography}
\bibliographystyle{unsrtnat}

\listoffigures
\doublespacing

\begin{appendices}

\end{appendices}

\end{document}